\documentclass[mnsc,nonblindrev]{informs3_hide} 

\OneAndAHalfSpacedXI 

\usepackage{natbib, multirow, multicol, xspace, enumitem, amsmath, pifont, subcaption}
\usepackage{hyperref}
\hypersetup{hidelinks}
\usepackage{accents}
\usepackage{mathrsfs}
\bibpunct[, ]{(}{)}{,}{a}{}{,}%
\def\bibfont{\small}%
\usepackage{bm}
\usepackage[normalem]{ulem}
\usepackage[dvipsnames, table]{xcolor}
\usepackage[all]{xy}

\usepackage{algorithm}
\usepackage[noend]{algpseudocode}

\algnewcommand{\algorithmicgoto}{\phantom{for} \textbf{go to} Line}%
\algnewcommand{\Goto}[1]{\State \algorithmicgoto~\ref{#1}}%

\usepackage{bigdelim}
\newcolumntype{C}[1]{>{\centering\let\newline\\\arraybackslash\hspace{0pt}}m{#1}}

\newcommand{\mathbbm}[1]{\text{\usefont{U}{bbm}{m}{n}#1}} 

\newcommand{\eps}{\varepsilon}
\newcommand{\bI}{\mathbbm{1}}

\newcommand{\bR}{\mathbb{R}}
\newcommand{\bN}{\mathbb{N}}

\newcommand{\cY}{\mathcal{Y}}
\newcommand{\cN}{\mathcal{N}}

\newcommand{\Var}{\mathrm{Var}}

\newcommand{\bE}{\mathrm{E}}

\newcommand{\cF}{\mathcal{F}}
\newcommand{\cE}{\mathcal{E}}
\newcommand{\sF}{\mathscr{F}}
\newcommand{\sP}{\mathscr{P}}

\newcommand{\cT}{\mathcal{T}}
\newcommand{\cH}{\mathcal{H}}
\newcommand{\Ber}{\mathrm{Ber}}

\TheoremsNumberedThrough     
\ECRepeatTheorems

\newtheorem{subassumption}{Assumption\normalfont}[assumption]

\newenvironment{assumption+}[1]
 {\renewcommand{\thesubassumption}{#1}\subassumption}
 {\endsubassumption}

\EquationsNumberedThrough    

\MANUSCRIPTNO{}

\begin{document}


\RUNAUTHOR{}

\RUNTITLE{Adaptive Neyman Allocation}

\TITLE{Adaptive Neyman Allocation}

\ARTICLEAUTHORS{
\AUTHOR{Jinglong Zhao}\AFF{Boston University, Questrom School of Business, Boston, MA, 02215,
\EMAIL{jinglong@bu.edu}}
}

\ABSTRACT{
In the experimental design literature, Neyman allocation refers to the practice of allocating units into treated and control groups, potentially in unequal numbers proportional to their respective standard deviations, with the objective of minimizing the variance of the treatment effect estimator. This widely recognized approach increases statistical power in scenarios where the treated and control groups have different standard deviations, as is often the case in social experiments, clinical trials, marketing research, and online A/B testing. However, Neyman allocation cannot be implemented unless the standard deviations are known in advance. Fortunately, the multi-stage nature of the aforementioned applications allows the use of earlier stage observations to estimate the standard deviations, which further guide allocation decisions in later stages. In this paper, we introduce a competitive analysis framework to study this multi-stage experimental design problem. We propose a simple adaptive Neyman allocation algorithm, which almost matches the information-theoretic limit of conducting experiments. We provide theory for estimation and inference using data collected from our adaptive Neyman allocation algorithm. We demonstrate the effectiveness of our adaptive Neyman allocation algorithm using both online A/B testing data from a social media site and synthetic data.
}

\HISTORY{\href{https://papers.ssrn.com/sol3/papers.cfm?abstract_id=4448249}{First draft: May 15, 2023.} This version: \today}

\maketitle

\section{Introduction}

Why are randomized controlled experiments usually conducted with half treated and half control?
One answer, dating back to \citet{neyman1934two}, is that experimenters usually believe the treated and control groups to have the same level of variability.
When the treated and control groups have different levels of variability, such as an intervention inducing heterogeneous responses or even polarization of the responses, the seminal work of \citet{neyman1934two} recommends unequal allocation: the sizes of treated and control groups should be proportional to their respective standard deviations.
This approach has later on been recognized as ``Neyman allocation.''

\begin{figure}[!tb]
\centering
\caption{Distributions of the number of clicks per million impressions at a social media site \citep{AB_testing_kaggle}}
\label{fig:polarization}
\includegraphics[width=0.6\textwidth]{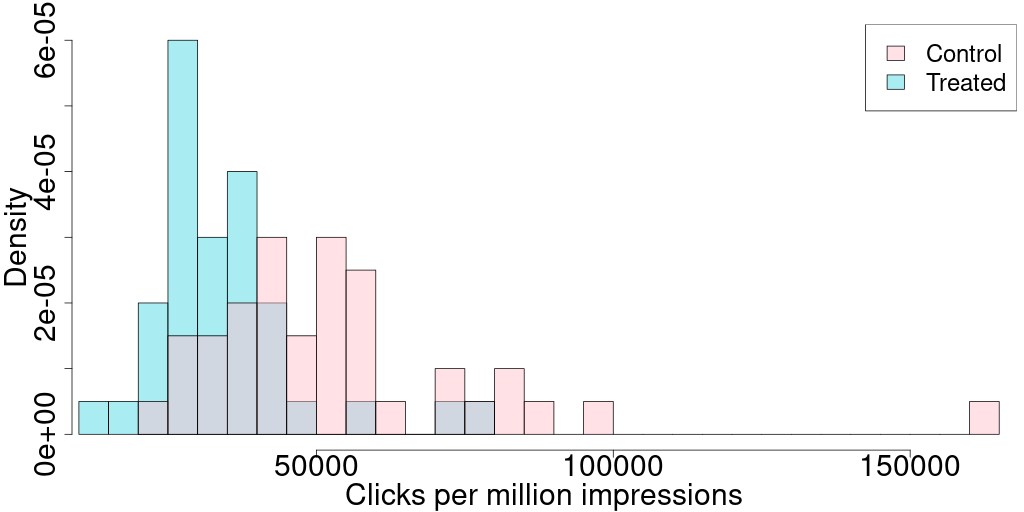}
\end{figure}

Neyman allocation has many desirable properties.
First, since it prescribes the sizes of the treated and control groups, it can be naturally combined with complete randomization \citep{cox2000theory, fisher1936design, imbens2015causal, wu2011experiments}.
Randomization then serves as the basis of validity for many randomized experiments \citep{cook2002experimental, deaton2018understanding}.
Second, it proves to minimize the variance of the widely used difference-in-means estimator, and increases the statistical power in scenarios where the the treated and control groups have different levels of variability \citep{neyman1934two}.
Consequently, it brings tremendous value to a wide range of applications whose treatment and control groups have different standard deviations, such as social experiments \citep{duflo2007using, karlan2008credit, mosleh2021shared}, clinical trials \citep{berry2006bayesian, hu2003optimality, rosenberger2015randomization}, marketing research \citep{rossi2003bayesian, sandor2001designing}, and online A/B testing \citep{bakshy2014designing, deng2013improving, kohavi2017online}.
For example, at a social media site who compares two advertisement strategies, the standard deviation of the treated group is much smaller than that of the control group;
see Figure~\ref{fig:polarization} for an illustration.

Albeit useful, a challenge in using Neyman allocation arises when the standard deviations of the treated and control groups are unknown in advance.
Fortunately, the multi-stage nature of the aforementioned applications allows the use of earlier stage observations to estimate the standard deviations.
If the earlier stage observations suggest a higher level of variability in one group, more experimental units will be allocated to the same group in the later stages, so that the confidence intervals of the average outcomes are roughly equal between the two groups.
We refer to this approach as ``adaptive Neyman allocation.''

In this paper, we study the optimal adaptive Neyman allocation problem.

To study this problem, we borrow the competitive analysis framework, a common optimization framework in the literature of decision making under uncertainty.
This framework minimizes the worst case ratio between a proposed algorithm and an optimal algorithm endowed with clairvoyant information.
This framework is scale-independent, ensuring that the ratio remains meaningful even on ``hard instances'' where both the proposed algorithm and the optimal algorithm perform poorly.
To the best of our knowledge, we are the first to introduce the competitive analysis framework into experimental designs.
In the single stage setup, an immediate implication of adopting this framework is that half-half allocations are optimal, without knowing the standard deviations of the treated and control groups, or any assumptions about these standard deviations.
In the multi-stage setup, this framework allows for meaningful comparisons across different problem instances, even if the standard deviations of the treated and control groups are different. 
This is in contrast to the conventional minimax framework or the regret minimization framework, as the objective values in such frameworks will change under re-scaling of the standard deviations.
To facilitate such comparisons, the minimax framework and the regret minimization framework need to assume the standard deviations being constants.

Another remarkable advantage of using the competitive analysis framework is that it facilitates a more precise examination of the second-order efficiency of experimental designs, which is different from the conventional emphasis on the first-order efficiency\footnote{First order efficiency in the context of experimental design is similar to semi-parametric efficiency in the context of observational study; see, e.g., \citet{hahn1998role, hirano2003efficient, robins1994estimation, robins1995semiparametric, scharfstein1999adjusting} and textbooks \citet{ding2024first, imbens2015causal, wager2024causal}.} such as in \citet{armstrong2022asymptotic} and \citet{hahn2011adaptive}.
More specifically, when a total of $T$ experimental units are enrolled over $M \geq 2$ stages, the adaptive Neyman allocation algorithm in this paper achieves $1 + O\big(T^{-\frac{M-1}{M}}\big)$ competitiveness against a hindsight benchmark that knew the standard deviations in advance.
In contrast, \citet{hahn2011adaptive} show that when there are $M = 2$ stages and when the first stage pilot experiment involves approximately $T^\alpha$ units, any value of $\alpha < 1$ is first-order efficient.
While two different parameterizations of $\alpha$ may both satisfy the first-order efficiency criterion, they can still lead to significantly different performances due to their second-order gap.
A more precise examination of the second-order efficiency is useful in determining which parameterization of $\alpha$ is optimal.

Our work presents how to use the notion of second-order efficiency to choose the sample size for each stage in an adaptive Neyman allocation algorithm.
In the $M = 2$ stage example above, the optimal sample size for the first stage pilot experiment should involve approximately $T^{\frac{1}{2}}$ units, i.e., $\alpha = \frac{1}{2}$.
In general, in an $M$ stage experiment, the optimal sample size for the $m$-th stage should involve approximately $T^{\frac{m}{M}}$ units, leading to an exponentially increasing number of units in the later stages of the experiment.
This exponentially increasing pattern may serve as a rule of thumb for practitioners who would like to conduct multi-stage experiments.

\begin{figure}[!tb]
\centering
\caption{Competitive ratios with respect to different numbers of stages}
\label{fig:Illustration:CR}
\includegraphics[width=0.75\textwidth]{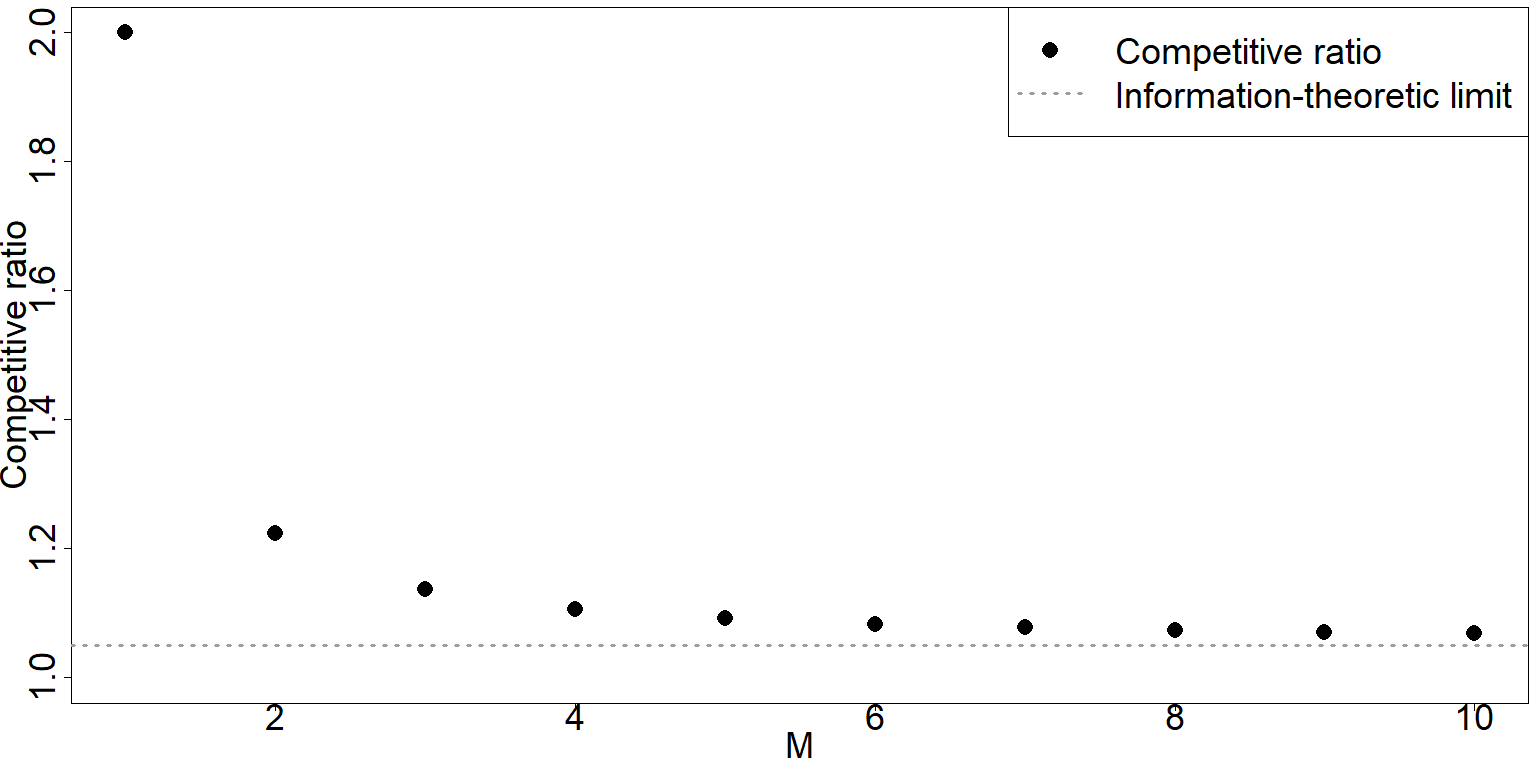}
\end{figure}

We also prove a novel information-theoretic $1 + O(T^{-1})$ competitive lower bound of conducting adaptive experiments.
Recall that the competitive ratio of the aforementioned adaptive Neyman allocation algorithm is $1 + O\big(T^{-\frac{M-1}{M}}\big)$, which quickly approaches $1 + O(T^{-1})$ when the number of stages is large.
Combining these two results, it shows that the adaptive Neyman allocation algorithm is second-order optimal when the number of stages is large.
See Figure~\ref{fig:Illustration:CR} for an illustration.
To the best of our knowledge, the best known result that studies the same question in the literature \citep{antos2010active, carpentier2011finite, grover2009active} translates into a $1 + O\big(T^{-\frac{1}{2}}\big)$ ratio (see Section~\ref{sec:Optimization} for details), and conjectures that this ratio is best possible.
Our work negates this conjecture by improving this ratio.

Our work has two practical implications. 
First, conducting a two-stage or three-stage experiment can be sufficiently efficient as long as the sample size in each stage approximately follows an exponentially increasing pattern. 
Even though the two-stage or three-stage experiment is not optimal, having the ability to adaptively adjust the allocation of units based on insights from earlier stages can greatly improve efficiency. 
Second, if there is existing experimental data available, practitioners can use it as the first stage experiment to estimate the levels of variability from the treated and control groups, and guide the allocation of units in later stages.

\subsection{Related Literature}

This paper bridges four different fields of literature, listed alphabetically below.
The subtle differences that distinguish these fields lie in the objective function and the underlying assumptions.
\begin{enumerate}
\item Active learning (theoretical computer science). 
In the active learning literature, prior works have adopted the same objective of minimizing the estimation error defined as the proxy mean squared error. 
But the optimization formulation is to minimize the worst case regret, defined as the difference between any proposed algorithm and the optimal algorithm endowed with clairvoyant information \citep{antos2010active, aznag2023active, carpentier2011finite, etore2010adaptive, etore2011adaptive, grover2009active, russac2021b}.

This literature usually assumes that the variances of outcomes in both treated and control groups are upper bounded by some constants. 
Under this assumption, any regret minimization result corresponds to a competitive ratio result that is comparable to our work.
After translation between the two types of results, our work improves the best known results from \citet{antos2010active, carpentier2011finite, grover2009active} even under a weaker assumption (Theorems~\ref{thm:2StageANA} and~\ref{thm:MStageANA}), negating the conjecture that existing results are best possible.
The key to this improvement lies in fully exploiting the uni-modal structure of the nonlinear objective function; whereas prior works make linear approximations.
There are two recent independent works.
\citet{aznag2023active} studies the same problem through regret minimization and proposes a fully adaptive algorithm that leads to a similar improvement as our work. 
\citet{dai2023clip} adopts an adversarial arrival model to study a similar problem.

Related to the active learning literature is the stochastic multi-armed bandit problem, with an objective of maximizing the cumulative rewards through balancing both exploration and exploitation.
We are unable to survey the rich literature on multi-armed bandits, but only point to \citet{chen2022elements, lattimore2020bandit, russo2018tutorial, slivkins2019introduction} for books and \citet{agrawal2012analysis, audibert2009exploration, auer2002finite, garivier2011kl, lai1985asymptotically, robbins1952some, russo2016information, simchi2023multi, thompson1933likelihood} for papers, and references therein.

\item Adaptive clinical trial (statistics and biostatistics). 
In the adaptive clinical trial literature, prior works have adopted a related but different objective of setting the proportion of treated and control units to asymptotically converge to a target proportion \citep{hu2004asymptotic, hu2006theory, jennison1999group, sverdlov2015modern}.
Stemming from the seminal work of the biased coin design \citep{efron1971forcing}, the literature mainly proposes two solutions: the Polya's urn design \citep{wei1978application, wei1979generalized} and the doubly adaptive biased coin design \citep{eisele1990adaptive, eisele1994doubly, eisele1995central, hu2004asymptotic, wei1978adaptive}.

The adaptively clinical trial literature usually assumes that the outcomes have bounded finite moments, which is the same as we assume in this work.
Some other works in the literature, such as \citet{azriel2014adaptive, melfi1998variablility, rosenberger2001optimal}, make a stronger assumption that the outcomes follow Bernoulli distributions.
This literature usually considers fully adaptive designs, which ensure that the proportion of treated and control units asymptotically converges to the proportion of Neyman allocation when the sample size is large.
Using a batched adaptive design, our adaptive Neyman allocation also ensures convergence (Corollaries~\ref{coro:2Stage:ArmPulls} and~\ref{coro:MStage:ArmPulls}), but at a very slow rate.

\item Adaptive experimental design (statistics and econometrics). 
In the adaptive experimental design literature, prior works have adopted a similar but slightly different objective of minimizing the estimation error defined as the variance of the estimator, and from a first-order efficiency perspective \citep{armstrong2022asymptotic, blackwell2022batch, cai2024performance, hahn2011adaptive}.
Intuitively, a first-order optimal design converges to the asymptotic variance lower bound when the sample size is large.
This literature has been further extended to incorporate adjustments in the presence of baseline covariates \citep{cytrynbaum2021optimal, li2024double, tabord2023stratification, wei2025adaptive}.
Although these works reveal many insights that guide the design of pilot experimental studies, these works do not precisely guide the selection of sample sizes, as any sub-linear sample size in the first stage is first-order optimal under the first-order efficiency framework.

In contrast, the objective of our work is to minimize the proxy mean squared error, and from a second-order efficiency perspective.
Intuitively, a second-order efficiency notion studies how fast a design converges to the proxy mean squared error lower bound as the sample size grows.
In our work, our adaptive Neyman allocation algorithm is both first-order efficient in minimizing the variance of the estimator (Theorem~\ref{thm:inference}), and second-order efficient in minimizing the proxy mean squared error (Theorems~\ref{thm:MStageANA} and~\ref{thm:ANA:LB}).
Additionally, the notion of second-order efficiency explicitly guides the selection of sample sizes in pilot experimental studies.

The adaptive experimental design literature also studies inference on adaptively collected data \citep{bowden2017unbiased, chen2025characterization, hirano2023asymptotic, khamaru2024inference, melfi2000estimation, nie2018adaptively, offer2021adaptive, shin2019bias, shin2019sample, zhang2020inference, zhang2021statistical}, with extensions to adjust for baseline covariates \citep{deshpande2018accurate, deshpande2019online, hadad2021confidence, xiong2019optimal, zhan2021off, zhan2023policy}.
For estimation, our work borrows ideas from \citet{xiong2019optimal} and shows that adaptive Neyman allocation, which adapts on the sample variance but not the sample mean, achieves finite-sample unbiasedness under a symmetric distribution assumption (Theorem~\ref{thm:estimation}).
It is different from the traditional adaptive experiments, where the unbiasedness property usually requires the sample size to be large.
For inference, our work borrows ideas from \citet{chen2025characterization, khamaru2024inference} and establishes a central limit theorem for adaptive Neyman allocation.

\item Ranking and selection (operations research and simulations). 
In the ranking and selection literature, prior works have adopted a related but different objective of maximizing the probability of correctly identifying the treatment with the largest mean outcome, usually involving more than two treatments \citep{bechhofer1954single, chick2001new, glynn2004large, hong2021review, hunter2017parallel}.
The literature has proposed various methods to allocate simulation budget to each treatment, such as the seminal optimal computing budget allocation (OCBA) method \citep{chen1996lower, chen2000simulation}.

The ranking and selection literature is also closely related to the best-arm identification literature, which essentially studies the same problem but under a different assumption about the outcomes \citep{adusumilli2022minimax, audibert2010best, kasy2021adaptive, kato2022best, mannor2004sample, russo2016simple}.
Ranking and selection usually assumes Gaussian distributions with unknown variances, whereas best-arm identification usually assumes sub-Gaussian distributions with constant upper bounds on the variances.
Compared with these two lines of literature, our work makes a weaker assumption.

In terms of algorithmic design, when there are more than two treatments, the difference between our adaptive Neyman allocation problem and these two lines of literature becomes apparent.
The optimal allocation in these two lines of literature usually follows some OCBA structure where the treatments with smaller mean outcomes are less explored than the optimal treatment.
In contrast, the optimal allocation in our problem, even if there were more than two treatments, follows the Neyman allocation structure where the mean outcomes are irrelevant.

When there are only two treatments, the adaptive Neyman allocation problem becomes similar to these two lines of literature.
If the outcomes of both treatments can be well-approximated by Gaussian distributions, such as in a small gap regime when the gap between the mean outcomes of both treatments decreases to zero \citep{adusumilli2022minimax, kato2022best, wager2021experimenting}, these two problems become equivalent to each other. 
In contrast, our work considers a fixed gap regime, and neither problem implies the other.
\end{enumerate}


\subsection*{Roadmap}
The paper is structured as follows. 
In Section~\ref{sec:Setup} we formally introduce adaptive Neyman allocation.
In Section~\ref{sec:Optimization} we introduce an optimization framework and show that the classical half-half allocation is optimal under this optimization framework.
In Sections~\ref{sec:TwoStage} and~\ref{sec:MultiStage} we study the two-stage and multi-stage adaptive Neyman allocation problem, respectively.
In Section~\ref{sec:Analysis} we study estimation and inference using adaptively collected data.
In Section~\ref{sec:InExpectation} we extend our high probability guarantees into in expectation guarantees.
In Sections~\ref{sec:Simulation} and~\ref{sec:Synthetic} we use online A/B testing data from a social media site and synthetic data to demonstrate the effectiveness of our adaptive Neyman allocation algorithm.
In Section~\ref{sec:Conclusions} we conclude the paper and point out some limitations and future research directions.
All mathematical details are deferred to the Online Appendix.

\section{Problem Setup}
\label{sec:Setup}

Consider the following problem.
There is a discrete, finite time horizon of $T \in \bN$ periods.
The time horizon $T$ stands for the size of the experiment, and is known to the experimenter before the start of the horizon.
At any time $t \in [T] := \{1,2,...,T\}$, one unit is involved in the experiment.
We interchangeably use unit $t$ to stand for the unit that arrives at time $t$.

Let there be two versions of treatments.
We use ``treatment'' and ``control'', or $1$ and $0$, respectively, to stand for these two versions of treatments.
Let $W_t \in \{0,1\}$ stand for the treatment assignment that unit $t$ receives.
Following convention, we use $W_t$ for a random treatment assignment, and $w_t$ for one realization.

Following the potential outcomes framework and under the Stable Unit Treatment Value Assumption \citep{rubin1974estimating, holland1986statistics, imbens2015causal}, each unit $t$ has a set of potential outcomes $Y_{t}(\cdot)$.
Each observed outcome is related to its respective potential outcomes $Y_t = Y_t(w)$, if $W_t=w$.
We assume the existence of a super-population \citep{abadie2020sampling}, such that each unit's potential outcomes $(Y_t(1), Y_t(0))$ are independent and identically distributed (i.i.d.) replicas of a pair of representative random variables $(Y(1), Y(0))$.
These random variables are drawn from a joint distribution of the super-population, i.e., $(Y(1), Y(0)) \sim \cF$.
We assume that $\cF$ belongs to $\sP$, the family of joint distributions where the first two moments exist. 
But we put no restrictions on the correlation between $Y(1)$ and $Y(0)$.

In this paper, we consider a multi-stage randomized experiment, which we refer to as ``adaptive Neyman allocation.''
The experiment is conducted in $M \in \bN$ stages.
In stage $m \in [M]$, the experimenter conducts a completely randomized experiment parameterized by $(T_m(1), T_m(0))$.
The size of the stage-$m$ experiment is $T_m = T_m(1)+T_m(0)$, and the experimenter randomly chooses exactly $T_m(1)$ units to receive treatment, and exactly $T_m(0)$ units to receive control.
After $M$ stages of experiments, the experimenter has assigned $T(1) = \sum_{m=1}^M T_m(1)$ units to receive treatment, and $T(0) = \sum_{m=1}^M T_m(0)$ units to receive control.
See Table~\ref{tbl:Notations} for a summary of notations.

\begin{table}[!tb]
\centering
\TABLE{Notations of the number of treated, control, and total units in each stage
\label{tbl:Notations}}
{\begin{tabular}{>{\centering}p{1.8cm}|>{\centering}p{1.8cm}>{\centering}p{1.8cm}>{\centering}p{1cm}>{\centering}p{1.8cm}>{\centering}p{1.8cm}c}
        & Stage $1$ & Stage $2$ & $\ldots$ & Stage $M$ & Total  & \\ \cline{1-6}
Treated & $T_1(1)$  & $T_2(1)$  & $\ldots$ & $T_M(1)$  & $T(1)$ & \\
Control & $T_1(0)$  & $T_2(0)$  & $\ldots$ & $T_M(0)$  & $T(0)$ & \\
Total   & $T_1$     & $T_2$     & $\ldots$ & $T_M$     & $T$    & 
\end{tabular}
}
{}
\end{table}

Formally, a design of $M$-stage adaptive experiment is defined as $\pi = (\cT_M, \phi_1, \phi_2, ..., \phi_T)$, where $\cT_M = \{T_1, T_2, ..., T_M\}$ is a sequence of sample sizes in the $M$ stages, and $\phi_t$ is a decision rule that decides the treatment probability of unit $t$.
Given $\cT_M$, let $m(t) \in [M]$ be the index of the current stage that contains $t$.
Let $\cH(t) = \{(W_s, Y_s) \vert s \leq \sum_{l=1}^{m(t)-1} T_l\}$ be the history of unit $t$, that is, a collection of treatment assignments and observed outcomes up to the end of stage $m(t)-1$, where stage $0$ stands for an empty set.
For each $t \in [T]$, $\phi_t: \cH(t) \to [0,1]$ maps from the space of histories to the space of treatment probabilities, such that $\Pr(W_t = 1) = \phi_t(\cH(t))$.
Let $\Pi_M$ be the family of $M$-stage adaptive experiments.
We have $\Pi_0 \subseteq \Pi_1 \subseteq ... \subseteq \Pi_T := \Pi$, where $\Pi_0$ stands for the family of non-adaptive experiments, and $\Pi$ stands for the family of fully adaptive experiments, or, simply, adaptive experiments.

The causal effect of interest is the average treatment effect of the super-population,
\begin{align*}
\tau = \bE[Y(1) - Y(0)],
\end{align*}
where the expectation is taken with respect to the joint distribution $\cF$.
After collecting data from the experiment, the experimenter uses the simple difference-in-means estimator to estimate the causal effect,
\begin{align}
\widehat{\tau} = \frac{1}{T(1)} \sum_{t: W_t = 1} Y_t - \frac{1}{T(0)} \sum_{t: W_t = 0} Y_t. \label{eqn:Estimator}
\end{align}
It is worth mentioning that $\widehat{\tau}$ may have two sources of randomness. 
The potential outcomes are random and the treatment assignments are also possibly random.

To evaluate the quality of the difference-in-means estimator, we consider the mean squared error of the estimator.
When $T(1)$ and $T(0)$ are fixed, the difference-in-means estimator is unbiased and the mean squared error is equivalent to the variance of the estimator, which could be further expressed as follows,
\begin{align}
\bE\left[(\widehat{\tau} - \tau)^2\right] = \Var(\widehat{\tau}) = \frac{1}{T(1)} \sigma^2(1) + \frac{1}{T(0)} \sigma^2(0), \label{eqn:MSE}
\end{align}
where $\sigma(1), \sigma(0) > 0$ stand for the standard deviations of the two representative random variables $Y(1)$ and $Y(0)$, respectively.
However, in an adaptive Neyman allocation algorithm, $T(1)$ and $T(0)$ are random in nature.
The number of treated and control units are adaptively determined by the observed outcomes in the previous stages.
Consequently, the mean squared error may not always have the same expression in \eqref{eqn:MSE} as if $T(1)$ and $T(0)$ were fixed quantities.

In this paper, we re-define expression \eqref{eqn:MSE} to be the proxy mean squared error,
\begin{align}
V(T(1), T(0)) = \frac{1}{T(1)} \sigma^2(1) + \frac{1}{T(0)} \sigma^2(0). \label{eqn:ProxyVariance}
\end{align}
The experimenter’s objective is then to minimize the proxy mean squared error as defined above.
We will show in Theorem~\ref{thm:inference} that, under appropriate allocation rules such as the adaptive Neyman allocation algorithms that we will introduce in this paper, the variance of estimator \eqref{eqn:Estimator} asymptotically converges to the proxy mean squared error \eqref{eqn:ProxyVariance}. 
So minimizing the proxy mean squared error \eqref{eqn:ProxyVariance} can be interpreted as minimizing the variance of estimator \eqref{eqn:Estimator} when $T$ is large.
See Section~\ref{sec:Analysis} for more discussions.

One benefit of using the proxy mean squared error as our objective is that the proxy mean squared error only depends on $(T(1), T(0))$ the numbers of treated and control units in total.
No matter how the experimenter adaptively chooses $(T_m(1), T_m(0))$ in each stage, the proxy mean squared error $V(T(1), T(0))$ is always well defined.
The multi-stage experiment enables the experimenter to make better choices for $(T(1), T(0))$ by appropriately selecting $(T_m(1), T_m(0))$ at each stage. 
In the following section, we present an optimization framework for making such decisions.

\section{An Optimization Framework}
\label{sec:Optimization}

If the experimenter was endowed with clairvoyant information about the standard deviations $\sigma(1)$ and $\sigma(0)$, the experimenter would allocate $(T(1), T(0))$ optimally in a single stage experiment to minimize $V(T(1), T(0))$.
Note that we do not take expectation for $V(T(1), T(0))$ in a single stage experiment because $T(1)$ and $T(0)$ are fixed.
The optimal solution can be explicitly calculated as,
\begin{align*}
T^*(1) = \ \frac{\sigma(1)}{\sigma(1) + \sigma(0)} T, && T^*(0) = \ \frac{\sigma(0)}{\sigma(1) + \sigma(0)} T,
\end{align*}
and the optimal proxy mean squared error is given by the following expression,
\begin{align}
V(T^*(1), T^*(0)) = \frac{1}{T}(\sigma(1) + \sigma(0))^2. \label{eqn:ClairvoyantOptimal}
\end{align}
This is what \citet{neyman1934two} suggests, and has been recognized as the Neyman allocation.
As the standard deviations were assumed given, the original work of Neyman allocation only focused on single stage experiments.

More often, the experimenter is not endowed with clairvoyant information about $\sigma(1)$ and $\sigma(0)$.
To solve this decision making under uncertainty problem, we introduce the competitive analysis framework to experimental design.
For any design $\pi \in \Pi$, let $(T^\pi(1), T^\pi(0))$ be the numbers of treated and control units assigned by policy $\pi$.
The competitive analysis framework suggests to solve the following problem,
\begin{align}
\inf_{\pi \in \Pi} \ \sup_{\cF \in \sP} \ \frac{\bE[V(T^\pi(1), T^\pi(0))]}{V(T^*(1), T^*(0))}. \label{eqn:Obj}
\end{align}
The optimal value to problem \eqref{eqn:Obj} is often referred to as the competitive ratio \citep{borodin2005online, buchbinder2009design}.

The above competitive analysis framework is similar to the minimax decision rule \citep{berger2013statistical, bickel2015mathematical, li1983minimaxity, wu1981robustness}, which solves
\begin{align*}
\inf_{\pi \in \Pi} \ \sup_{\cF \in \sP} \quad \bE[V(T^\pi(1), T^\pi(0))], 
\end{align*}
as well as the minimax regret decision rule \citep{lai1985asymptotically, manski2004statistical, robbins1952some, stoye2009minimax}, which solves
\begin{align*}
\inf_{\pi \in \Pi} \ \sup_{\cF \in \sP} \quad \bE[V(T^\pi(1), T^\pi(0))] - V(T^*(1), T^*(0)).
\end{align*}
But the above two decision rules are not directly applicable in our work because both objective values scale with the magnitudes of the potential outcomes or the variances.
Consequently, the active learning literature assumes that $\sigma(1)$ and $\sigma(0)$ are constants \citep{antos2010active, carpentier2011finite, grover2009active}.
Under this assumption, $V(T^*(1), T^*(0)) = \Theta(T^{-1})$. 
So any regret minimization result on the order of $O\big(T^{-1-\alpha}\big)$ corresponds to a competitive ratio result on the order of $1+O\big(T^{-\alpha}\big)$.

To illustrate the competitive analysis framework, consider the setup of the traditional single stage Neyman allocation, but with unknown standard deviations.
In the single stage experiment, the policy $\pi$ only determines one single and fixed pair of $(T(1), T(0))$.
We replace the policy $\pi$ with this pair of actions $(T(1), T(0))$ in \eqref{eqn:Obj}, and solve the new problem to optimal.
This yields the following result, the proof of which is deferred to Section~\ref{sec:proof:thm:OneStage} in the Online Appendix.

\begin{theorem}
\label{thm:OneStage}
The optimal solution to
\begin{align*}
\inf_{\pi \in \Pi_0} \ \sup_{\cF \in \sP} \ \frac{V(T(1), T(0))}{V(T^*(1), T^*(0))}
\end{align*}
is given by $T(1) = T(0) = T / 2$. 
The supremum of the inner optimization problem is achieved when either the treated group or the control group has zero variance, that is, $\sigma(1) = 0$ or $\sigma(0) = 0$.
\end{theorem}

Theorem~\ref{thm:OneStage} reproduces the classical result that the optimal design involves an equal number of treated and control units \citep{neyman1934two}.
But Theorem~\ref{thm:OneStage} does not require any knowledge of the standard deviations of the treatment or control populations.
More importantly, Theorem~\ref{thm:OneStage} does not even require any assumption about the data generating process, such as the treatment or control populations having the same support (see, e.g., \citet{bojinov2023design, ni2023design}), or the treatment effects being additive, which implies that the standard deviations are the same (see, e.g., \citet{xiong2019optimal}), or permutation invariance (see, e.g., \citet{bai2023randomize, basse2023minimax, wu1981robustness}).

In other experimental design literature, Theorem~\ref{thm:OneStage} is often presented as an assumption and serves as the basis for designing optimal experiments \citep{bai2022optimality, candogan2021near, greevy2004optimal, harshaw2019balancing, lu2011optimal, rosenbaum1989optimal, xiong2019optimal, zhao2022pigeonhole}.
In contrast, by using the competitive analysis framework, Theorem~\ref{thm:OneStage} establishes the credibility of such an assumption.

In the following sections, we will use this competitive analysis framework to study adaptive Neyman allocation.
We will start with the two-stage adaptive Neyman allocation to introduce the basic estimation ideas and build some intuitions in Section~\ref{sec:TwoStage}.
We will then introduce the more general multi-stage adaptive Neyman allocation in Section~\ref{sec:MultiStage}.

\section{Two-Stage Adaptive Neyman Allocation}
\label{sec:TwoStage}

In this section, we focus on the $M=2$ case, which we refer to as the two-stage adaptive Neyman allocation.
When there are two stages, the experimental data collected during the first stage reveals information about the magnitudes of $\sigma(1)$ and $\sigma(0)$, which can be used to guide the design of the second stage experiment.

\subsubsection*{Algorithm.}
Recall that $(T_1(1), T_1(0))$ stand for the numbers of treated and control units in the first stage, respectively, and that $T_1 = T_1(1) + T_1(0)$ stands for the total number of units in the first stage.
We consider the following sample variance estimators at the end of the first stage,
\begin{subequations}
\begin{align}
\widehat{\sigma}^2_1(1) = & \ \frac{1}{T_1(1) - 1}\sum_{\substack{1 \leq t \leq T_1 \\ t: W_t = 1}} \bigg( Y_t - \frac{1}{T_1(1)} \sum_{\substack{1 \leq t \leq T_1 \\ t: W_t = 1}} Y_t \bigg)^2, \label{eqn:SampleVariance1}\\
\widehat{\sigma}^2_1(0) = & \ \frac{1}{T_1(0) - 1}\sum_{\substack{1 \leq t \leq T_1 \\ t: W_t = 0}} \bigg( Y_t - \frac{1}{T_1(0)} \sum_{\substack{1 \leq t \leq T_1 \\ t: W_t = 0}} Y_t \bigg)^2. \label{eqn:SampleVariance0}
\end{align}
\end{subequations}

After obtaining the sample variance estimators, it is natural to use such sample variance estimators to guide the Neyman allocation in the second stage.
The allocation of treated and control units should roughly follow the ones suggested by \eqref{eqn:ClairvoyantOptimal}, but the estimated standard deviations from the first stage will be used instead of the true standard deviations, i.e.\footnote{When $\widehat{\sigma}(1) = \widehat{\sigma}(0) = 0$, we abuse the notation and denote $\frac{0}{0+0} = \frac{1}{2}$.},
\begin{align}
T(1) = \ \frac{\widehat{\sigma}_1(1)}{\widehat{\sigma}_1(1) + \widehat{\sigma}_1(0)} T, && T(0) = \ \frac{\widehat{\sigma}_1(0)}{\widehat{\sigma}_1(1) + \widehat{\sigma}_1(0)} T. \label{eqn:EstimatedOPT}
\end{align}
Based on this natural intuition, we define the two-stage adaptive Neyman allocation in Algorithm~\ref{alg:2StageANA}.

\begin{algorithm}[!tb]
\caption{Two-stage adaptive Neyman allocation}
\label{alg:2StageANA}
\small 
\textbf{Input}: Tuning parameter $\beta$.
\begin{algorithmic}[1]
\State \textbf{Initialize}: $(T_1(1), T_1(0)) \gets (\frac{\beta}{2} \sqrt{T}, \frac{\beta}{2} \sqrt{T})$.
\State Conduct a completely randomized experiment parameterized by $(T_1(1), T_1(0))$; \Comment{Stage 1 experiment}
\State Calculate two estimators $\widehat{\sigma}^2_1(1)$ and $\widehat{\sigma}^2_1(0)$ as defined in \eqref{eqn:SampleVariance1} and \eqref{eqn:SampleVariance0}.
\State \textbf{Case 1}: $\frac{\widehat{\sigma}_1(1)}{\widehat{\sigma}_1(1) + \widehat{\sigma}_1(0)} T > \frac{\beta}{2} \sqrt{T}$ and $\frac{\widehat{\sigma}_1(0)}{\widehat{\sigma}_1(1) + \widehat{\sigma}_1(0)} T > \frac{\beta}{2} \sqrt{T}$
\State \phantom{for} $(T_2(1), T_2(0)) \gets (\frac{\widehat{\sigma}_1(1)}{\widehat{\sigma}_1(1) + \widehat{\sigma}_1(0)} T - \frac{\beta}{2}\sqrt{T}, \frac{\widehat{\sigma}_1(0)}{\widehat{\sigma}_1(1) + \widehat{\sigma}_1(0)} T - \frac{\beta}{2}\sqrt{T})$.
\State \textbf{Case 2}: $\frac{\widehat{\sigma}_1(1)}{\widehat{\sigma}_1(1) + \widehat{\sigma}_1(0)} T \leq \frac{\beta}{2} \sqrt{T}$
\State \phantom{for} $(T_2(1), T_2(0)) \gets (0, T - \beta\sqrt{T})$.
\State \textbf{Case 3}: $\frac{\widehat{\sigma}_1(0)}{\widehat{\sigma}_1(1) + \widehat{\sigma}_1(0)} T \leq \frac{\beta}{2} \sqrt{T}$
\State \phantom{for} $(T_2(1), T_2(0)) \gets (T - \beta\sqrt{T}, 0)$.
\State Conduct a completely randomized experiment parameterized by $(T_2(1), T_2(0))$; \Comment{Stage 2 experiment}
\end{algorithmic}
\end{algorithm}

In words, the experiment consists of two stages.
In the first stage, the experiment has a total size of $\sqrt{T}$, and assigns half units to treated and the other half to control.
Then we calculate the sample variance estimators $\widehat{\sigma}^2_1(1)$ and $\widehat{\sigma}^2_1(0)$.
If neither $\widehat{\sigma}_1(1)$ or $\widehat{\sigma}_1(0)$ is too small, the second stage experiment roughly mimics the Neyman allocation by using the estimated variances.
If $\widehat{\sigma}_1(1)$ or $\widehat{\sigma}_1(0)$ is too small, the second stage experiment assigns all units to control or treated, respectively.

In essence, the two-stage adaptive Neyman allocation is similar to the design in \citet{hahn2011adaptive}.
But there are two differences.
First and more importantly, we specify the optimal size for the first-stage experiment.
Second, we use equation \eqref{eqn:EstimatedOPT} to guide the allocation of treated and control units for the entire horizon; whereas \citet{hahn2011adaptive} uses equation \eqref{eqn:EstimatedOPT} to guide the allocation for the second stage.

\subsubsection*{Analysis.}
We now present a formal analysis about the quality of the two-stage adaptive Neyman allocation.
Recall that the sample variance estimators are unbiased, i.e., $\bE[\widehat{\sigma}^2_1(1)] = \sigma^2(1)$ and $\bE[\widehat{\sigma}^2_1(0)] = \sigma^2(0)$.
To ensure that the distributions of the sample variance estimators are concentrated enough around the true variances $\sigma^2(1)$ and $\sigma^2(0)$, we make the following assumption.

\begin{assumption}
\label{asp:kurtosis}
There exist two constants $\kappa(1), \kappa(0) < \infty$ which do not depend on $T$, such that
\begin{align*}
\kappa(1) = \frac{\bE\left[(Y(1) - \bE Y(1))^4\right]}{\sigma^4(1)}, && \kappa(0) = \frac{\bE\left[(Y(0) - \bE Y(0))^4\right]}{\sigma^4(0)}.
\end{align*}
\end{assumption}

Assumption~\ref{asp:kurtosis} asserts that the representative random variables $Y(1), Y(0)$ are sufficiently light-tailed, in the sense that their respective kurtosis values $\kappa(1), \kappa(0)$ exist.
It is worth noting that the kurtosis values are always greater than $1$, i.e., $\kappa(1), \kappa(0) > 1$.
For a Gaussian random variable $Z$, its kurtosis is equal to $3$, i.e., $\bE\left[(Z - \bE Z)^4\right] / \sigma^4(Z) = 3$.
The kurtosis will be larger for distributions with heavier tails, and smaller for those with lighter tails.
In other papers, instead of assuming Assumption~\ref{asp:kurtosis}, either sub-Gaussianity or boundedness is often assumed \citep{lattimore2020bandit, slivkins2019introduction}. 
Let $\sP^{[\kappa]}$ be the family of joint distributions that satisfies Assumption~\ref{asp:kurtosis}. 

With the above assumption, we now show the quality of the two-stage adaptive Neyman allocation, as measured by the competitive ratio, in Theorem~\ref{thm:2StageANA}.

\begin{theorem}
\label{thm:2StageANA}
Let $T \geq 16$ and $\eps \in \left(0, \frac{1}{8}\right)$.
Let $\beta = 1$ in Algorithm~\ref{alg:2StageANA}.
Let $(T(1), T(0))$ be the number of total treated and control units from Algorithm~\ref{alg:2StageANA}, respectively. Under Assumption~\ref{asp:kurtosis}, there exists an event that happens with probability at least $1 - (\kappa(1) + \kappa(0)) T^{-\eps}$, conditional on which
\begin{align*}
\sup_{\cF \in \sP^{[\kappa]}} \ \frac{V(T(1), T(0))}{V(T^*(1), T^*(0))} \leq 1 + T^{-\frac{1}{2} + \eps}.
\end{align*}
\end{theorem}

Theorem~\ref{thm:2StageANA} presents a high probability bound.
Using an assumption that enables us to show exponentially small probability, we will be able to show that a similar bound (up to some logarithm factors) holds in expectation.
See Corollary~\ref{coro:2StageANA} in Section~\ref{sec:InExpectation}.

Results that study the quality of adaptive allocation policies also frequently appear in the active learning literature \citep{antos2010active, carpentier2011finite, grover2009active, russac2021b}.
This literature adopts a minimax regret framework, which differs from the competitive analysis framework.
The minimax regret framework focuses on the difference between the variance of any policy and the optimal variance, rather than the ratio between them.
As the magnitudes of $\sigma(1)$ and $\sigma(0)$ directly impact the objective value, the minimax regret framework typically assumes $\sigma(1)$ and $\sigma(0)$ are constants and that they are on the same order.
In contrast, this paper allows $\sigma(1)$ and $\sigma(0)$ to differ significantly, with one potentially much larger than the other.
After translating into the framework of this paper, the best competitive ratio suggested by the literature is on the order of $1 + O(T^{-\frac{1}{2}})$, using much more complicated and fully adaptive experimental designs such as upper confidence bound approaches.
In contrast, Theorem~\ref{thm:2StageANA} shows that a simple two-stage adaptive Neyman allocation can achieve the same competitive ratio, by adapting only once.

In Section~\ref{sec:MultiStage}, we will show that conducting experiments in more than two stages can improve the competitive ratio, to an extent that almost matches the information-theoretic limit of conducting adaptive experiments.

To conclude this section, we sketch some unrigorous intuitions behind the proof of Theorem~\ref{thm:2StageANA} below, and defer the complete proof to Section~\ref{sec:proof:thm:2StageANA} in the Online Appendix.

\proof{Sketch proof of Theorem~\ref{thm:2StageANA}.}
Denote $\rho = \frac{\sigma(1)}{\sigma(0)}$.
Without loss of generality assume $\rho \geq 1$.
Suppose the length of the first stage is parameterized by $T^\alpha$ (we ignore $\beta$ in this unrigorous sketch proof).
We aim to find the optimal length $T^\alpha$ of the first stage.

Case 1: $\rho > \frac{T - T^{\alpha}}{T^{\alpha}}$. 
Then with high probability, the first stage reveals this condition and Algorithm~\ref{alg:2StageANA} stops allocating units to the control group in the second stage.
In this case, we will show that
\begin{align*}
\frac{V(T(1), T(0))}{V(T^*(1), T^*(0))} = \frac{T}{T-T^{\alpha}} \cdot \frac{\rho^2}{(\rho+1)^2} + \frac{T}{T^{\alpha}} \cdot \frac{1}{(\rho+1)^2} \leq \frac{T}{T-T^{\alpha}} \approx 1 + T^{\alpha - 1}.
\end{align*}

Case $2$: $1 \leq \rho \leq \frac{T - T^{\alpha}}{T^{\alpha}}$.
Then with high probability, the first stage reveals this condition and Algorithm~\ref{alg:2StageANA} mimics the Neyman allocation by using the estimated variances.
In this case, the estimation errors of estimating $\sigma(1)$ and $\sigma(0)$ are on the order of $T^{-\frac{\alpha}{2}}$. 
So the estimation error of estimating $\rho$ is on the order of $T^{-\alpha}$. 
We will show that
\begin{align*}
\frac{V(T(1), T(0))}{V(T^*(1), T^*(0))} \leq 1 + \frac{\rho}{(\rho+1)^2} \cdot T^{-\alpha} \approx 1 + T^{-\alpha}.
\end{align*}

Combining both cases, we set $\alpha-1 = -\alpha$ and obtain $\alpha = \frac{1}{2}$, which leads to competitive ratio $\frac{V(T(1), T(0))}{V(T^*(1), T^*(0))} \approx 1 + T^{-\frac{1}{2}}.$
\hfill \halmos
\endproof

\section{Multi-Stage Adaptive Neyman Allocation}
\label{sec:MultiStage}

This section presents an extension of the two-stage adaptive Neyman allocation to multiple stages. 
We provide a formal analysis of the competitive ratio of the multi-stage adaptive Neyman allocation, and show that such a competitive ratio is nearly optimal.

\subsubsection*{Algorithm.} 
The multi-stage adaptive Neyman allocation algorithm uses the following sample variance estimators at the end of each stage.
Recall that $(T_m(1), T_m(0))$ stand for the numbers of treated and control units in stage $m$, respectively, and that $T_m = T_m(1) + T_m(0)$ stands for the total number of units in stage $m$.
At the end of stage $m$, define the following sample variance estimators,
\begin{subequations}
\begin{align}
\widehat{\sigma}^2_m(1) = & \ \frac{1}{\sum_{l=1}^m T_l(1) - 1}\sum_{\substack{1 \leq t \leq \sum_{l=1}^m T_l \\ t: W_t = 1}} \bigg( Y_t - \frac{1}{\sum_{l=1}^m T_l(1)} \sum_{\substack{1 \leq t \leq \sum_{l=1}^m T_l \\ t: W_t = 1}} Y_t \bigg)^2, \label{eqn:mVariance1} \\
\widehat{\sigma}^2_m(0) = & \ \frac{1}{\sum_{l=1}^m T_l(0) - 1}\sum_{\substack{1 \leq t \leq \sum_{l=1}^m T_l \\ t: W_t = 0}} \bigg( Y_t - \frac{1}{\sum_{l=1}^m T_l(0)} \sum_{\substack{1 \leq t \leq \sum_{l=1}^m T_l \\ t: W_t = 0}} Y_t \bigg)^2. \label{eqn:mVariance0} 
\end{align}
\end{subequations}
Using the above sample variance estimators \eqref{eqn:mVariance1} and \eqref{eqn:mVariance0}, we could update the allocation of treated and control units following the Neyman allocation, i.e.,
\begin{align}
T(1) = \ \frac{\widehat{\sigma}_m(1)}{\widehat{\sigma}_m(1) + \widehat{\sigma}_m(0)} T, && T(0) = \ \frac{\widehat{\sigma}_m(0)}{\widehat{\sigma}_m(1) + \widehat{\sigma}_m(0)} T. \label{eqn:EstimatedOPT:M}
\end{align}
We can update the above Neyman allocation as defined in \eqref{eqn:EstimatedOPT:M} using the estimated variances at the end of each stage $m$.
Using \eqref{eqn:EstimatedOPT:M} we define the $M$-stage adaptive Neyman allocation.
See Algorithm~\ref{alg:MStageANA} for Pseudo-codes.

\begin{algorithm}[!htb]
\caption{$M$-stage adaptive Neyman allocation}
\label{alg:MStageANA}
\small
\textbf{Inputs}: Tuning parameters $\beta_1, \beta_2, ..., \beta_{M-1}$. \Comment{There are $M$ pre-determined stages $[0,\beta_1 T^{\frac{1}{M}}]$, $(\beta_1 T^{\frac{1}{M}}$, $\beta_2 T^{\frac{2}{M}}]$, ..., $(\beta_{M-1} T^{\frac{M-1}{M}}, T]$} 
\begin{algorithmic}[1]
\State \textbf{Initialize:} $(T_1(1), T_1(0)) \gets (\frac{\beta_1}{2} T^{\frac{1}{M}}, \frac{\beta_1}{2} T^{\frac{1}{M}})$;
\For{$m = 1, 2, ..., M-2$} \Comment{The $m$-th stage experiment}
\State Conduct a completely randomized experiment parameterized by $(T_m(1), T_m(0))$;
\State Estimate $\widehat{\sigma}^2_m(1)$ and $\widehat{\sigma}^2_m(0)$ as in \eqref{eqn:mVariance1} and \eqref{eqn:mVariance0} using data collected during stages $1 \sim m$;
\State \textbf{Case 1}: $\frac{\widehat{\sigma}_m(0)}{\widehat{\sigma}_m(1) + \widehat{\sigma}_m(0)} T < \frac{\beta_m}{2} T^{\frac{m}{M}}$ \label{mrk:Case1}
\State \phantom{for} For any $l \geq m+1$, $(T_l(1), T_l(0)) \gets (\beta_l T^{\frac{l}{M}} - \beta_{l-1} T^{\frac{l-1}{M}}, 0)$;
\Goto{marker};
\State \textbf{Case 2}: $\frac{\beta_{m}}{2} T^{\frac{m}{M}} \leq \frac{\widehat{\sigma}_m(0)}{\widehat{\sigma}_m(1) + \widehat{\sigma}_m(0)} T < \frac{\beta_{m+1}}{2} T^{\frac{m+1}{M}}$ \label{mrk:Case2}
\State \phantom{for} $(T_{m+1}(1), T_{m+1}(0)) \gets (\beta_{m+1}T^{\frac{m+1}{M}}-\frac{\widehat{\sigma}_m(0)}{\widehat{\sigma}_m(1)+\widehat{\sigma}_m(0)}T - \frac{\beta_m}{2}T^{\frac{m}{M}}, \frac{\widehat{\sigma}_m(0)}{\widehat{\sigma}_m(1)+\widehat{\sigma}_m(0)}T - \frac{\beta_m}{2}T^{\frac{m}{M}})$;
\State \phantom{for} For any $l \geq m+2$, $(T_l(1), T_l(0)) \gets (\beta_l T^{\frac{l}{M}} - \beta_{l-1} T^{\frac{l-1}{M}}, 0)$; 
\Goto{marker};
\State \textbf{Case 3}: $\frac{\widehat{\sigma}_m(0)}{\widehat{\sigma}_m(1) + \widehat{\sigma}_m(0)} T \geq \frac{\beta_{m+1}}{2} T^{\frac{m+1}{M}}$ and $\frac{\widehat{\sigma}_m(1)}{\widehat{\sigma}_m(1) + \widehat{\sigma}_m(0)} T \geq \frac{\beta_{m+1}}{2} T^{\frac{m+1}{M}}$ \label{mrk:Case3}
\State \phantom{for} $(T_{m+1}(1), T_{m+1}(0)) \gets (\frac{\beta_{m+1}}{2}T^{\frac{m+1}{M}} - \frac{\beta_{m}}{2}T^{\frac{m}{M}}, \frac{\beta_{m+1}}{2}T^{\frac{m+1}{M}} - \frac{\beta_{m}}{2}T^{\frac{m}{M}})$; \Comment{Note: there is no ``go to''}
\State \textbf{Case 4}: $\frac{\beta_{m}}{2} T^{\frac{m}{M}} \leq \frac{\widehat{\sigma}_m(1)}{\widehat{\sigma}_m(1) + \widehat{\sigma}_m(0)} T < \frac{\beta_{m+1}}{2} T^{\frac{m+1}{M}}$ \label{mrk:Case4}
\State \phantom{for} $(T_{m+1}(1), T_{m+1}(0)) \gets (\frac{\widehat{\sigma}_m(1)}{\widehat{\sigma}_m(1)+\widehat{\sigma}_m(0)}T - \frac{\beta_m}{2}T^{\frac{m}{M}}, \beta_{m+1}T^{\frac{m+1}{M}}-\frac{\widehat{\sigma}_m(1)}{\widehat{\sigma}_m(1)+\widehat{\sigma}_m(0)}T - \frac{\beta_m}{2}T^{\frac{m}{M}})$;
\State \phantom{for} For any $l \geq m+2$, $(T_l(1), T_l(0)) \gets (\beta_l T^{\frac{l}{M}} - \beta_{l-1} T^{\frac{l-1}{M}}, 0)$; 
\Goto{marker};
\State \textbf{Case 5}: $\frac{\widehat{\sigma}_m(1)}{\widehat{\sigma}_m(1) + \widehat{\sigma}_m(0)} T < \frac{\beta_m}{2} T^{\frac{m}{M}}$ \label{mrk:Case5}
\State \phantom{for} For any $l \geq m+1$, $(T_l(1), T_l(0)) \gets (0, \beta_l T^{\frac{l}{M}} - \beta_{l-1} T^{\frac{l-1}{M}})$;
\Goto{marker};
\EndFor

\If{$m = M-1$} \Comment{The $(M-1)$-th stage experiment} \label{mrk:LastStage}
\State Conduct a completely randomized experiment parameterized by $(T_{M-1}(1), T_{M-1}(0))$;
\State Estimate $\widehat{\sigma}^2_{M-1}(1)$ and $\widehat{\sigma}^2_{M-1}(0)$ as in \eqref{eqn:mVariance1} and \eqref{eqn:mVariance0} using data collected during stages $1 \sim (M-1)$;
\State \textbf{Case 1}: $\frac{\widehat{\sigma}_{M-1}(0)}{\widehat{\sigma}_{M-1}(1) + \widehat{\sigma}_{M-1}(0)} T < \frac{\beta_{M-1}}{2} T^{\frac{M-1}{M}}$ \label{mrk:LastStage:Case1}
\State \phantom{for} $(T_M(1), T_M(0)) \gets (T - \beta_{M-1} T^{\frac{M-1}{M}}, 0)$;
\State \textbf{Case 2}: $\frac{\widehat{\sigma}_{M-1}(0)}{\widehat{\sigma}_{M-1}(1) + \widehat{\sigma}_{M-1}(0)} T \geq \frac{\beta_{M-1}}{2} T^{\frac{M-1}{M}}$ and $\frac{\widehat{\sigma}_{M-1}(1)}{\widehat{\sigma}_{M-1}(1) + \widehat{\sigma}_{M-1}(0)} T \geq \frac{\beta_{M-1}}{2} T^{\frac{M-1}{M}}$ \label{mrk:LastStage:Case2}
\State \phantom{for} $(T_M(1), T_M(0)) \gets (\frac{\widehat{\sigma}_{M-1}(1)}{\widehat{\sigma}_{M-1}(1) + \widehat{\sigma}_{M-1}(0)} T - \frac{\beta_{M-1}}{2} T^{\frac{M-1}{M}}, \frac{\widehat{\sigma}_{M-1}(0)}{\widehat{\sigma}_{M-1}(1) + \widehat{\sigma}_{M-1}(0)} T - \frac{\beta_{M-1}}{2} T^{\frac{M-1}{M}})$;
\State \textbf{Case 3}: $\frac{\widehat{\sigma}_{M-1}(1)}{\widehat{\sigma}_{M-1}(1) + \widehat{\sigma}_{M-1}(0)} T < \frac{\beta_{M-1}}{2} T^{\frac{M-1}{M}}$ \label{mrk:LastStage:Case3}
\State \phantom{for} $(T_M(1), T_M(0)) \gets (0, T - \beta_{M-1} T^{\frac{M-1}{M}})$;
\EndIf

\For{$m' = m+1, ..., M$} \Comment{A sub-routine for experiments in the remaining stages} \label{marker}
\State Conduct a completely randomized experiment parameterized by $(T_{m'}(1), T_{m'}(0))$;
\EndFor
\end{algorithmic}
\clearpage
\end{algorithm}

The $M$-stage adaptive Neyman allocation generalizes the idea of two-stage adaptive Neyman allocation: we use the observations in the earlier stages to estimate the variances, and use the estimated variances to guide the allocation in the later stages.
Initially, an equal number of treated and control units are allocated in the first stage. 
At the end of each stage, sample variances are estimated using \eqref{eqn:mVariance1} and \eqref{eqn:mVariance0}, and the number of treated and control units is determined using the Neyman allocation formula, as shown in \eqref{eqn:EstimatedOPT:M}.

There are three major cases that will happen.
First, the estimated standard deviations $\widehat{\sigma}_m(1)$ and $\widehat{\sigma}_m(0)$ indicate that there is already an excessive allocation to either the treated or control group by the end of stage $m$.
In this case, we immediately stop allocating units to that group in the subsequent stages.
See Case~1 (Line~\ref{mrk:Case1}) and Case~5 (Line~\ref{mrk:Case5}) in Algorithm~\ref{alg:MStageANA}.
Intuitively, we are pruning the corner cases: once we have used a small number of stages to identify that the standard deviation $\sigma(1)$ or $\sigma(0)$ is very small, we stop allocating units to that group.

Second, the estimated standard deviations $\widehat{\sigma}_m(1)$ and $\widehat{\sigma}_m(0)$ indicate that we have not allocated too many units to both groups by the end of stage $m$, but an equal allocation in the next stage $(m+1)$ would result in an excessive allocation to either the treated or control group.
In this case, we follow the Neyman allocation in the next stage $(m+1)$ only.
We then stop allocating units to that treated or control group in the subsequent stages after the next stage.
See Case~2 (Line~\ref{mrk:Case2}) and Case~4 (Line~\ref{mrk:Case4}) in Algorithm~\ref{alg:MStageANA}.
This is the non-trivial generalization from Algorithm~\ref{alg:2StageANA} in the two-stage adaptive Neyman allocation.
Intuitively, we are pruning the corner cases as early as possible: now that we have identified that the standard deviation $\sigma(1)$ or $\sigma(0)$ is small enough, we do not spend an extra stage to allocate more units than necessary and convince ourselves that they are small.
Instead, we follow the Neyman allocation in the next stage, and, without even updating the estimators $\widehat{\sigma}_{m+1}(1)$ and $\widehat{\sigma}_{m+1}(0)$, directly stop allocating future units to that group.

Third, the estimated standard deviations $\widehat{\sigma}_m(1)$ and $\widehat{\sigma}_m(0)$ indicate that even with an equal allocation in the next stage, we will not have allocated too many units to both groups.
In this case, we keep an equal allocation in the next stage.
See Case~3 (Line~\ref{mrk:Case3}) in Algorithm~\ref{alg:MStageANA}. 
Intuitively, we have not identified a significant difference between the standard deviation $\sigma(1)$ or $\sigma(0)$, so we keep a balanced exploration. 
After collecting data from the next stage, the above procedure is repeated.

\subsubsection*{Analysis.} Such a simple idea leads to an effective improvement over the two-stage adaptive Neyman allocation.
We show the quality of the multi-stage adaptive Neyman allocation, as measured by the competitive ratio, in Theorem~\ref{thm:MStageANA} below.

\begin{theorem}
\label{thm:MStageANA}
Let $M \geq 3$, $T \geq 16$, and $0 < \eps \leq \min\{\frac{1}{M}, \frac{1}{100}\}$.
Let the tuning parameters from Algorithm~\ref{alg:MStageANA} be defined as $\beta_m = 6 \cdot 15^{-\frac{m}{M}}$.
Under these parameters, Algorithm~\ref{alg:MStageANA} is feasible, i.e., $1 < \beta_1 T^{\frac{1}{M}} < ... < \beta_{M-1} T^{\frac{M-1}{M}} < T$.
Furthermore, let $(T(1), T(0))$ be the total number of treated and control units from Algorithm~\ref{alg:MStageANA}, respectively. 
Under Assumption~\ref{asp:kurtosis}, there exists an event that happens with probability at least $1 - (M-1)(\kappa(1) + \kappa(0)) T^{-\eps}$, conditional on which
\begin{align*}
\sup_{\cF \in \sP^{[\kappa]}} \ \frac{V(T(1), T(0))}{V(T^*(1), T^*(0))} \leq 1 + 4 \cdot 15^{-\frac{1}{M}} T^{-\frac{M-1}{M} + \eps}.
\end{align*}
\end{theorem}

Theorem~\ref{thm:MStageANA} presents a high probability bound.
Using an assumption that enables us to show exponentially small probability, we will be able to show that a similar bound (up to some logarithm factors) holds in expectation.
See Corollary~\ref{coro:MStageANA} in Section~\ref{sec:InExpectation}.
This result improves the best existing results in the literature \citep{antos2010active, carpentier2011finite, grover2009active}, and negates the conjecture that the competitive ratio is lower bounded by $1 + \Omega(T^{-\frac{1}{2}})$.
We sketch some unrigorous intuitions behind the proof of Theorem~\ref{thm:MStageANA} below. 
The proof borrows ideas from \citet{perchet2016batched}.
We defer the complete proof to Section~\ref{sec:proof:thm:MStageANA} in the Online Appendix.

\proof{Sketch proof of Theorem~\ref{thm:MStageANA}.}
Denote $\rho = \frac{\sigma(1)}{\sigma(0)}$.
Without loss of generality assume $\rho \geq 1$.
Suppose there are $(M-1)$ constants $0 \leq \alpha_1 \leq \alpha_2 \leq \ldots \leq \alpha_{M-1} \leq 1$, such that we can choose the lengths of the $M$ stages to be roughly in the following order: $[0,T^{\alpha_1}]$, $(T^{\alpha_1}, T^{\alpha_2}]$, ..., $(T^{\alpha_{M-1}}, T]$.

Case 1: $\rho > \frac{T - T^{\alpha_1}}{T^{\alpha_1}}$. 
Then with high probability, the first stage reveals this condition and the algorithm stops allocating units to the control group from the second stage.
In this case, we will show that
\begin{align*}
\frac{V(T(1), T(0))}{V(T^*(1), T^*(0))} = \frac{T}{T-T^{\alpha_1}} \cdot \frac{\rho^2}{(\rho+1)^2} + \frac{T}{T^{\alpha_1}} \cdot \frac{1}{(\rho+1)^2} \leq \frac{T}{T-T^{\alpha_1}} \approx 1 + T^{\alpha_1 - 1}.
\end{align*}

Case $m$ $(2 \leq m \leq M-1)$: $\frac{T - T^{\alpha_{m}}}{T^{\alpha_{m}}} < \rho \leq \frac{T - T^{\alpha_{m-1}}}{T^{\alpha_{m-1}}}$.
Then with high probability, this condition is not revealed until the end of the $(m-1)$-th stage.
Once this condition is revealed, the algorithm allocates a few units to the control group in the $m$-th stage, and stops allocating units to the control group from the $(m+1)$-th stage.
In this case, the estimation errors of estimating $\sigma(1)$ and $\sigma(0)$ are on the order of $T^{-\frac{\alpha_{m-1}}{2}}$. 
So the estimation error of estimating $\rho$ is on the order of $T^{-\alpha_{m-1}}$. 
We will show that
\begin{align*}
\frac{V(T(1), T(0))}{V(T^*(1), T^*(0))} \leq 1 + \frac{\rho}{(\rho+1)^2} \cdot T^{-\alpha_{m-1}} \approx 1 + T^{\alpha_{m}-\alpha_{m-1}-1}.
\end{align*}

Case $M$: $1 \leq \rho \leq \frac{T - T^{\alpha_{M-1}}}{T^{\alpha_{M-1}}}$.
Then with high probability, this condition is not revealed until the end of the $(M-1)$-th stage.
In the last stage, the algorithm mimics Neyman allocation by using the estimated variances.
In this case, the estimation errors of estimating $\sigma(1)$ and $\sigma(0)$ are on the order of $T^{-\frac{\alpha_{M-1}}{2}}$. 
So the estimation error of estimating $\rho$ is on the order of $T^{-\alpha_{M-1}}$. 
We will show that
\begin{align*}
\frac{V(T(1), T(0))}{V(T^*(1), T^*(0))} \leq 1 + \frac{\rho}{(\rho+1)^2} \cdot T^{-\alpha_{M-1}} \approx 1 + T^{-\alpha_{M-1}}.
\end{align*}

Combining all cases, we solve
\begin{align*}
\min_{\alpha_1, ..., \alpha_{M-1}} \max\{\alpha_1 - 1, \alpha_2 - \alpha_1 - 1, ..., -\alpha_{M-1}\}
\end{align*}
and obtain $\alpha_m = \frac{m}{M}$, which leads to a competitive ratio $\frac{V(T(1), T(0))}{V(T^*(1), T^*(0))} \approx 1 + T^{-\frac{M-1}{M}}.$
\hfill \halmos
\endproof

\subsubsection*{Information-theoretic limit.}
Next, we present an information-theoretic limit of such experiments, as measured by the competitive ratio, in Theorem~\ref{thm:ANA:LB} below.

\begin{theorem}
\label{thm:ANA:LB}
Let $T \geq 4$. 
For any adaptive design of experiment $\pi \in \Pi$, let $(T^\pi(1), T^\pi(0))$ be the total number of treated and control units from $\pi$, respectively.
There exists a problem instance such that on this problem instance, for any adaptive design of experiment $\pi \in \Pi$,
\begin{align*}
\frac{\bE[V(T^\pi(1), T^\pi(0))]}{V(T^*(1), T^*(0))} \geq 1 + \frac{1}{480} T^{-1}.
\end{align*}
\end{theorem}

We sketch some unrigorous intuitions behind the proof of Theorem~\ref{thm:ANA:LB} as follows, and defer the complete information-theoretic proof of Theorem~\ref{thm:ANA:LB} to Section~\ref{sec:proof:thm:LB} in the Online Appendix.

\proof{Sketch proof of Theorem~\ref{thm:ANA:LB}.}
To prove Theorem~\ref{thm:ANA:LB}, we construct two probability distributions $\nu$ and $\nu'$ that are challenging to distinguish.
Intuitively, we know that $Y(0)$ and $Y(1)$ follow $\nu$ and $\nu'$, but it is challenging to distinguish which outcome corresponds to which distribution.

Now define $\eps = \frac{1}{3T^{\frac{1}{2}}}.$
Both distributions have three discrete supports $\{-1,0,1\}$.
The probability mass for distribution $\nu$ is given by 
\begin{align*}
p_{-1} = \frac{1}{3}, && p_{0} = \frac{1}{3}, && p_{1} = \frac{1}{3}.
\end{align*}
The probability mass for distribution $\nu'$ is given by 
\begin{align*}
p'_{-1} = \frac{1}{3}+\frac{\eps}{2}, && p'_{0} = \frac{1}{3}-\eps, && p'_{1} = \frac{1}{3}+\frac{\eps}{2}.
\end{align*}
Then we bound the KL-divergences of these two probability distributions,
\begin{align*}
D_{KL}(\nu \vert\vert \nu') \leq \frac{9}{2} \eps^2 = \frac{1}{2T}, && D_{KL}(\nu' \vert\vert \nu) \leq \frac{9}{2} \eps^2 = \frac{1}{2T}.
\end{align*}

Intuitively, it is challenging to distinguish the above two probability distributions within $T$ rounds.
This means that, any policy can not distinguish the above two probability distributions until the end of horizon.
Since the two probability distributions are not distinguishable, the best policy in this situation has to follow the half-half allocation, which leads to a competitive ratio of $\frac{\bE[V(T^\pi(1), T^\pi(0))]}{V(T^*(1), T^*(0))} \approx 1 + T^{-1}.$
\hfill \halmos
\endproof

By comparing Theorems~\ref{thm:MStageANA} and~\ref{thm:ANA:LB}, we see that when the number of stages $M$ is large, the two results are close to each other.
When there are $\log(T)$ many stages, the two results almost match with each other, suggesting that the multi-stage adaptive Neyman allocation is the optimal design of experiments, whose competitive ratio almost matches the information-theoretic limit of conducting adaptive experiments.

\section{Post-Experiment Analysis Using Adaptively Collected Data}
\label{sec:Analysis}

In this section, we establish estimation and inference results using data collected via our adaptive Neyman allocation algorithms. 
Generally speaking, analyzing data collected by adaptive experiments could be challenging. 
However, our proposed adaptive Neyman allocation algorithms enjoy the key property of adapting on the sample variance, not on the sample mean.
This property makes estimation and inference easy.
The estimation result in this section borrows ideas from \citet{xiong2019optimal}, and the inference result in this section borrows ideas from \citet{chen2025characterization, khamaru2024inference}.

\subsubsection*{Estimation.}
We start with establishing an unbiased estimation result that holds in finite sample.
Recall that the pair of potential outcomes $(Y(1), Y(0)) \sim \cF$ is sampled from the joint distribution $\cF$.
We show that the difference-in-means estimator is unbiased, as long as distribution $\cF$ satisfies Assumption~\ref{asp:symmetric} below.
Assumption~\ref{asp:symmetric} borrows ideas from \citet{xiong2019optimal}, who study a similar assumption and show unbiasedness of least squares estimators.

\begin{assumption}[Symmetric Distribution]
\label{asp:symmetric}
Let $(Y(1), Y(0))$ be a pair of random variables sampled from $\cF$.
Assume that the joint probability distributions of 
\begin{align*}
& \Big(Y(1) - \bE[Y(1)], \ Y(0) - \bE[Y(0)]\Big) & & \text{and} & & \Big(\bE[Y(1)] - Y(1), \ \bE[Y(0)] - Y(0)\Big)
\end{align*}
are identical.
\end{assumption}

The symmetric distribution is satisfied by many families of distributions, such as the family of joint normal distributions.
Intuitively, because the family of joint normal distributions is symmetric and has two parameters, adapting to the sample variance does not bias the mean estimation.
On the other hand, Assumption~\ref{asp:symmetric} does not hold for the family of Bernoulli distributions.
The family of Bernoulli distributions is generally not symmetric and has only one single parameter, and adapting to the estimated variance biases the mean estimation. 
We provide a simplified toy example below to illustrate the intuitions.

\begin{example}[Symmetric Distribution Implies No Conditioning Bias]
\label{exa:Symmetric}
We consider a simplified, single-dimensional example.
Suppose we have two scalar random variables $Z_1$ and $Z_2$ that are i.i.d. sampled from the same distribution $\cF$.
We consider the following conditional expectation for any $a \geq 0$,
\begin{align*}
\bE\Big[ Z_1 \Big\vert \vert Z_1 - Z_2 \vert = a \Big].
\end{align*}
This conditional expectation reflects estimating the mean value of $Z_1$ conditioning on observing the sample variance of the two samples, because the sample variane is equal to $\frac{1}{2}(Z_1 - Z_2)^2$.
Next we consider two distributions.

First, we consider a normal distribution $\cN(\mu, \sigma^2)$.
Because normal distribution is symmetric and $Z_1$ and $Z_2$ are independent, $(Z_1, Z_2)$ and $(2\mu-Z_1, 2\mu-Z_2)$ follow the same distribution.
Consequently,
\begin{align*}
\bE\Big[ Z_1 \Big\vert \vert Z_1 - Z_2 \vert = a \Big] = \bE\Big[ 2\mu - Z_1 \Big\vert \vert (2\mu - Z_1) - (2\mu - Z_2) \vert = a \Big] = 2\mu - \bE\Big[ Z_1 \Big\vert \vert Z_1 - Z_2 \vert = a \Big],
\end{align*}
which yields $\bE\Big[ Z_1 \Big\vert \vert Z_1 - Z_2 \vert = a \Big] = \mu = \bE\big[Z_1\big]$. 
This example shows that, for normal distributions, conditioning on the sample variance does not bias the mean estimation.

Second, we consider a Bernoulli distribution $\Ber(p)$.
When $p \ne \frac{1}{2}$, the Bernoulli distribution is not symmetric.
Because $Z_1$ and $Z_2$ are independent, we can calculate that $\Pr(Z_1 = 1, Z_2 = 0) = \Pr(Z_1 = 0, Z_2 = 1) = p(1-p)$.
Consequently, when $a = 1$,
\begin{align*}
\bE\Big[ Z_1 \Big\vert \vert Z_1 - Z_2 \vert = 1 \Big] = \frac{1 \cdot \Pr(Z_1 = 1, Z_2 = 0) + 0 \cdot \Pr(Z_1 = 0, Z_2 = 1)}{\Pr(Z_1 = 1, Z_2 = 0) + \Pr(Z_1 = 0, Z_2 = 1)} = \frac{1}{2} \ne p = \bE\big[Z_1\big].
\end{align*}
This example shows that, for Bernoulli distributions with $p \ne \frac{1}{2}$, conditioning on the sample variance biases the mean estimation.
\hfill \halmos
\end{example}

We formalize the intuitions built in Example~\ref{exa:Symmetric} into the following theorem.

\begin{theorem}[Finite Sample Unbiasedness]
\label{thm:estimation}
When $M=2$, use Algorithm~\ref{alg:2StageANA}. 
When $M \geq 3$, use Algorithm~\ref{alg:MStageANA}.
Under Assumption~\ref{asp:symmetric}, the difference-in-means estimator as defined in \eqref{eqn:Estimator} is unbiased, that is,
\begin{align*}
\bE\big[\widehat{\tau}\big] = \tau.
\end{align*}
\end{theorem}

We prove Theorem~\ref{thm:estimation} in Section~\ref{sec:proof:thm:estimation} in the Online Appendix.
It is worth noting that Theorem~\ref{thm:estimation} is a non-asymptotic result.
This non-asymptotic nature is unique to our adaptive Neyman allocation algorithms, which adapts on the sample variance but not the sample mean. 
It is different from the traditional adaptive experiments literature, where the unbiasedness property usually requires the sample size $T$ to be large \citep{bowden2017unbiased, chen2025characterization, hadad2021confidence, hirano2023asymptotic, khamaru2024inference, melfi2000estimation, nie2018adaptively, offer2021adaptive, shin2019bias, shin2019sample, zhan2021off, zhan2023policy, zhang2020inference, zhang2021statistical}.
In what follows, we provide a standard asymptotic result that does not require Assumption~\ref{asp:symmetric} but requires the sample size $T$ to be large.

\subsubsection*{Inference.}
Now we turn our attention to inference when the sample size $T$ is large.
Making inference for adaptively collected data is an active literature.
As long as some notion of ``stability condition'' holds, one can establish central limit theorems for the sample means \citep{chen2025characterization, khamaru2024inference, melfi2000estimation}.
We borrow this technique and provide a central limit theorem that allows us to construct confidence intervals under asymptotic normality. 

\begin{theorem}[Asymptotic Normality]
\label{thm:inference}
When $M=2$, use Algorithm~\ref{alg:2StageANA}. 
When $M \geq 3$, use Algorithm~\ref{alg:MStageANA}.
Recall that $T(1)$ and $T(0)$ stand for the numbers of treated and control units, respectively.
Under Assumption~\ref{asp:kurtosis}, we have
\begin{align*}
\lim_{T \to +\infty} \begin{pmatrix}
\frac{1}{\sqrt{T(1)}} \sum_{t=1}^T (Y_t - \bE[Y(1)]) \bI\{W_t=1\}\\
\frac{1}{\sqrt{T(0)}} \sum_{t=1}^T (Y_t - \bE[Y(0)]) \bI\{W_t=0\}
\end{pmatrix} 
\xrightarrow{d} \cN\left( 
\begin{pmatrix}
0 \\
0
\end{pmatrix},
\begin{bmatrix}
\sigma^2(1) & 0 \\
0 & \sigma^2(0)
\end{bmatrix}
\right),
\end{align*}
where $\xrightarrow{d}$ stands for convergence in distribution.
\end{theorem}

We can use Theorem~\ref{thm:inference} to construct confidence intervals using any consistent estimator of the true variance.
It turns out that the sample variance estimators using all the data,
\begin{subequations}
\begin{align}
\widehat{\sigma}^2(1) = & \ \frac{1}{T(1)-1} \sum_{t: W_t=1} \bigg( Y_t - \frac{1}{T(1)} \sum_{t: W_t=1} Y_t \bigg)^2, \label{eqn:AllVariance1} \\
\widehat{\sigma}^2(0) = & \ \frac{1}{T(0)-1} \sum_{t: W_t=0} \bigg( Y_t - \frac{1}{T(0)} \sum_{t: W_t=0} Y_t \bigg)^2, \label{eqn:AllVariance0}
\end{align}
\end{subequations}
are consistent in estimating the true variances $\sigma^2(1)$ and $\sigma^2(0)$.
This consistency result follows arguments similar to those in \citet{melfi2000estimation, shin2019bias, khamaru2024inference}, and is a consequence of the law of large numbers.

\begin{proposition}[Sample Variance Estimator Consistency]
\label{prop:SampleVarianceConsistent}
When $M=2$, use Algorithm~\ref{alg:2StageANA}. 
When $M \geq 3$, use Algorithm~\ref{alg:MStageANA}.
Under Assumption~\ref{asp:kurtosis}, the sample variance estimators as defined in \eqref{eqn:AllVariance1} and \eqref{eqn:AllVariance0} are consistent estimators of the true variances $\sigma^2(1)$ and $\sigma^2(0)$, that is, as $T \to +\infty$,
\begin{align*}
\widehat{\sigma}^2(1) \xrightarrow{p} \sigma^2(1), && \widehat{\sigma}^2(0) \xrightarrow{p} \sigma^2(0),
\end{align*}
where $\xrightarrow{p}$ stands for convergence in probability.
\end{proposition}

Combining Theorem~\ref{thm:inference} and Proposition~\ref{prop:SampleVarianceConsistent}, we can construct the classical asymptotically valid $1-\alpha$ confidence interval as follows,
\begin{align*}
\Big[\widehat{\tau} - z_{1-\frac{\alpha}{2}} \sqrt{\frac{\widehat{\sigma}^2(1)}{T(1)} + \frac{\widehat{\sigma}^2(0)}{T(0)}}, \ \widehat{\tau} + z_{1-\frac{\alpha}{2}} \sqrt{\frac{\widehat{\sigma}^2(1)}{T(1)} + \frac{\widehat{\sigma}^2(0)}{T(0)}}\Big],
\end{align*}
where $z_{1-\frac{\alpha}{2}}$ is the $1-\frac{\alpha}{2}$ quantile of the standard normal distribution.

We sketch the intuitions behind the proof of Theorem~\ref{thm:inference} as follows, and defer the complete proof of Theorem~\ref{thm:inference} to Section~\ref{sec:proof:thm:inference} in the Online Appendix.
The proof of Proposition~\ref{prop:SampleVarianceConsistent} is simple, and we defer the self-contained proof to Section~\ref{sec:proof:prop:SampleVarianceConsistent} in the Online Appendix.

\proof{Sketch proof of Theorem~\ref{thm:inference}.}
The proof of Theorem~\ref{thm:inference} relies on a standard martingale central limit theorem.
To appropriately apply the martingale central limit theorem, we will identify the martingale sequence.
Apparently, the following sequence of (re-centered) sample mean estimators
\begin{align*}
\bigg\{\frac{1}{T(1)}\sum_{t=1}^T\big(Y_t(1)-\bE[Y(1)]\big)\bI\{W_t=1\}\bigg\}_{T=1,2,...}
\end{align*}
is not a martingale sequence, because the denominator $T(1)$ is a random variable.
To overcome this issue, note that under our adaptive Neyman allocation algorithms, the denominator $T(1)$ converges to a deterministic quantity $T^*(1)$ in probability, that is, $T(1) \xrightarrow{p} T^*(1)$ as $T \to +\infty$.
Consequently, 
\begin{align*}
\frac{1}{T(1)}\sum_{t=1}^T\big(Y_t(1)-\bE[Y(1)]\big)\bI\{W_t=1\} \quad \xrightarrow{p} \quad \frac{1}{T^*(1)}\sum_{t=1}^T\big(Y_t(1)-\bE[Y(1)]\big)\bI\{W_t=1\}
\end{align*}
as $T \to +\infty$.
We can now show that the sequence 
\begin{align*}
\bigg\{\frac{1}{T^*(1)}\sum_{t=1}^T\big(Y_t(1)-\bE[Y(1)]\big)\bI\{W_t=1\}\bigg\}_{T=1,2,...}
\end{align*}
is a martingale sequence, and establish a martingale central limit theorem.
\hfill \halmos
\endproof

In the above sketch proof, the condition that there exists a deterministic quantity $T^*(1)$ such that $T(1) \xrightarrow{p} T^*(1)$ is referred to as the stability condition in \citet{chen2025characterization, khamaru2024inference}. 
This stability condition is naturally satisfied by our adaptive Neyman allocation algorithms.

\subsubsection*{Implications.}
First, Theorem~\ref{thm:inference} is an asymptotic result that requires $T$ to be large, but does not require Assumption~\ref{asp:symmetric}.
An immediate implication of Theorem~\ref{thm:inference} is that the difference-in-means estimator is asymptotically unbiased.
This is in contrast to Theorem~\ref{thm:estimation} which suggests that the difference-in-means estimator is unbiased in finite sample but requires Assumption~\ref{asp:symmetric}.

Second, recall from Section~\ref{sec:Setup} that the objective function of our adaptive Neyman allocation algorithms is the proxy mean squared error, which is not equal to the variance of the difference-in-means estimator \eqref{eqn:Estimator} because the data is adaptively collected.
But Theorem~\ref{thm:inference} implies that using this proxy is a reasonable idea.
More specifically, Theorem~\ref{thm:inference} implies that the difference-in-means estimator \eqref{eqn:Estimator} using adaptively collected data has an asymptotic variance of
\begin{align*}
\frac{\sigma^2(1)}{T(1)} + \frac{\sigma^2(0)}{T(0)},
\end{align*}
which is exactly the same expression as the proxy mean squared error \eqref{eqn:ProxyVariance}.

The fact that the proxy mean squared error \eqref{eqn:ProxyVariance} is not equal to the variance of estimator \eqref{eqn:Estimator} in finite sample has been well recognized in the literature across three different fields:
the active learning literature where people directly use the proxy mean squared error as an objective function \citep{antos2010active, carpentier2011finite, grover2009active}, 
the experimental design literature where people analyze adaptively collected data \citep{chen2025characterization, hahn2011adaptive, hirano2023asymptotic, khamaru2024inference, li2024double, nie2018adaptively, offer2021adaptive, shin2019bias, shin2019sample, zhang2020inference, zhang2021statistical},
and the simulations literature where people discuss the bias in estimating confidence intervals \citep{ross2013simulation}.
Theorem~\ref{thm:inference} aligns with the same message from \citet{hahn2011adaptive, li2024double} that, under properly designed adaptive Neyman allocation algorithms, the variance of estimator \eqref{eqn:Estimator} converges to the proxy mean squared error asymptotically.
We conduct simulations in Section~\ref{sec:Synthetic} to verify that the gap between these two quantities is small.
However, the magnitude of the non-asymptotic gap between these two quantities remains unknown, which we discuss as a future research direction in Section~\ref{sec:Conclusions}.

\section{Extensions}
\label{sec:InExpectation}

In the previous sections, we have seen that Theorems~\ref{thm:2StageANA} and~\ref{thm:MStageANA} provide high probability bounds for the competitive ratios of using adaptive Neyman allocation.
But with low probability, the competitive ratios might be much larger.
In this section, we show that similar bounds as in Theorems~\ref{thm:2StageANA} and~\ref{thm:MStageANA} still hold in expectation.

We will need a stronger assumption to show that the low probability events happen with exponentially small probability.
But this stronger assumption is still weaker than the standard modeling assumptions that commonly appear in the active learning literature, where existing works assume that both the standard deviations and the supports are bounded \citep{antos2010active, carpentier2011finite, grover2009active}.

\begin{assumption}
\label{asp:bounded}
There exists a constant $C < \infty$ which does not depend on $T$, such that
\begin{align*}
\vert Y(1) \vert \leq C \sigma(1), && \vert Y(0) \vert \leq C \sigma(0). 
\end{align*}
\end{assumption}

Assumption~\ref{asp:bounded} asserts that the representative random variables $Y(1), Y(0)$ have bounded supports that depend on the variances.
In contrast, the traditional literature sometimes assumes that the bounded support is between $[0,1]$, and that the variances are constants.
To illustrate Assumption~\ref{asp:bounded}, consider the following example.
Consider a three-point distribution $Y$, such that with probability $1-2p$, $Y=0$; with probability $p$, $Y=1$; and with probability $p$, $Y=-1$. 
In this example, $\vert Y \vert \leq 1$ and $\sigma(Y) = \sqrt{2p}$.
If $\lim_{T \to +\infty} p \to 0$, Assumption~\ref{asp:bounded} does not hold.
On the other hand, if $p$ is a constant, Assumption~\ref{asp:bounded} holds.
For example, if $p = \frac{1}{3}$, then Assumption~\ref{asp:bounded} holds with constant $C = \sqrt{\frac{3}{2}}$.
Let $\sP^{[C]}$ be the family of joint distributions that satisfies Assumption~\ref{asp:bounded}. 

Under Assumption~\ref{asp:bounded}, we are able to show Corollary~\ref{coro:2StageANA} as an extension of Theorem~\ref{thm:2StageANA}, and Corollary~\ref{coro:MStageANA} as an extension of Theorem~\ref{thm:MStageANA}.

\begin{corollary}
\label{coro:2StageANA}
Let $T \geq 320^\frac{5}{4} C^5$.
Let the tuning parameter from Algorithm~\ref{alg:2StageANA} be defined as $\beta = 4C^2 (\log{T})^{\frac{1}{2}}$.
Algorithm~\ref{alg:MStageANA} is feasible under $\beta$.

Furthermore, let $(T(1), T(0))$ be the total number of treated and control units from Algorithm~\ref{alg:2StageANA}, respectively. 
Under Assumption~\ref{asp:bounded},
\begin{align*}
\sup_{\cF \in \sP^{[C]}} \ \frac{\bE[V(T(1), T(0))]}{V(T^*(1), T^*(0))} < 1 + 5 C^2 T^{-\frac{1}{2}} (\log{T})^{\frac{1}{2}}.
\end{align*}
\end{corollary}

\begin{corollary}
\label{coro:MStageANA}
Let $M \geq 3$ and $T \geq (\frac{5000}{3})^{\frac{5}{4}} C^5$.
Let the tuning parameters from Algorithm~\ref{alg:MStageANA} be defined as $\beta_m = \frac{400}{3} C^4 \log{T} \cdot (\frac{1000}{3} C^4 \log{T})^{-\frac{m}{M}}$.
Under these parameters, Algorithm~\ref{alg:MStageANA} is feasible, i.e., $1 < \beta_1 T^{\frac{1}{M}} < ... < \beta_{M-1} T^{\frac{M-1}{M}} < T$.

Furthermore, let $(T(1), T(0))$ be the total number of treated and control units from Algorithm~\ref{alg:MStageANA}, respectively. 
Under Assumption~\ref{asp:bounded},
\begin{align*}
\sup_{\cF \in \sP^{[C]}} \ \frac{\bE[V(T(1), T(0))]}{V(T^*(1), T^*(0))} < 1 + 97 \left(\frac{1000}{3}\right)^{-\frac{1}{M}} C^{\frac{4(M-1)}{M}} T^{-\frac{M-1}{M}} (\log{T})^{\frac{M-1}{M}}.
\end{align*}
\end{corollary}

\section{Simulations Using Online A/B Testing Data}
\label{sec:Simulation}

In this section, we conduct simulations using online A/B testing data from a social media site \citet{AB_testing_kaggle}.
We combine this online A/B testing data with a resampling process which generates the sequence of experiments.
Following each trajectory of the generated sequence, we calculate the difference-in-means estimator as in \eqref{eqn:Estimator} under the adaptive Neyman allocation and the half-half allocation, respectively.
By drawing different trajectories from the resampling process, we are able to compare the adaptive Neyman allocation and the half-half allocation.
Overall, the simulation suggests a $\sim 10\%$ reduction in the variance.
We describe the details below.

\subsubsection*{Raw data and pre-processing.}
This social media site has conducted an online A/B test to compare two advertisement strategies, which they refer to as the average bidding strategy and the maximum bidding strategy.
The true label of treatment and control is masked from the data. 
We only know that they refer to two different advertisement strategies: average bidding and maximum bidding.
This social media site is interested in understanding which bidding strategy generates more conversion, that is, user clicks.

Unlike traditional user click data which documents the binary user click records from one single experiment, this data set documents a total of $80$ experiments, $40$ treated and $40$ control.
Each experiment is stored in one row, which documents aggregate data of the number of impressions (one impression refers to one view of the advertisement) and the number of clicks.
We normalize the number of clicks by the number of impressions, and use the number of clicks per million impressions as the unit of measurement.
We denote the numbers of clicks from these two groups as $\cY(1)$ and $\cY(0)$, respectively.
See Table~\ref{tbl:SummaryStatistics} for the summary statistics of these two groups.

\begin{table}[!tb]
\centering
\TABLE{Summary statistics of the number of clicks per million impressions at a social media site \citep{AB_testing_kaggle}
\label{tbl:SummaryStatistics}}
{\begin{tabular}{>{\centering}p{2.8cm}>{\centering}p{2cm}>{\centering}p{2cm}>{\centering}p{2cm}>{\centering}p{2cm}>{\centering}p{2cm}c}
        & Mean    & St dev  & Min     & Median  & Max      & \\ \hline
Treated & $34176$ & $12256$ & $14732$ & $31358$ & $75752$  & \\
Control & $53618$ & $24850$ & $20757$ & $48796$ & $162068$ & 
\end{tabular}
}
{\textit{Note:} In this data, the labels of treatment and control are masked; we only know that they refer to two different advertisement strategies.}
\end{table}

\subsubsection*{Resampling process.}
Since the data does not show us the sequence of experiments, we use a resampling process to generate the sequence.
In the resampling process, we consider $T = 1000$.
For each $t \in [T]$, we generate the potential outcomes $Y_t(1)$ and $Y_t(0)$ by sampling from the two groups $\cY(1)$ and $\cY(0)$ with replacement.
We refer to the data generated above as one trajectory, and following each trajectory we calculate the difference-in-means estimator \eqref{eqn:Estimator} under different experients.
By drawing a total of $1,000,000$ different trajectories, we obtain the distributions of the same difference-in-means estimator and compare the performances across different experiments.

Since we generate the potential outcomes, we can calculate the average treatment effect of the super-population as $\tau = \bE[Y(1) - Y(0)] = -19442$, where the expectation is taken over the resampling process.

\subsubsection*{Experimental design and results.}
We consider the following three designs of experiments.
\begin{enumerate}
\item Half-half allocation: a completely randomized experiment with half units in the treated group and half units in the control group.
\item Two stage adaptive Neyman allocation: the two stage adaptive experiment as described in Algorithm~\ref{alg:2StageANA}, using parameter $\beta = 1$ as suggested in Theorem~\ref{thm:2StageANA}.
\item M stage adaptive Neyman allocation: the $M$ stage adaptive experiment as described in Algorithm~\ref{alg:MStageANA}, using parameters $\beta_m = 6 \cdot 15^{-\frac{m}{M}}$. as suggested in Theorem~\ref{thm:MStageANA}.
\end{enumerate}

\begin{figure}[!tb]
\centering
\caption{Simulated variances of experiments under different numbers of stages}
\label{fig:Variances}
\includegraphics[width=0.65\textwidth]{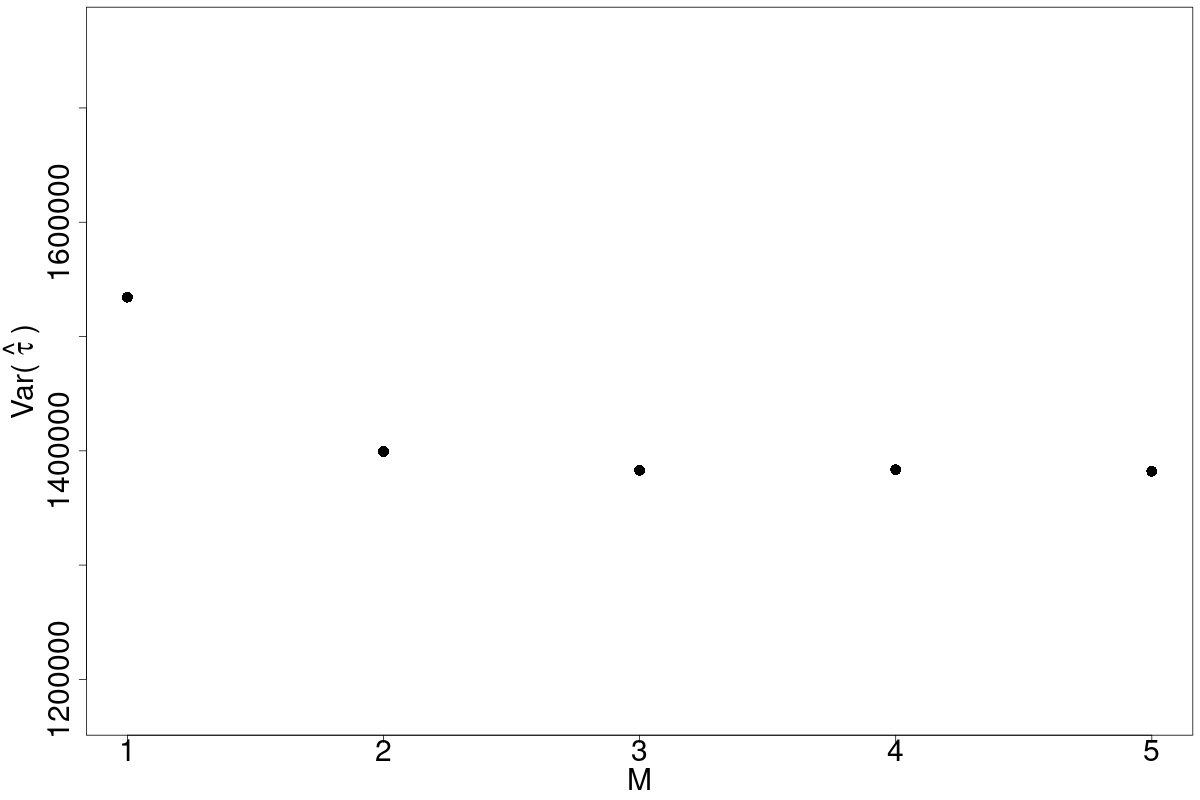}
\end{figure}

\begin{figure}[!tb]
\centering
\caption{Simulated distributions of experiments under different numbers of stages}
\label{fig:Histogram}
\includegraphics[width=0.65\textwidth]{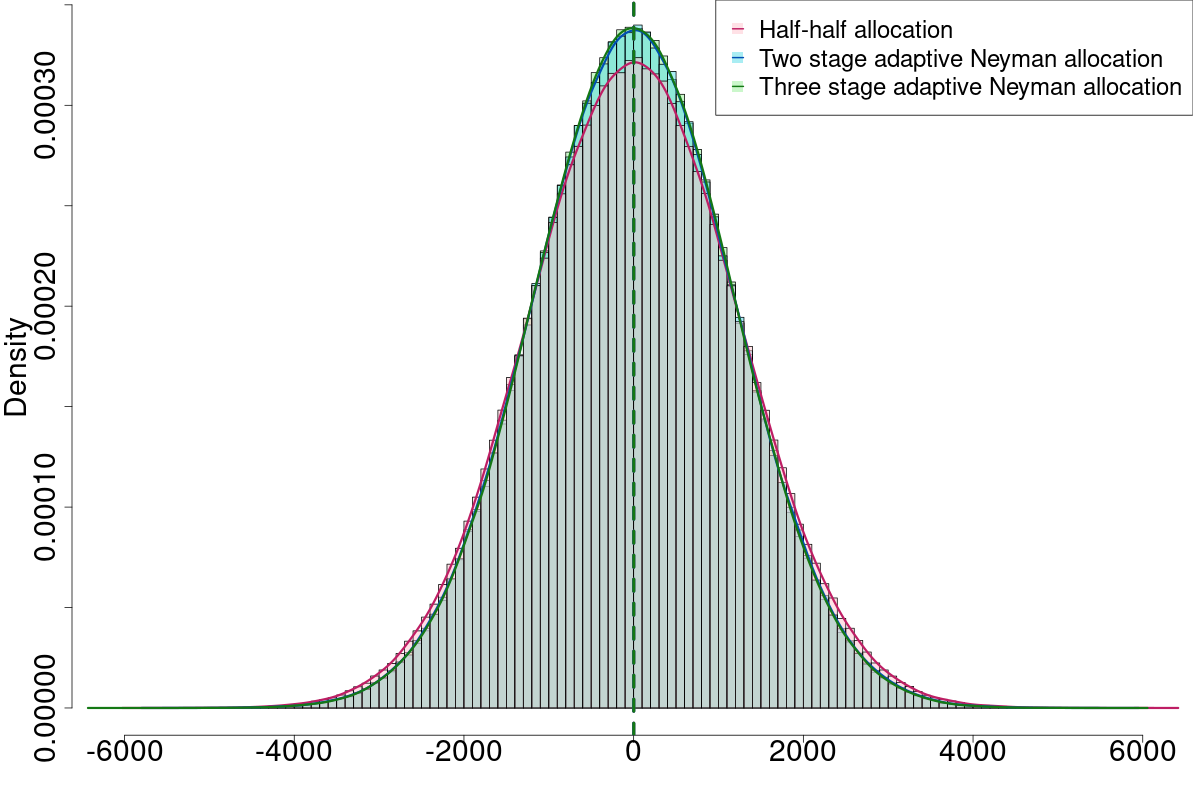}
\end{figure}

For each design, we conduct the experiment and calculate the difference-in-means estimator $\widehat{\tau}$.
We compare the variances of the estimators in Figure~\ref{fig:Variances} and compare the distributions of the estimators in Figure~\ref{fig:Histogram}.
Note that, there are two sources of randomness in Figures~\ref{fig:Variances} and~\ref{fig:Histogram}. 
First, the resampling process draws random samples when generating the potential outcomes; second, the experiments are randomized experiments when determining the treatment assignments.

In Figure~\ref{fig:Variances}, we simulate the variances of the estimator $\widehat{\tau}$.
Figure~\ref{fig:Variances} shows a significant reduction in variance when increasing the number of stages from $M=1$ to $M=2$. 
However, the marginal benefit of increasing the number of stages from $M=2$ to $M=3$ is substantially smaller.
Further increasing the number of stages beyond this point leads to negligible improvements or even a slight increase in variance.
We recommend using a small value for $M$, such as $2$ or $3$, to have the best the numerical performance.

In Figure~\ref{fig:Histogram}, we simulate the distributions of $\widehat{\tau} - \tau$, the difference-in-means estimator subtracted by the average treatment effect of the super-population.
The vertical dashed lines indicate the mean of each distribution, while the solid curves represent the respective kernel density estimates.
As shown in Figure~\ref{fig:Histogram}, the three dashed lines are virtually indistinguishable at zero, suggesting that all three estimators are unbiased.
Furthermore, the blue and green density curves almost coincide, and both of them are taller than the red density curve.
They suggest that the two stage and three stage adaptive Neyman allocation algorithms have similar performances, and both of them outperform the half-half allocation benchmark.

Our simulations in Figures~\ref{fig:Variances} and~\ref{fig:Histogram} suggest that, on this user click data from a social media site, adaptive Neyman allocation leads to a $\sim 10\%$ reduction in variance compared to half-half allocations.
This will lead to faster business decisions as the experimenter would require less samples to draw the same causal conclusion.

\section{Simulations Using Synthetic Data}
\label{sec:Synthetic}

In this section, we conduct extensive simulations using synthetic data.
The main purposes of this section are two fold.
First, we compare the performances of the adaptive Neyman allocation algorithms we propose in this paper with a series of benchmarks in the literature.
Second, we numerically study the gap between the proxy mean squared error \eqref{eqn:ProxyVariance} and the variance of estimator \eqref{eqn:Estimator} when the sample size is relatively small.

\subsubsection*{Simulation Setup.}
We change the values of $T$ from $\{100, 200, ..., 1000\}$.
For each $t \in [T]$, we set the potential outcomes $Y_t(1)$ and $Y_t(0)$ to be independent normal $\cN(1,\sigma^2(1))$ and $\cN(0,1)$ random variables, respectively, where we set $\sigma(1) = 5$.
We also study other values of $\sigma(1)$ in Section~\ref{sec:append:AdditionalSynthetic}.

\subsubsection*{Comparing the performances of different algorithms.}
We study the performances of eight algorithms that fall into the six types below.
For each algorithm, we normalize the mean squared error by the theoretical value of the mean squared error under optimal allocation, which is derived in \eqref{eqn:ClairvoyantOptimal}.
Such a normalization enables us to focus on the relative performance of the eight algorithms.

\begin{enumerate}
\item \textit{Optimal}: Optimal Neyman allocation if $\sigma(1)$ and $\sigma(0)$ are given. This algorithm requires knowledge of $\sigma(1)$ and $\sigma(0)$ as input.
\item \textit{HalfHalf}: Half-half allocation, which is the optimal solution from Theorem~\ref{thm:OneStage}.
\item \textit{ANA(M)}: Adaptive Neyman allocation when the number of stages is $M$. We consider two cases: when $M=2$, we implement Algorithm~\ref{alg:2StageANA} with $\beta=1$; when $M=3$, we implement Algorithm~\ref{alg:MStageANA} with $\beta_m = 6 \cdot 15^{-\frac{m}{M}}, \forall m \leq M-1$. The performance of adaptive Neyman allocation depends on the choice of these tuning parameters. Sometimes a slightly larger tuning parameter in the earlier stages may improve the performance.
\item \textit{Discard(M)}: A sample-discarding batched algorithm which borrows ideas from \citet{perchet2016batched} to discard earlier stage data, when the number of stages is $M$. At the end of each stage, we first estimate the sample variances using data from this stage only, and then determine the number of treated and control units in the next stage following Neyman allocation using the estimated variances. At the end of the experiment, we only use data from the last stage to estimate $\widehat{\tau}$. Because we ``discard'' data from the first $M-1$ stages, $\widehat{\tau}$ is estimated using i.i.d. data. We consider two cases $M=2$ and $M=3$.
\item \textit{DBCD}: Doubly adaptive biased coin design. We implement the algorithm from \citet{hu2004asymptotic} with the following specification, $g(x,y) = \bI\{x \leq y\}$, where $\bI\{x \leq y\}$ is an indicator that takes value $1$ when $x \leq y$. In other words, this algorithm assigns unit $(t+1)$ to the treated group if and only if $\frac{T_t(1)}{t} \leq \frac{\widehat{\sigma}_t(1)}{t}$, where $T_t(1)$ stands for the number of treated units among the first $t$ units, and $\widehat{\sigma}_t(1)$ stands for the estimated standard deviation using data collected up to unit $t$. We initialize this algorithm by randomly assigning a half of the units among the first $T^{\frac{1}{2}}$ into treated and control, respectively. It is worth noting that, our specification $g(x,y) = \bI\{x \leq y\}$ does not satisfy the joint continuity condition in \citet[p.~272, condition (i)]{hu2004asymptotic}. So this specification is beyond the family of DBCDs considered in \citet{hu2004asymptotic}. It is unclear how to construct a DBCD satisfying the conditions in \citet{hu2004asymptotic} that is also comparable to the adaptive Neyman allocation algorithms in this paper.
\item \textit{UCB}: Upper confidence bound. We implement the algorithm from \citet{carpentier2011finite} with the following specification, $c_1 = 1$. Note that this algorithm requires knowledge of an upper bound of $\sigma(1)$ and $\sigma(0)$ as input. The specification that we choose, $c_1 = 1$, is one that does not use such knowledge.
\end{enumerate}

\begin{figure}[!tb]
\centering
\caption{Normalized mean squared error with respect to sample size when $\sigma(1) / \sigma(0)=5$}
\label{fig:MSE:sigma=5}
\includegraphics[width=0.75\textwidth]{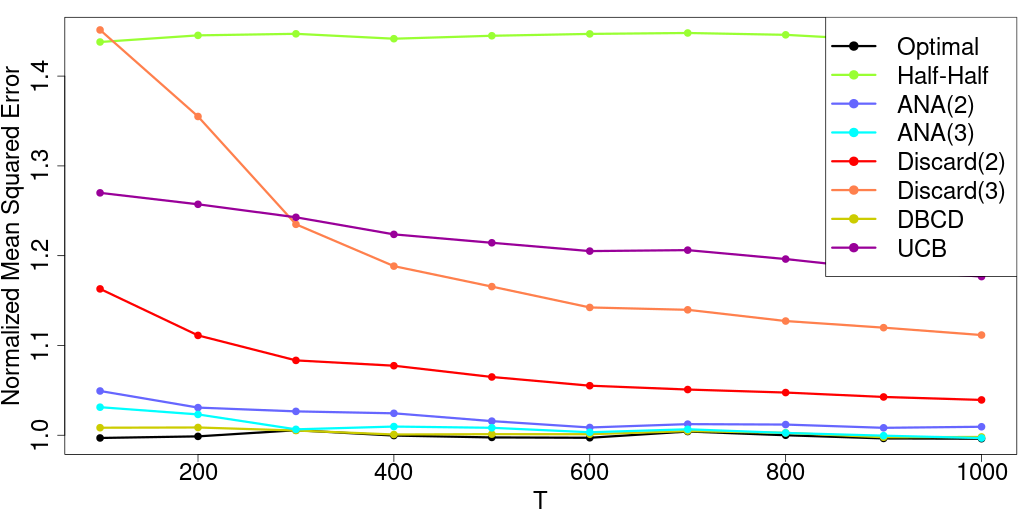}
\end{figure}

\begin{figure}[!tb]
\centering
\caption{Normalized proxy mean squared error with respect to sample size when $\sigma(1) / \sigma(0)=5$}
\label{fig:proxyMSE:sigma=5}
\includegraphics[width=0.75\textwidth]{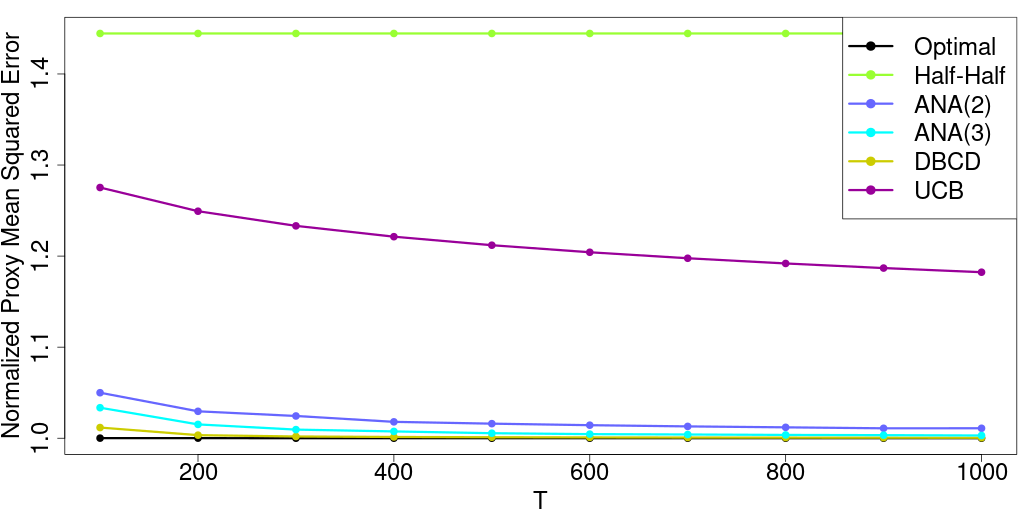}
\end{figure}

We first compare the performances of the above eight algorithms and report their mean squared errors in Figure~\ref{fig:MSE:sigma=5}.
As we see from Figure~\ref{fig:MSE:sigma=5}, the normalized mean square error of the optimal Neyman allocation stays at $1$, suggesting that the simulation performance of the optimal Neyman allocation closely mimics the theoretical calculation of expression~\eqref{eqn:ClairvoyantOptimal}.
The normalized mean square error of half-half allocation, on the other hand, stays at the theoretical calculation of $\frac{2(\sigma^2(1)+\sigma^2(0))}{(\sigma(1)+\sigma(0))^2} \approx 1.44$, and does not change as $T$ changes.
The two adaptive Neyman allocation algorithms as we proposed in Theorems~\ref{thm:2StageANA} and~\ref{thm:MStageANA} perform well, with the three-stage adaptive Neyman allocation algorithm slightly outperforming the two-stage one.
The two sample-discarding variants of batched Neyman allocation algorithm perform worse than the adaptive Neyman allocation algorithms, because they do not fully utilize all the samples.
The three-stage sample-discarding algorithm performs even worse, because it discards even more samples, on the order of $T^{\frac{2}{3}}$, than compared with the two-stage algorithm.
The doubly adaptive biased coin design has the best performance. 
This is not surprising because this algorithm is fully adaptive, whereas the adaptive Neyman allocation algorithms run in batches.
Finally, the upper confidence bound algorithm performs worse than the adaptive Neyman allocation algorithms.
This is probably because elimination-based algorithms, which are the algorithmic ideas behind our adaptive Neyman allocation algorithms, outperform the upper confidence bound algorithm when the sub-optimal treatment is easy to identify, which is the case in our simulation when $\sigma(1)=5$.
In Section~\ref{sec:append:AdditionalSynthetic} in the Online Appendix, we will see that the upper confidence bound algorithm performs better when $\sigma(1)=1$.

We also report the proxy mean squared errors as defined in \eqref{eqn:ProxyVariance} in Figure~\ref{fig:proxyMSE:sigma=5}.
In this comparison, we omit the two curves of the sample-discarding algorithms because they are not defined for the proxy mean squared error.
Among the remaining six algorithms, the proxy mean squared error serves as the objective function of both the adaptive Neyman allocation algorithms and the upper confidence bound algorithm.
The performances of the six algorithms follow the same pattern as we have observed in Figure~\ref{fig:MSE:sigma=5}.

\subsubsection*{Gaps between $\Var(\widehat{\tau})$, $\bE[(\widehat{\tau} - \tau)^2]$, and $\bE[V(T(1),T(0))]$.}

In this simulation, we focus on the following three quantities: the proxy mean squared error $\bE[V(T(1),T(0))]$ as in \eqref{eqn:ProxyVariance}, the variance of estimator \eqref{eqn:Estimator}, $\Var(\widehat{\tau})$, and the mean squared error of estimator \eqref{eqn:Estimator}, $\bE\left[(\widehat{\tau} - \tau)^2\right]$.
We conduct the simulation under two cases when $M=2$ and $M=3$.

\begin{figure}[!tb]
\centering
\caption{Gap between $\bE[V(T(1),T(0))]$, $\Var(\widehat{\tau})$, and $\bE[(\widehat{\tau} - \tau)^2]$ when $\sigma(1) / \sigma(0)=5$}
\label{fig:Gap:sigma=5}
\includegraphics[width=0.75\textwidth]{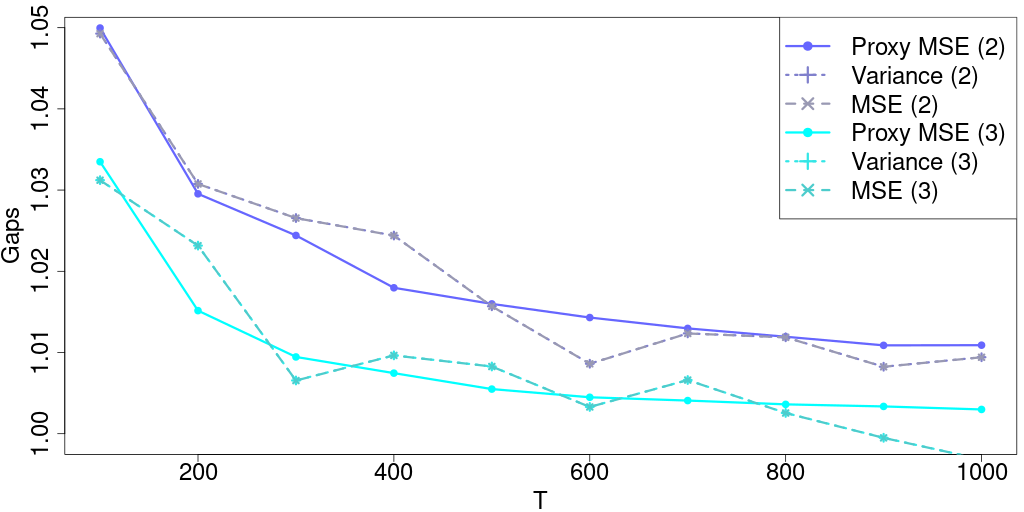}
\end{figure}

We report the simulation results in Figure~\ref{fig:Gap:sigma=5}. 
First, we compare the variance of estimator \eqref{eqn:Estimator}, $\Var(\widehat{\tau})$, and the mean squared error of estimator \eqref{eqn:Estimator}, $\bE\left[(\widehat{\tau} - \tau)^2\right]$.
They seem indistinguishable in this figure.
This is partly because we generate the potential outcomes from normal distributions, which satisfies Assumption~\ref{asp:symmetric}.
So the estimator \eqref{eqn:Estimator} is unbiased.
Second, we compare the proxy mean squared error $\bE[V(T(1),T(0))]$ with $\Var(\widehat{\tau})$ and $\bE[(\widehat{\tau} - \tau)^2]$.
The gap between the proxy mean squared error $\bE[V(T(1),T(0))]$ and $\Var(\widehat{\tau})$ or $\bE[(\widehat{\tau} - \tau)^2]$ seems to be small.
Third, the proxy mean squared error $\bE[V(T(1),T(0))]$ is relatively stable compared with $\Var(\widehat{\tau})$ or $\bE[(\widehat{\tau} - \tau)^2]$.

\section{Conclusions}
\label{sec:Conclusions}

In this paper, we present a competitive analysis framework to study the optimal multi-stage experimental design problem. 
We propose an adaptive Neyman allocation algorithm that is nearly optimal and almost matches the information-theoretic limit of conducting experiments. 
Our algorithm allows for efficient allocation of units into treated and control groups in multi-stage experiments, and can guide researchers towards the best allocation decisions when standard deviations are unknown in advance. 
Overall, our approach offers a solution for researchers seeking to optimize their experimental designs and increase statistical power, particularly in cases where the treated and control groups have different standard deviations, such as in social experiments, clinical trials, marketing research, and online A/B testing.

We conclude this paper with three potential limitations that should serve as cautionary notes for practitioners.
First, while adaptive Neyman allocation as described in this paper is suitable for sequential experimental design with a limited sample size, it still requires a minimum amount of sample size on the scale of at least several hundreds, to have reasonable performance.
In cases where a social experiment only involves a very small number of units, such as $\sim 30$ districts in a developing economy \citep{gibson2023redesigning}, and especially when there is a constraint that limits the size of the treated group to be only $2$ or $3$, we do not recommend the usage of adaptive Neyman allocation, or any randomized experiment design method.
Instead, we recommend conducting non-randomized experiments using similar ideas as the synthetic control method; see, e.g., \citet{abadie2021synthetic, doudchenko2021synthetic}.

Second, we have used the proxy mean squared error as the primary objective, instead of using the variance of the estimator $\widehat{\tau}$.
Since there is a gap between the proxy mean squared error and the variance, the confidence intervals derived based on the proxy mean squared error may suffer from undercoverage issues when the sample size is small.
In the simulations literature \citep{asmussen2007stochastic, glasserman2004monte, ross2013simulation}, this gap could be estimated if the outcomes are assumed to come from known parametric distribution families.
Yet there is no general method that estimates such a gap, not even the magnitude of such a gap when the sample size is small.
We leave this as a future research direction.

Third, we have shown both high probability guarantees (Theorems~\ref{thm:2StageANA} and~\ref{thm:MStageANA}) and in expectation guarantees (Corollaries~\ref{coro:2StageANA} and~\ref{coro:MStageANA}) in this paper.
However, with a small probability, the estimation error can still be very large.
This is usually referred to as the ``tail risk'' of an adaptive algorithm, which we do not discuss in this paper.
We refer to \citet{fan2024fragility, kalvit2021closer} for more details, and leave this as a future research direction.

\ACKNOWLEDGMENT{
The author would like to thank Alberto Abadie, Yilun Chen, Dean Eckles, Ivan Fernandez-Val, Christopher Harshaw, Susan Hunter, Ramesh Johari, Menglong Li, Tesary Lin, Tu Ni, Erol Pekoz, Pengyu Qian, Chao Qin, Philippe Rigollet, Daniel Russo, Nian Si, Stefan Wager, Chonghuan Wang, Yunzong Xu, Ruoxuan Xiong, Ruohan Zhan, and Zijie Zhou
for their insightful comments that have greatly improved this paper. 
In particular, the literature review on small gap regime versus fixed gap regime drew from the wisdom of Daniel Russo.
}

\bibliographystyle{informs2014} 
\bibliography{bibliography} 

\ECSwitch


\ECHead{Online Appendix}

\section{Intuitions Behind Algorithm Design}
\label{sec:append:AlgIntuitions}

In this section, we discuss some unrigorous intuitions behind the design of Algorithm~\ref{alg:MStageANA}.
Intuitively, at the end of each stage, Algorithm~\ref{alg:MStageANA} considers three different cases: when the current allocation is significantly different from the estimated Neyman allocation, or when it is moderately different, or when it is not very different.
A very natural idea is to directly extend Algorithm~\ref{alg:2StageANA} and consider only two cases: when the current allocation is significantly different from the estimated Neyman allocation, or when it is not very different. 
This direct extension will lead to Algorithm~\ref{alg:MStageNaive} as follows.

\begin{algorithm}[!htb]
\caption{$M$-stage adaptive Neyman allocation directly extended from Algorithm~\ref{alg:2StageANA}}
\label{alg:MStageNaive}
\small 
\textbf{Inputs}: Tuning parameters $\beta_1, \beta_2, ..., \beta_{M-1}$. \Comment{There are $M$ pre-determined stages $[0,\beta_1 T^{\frac{1}{M}}]$, $(\beta_1 T^{\frac{1}{M}}$, $\beta_2 T^{\frac{2}{M}}]$, ..., $(\beta_{M-1} T^{\frac{M-1}{M}}, T]$} 
\begin{algorithmic}[1]
\State \textbf{Initialize:} $(T_1(1), T_1(0)) \gets (\frac{\beta_1}{2} T^{\frac{1}{M}}, \frac{\beta_1}{2} T^{\frac{1}{M}})$;
\For{$m = 1, 2, ..., M-1$} \Comment{The $m$-th stage experiment}
\State Conduct a completely randomized experiment parameterized by $(T_m(1), T_m(0))$;
\State Estimate $\widehat{\sigma}^2_m(1)$ and $\widehat{\sigma}^2_m(0)$ as in \eqref{eqn:mVariance1} and \eqref{eqn:mVariance0} using data collected during stages $1 \sim m$;
\State \textbf{Case 1}: $\frac{\widehat{\sigma}_m(0)}{\widehat{\sigma}_m(1) + \widehat{\sigma}_m(0)} T < \frac{\beta_m}{2} T^{\frac{m}{M}}$ 
\State \phantom{for} For any $l \geq m+1$, $(T_l(1), T_l(0)) \gets (\beta_l T^{\frac{l}{M}} - \beta_{l-1} T^{\frac{l-1}{M}}, 0)$;
\Goto{markerNaive};
\State \textbf{Case 2}: $\frac{\widehat{\sigma}_m(0)}{\widehat{\sigma}_m(1) + \widehat{\sigma}_m(0)} T \geq \frac{\beta_{m}}{2} T^{\frac{m}{M}}$ and $\frac{\widehat{\sigma}_m(1)}{\widehat{\sigma}_m(1) + \widehat{\sigma}_m(0)} T \geq \frac{\beta_{m}}{2} T^{\frac{m}{M}}$ 
\State \phantom{for} $(T_{m+1}(1), T_{m+1}(0)) \gets (\frac{\beta_{m+1}}{2}T^{\frac{m+1}{M}} - \frac{\beta_{m}}{2}T^{\frac{m}{M}}, \frac{\beta_{m+1}}{2}T^{\frac{m+1}{M}} - \frac{\beta_{m}}{2}T^{\frac{m}{M}})$; \Comment{Note: there is no ``go to''}
\State \textbf{Case 3}: $\frac{\widehat{\sigma}_m(1)}{\widehat{\sigma}_m(1) + \widehat{\sigma}_m(0)} T < \frac{\beta_m}{2} T^{\frac{m}{M}}$ 
\State \phantom{for} For any $l \geq m+1$, $(T_l(1), T_l(0)) \gets (0, \beta_l T^{\frac{l}{M}} - \beta_{l-1} T^{\frac{l-1}{M}})$;
\Goto{markerNaive};
\EndFor
\For{$m' = m+1, ..., M$} \Comment{A sub-routine for experiments in the remaining stages} \label{markerNaive}
\State Conduct a completely randomized experiment parameterized by $(T_{m'}(1), T_{m'}(0))$;
\EndFor
\end{algorithmic}
\end{algorithm}

However, Algorithm~\ref{alg:MStageNaive} does not lead to the competitive ratio $\frac{V(T(1), T(0))}{V(T^*(1), T^*(0))} \approx 1 + T^{-\frac{M-1}{M}}$ as we were able to show in Theorem~\ref{thm:MStageANA}.

To see this, consider the following example when $M=3$, 
Denote $\rho = \frac{\sigma(1)}{\sigma(0)}$.
Consider the case when $\rho = \frac{T - T^{\frac{1}{3}}}{T^{\frac{1}{3}}} - \eps$, where $\eps > 0$ is a small number.
So $\rho$ falls into the following case $\frac{T - T^{\frac{2}{3}}}{T^{\frac{2}{3}}} < \rho \leq \frac{T - T^{\frac{1}{3}}}{T^{\frac{1}{3}}}$.
Then with high probability, we will stick with equal allocation in the first $2$ stages, and then in the last stage only allocate units into the treated group.
This means that we will allocate a total of $T^{\frac{2}{3}}$ units into the control group, and $T - T^{\frac{2}{3}}$ units into the treated group.

In this case, 
\begin{multline*}
\frac{V(T(1), T(0))}{V(T^*(1), T^*(0))} = \frac{T}{T-T^{\frac{2}{3}}} \cdot \frac{\rho^2}{(\rho+1)^2} + \frac{T}{T^{\frac{2}{3}}} \cdot \frac{1}{(\rho+1)^2} \\
\approx \frac{T}{T-T^{\frac{2}{3}}} \bigg(\frac{T-T^{\frac{1}{3}}}{T}\bigg)^2 + \frac{T}{T^{\frac{2}{3}}} \bigg(\frac{T^{\frac{1}{3}}}{T}\bigg)^2 = 1 + \frac{1}{T} + \frac{T^{\frac{5}{3}}-2T^{\frac{4}{3}}+T^{\frac{2}{3}}}{T^2-T^{\frac{5}{3}}} \approx 1 + T^{-\frac{1}{3}},
\end{multline*}
which is larger than the $1 + T^{-\frac{2}{3}}$ competitive ratio result that we were able to prove in Theorem~\ref{thm:MStageANA}.

Through this counterexample, we see that Algorithm~\ref{alg:MStageNaive}, the direct extension of Algorithm~\ref{alg:2StageANA}, does not yield a desirable competitive ratio. 
This is the reason why we need the many different cases as in Algorithm~\ref{alg:MStageANA}.
However, it is still unclear if there are other simpler algorithms that can yield the same competitive ratio as in Algorithm~\ref{alg:MStageANA}.

\section{Further Extensions}
\label{sec:ArmPulls}

In this section, we examine the number of treated and control units after we run Algorithms~\ref{alg:2StageANA} and~\ref{alg:MStageANA}.
We show that the number of treated and control units converge to the optimal allocation, although the rate that we provide below may not be optimal.
This results are implications of Theorems~\ref{thm:2StageANA} and~\ref{thm:MStageANA}.

\begin{corollary}
\label{coro:2Stage:ArmPulls}
Let $(T(1), T(0))$ be the total number of treated and control units from Algorithm~\ref{alg:2StageANA}, respectively. 
Under Assumption~\ref{asp:kurtosis}, there exists an event that happens with probability at least $1 - (\kappa(1) + \kappa(0)) T^{-\eps}$, conditional on which
\begin{align*}
\left|\frac{T(1)}{T(1)+T(0)}-\frac{\sigma_{1}}{\sigma_{1}+\sigma_{0}}\right| = O\Big(T^{-\frac{1}{4} + \frac{\eps}{2}}\Big).
\end{align*}
\end{corollary}

\begin{corollary}
\label{coro:MStage:ArmPulls}
Let $(T(1), T(0))$ be the total number of treated and control units from Algorithm~\ref{alg:MStageANA}, respectively. 
Under Assumption~\ref{asp:kurtosis}, there exists an event that happens with probability at least $1 - (M-1)(\kappa(1) + \kappa(0)) T^{-\eps}$, conditional on which
\begin{align*}
\left|\frac{T(1)}{T(1)+T(0)}-\frac{\sigma_{1}}{\sigma_{1}+\sigma_{0}}\right| = O\Big(T^{-\frac{1}{2M} + \frac{\eps}{2}}\Big).
\end{align*}
\end{corollary}

We defer the proof of Corollary~\ref{coro:2Stage:ArmPulls} to Section~\ref{sec:proof:coro:2Stage:ArmPulls} and the proof of Corollary~\ref{coro:MStage:ArmPulls} to Section~\ref{sec:proof:coro:MStage:ArmPulls} in the Online Appendix.

\section{Useful Lemmas}
\label{sec:UsefulLemmas}

\subsection{Martingale Central Limit Theorem}

We state Theorem~2 from \citet{brown1971martingale} below without a proof.

\begin{lemma}[Theorem~2, \citet{brown1971martingale}]
\label{lem:martingaleCLT}
Let $\{X_t, \sF_{t} \}_{t=1,2,...}$ be a martingale difference sequence on the probability space $(\Omega, \sF, P)$ such that $\bE[X_t \vert \sF_{t-1}] = 0$.
Let $\xrightarrow{d}$ and $\xrightarrow{p}$ stand for convergence in distribution and convergence in probability, respectively.
If the following two conditions hold,
\begin{enumerate}[label=(\roman*)]
\item Bounded variance. As $T \to +\infty$,
\begin{align*}
\sum_{t=1}^T \bE\Big[X_t^2 \Big\vert \sF_{t-1}\Big] \xrightarrow{p} s^2,
\end{align*}
\item Lindeberg condition. There exists some $\eps > 0$, such that as $T \to +\infty$,
\begin{align*}
\sum_{t=1}^T \bE\Big[X_t^2 \bI\{\vert X_t \vert \geq \eps s\} \Big\vert \sF_{t-1}\Big] \xrightarrow{p} 0,
\end{align*}
\end{enumerate}
Then,
\begin{align*}
\lim_{T \to +\infty} \sum_{t=1}^T X_t \xrightarrow{d} \cN(0,s^2).
\end{align*}
\end{lemma}

\subsection{Law of Large Numbers with Random Indices}

We state Theorem~2.2 from \citet{gut2009stopped} below without a proof.
Note that, this result does not require that the sequence of random variables $\big\{Y_n, n \geq 1\big\}$ and the family of random variables $\big\{N(t), t \geq 0\big\}$ are independent.

\begin{lemma}[Theorem~2.2, \citet{gut2009stopped}]
\label{lem:LLNRandomIndices}
Let $\big\{Y_n, n \geq 1\big\}$ be a sequence of random variables and $\big\{N(t), t \geq 0\big\}$ be a family of positive, integer-valued random variables.
Suppose that as $n \to +\infty$,
\begin{align*}
Y_t \xrightarrow{a.s.} Y,
\end{align*}
where $\xrightarrow{a.s.}$ stands for almost sure convergence, and as $t \to +\infty$,
\begin{align*}
N(t) \xrightarrow{p} +\infty.
\end{align*}
Then, as $t \to +\infty$,
\begin{align*}
Y_{N(t)} \xrightarrow{p} Y.
\end{align*}
\end{lemma}

\subsection{Algebraic Inequalities}

\begin{lemma}
\label{lem:g:rho}
Let $G_1, G_2 > 0$ be positive.
Let $g: \bR^+ \to \bR^+$ be a univariate function defined by 
\begin{align*}
g(\rho) = \frac{1}{G_1} \frac{\rho^2}{(\rho+1)^2} + \frac{1}{G_2} \frac{1}{(\rho+1)^2}.
\end{align*}
Then, 
\begin{enumerate}
\item $g(\rho)$ is decreasing when $\rho < \frac{G_1}{G_2}$, and increasing when $\rho > \frac{G_1}{G_2}$.
\item $g(\rho) \leq \max\{\frac{1}{G_1}, \frac{1}{G_2}\}$.
\end{enumerate}
\end{lemma}

\begin{lemma}
\label{lem:h:rhohat}
Let $\sigma(1), \sigma(0) > 0$ be positive.
Let $h: \bR^+ \to \bR^+$ be a univariate function defined by 
\begin{align*}
h(\widehat{\rho}) = \frac{1}{\widehat{\rho}} \ \sigma^2(1) + \widehat{\rho} \ \sigma^2(0).
\end{align*}
Then,
\begin{enumerate}
\item $h(\widehat{\rho})$ is decreasing when $\widehat{\rho} < \frac{\sigma(1)}{\sigma(0)}$, and increasing when $\widehat{\rho} > \frac{\sigma(1)}{\sigma(0)}$.
\item $h(\widehat{\rho})$ is a convex function.
\item Let $\zeta \in (0,1)$. When $\frac{\sigma(1)}{\sigma(0)}\sqrt{\frac{1-\zeta}{1+\zeta}} \leq \widehat{\rho} \leq \frac{\sigma(1)}{\sigma(0)}\sqrt{\frac{1+\zeta}{1-\zeta}}$, $h(\widehat{\rho}) \leq \sigma(1) \sigma(0) (\sqrt{\frac{1-\zeta}{1+\zeta}} + \sqrt{\frac{1+\zeta}{1-\zeta}})$.
\end{enumerate}
\end{lemma}

\begin{lemma}
\label{lem:AlgebraicTrick1}
When $T \geq 16$, $\eps \in \left(0, \frac{1}{8}\right)$, the following inequality holds,
\begin{align*}
\left( \frac{T - \frac{1}{2}T^{\frac{1}{2}}}{\frac{1}{2}T^{\frac{1}{2}}} \right)^4 > \frac{ 1 + 2^{\frac{1}{2}}T^{-\frac{1}{4} + \frac{\eps}{2}} }{ 1 - 2^{\frac{1}{2}}T^{-\frac{1}{4} + \frac{\eps}{2}} }.
\end{align*}
\end{lemma}

\begin{lemma}
\label{lem:AlgebraicTrick2}
Let $M \geq 3$ and $T \geq 16$, and $0 < \eps \leq \min\{\frac{1}{M}, \frac{1}{100}\}$.
For any $m \leq M-1$, let $\beta_m = 6 \cdot 15^{-\frac{m}{M}}$.
Then we have, for any $m \leq M-1$,
\begin{align*}
(1 - 2 \beta_m^{-1} T^{-\frac{m}{M} + \eps})^{-\frac{1}{2}} \leq 1 + 2 \beta_m^{-1} T^{-\frac{m}{M} + \eps}. 
\end{align*}
\end{lemma}

\begin{lemma}
\label{lem:AlgebraicTrick3}
Let $M \geq 3$, $T \geq 16$, and $0 < \eps \leq \min\{\frac{1}{M}, \frac{1}{100}\}$.
For any $m \leq M-1$, let $\beta_m = 6 \cdot 15^{-\frac{m}{M}}$.
Then we have, for any $m \leq M-1$,
\begin{align*}
\sqrt{\frac{1-2^{\frac{1}{2}}\beta_m^{-\frac{1}{2}}T^{-\frac{m}{2M}+\frac{\eps}{2}}}{1+2^{\frac{1}{2}}\beta_m^{-\frac{1}{2}}T^{-\frac{m}{2M}+\frac{\eps}{2}}}} > \frac{1}{2}.
\end{align*}
\end{lemma}

\begin{lemma}
\label{lem:AlgebraicTrick4}
Let $M \geq 3$ and $T \geq 16$.
For any $m \leq M-1$, let $\beta_m = 6 \cdot 15^{-\frac{m}{M}}$.
Then we have, for any $m \leq M-1$,
\begin{align*}
\frac{T - \frac{1}{2} \beta_m T^{\frac{m}{M}}}{\frac{1}{2} \beta_m T^{\frac{m}{M}}} \geq 4 > 2.
\end{align*}
\end{lemma}

\begin{lemma}
\label{lem:AlgebraicTrick5}
Let $M \geq 3$ and $T \geq 16$.
Let $\beta_1 = 6 \cdot 15^{-\frac{1}{M}}$.
Then we have
\begin{align*}
\frac{\frac{1}{2} \beta_1 T^{\frac{1}{M}}}{T - \frac{1}{2} \beta_1 T^{\frac{1}{M}}} < 4 \cdot 15^{-\frac{1}{M}} \cdot T^{-\frac{M-1}{M}}.
\end{align*}
\end{lemma}

\begin{lemma}
\label{lem:AlgebraicTrick6}
Let $0 < \eps \leq \frac{1}{6}$.
Then, 
\begin{align*}
\sqrt{1+\frac{3}{2}\eps} \leq 1 + \frac{3}{4} \eps - \frac{9}{64} \eps^2 < 1 + \frac{3}{4} \eps.
\end{align*}
\end{lemma}

\begin{lemma}
\label{lem:AlgebraicTrick7}
Let $0 < \eps \leq \frac{1}{6}$.
Then, 
\begin{align*}
\frac{1}{1-\frac{27}{4}\eps^2 - \frac{27}{4}\eps^3} \leq 1 + \frac{27}{2} \eps^2.
\end{align*}
\end{lemma}

\subsection{Extensions of Algebraic Inequalities}

\begin{lemma}
\label{lem:AlgebraicTrick:Basic:1}
Let $T \geq 320^\frac{5}{4} C^5$.
Then,
\begin{align*}
T \geq 64 C^4 \log{T}.
\end{align*}
\end{lemma}

\begin{lemma}
\label{lem:AlgebraicTrick:Basic}
Let $T \geq (\frac{5000}{3})^{\frac{5}{4}} C^5$.
Then,
\begin{align*}
T \geq \frac{1000}{3} C^4 \log{T}.
\end{align*}
\end{lemma}

\begin{lemma}
\label{lem:AlgebraicTrick1:Refined}
Let $T \geq 320^\frac{5}{4} C^5$.
Then, we have
\begin{enumerate}[label=(\roman*)]
\item \begin{align*}
4 C^2 T^{-\frac{1}{2}} (\log{T})^{\frac{1}{2}} \leq \frac{1}{2}.
\end{align*}
\item \begin{align*}
\left( \frac{T - 2 C^2 T^{\frac{1}{2}} (\log{T})^{\frac{1}{2}}}{2 C^2 T^{\frac{1}{2}} (\log{T})^{\frac{1}{2}}} \right)^4 > \frac{ 1 + 2 C T^{-\frac{1}{4}} (\log{T})^{\frac{1}{4}} }{ 1 - 2 C T^{-\frac{1}{4}} (\log{T})^{\frac{1}{4}} }.
\end{align*}
\item \begin{align*}
\left(1 - 4 C^2 T^{-\frac{1}{2}} (\log{T})^{\frac{1}{2}}\right)^{-\frac{1}{2}} \leq 4 C^2 T^{-\frac{1}{2}} (\log{T})^{\frac{1}{2}}
\end{align*}
\end{enumerate}
\end{lemma}

\begin{lemma}
\label{lem:AlgebraicTrick2:Refined}
Let $M \geq 3$, $T \geq (\frac{5000}{3})^{\frac{5}{4}} C^5$.
Let $\beta_m = \frac{400}{3} C^4 \log{T} \cdot (\frac{1000}{3} C^4 \log{T})^{-\frac{m}{M}}$ for any $m \leq M-1$.
Then we have, for any $m \leq M-1$,
\begin{align*}
(1 - 48 C^4\beta_m^{-1}T^{-\frac{m}{M}}\log{T})^{-\frac{1}{2}} \leq 1 + 48 C^4\beta_m^{-1}T^{-\frac{m}{M}}\log{T}. 
\end{align*}
\end{lemma}

\begin{lemma}
\label{lem:AlgebraicTrick3:Refined}
Let $M \geq 3$, $T \geq (\frac{5000}{3})^{\frac{5}{4}} C^5$.
Let $\beta_m = \frac{400}{3} C^4 \log{T} \cdot (\frac{1000}{3} C^4 \log{T})^{-\frac{m}{M}}$ for any $m \leq M-1$.
Then we have, for any $m \leq M-1$,
\begin{align*}
\sqrt{\frac{1-48^{\frac{1}{2}} C^2 \beta_m^{-\frac{1}{2}} T^{-\frac{m}{2M}} (\log{T})^{\frac{1}{2}}}{1+48^{\frac{1}{2}} C^2 \beta_m^{-\frac{1}{2}} T^{-\frac{m}{2M}} (\log{T})^{\frac{1}{2}}}} > \frac{1}{2}.
\end{align*}
\end{lemma}

\begin{lemma}
\label{lem:AlgebraicTrick4:Refined}
Let $M \geq 3$, $T \geq (\frac{5000}{3})^{\frac{5}{4}} C^5$.
Let $\beta_m = \frac{400}{3} C^4 \log{T} \cdot (\frac{1000}{3} C^4 \log{T})^{-\frac{m}{M}}$ for any $m \leq M-1$.
Then we have, for any $m \leq M-1$,
\begin{align*}
\frac{T - \frac{1}{2} \beta_m T^{\frac{m}{M}}}{\frac{1}{2} \beta_m T^{\frac{m}{M}}} \geq 4 > 2.
\end{align*}
\end{lemma}

\begin{lemma}
\label{lem:AlgebraicTrick5:Refined}
Let $M \geq 3$, $T \geq (\frac{5000}{3})^{\frac{5}{4}} C^5$.
Let $\beta_1 = \frac{400}{3} C^4 \log{T} \cdot (\frac{1000}{3} C^4 \log{T})^{-\frac{1}{M}}$.
Then we have
\begin{align*}
\frac{\frac{1}{2} \beta_1 T^{\frac{1}{M}}}{T - \frac{1}{2} \beta_1 T^{\frac{1}{M}}} < 4 \cdot 15^{-\frac{1}{M}} C^{\frac{2(M-1)}{M}} \cdot T^{-\frac{M-1}{M}} (\log{T})^{\frac{M-1}{M}}.
\end{align*}
\end{lemma}

\subsection{Probability Inequalities}

\begin{lemma}
\label{lem:EC:Kurtosis}
Let $Y_1, ..., Y_n$ be $n$ identical and independent copies of some random variable $Y$.
Let $\sigma^2$ be the variance of $Y$, and let $\widehat{\sigma}^2 = \frac{1}{n-1}\sum_{i=1}^n \left(Y_i - \frac{1}{n} \sum_{i=1}^n Y_i\right)^2$ be the sample variance estimator. The variance of the sample variance estimator can be expressed as
\begin{align*}
\bE\left[\left(\widehat{\sigma}^2\right)^2\right] = \frac{1}{n} \bE\left[(Y - \bE Y)^4\right] + \frac{n^2-2n+3}{n(n-1)} \sigma^4.
\end{align*}
\end{lemma}

\begin{lemma}
\label{lem:LightTail}
At the end of stage $m$, consider the sample variance estimators as defined in \eqref{eqn:mVariance1} and \eqref{eqn:mVariance0}.
Under Assumption~\ref{asp:kurtosis}, for any $m \in [M]$ and for any $\delta > 0$, if $\sum_{l=1}^m T_l(1) \geq 3$, then
\begin{align*}
\Pr\left( \vert \widehat{\sigma}^2_m(1) - \sigma^2(1) \vert \geq \delta \right) \leq & \ \frac{\kappa(1) \sigma^4(1)}{\delta^2 \sum_{l=1}^m T_l(1)}.
\end{align*}
If $\sum_{l=1}^m T_l(0) \geq 3$, then
\begin{align*}
\Pr\left( \vert \widehat{\sigma}^2_m(0) - \sigma^2(0) \vert \geq \delta \right) \leq & \ \frac{\kappa(0) \sigma^4(0)}{\delta^2 \sum_{l=1}^m T_l(0)},
\end{align*}
where $\kappa(1)$ and $\kappa(0)$ are defined in Assumption~\ref{asp:kurtosis}.
\end{lemma}

\begin{lemma}
\label{lem:ExponentialTail}
Consider, either the sample variance estimators as defined in \eqref{eqn:SampleVariance1} and \eqref{eqn:SampleVariance0} at the end of the first stage, or the sample variance estimators as defined in \eqref{eqn:mVariance1} and \eqref{eqn:mVariance0} at the end of stage $m$.
Under Assumption~\ref{asp:bounded}, for any $m \leq M-1$ and for any $\delta > 0$,
\begin{align*}
\Pr\left( \vert \widehat{\sigma}^2_m(1) - \sigma^2(1) \vert \geq \delta \right) \leq & \ 2 \exp\left\{-\frac{\delta^2 \sum_{l=1}^m T_l(1)}{8 C^4 \sigma^4(1)}\right\}, \\ 
\Pr\left( \vert \widehat{\sigma}^2_m(0) - \sigma^2(0) \vert \geq \delta \right) \leq & \ 2 \exp\left\{-\frac{\delta^2 \sum_{l=1}^m T_l(0)}{8 C^4 \sigma^4(0)}\right\},
\end{align*}
where $C$ is defined in Assumption~\ref{asp:bounded}.
\end{lemma}

\subsection{Probability Equality}

It is well known that the sample variance can be expressed as a sum of squares.
\begin{lemma}
\label{lem:CommonKnowledge:SumOfSquares}
Let there be $n$ i.i.d. samples $X_1, X_2, ..., X_n$ of the same distribution.
The sample variance $\widehat{\sigma}^2$ can be expressed as follows,
\begin{align*}
\widehat{\sigma}^2 = \frac{1}{2n(n-1)} \sum_{i=1}^n \sum_{j=1}^n (X_i - X_j)^2.
\end{align*}
\end{lemma}

\section{Missing Proofs}

\subsection{Proofs of Lemmas from Section~\ref{sec:UsefulLemmas}}

\subsubsection{Proof of Lemma~\ref{lem:g:rho}}
\proof{Proof of Lemma~\ref{lem:g:rho}.}
Taking first order derivative, we have
\begin{align*}
g'(\rho) = & \ \frac{1}{G_1} \cdot \frac{2 \rho}{(\rho+1)^3} - \frac{1}{G_2} \cdot \frac{2}{(\rho+1)^3} \\
= & \ \frac{2}{(\rho+1)^3} \left( \frac{\rho}{G_1} - \frac{1}{G_2} \right).
\end{align*}
When $\rho < \frac{G_1}{G_2}$, $g'(\rho) < 0$ so $g(\rho)$ is decreasing;
when $\rho > \frac{G_1}{G_2}$, $g'(\rho) > 0$ so $g(\rho)$ is increasing.

Using the above, we have that 
\begin{align*}
g(\rho) \leq \max\left\{\lim_{\rho \to +\infty} g(\rho), \lim_{\rho \to 0^+} g(\rho)\right\} = \max\left\{\frac{1}{G_1}, \frac{1}{G_2}\right\}.
\end{align*}
\hfill \halmos
\endproof

\subsubsection{Proof of Lemma~\ref{lem:h:rhohat}}
\proof{Proof of Lemma~\ref{lem:h:rhohat}.}
Taking first order derivative, we have
\begin{align*}
h'(\widehat{\rho}) = \sigma^2(0) - \frac{1}{\widehat{\rho}^2} \ \sigma^2(1).
\end{align*}
When $\widehat{\rho} < \frac{\sigma(1)}{\sigma(0)}$, $h'(\widehat{\rho}) < 0$, so $h(\widehat{\rho})$ is decreasing in $\widehat{\rho}$; when $\widehat{\rho} > \frac{\sigma(1)}{\sigma(0)}$, $h'(\widehat{\rho}) > 0$, so $h(\widehat{\rho})$ is increasing in $\widehat{\rho}$.

Next, taking second order derivative, we have
\begin{align*}
h'(\widehat{\rho}) = \frac{2}{\widehat{\rho}^3} \ \sigma^2(1) > 0.
\end{align*}
So $h(\widehat{\rho})$ is a convex function.

Combing above, we know that $h(\widehat{\rho})$ is a convex function, increasing when $\widehat{\rho} > \frac{\sigma(1)}{\sigma(0)}$ and decreasing when $\widehat{\rho} < \frac{\sigma(1)}{\sigma(0)}$.
When $\frac{\sigma(1)}{\sigma(0)}\sqrt{\frac{1-\zeta}{1+\zeta}} \leq \widehat{\rho} \leq \frac{\sigma(1)}{\sigma(0)}\sqrt{\frac{1+\zeta}{1-\zeta}}$, the maximum is taken on the boundaries, i.e., 
\begin{align*}
h(\widehat{\rho}) \leq \max\left\{ h\left(\frac{\sigma(1)}{\sigma(0)}\sqrt{\frac{1-\zeta}{1+\zeta}}\right), h\left(\frac{\sigma(1)}{\sigma(0)}\sqrt{\frac{1+\zeta}{1-\zeta}}\right) \right\} = \sigma(1) \sigma(0) \left(\sqrt{\frac{1-\zeta}{1+\zeta}} + \sqrt{\frac{1+\zeta}{1-\zeta}}\right).
\end{align*}
\hfill \halmos
\endproof

\subsubsection{Proof of Lemma~\ref{lem:AlgebraicTrick1}}
\proof{Proof of Lemma~\ref{lem:AlgebraicTrick1}.}
When $T \geq 16$, we have
\begin{align}
\frac{T - \frac{1}{2}T^{\frac{1}{2}}}{\frac{1}{2}T^{\frac{1}{2}}} > 3 > \left( \frac{1 + 2^{-\frac{1}{4}}}{1 - 2^{-\frac{1}{4}}} \right)^4 \approx 1.84. \label{eqn:AlgebraicTrick:proof:1}
\end{align}
On the other hand, when $T \geq 16$ and $\eps \in \left(0, \frac{1}{8}\right)$, we have $T^{1-2\eps} \geq 8 = \left( 2^{\frac{3}{4}} \right)^4$, which then suggests $2^{-\frac{1}{2}}T^{\frac{1}{4}-\frac{\eps}{2}} \geq 2^{\frac{1}{4}} > 0$.
So 
\begin{align*}
0 < 2^{\frac{1}{2}} T^{-\frac{1}{4}+\frac{\eps}{2}} \leq 2^{-\frac{1}{4}}.
\end{align*}
Since $\frac{1+x}{1-x}$ is an increasing function on $x > 0$, we have
\begin{align}
\frac{1 + 2^{-\frac{1}{4}}}{1 - 2^{-\frac{1}{4}}} \geq \frac{1 + 2^{\frac{1}{2}} T^{-\frac{1}{4}+\frac{\eps}{2}}}{1 - 2^{\frac{1}{2}} T^{-\frac{1}{4}+\frac{\eps}{2}}}. \label{eqn:AlgebraicTrick:proof:2}
\end{align}
Combining \eqref{eqn:AlgebraicTrick:proof:1} and \eqref{eqn:AlgebraicTrick:proof:2} we finish the proof.
\hfill \halmos
\endproof

\subsubsection{Proof of Lemma~\ref{lem:AlgebraicTrick2}}
\proof{Proof of Lemma~\ref{lem:AlgebraicTrick2}.}
When $0 \leq x \leq \frac{1}{2}$, we have $x + x^2 \leq 1$. 
Then, since $x \geq 0$, we have $1 \leq 1 + x - x^2 - x^3 = (1+x)^2 (1-x)$.
Since $1-x > 0$, this leads to $0 < \frac{1}{1-x} \leq (1+x)^2$.
Taking square root we have
\begin{align}
(1-x)^{-\frac{1}{2}} \leq 1+x. \label{eqn:proof:Trick2:intermediate}
\end{align}

Next we show that $\beta_m^{-1} T^{-\frac{m}{M} + \eps} \leq \frac{1}{4}$.
To see this, we use the definition of $\beta_m = 6 \cdot 15^{-\frac{m}{M}}$.
\begin{align*}
\beta_m^{-1} T^{-\frac{m}{M} + \eps} = \frac{1}{6} \cdot \left(\frac{1}{15}\right)^{-\frac{m}{M}} T^{-\frac{m}{M} + \eps} = \frac{15^{\eps}}{6} \cdot \left(\frac{T}{15}\right)^{-\frac{m}{M} + \eps} \leq \frac{15^{\eps}}{6} \leq \frac{15^{\frac{1}{100}}}{6} \approx 0.1713 < \frac{1}{4}.
\end{align*}
where the first inequality is because $T > 15$ and $-\frac{m}{M}+\eps \leq 0$; the second inequality is because $0 < \eps \leq \frac{1}{100}$.
Replacing $x = 2 \beta_m^{-1} T^{-\frac{m}{M} + \eps}$ into \eqref{eqn:proof:Trick2:intermediate} we finish the proof.
\hfill \halmos
\endproof

\subsubsection{Proof of Lemma~\ref{lem:AlgebraicTrick3}}
\proof{Proof of Lemma~\ref{lem:AlgebraicTrick3}.}
First we focus on $\beta_m^{-1} T^{-\frac{m}{M} + \eps}$.
Using the definition of $\beta_m = 6 \cdot 15^{-\frac{m}{M}}$,
\begin{align}
\beta_m^{-1} T^{-\frac{m}{M} + \eps} = \frac{1}{6} \cdot \left(\frac{1}{15}\right)^{-\frac{m}{M}} T^{-\frac{m}{M} + \eps} = \frac{15^{\eps}}{6} \cdot \left(\frac{T}{15}\right)^{-\frac{m}{M} + \eps} \leq \frac{15^{\eps}}{6} \leq \frac{15^{\frac{1}{100}}}{6} \approx 0.1713 < \frac{9}{50}. \label{eqn:proof:Trick3:key}
\end{align}
where the first inequality is because $T > 15$ and $-\frac{m}{M}+\eps \leq 0$; the second inequality is because $0 < \eps \leq \frac{1}{100}$.

Using \eqref{eqn:proof:Trick3:key} as above, we have
\begin{align*}
0 < 2^{\frac{1}{2}} \beta_m^{-\frac{1}{2}} T^{-\frac{m}{2M} + \frac{\eps}{2}} < \frac{3}{5} < 1.
\end{align*}
Since $\frac{1-x}{1+x}$ is a decreasing function in $x$ when $0<x<1$, 
\begin{align*}
\frac{1-2^{\frac{1}{2}} \beta_m^{-\frac{1}{2}} T^{-\frac{m}{2M}+\frac{\eps}{2}}}{1+2^{\frac{1}{2}} \beta_m^{-\frac{1}{2}} T^{-\frac{m}{2M}+\frac{\eps}{2}}} > \frac{1-\frac{3}{5}}{1+\frac{3}{5}} = \frac{1}{4}.
\end{align*}
Taking square root finishes the proof.
\hfill \halmos
\endproof

\subsubsection{Proof of Lemma~\ref{lem:AlgebraicTrick4}}
\proof{Proof of Lemma~\ref{lem:AlgebraicTrick4}.}
Using the definition of $\beta_m = 6 \cdot 15^{-\frac{m}{M}}$,
\begin{align*}
0 < \frac{5}{2} \beta_m T^{\frac{m}{M}} = 15 \cdot \left( \frac{T}{15} \right)^{\frac{m}{M}} \leq 15 \cdot \left( \frac{T}{15} \right) = T,
\end{align*}
where the inequality is due to $m < M$.
Then we have 
\begin{align*}
\frac{T - \frac{1}{2} \beta_m T^{\frac{m}{M}}}{\frac{1}{2} \beta_m T^{\frac{m}{M}}} > \frac{\frac{5}{2} \beta_m T^{\frac{m}{M}} - \frac{1}{2} \beta_m T^{\frac{m}{M}}}{\frac{1}{2} \beta_m T^{\frac{m}{M}}} = 4,
\end{align*}
which finishes the proof.
\hfill \halmos
\endproof

\subsubsection{Proof of Lemma~\ref{lem:AlgebraicTrick5}}
\proof{Proof of Lemma~\ref{lem:AlgebraicTrick5}.}
Using the definition of $\beta_1 = 6 \cdot 15^{-\frac{1}{M}}$,
\begin{align*}
\frac{1}{4} T = \left( \frac{1}{2} \beta_1 T^{\frac{1}{M}} \right) \cdot \left( \frac{1}{12} 15^{\frac{1}{M}} T^{\frac{M-1}{M}} \right) > \left( \frac{1}{2} \beta_1 T^{\frac{1}{M}} \right) \cdot \left( \frac{1}{12} 15^{\frac{1}{M}} 15^{\frac{M-1}{M}} \right) > \frac{1}{2} \beta_1 T^{\frac{1}{M}}.
\end{align*}
Then, replacing $\frac{1}{2} \beta_1 T^{\frac{1}{M}}$ with $\frac{1}{4}T$ in the denominator, we have
\begin{align*}
\frac{\frac{1}{2} \beta_1 T^{\frac{1}{M}}}{T - \frac{1}{2} \beta_1 T^{\frac{1}{M}}} < \frac{\frac{1}{2} \beta_1 T^{\frac{1}{M}}}{\frac{3}{4}T} = \frac{2}{3} \beta_1 T^{-\frac{M-1}{M}} = 4 \cdot 15^{-\frac{1}{M}} \cdot T^{-\frac{M-1}{M}}.
\end{align*}
\hfill \halmos
\endproof

\subsubsection{Proof of Lemma~\ref{lem:AlgebraicTrick6}}
\proof{Proof of Lemma~\ref{lem:AlgebraicTrick6}.}
From $\eps < 1$, we have
\begin{align*}
\frac{3}{4} \eps - \frac{9}{128} \eps^2 < \frac{3}{4} \eps < 1.
\end{align*}
Since $\eps > 0$,
\begin{align*}
\frac{27}{128} \eps^3 < \frac{9}{32}\eps^2 + \frac{81}{4096} \eps^4.
\end{align*}
Then we have,
\begin{align*}
0 < 1 + \frac{3}{2} \eps \leq 1 + \frac{3}{2} \eps + \frac{9}{16} \eps^2 - \frac{9}{32} \eps^2 - \frac{27}{128} \eps^3 + \frac{81}{4096} \eps^4 = \left(1 + \frac{2}{4}\eps - \frac{9}{64}\eps^2 \right)^2.
\end{align*}
Taking square root we finish the proof.
\hfill \halmos
\endproof

\subsubsection{Proof of Lemma~\ref{lem:AlgebraicTrick7}}
\proof{Proof of Lemma~\ref{lem:AlgebraicTrick7}.}
From $\eps < \frac{1}{6}$, we have
\begin{align*}
\eps + \frac{27}{2} \eps^2 + \frac{27}{2} \eps^3 \leq \frac{1}{6} + \frac{27}{72} + \frac{27}{432} < 1.
\end{align*}
Since $\eps > 0$,
\begin{align*}
\frac{27}{4} \cdot (\eps^3 + \frac{27}{2} \eps^4 + \frac{27}{2} \eps^5) < \frac{27}{4} \eps^2.
\end{align*}
Then we have,
\begin{align*}
1 \leq 1 - \frac{27}{4} \eps^2 - \frac{27}{4} \eps^3 + \frac{27}{2} \eps^2 - \frac{27^2}{8} \eps^4 - \frac{27^2}{8} \eps^5 = (1 + \frac{27}{2}\eps^2) \cdot \left(1-\frac{27}{4}\eps^2 - \frac{27}{4}\eps^3\right).
\end{align*}
Since $1-\frac{27}{4}\eps^2 - \frac{27}{4}\eps^3 > 0$, moving it to the left hand side finishes the proof.
\hfill \halmos
\endproof

\subsubsection{Proof of Lemma~\ref{lem:AlgebraicTrick:Basic:1}}
\proof{Proof of Lemma~\ref{lem:AlgebraicTrick:Basic:1}.}
To prove the first claim, note that $\frac{T}{\log{T}}$ is an increasing function, and that $T \geq 320^{\frac{5}{4}} C^5$, so we have
\begin{align*}
\frac{T}{64 C^4 \log{T}} > \frac{320^{\frac{5}{4}} C^5}{64 C^4 \log{(320^{\frac{5}{4}} C^5)}} = \frac{320^{\frac{5}{4}} C^5}{64 C^4 \cdot 5 \cdot \log{(320^{\frac{1}{4}}C)}} = \frac{320^{\frac{1}{4}}C}{\log{(320^{\frac{1}{4}}C)}} \geq 1,
\end{align*}
which finishes the proof.
\hfill \halmos
\endproof

\subsubsection{Proof of Lemma~\ref{lem:AlgebraicTrick:Basic}}
\proof{Proof of Lemma~\ref{lem:AlgebraicTrick:Basic}.}
Note that $\frac{T}{\log{T}}$ is an increasing function, and that $T \geq (\frac{5000}{3})^{\frac{5}{4}} C^5$, so we have
\begin{align*}
\frac{T}{\frac{1000}{3} C^4 \log{T}} > \frac{(\frac{5000}{3})^{\frac{5}{4}} C^5}{\frac{1000}{3} C^4 \log{((\frac{5000}{3})^{\frac{5}{4}} C^5)}} = \frac{(\frac{5000}{3})^{\frac{5}{4}} C^5}{\frac{1000}{3} C^4 \cdot 5 \cdot \log{((\frac{5000}{3})^{\frac{1}{4}}C)}} = \frac{(\frac{5000}{3})^{\frac{1}{4}}C}{\log{((\frac{5000}{3})^{\frac{1}{4}}C)}} \geq 1,
\end{align*}
which finishes the proof.
\hfill \halmos
\endproof

\subsubsection{Proof of Lemma~\ref{lem:AlgebraicTrick1:Refined}}
\proof{Proof of Lemma~\ref{lem:AlgebraicTrick1:Refined}.}
To prove the first claim, we re-arrange terms from Lemma~\ref{lem:AlgebraicTrick:Basic:1} and obtain
\begin{align*}
4 C^2 T^{-\frac{1}{2}} (\log{T})^{\frac{1}{2}} \leq \frac{1}{2}.
\end{align*}

To prove the second claim, note that from above, we have $T \geq 8 C^2 T^{\frac{1}{2}} (\log{T})^{\frac{1}{2}}$, so that
\begin{align*}
\left(\frac{T - 2 C^2 T^{\frac{1}{2}} (\log{T})^{\frac{1}{2}}}{2 C^2 T^{\frac{1}{2}} (\log{T})^{\frac{1}{2}}}\right)^4 \geq 3^4
\end{align*}
We also have $2 C T^{-\frac{1}{4}} (\log{T})^{\frac{1}{4}} \leq \frac{\sqrt{2}}{2}$, so that
\begin{align*}
\frac{1 + 2 C T^{-\frac{1}{4}} (\log{T})^{\frac{1}{4}}}{1 - 2 C T^{-\frac{1}{4}} (\log{T})^{\frac{1}{4}}} \leq \frac{1 + \frac{\sqrt{2}}{2}}{1 - \frac{\sqrt{2}}{2}} < 3^4,
\end{align*}
which concludes the proof of the second claim.

To prove the third claim, note that when $0 \leq x \leq \frac{1}{2}$, we have $x + x^2 \leq 1$. 
Then, since $x \geq 0$, we have $1 \leq 1 + x - x^2 - x^3 = (1+x)^2 (1-x)$.
Since $1-x > 0$, this leads to $0 < \frac{1}{1-x} \leq (1+x)^2$.
Taking square root we have
\begin{align*}
(1-x)^{-\frac{1}{2}} \leq 1+x.
\end{align*}
Replacing $x = 4 C^2 T^{-\frac{1}{2}} (\log{T})^{\frac{1}{2}}$ we conclude the proof of the third claim.
\hfill \halmos
\endproof

\subsubsection{Proof of Lemma~\ref{lem:AlgebraicTrick2:Refined}}
\proof{Proof of Lemma~\ref{lem:AlgebraicTrick2:Refined}.}
When $0 \leq x \leq \frac{1}{2}$, we have $x + x^2 \leq 1$. 
Then, since $x \geq 0$, we have $1 \leq 1 + x - x^2 - x^3 = (1+x)^2 (1-x)$.
Since $1-x > 0$, this leads to $0 < \frac{1}{1-x} \leq (1+x)^2$.
Taking square root we have
\begin{align*}
(1-x)^{-\frac{1}{2}} \leq 1+x.
\end{align*}

Using the definition of $\beta_m = \frac{400}{3} C^4 \log{T} \cdot (\frac{1000}{3} C^4 \log{T})^{-\frac{m}{M}}$,
\begin{align*}
48 C^4\beta_m^{-1}T^{-\frac{m}{M}}\log{T} = \frac{9}{25} \left( \frac{T}{\frac{1000}{3} C^4 \log{T}} \right)^{-\frac{m}{M}} \leq \frac{9}{25} < \frac{1}{2},
\end{align*}
where the first inequality is due to Lemma~\ref{lem:AlgebraicTrick:Basic} and $m \geq 1$.

Replacing $x = 48 C^4\beta_m^{-1}T^{-\frac{m}{M}}\log{T}$ finishes the proof.
\hfill \halmos
\endproof

\subsubsection{Proof of Lemma~\ref{lem:AlgebraicTrick3:Refined}}
\proof{Proof of Lemma~\ref{lem:AlgebraicTrick3:Refined}.}
Using the definition of $\beta_m = \frac{400}{3} C^4 \log{T} \cdot (\frac{1000}{3} C^4 \log{T})^{-\frac{m}{M}}$,
\begin{align}
48 C^4\beta_m^{-1}T^{-\frac{m}{M}}\log{T} = \frac{9}{25} \left( \frac{T}{\frac{1000}{3} C^4 \log{T}} \right)^{-\frac{m}{M}} \leq \frac{9}{25}, \label{eqn:proof:Trick3:key:Refined}
\end{align}
where the first inequality is due to Lemma~\ref{lem:AlgebraicTrick:Basic} and $m \geq 1$.

Using \eqref{eqn:proof:Trick3:key:Refined} as above, we have
\begin{align*}
0 < 48^{\frac{1}{2}} C^2 \beta_m^{-\frac{1}{2}} T^{-\frac{m}{2M}} (\log{T})^{\frac{1}{2}} < \frac{3}{5} < 1.
\end{align*}
Since $\frac{1-x}{1+x}$ is a decreasing function in $x$ when $0<x<1$, 
\begin{align*}
\frac{1-48^{\frac{1}{2}} C^2 \beta_m^{-\frac{1}{2}} T^{-\frac{m}{2M}} (\log{T})^{\frac{1}{2}}}{1+48^{\frac{1}{2}} C^2 \beta_m^{-\frac{1}{2}} T^{-\frac{m}{2M}} (\log{T})^{\frac{1}{2}}} > \frac{1-\frac{3}{5}}{1+\frac{3}{5}} = \frac{1}{4}.
\end{align*}
Taking square root finishes the proof.
\hfill \halmos
\endproof

\subsubsection{Proof of Lemma~\ref{lem:AlgebraicTrick4:Refined}}
\proof{Proof of Lemma~\ref{lem:AlgebraicTrick4:Refined}.}
Using the definition of $\beta_m = \frac{400}{3} C^4 \log{T} \cdot (\frac{1000}{3} C^4 \log{T})^{-\frac{m}{M}}$,
\begin{align*}
0 < \frac{5}{2} \beta_m T^{\frac{m}{M}} = \frac{1000}{3} C^4 \log{T} \cdot \left( \frac{T}{\frac{1000}{3} C^4 \log{T}} \right)^{\frac{m}{M}} < \frac{1000}{3} C^4 \log{T} \cdot \left( \frac{T}{\frac{1000}{3} C^4 \log{T}} \right) = T,
\end{align*}
where the inequality is due to Lemma~\ref{lem:AlgebraicTrick:Basic} and $m < M$.
Then we have 
\begin{align*}
\frac{T - \frac{1}{2} \beta_m T^{\frac{m}{M}}}{\frac{1}{2} \beta_m T^{\frac{m}{M}}} > \frac{\frac{5}{2} \beta_m T^{\frac{m}{M}} - \frac{1}{2} \beta_m T^{\frac{m}{M}}}{\frac{1}{2} \beta_m T^{\frac{m}{M}}} = 4,
\end{align*}
which finishes the proof.
\hfill \halmos
\endproof

\subsubsection{Proof of Lemma~\ref{lem:AlgebraicTrick5:Refined}}
\proof{Proof of Lemma~\ref{lem:AlgebraicTrick5:Refined}.}
Using the definition of $\beta_1 = \frac{400}{3} C^4 \log{T} \cdot (\frac{1000}{3} C^4 \log{T})^{-\frac{1}{M}}$,
\begin{align*}
\frac{1}{2} \beta_1 T^{\frac{1}{M}} = \frac{200}{3} C^4 \log{T} \cdot \left(\frac{T}{\frac{1000}{3} C^4 \log{T}}\right)^{\frac{1}{M}} \leq \frac{200}{3} C^4 \log{T} \cdot \left(\frac{T}{\frac{1000}{3} C^4 \log{T}}\right) = \frac{1}{5} T,
\end{align*}
where the inequality is due to Lemma~\ref{lem:AlgebraicTrick:Basic}.
Then, replacing $\frac{1}{2} \beta_1 T^{\frac{1}{M}}$ with $\frac{1}{5}T$ in the denominator, we have
\begin{multline*}
\frac{\frac{1}{2} \beta_1 T^{\frac{1}{M}}}{T - \frac{1}{2} \beta_1 T^{\frac{1}{M}}} < \frac{\frac{1}{2} \beta_1 T^{\frac{1}{M}}}{\frac{4}{5}T} = \frac{250}{3} C^4 \log{T} \cdot \left(\frac{T}{\frac{1000}{3} C^4 \log{T}}\right)^{\frac{1}{M}} \cdot T^{-1} \\
= 96 \cdot \left(\frac{1000}{3}\right)^{-\frac{1}{M}} C^{\frac{4(M-1)}{M}} \cdot T^{-\frac{M-1}{M}} (\log{T})^{\frac{M-1}{M}},
\end{multline*}
where the last inequality is re-arranging terms, and using the fact that $\frac{250}{3} < 96$.
\hfill \halmos
\endproof

\subsubsection{Proof of Lemma~\ref{lem:EC:Kurtosis}}
\proof{Proof of Lemma~\ref{lem:EC:Kurtosis}.}
Note that we can re-write the sample variance estimator as 
\begin{align*}
\widehat{\sigma}^2 = \frac{1}{n(n-1)} \left( n \cdot \sum_{i=1}^n Y_i^2 - (\sum_{i=1}^n Y_i)^2 \right).
\end{align*}

We now expand the variance of the sample variance estimator.
\begin{align*}
(\widehat{\sigma}^2)^2 = \frac{1}{n^2(n-1)^2} \left( n^2 (\sum_{i=1}^n Y_i^2)^2 - 2n ( \sum_{i=1}^n Y_i^2 ) (\sum_{i=1}^n Y_i)^2 + (\sum_{i=1}^n Y_i)^4 \right)
\end{align*}

Note that, the first term after taking expectation is
\begin{align*}
\bE\left[(\sum_{i=1}^n Y_i^2)^2\right] = n \bE\left[Y^4\right] + n(n-1) (\bE[Y^2])^2.
\end{align*}
The second term is
\begin{multline*}
\bE\left[(\sum_{i=1}^n Y_i^2) (\sum_{i=1}^n Y_i)^2\right] = n \bE\left[Y^4\right] + n(n-1) (\bE[Y^2])^2 + 2n(n-1) \bE\left[Y^3\right] \bE[Y] \\
+ n(n-1)(n-2) \bE\left[Y^2\right] (\bE[Y])^2.
\end{multline*}
The third term is
\begin{multline*}
\bE\left[(\sum_{i=1}^n Y_i)^4\right] = n \bE\left[Y^4\right] + 3 n(n-1) (\bE[Y^2])^2 + 4n(n-1) \bE\left[Y^3\right] \bE[Y] \\
+ 6n(n-1)(n-2) \bE\left[Y^2\right] (\bE[Y])^2 + n(n-1)(n-2)(n-3) (\bE[Y])^4.
\end{multline*}

Due to linearity of expectations and merging common terms,
\begin{multline}
\bE\left[(\widehat{\sigma}^2)^2\right] = \frac{1}{n^2(n-1)^2} \Bigg( n(n-1)^2 \bE\left[Y^4\right] - 4n(n-1)^2 \bE\left[Y^3\right] \bE[Y] + n(n-1)(n^2-2n+3) (\bE[Y^2])^2 \\
- 2n(n-1)(n-2)(n-3) \bE\left[Y^2\right] (\bE[Y])^2 + n(n-1)(n-2)(n-3) (\bE[Y])^4 \Bigg) \label{eqn:EC:ProofForKurtosis}
\end{multline}

Note that, 
\begin{align*}
\bE\left[ (Y - \bE[Y])^4 \right] = \ \bE\left[Y^4\right] - 4 \bE\left[Y^3\right] \bE[Y] + 6 \left[Y^2\right] (\bE[Y])^2 - 3 \left(\bE[Y]\right)^4,
\end{align*}
and
\begin{align*}
\left(\bE\left[ (Y - \bE[Y])^2 \right]\right)^2 := \sigma^4 = \ \bE\left[Y^2\right]^2 - \bE\left[Y^2\right] (\bE[Y])^2 + \left(\bE[Y]\right)^4.
\end{align*}

Putting the above two expressions into \eqref{eqn:EC:ProofForKurtosis} we have
\begin{align*}
\bE\left[(\widehat{\sigma}^2)^2\right] = & \ \frac{1}{n^2(n-1)^2} \Bigg( n(n-1)^2 \bE\left[ (Y - \bE[Y])^4 \right] + n(n-1)(n^2-2n+3) \left(\bE\left[ (Y - \bE[Y])^2 \right]\right)^2 \Bigg) \\
= & \ \frac{1}{n} \bE\left[ (Y - \bE[Y])^4 \right] + \frac{n^2-2n+3}{n(n-1)} \sigma^4
\end{align*}
which finishes the proof.
\hfill \halmos
\endproof

\subsubsection{Proof of Lemma~\ref{lem:LightTail}}
\proof{Proof of Lemma~\ref{lem:LightTail}.}
We prove the first inequality, and the second follows similarly.

Due to Chebyshev inequality,
\begin{align}
\Pr\left( \vert \widehat{\sigma}^2_m(1) - \sigma^2(1) \vert \geq \delta \right) \leq \frac{\bE\left[ (\widehat{\sigma}^2_m(1) - \sigma^2(1))^2 \right]}{\delta^2}. \label{eqn:Chebyshev}
\end{align}

Note that,
\begin{align*}
\bE\left[ (\widehat{\sigma}^2_m(1) - \sigma^2(1))^2 \right] = & \ \bE\left[ (\widehat{\sigma}^2_m(1))^2 \right] - \bE[\sigma^4(1)].
\end{align*}

Using Lemma~\ref{lem:EC:Kurtosis}, the above can be expressed as
\begin{align*}
\bE\left[ (\widehat{\sigma}^2_m(1) - \sigma^2(1))^2 \right] = & \ \frac{1}{\sum_{l=1}^m T_l(1)} \kappa(1) \sigma^4(1) - \frac{\sum_{l=1}^m T_l(1) - 3}{(\sum_{l=1}^m T_l(1))(\sum_{l=1}^m T_l(1) - 1)} \sigma^4(1) \\
= & \ \frac{\kappa(1) - 1}{\sum_{l=1}^m T_l(1)} \sigma^4(1) + \frac{2}{(\sum_{l=1}^m T_l(1))(\sum_{l=1}^m T_l(1) - 1)} \sigma^4(1) \\
\leq & \ \frac{\kappa(1) - 1}{\sum_{l=1}^m T_l(1)} \sigma^4(1) + \frac{3}{(\sum_{l=1}^m T_l(1))^2} \sigma^4(1) \\
\leq & \ \frac{\kappa(1)}{\sum_{l=1}^m T_l(1)} \sigma^4(1),
\end{align*}
where the last two inequalities are due to $\sum_{l=1}^m T_l(1) \geq 3$.
Putting this inequality back to \eqref{eqn:Chebyshev} we finish the proof.
\hfill \halmos
\endproof

\subsubsection{Proof of Lemma~\ref{lem:ExponentialTail}}
\proof{Proof of Lemma~\ref{lem:ExponentialTail}.}
The proof is by applying the bounded difference inequality.

First, denote $N = \sum_{l=1}^m T_l(1)$ as a short-hand notion.
Denote $\phi(Y_1, ..., Y_{N}) = \widehat{\sigma}^2_m(1)$ to emphasize the dependence on all the potential outcomes up to $N$.
Conditional on $W_1, ..., W_{N}$, we distinguish between two cases.
If $W_i = 0$, then 
\begin{align*}
\phi(Y_1, ..., Y_i, ..., Y_{N}) - \phi(Y_1, ..., Y'_i, ..., Y_{N}) = 0.
\end{align*}
If $W_i = 1$, then
\begin{align}
& \phi(Y_1, ..., Y_i, ..., Y_{N}) - \phi(Y_1, ..., Y'_i, ..., Y_{N}) \nonumber \\
= & \ \frac{1}{N-1} \sum_{\substack{t: W_t = 1 \\ t \ne i}} \bigg( Y_t - \frac{1}{N} \sum_{\substack{t': W_{t'} = 1 \\ t' \ne i}} Y_{t'} - \frac{1}{N} Y_i \bigg)^2 + \frac{1}{N-1} \bigg( Y_i - \frac{1}{N} \sum_{\substack{t': W_{t'} = 1 \\ t' \ne i}} Y_{t'} - \frac{1}{N} Y_i \bigg)^2 \nonumber \\
& \quad - \frac{1}{N-1} \sum_{\substack{t: W_t = 1 \\ t \ne i}} \bigg( Y_t - \frac{1}{N} \sum_{\substack{t': W_{t'} = 1 \\ t' \ne i}} Y_{t'} - \frac{1}{N} Y'_i \bigg)^2 - \frac{1}{N-1} \bigg( Y'_i - \frac{1}{N} \sum_{\substack{t': W_{t'} = 1 \\ t' \ne i}} Y_{t'} - \frac{1}{N} Y'_i \bigg)^2 \nonumber \\
= & \ \frac{1}{N-1} \sum_{\substack{t: W_t = 1 \\ t \ne i}} \bigg\{ \frac{2}{N}\bigg( Y_t-\frac{1}{N}\sum_{\substack{t': W_{t'} = 1 \\ t' \ne i}} Y_{t'}\bigg)\left(Y'_i-Y_i\right) + \frac{1}{N^2} \left(Y_i^2-(Y'_i)^2\right) \bigg\} \label{eqn:EC:ExponentialSmall:TwoParts1} \\
& \quad + \frac{1}{N-1} \bigg\{ \bigg(\frac{N-1}{N} Y_i - \frac{1}{N}\sum_{\substack{t': W_{t'} = 1 \\ t' \ne i}} Y_{t'} \bigg)^2 - \bigg(\frac{N-1}{N} Y'_i - \frac{1}{N}\sum_{\substack{t': W_{t'} = 1 \\ t' \ne i}} Y_{t'} \bigg)^2 \bigg\}. \label{eqn:EC:ExponentialSmall:TwoParts2}
\end{align}

To start with \eqref{eqn:EC:ExponentialSmall:TwoParts1}, we see that it is equal to 
\begin{align*}
& \ \frac{1}{N-1} \sum_{\substack{t: W_{t} = 1 \\ t \ne i}} \frac{2}{N} (Y'_i - Y_i) \bigg( Y_t - \frac{1}{N} \sum_{\substack{t': W_{t'} = 1 \\ t' \ne i}} Y_{t'} \bigg) + \frac{1}{N-1} \frac{N-1}{N^2} \left(Y_i^2-(Y'_i)^2\right) \\
= & \ \frac{2}{N(N-1)} (Y'_i - Y_i) \bigg( \sum_{\substack{t: W_{t} = 1 \\ t \ne i}} Y_{t} - \frac{N-1}{N} \sum_{\substack{t': W_{t'} = 1 \\ t' \ne i}} Y_{t'} \bigg) + \frac{1}{N^2} \left(Y_i^2-(Y'_i)^2\right) \\
= & \ \frac{2}{N^2(N-1)} (Y'_i - Y_i) \sum_{\substack{t: W_{t} = 1 \\ t \ne i}} Y_{t} + \frac{1}{N^2} \left(Y_i^2-(Y'_i)^2\right).
\end{align*}

Next, focusing on \eqref{eqn:EC:ExponentialSmall:TwoParts2}, we see that it is equal to 
\begin{align*}
& \ \frac{1}{N-1} \bigg\{ \frac{(N-1)^2}{N^2} \left(Y_i^2-(Y'_i)^2\right) + 2 \frac{N-1}{N} \frac{1}{N} (Y'_i - Y_i) \sum_{\substack{t': W_{t'} = 1 \\ t' \ne i}} Y_{t'} \bigg\} \\
= & \ \frac{N-1}{N^2} \left(Y_i^2-(Y'_i)^2\right) + \frac{2}{N^2} (Y'_i - Y_i) \sum_{\substack{t: W_{t} = 1 \\ t \ne i}} Y_{t}.
\end{align*}

Combining both parts, we have
\begin{align*}
\vert \phi(Y_1, ..., Y_i, ..., Y_{N}) - \phi(Y_1, ..., Y'_i, ..., Y_{N}) \vert = & \ \bigg\vert \frac{1}{N} \left(Y_i^2-(Y'_i)^2\right) + \frac{2}{N(N-1)} (Y'_i - Y_i) \sum_{\substack{t: W_{t} = 1 \\ t \ne i}} Y_{t} \bigg\vert \\
= & \ \frac{1}{N} \bigg\vert (Y'_i - Y_i) \bigg( \frac{2}{N-1} \sum_{\substack{t: W_{t} = 1 \\ t \ne i}} Y_{t} - (Y'_i+Y_i) \bigg) \bigg\vert.
\end{align*}

Note that for any $x,y,z \in [-V,V]$, we have
\begin{align*}
\vert (x-y)(2z - (x+y)) \vert \leq & \ \max\left\{ \vert (x-y)(2V-(x+y)) \vert, \vert (x-y)(-2V-(x+y)) \vert \right\} \\
= & \ \max\left\{ \vert 2(x-y)V-(x^2-y^2) \vert, \vert 2(x-y)V+(x^2-y^2) \vert \right\} \\
\leq & \ 4 V^2,
\end{align*}
where the first inequality is because the function is monotone with respect to $z$;
the last inequality is because both functions are monotone with respect to $x$ and $y$.
Replacing $x = Y'_i$, $y = Y_i$, $z = \frac{1}{N-1} \sum_{\substack{t: W_{t} = 1 \\ t \ne i}} Y_{t}$, and $V = C \sigma(1)$ into the above inequality, we have
\begin{align*}
\vert \phi(Y_1, ..., Y_i, ..., Y_{N}) - \phi(Y_1, ..., Y'_i, ..., Y_{N}) \vert \leq \frac{4 C^2 \sigma^2(1)}{N},
\end{align*}
which finishes discussing the case of $W_i=1$.

Using the bounded difference inequality \citep{boucheron2013concentration, mcdiarmid1989method}, 
\begin{multline*}
\Pr\left( \vert \widehat{\sigma}^2_m(1) - \sigma^2(1) \vert \geq \delta \right) \leq 2 \exp\left\{-\frac{2 \delta^2}{\sum_{t: W_t=1} \frac{16 C^4 \sigma^4(1)}{N^2}}\right\} \\
= 2 \exp\left\{-\frac{\delta^2 N}{8 C^4 \sigma^4(1)}\right\} = 2 \exp\left\{-\frac{\delta^2 \left(\sum_{l=1}^m T_l(1)\right)}{8 C^4 \sigma^4(1)}\right\}.
\end{multline*}

Similarly, we can show
\begin{align*}
\Pr\left( \vert \widehat{\sigma}^2_m(0) - \sigma^2(0) \vert \geq \delta \right) \leq 2 \exp\left\{-\frac{\delta^2 \left(\sum_{l=1}^m T_l(0)\right)}{8 C^4 \sigma^4(0)}\right\},
\end{align*}
which finishes the proof.
\hfill \halmos
\endproof

\subsubsection{Proof of Lemma~\ref{lem:CommonKnowledge:SumOfSquares}}

Lemma~\ref{lem:CommonKnowledge:SumOfSquares} is very common knowledge. 
We provide a proof below for completeness.

\proof{Proof of Lemma~\ref{lem:CommonKnowledge:SumOfSquares}.}
Denote $\bar{X} = \frac{1}{n} \sum_{i=1}^n X_i$.
\begin{align*}
\sum_{i=1}^n \sum_{j=1}^n (X_i - X_j)^2 & = \sum_{i=1}^n \sum_{j=1}^n \Big( (X_i - \bar{X}) - (X_j - \bar{X}) \Big)^2 \\
& = \sum_{i=1}^n \sum_{j=1}^n \Big( (X_i - \bar{X})^2 + (X_j - \bar{X})^2 - 2 (X_i - \bar{X}) (X_j - \bar{X}) \Big) \\
& = \sum_{i=1}^n \Big( n (X_i - \bar{X})^2 + \sum_{j=1}^n (X_j - \bar{X})^2 - 2 (X_i - \bar{X}) (\sum_{j=1}^n X_j - n \bar{X}) \Big) \\
& = \sum_{i=1}^n n (X_i - \bar{X})^2 + n \sum_{j=1}^n (X_j - \bar{X})^2 \\
& = 2n \sum_{i=1}^n (X_i - \bar{X})^2.
\end{align*}
\hfill \halmos

\subsection{Derivations of Equations in Sections~\ref{sec:Setup} and~\ref{sec:Optimization}}

In the main paper, we did not provide proofs to \eqref{eqn:ProxyVariance} and \eqref{eqn:ClairvoyantOptimal} because they are very well-known.
For completeness, we provide proofs to the derivation of expressions \eqref{eqn:ProxyVariance} and \eqref{eqn:ClairvoyantOptimal} here.

\subsubsection*{Derivation of \eqref{eqn:ProxyVariance}.}
Consider the case when $T(1)$ and $T(0)$ are fixed.
Note that there are two sources of randomness: the treatment assignments are random, and the potential outcomes are also random.
Using the law of total variance,
\begin{align*}
\Var(\widehat{\tau}) = \bE\left[ \Var\left( \widehat{\tau} \vert W_1, ..., W_T \right) \right] + \Var\left( \bE\left[ \widehat{\tau} \vert W_1, ..., W_T \right] \right).
\end{align*}

We derive both terms separately.
First,
\begin{align*}
\Var\left(\widehat{\tau} \vert W_1, ..., W_T\right) = & \frac{1}{T(1)^2} \Var\left( \sum_{t: W_t = 1} Y_t(1) \right) + \frac{1}{T(0)^2} \Var\left( \sum_{t: W_t = 0} Y_t(0) \right) \\
= & \frac{1}{T(1)^2} \cdot T(1) \cdot \Var(Y(1)) + \frac{1}{T(0)^2} \cdot T(0) \cdot \Var(Y(0)) \\
= & \frac{1}{T(1)} \sigma^2(1) + \frac{1}{T(0)} \sigma^2(0).
\end{align*}
Since the expression of $\Var\left(\widehat{\tau} \vert W_1, ..., W_T\right)$ only directly depends on $T(1)$ and $T(0)$ but not directly on $W_1, ..., W_T$,
\begin{align*}
\bE\left[ \Var\left( \widehat{\tau} \vert W_1, ..., W_T \right) \right] = & \frac{1}{T(1)} \sigma^2(1) + \frac{1}{T(0)} \sigma^2(0).
\end{align*}

Second,
\begin{align*}
\bE\left[\widehat{\tau} \vert W_1, ..., W_T\right] = & \frac{1}{T(1)} \bE\left[ \sum_{t: W_t = 1} Y_t(1) \right] - \frac{1}{T(0)} \bE\left[ \sum_{t: W_t = 0} Y_t(0) \right] \\
= & \bE[Y(1)] - \bE[Y(0)].
\end{align*}
Since the expression of $\bE\left[\widehat{\tau} \vert W_1, ..., W_T\right]$ does not depend on $W_1, ..., W_T$,
\begin{align*}
\Var\left( \bE\left[ \widehat{\tau} \vert W_1, ..., W_T \right] \right) = & 0.
\end{align*}

Combining both parts, 
\begin{align*}
\Var(\widehat{\tau}) = \frac{1}{T(1)} \sigma^2(1) + \frac{1}{T(0)} \sigma^2(0).
\end{align*}

\subsubsection*{Derivation of \eqref{eqn:ClairvoyantOptimal}.}
Consider the following problem:
\begin{align*}
\inf_{0 < x < T} & \ \frac{1}{T - x} \sigma^2(1) + \frac{1}{x} \sigma^2(0).
\end{align*}

Consider the first order condition, which leads to
\begin{align*}
\frac{1}{(T-x)^2} \sigma^2(1) - \frac{1}{x^2} \sigma^2(0) = 0.
\end{align*}
Simplifying terms this reduces to
\begin{align*}
x = \frac{\sigma^(0)}{\sigma(1) + \sigma(0)} T.
\end{align*}
And the optimal objective value is
\begin{align*}
\frac{1}{T - \frac{\sigma^(0)}{\sigma(1) + \sigma(0)} T} \sigma^2(1) + \frac{1}{\frac{\sigma^(0)}{\sigma(1) + \sigma(0)} T} \sigma^2(0) = \ \frac{1}{T} \cdot (\sigma(1) + \sigma(0))^2.
\end{align*}

\subsection{Proof of Theorem~\ref{thm:OneStage}}
\label{sec:proof:thm:OneStage}

\proof{Proof of Theorem~\ref{thm:OneStage}.}
Since this is a single stage experiment, we use $V(T(1), T(0))$ instead of $\bE[V(T(1), T(0))]$.
Suppose the optimal solution is not $T(1) = T(0) = T/2$.
Without loss of generality, assume the optimal solution is such that $T(1) > T(0) > 0$.
Then for any $(T(1), T(0))$, the worst case $\sigma(1), \sigma(0)$ should solve the following problem,
\begin{align}
\sup_{\cF \in \sP} \ \frac{V(T(1), T(0))}{V(T^*(1), T^*(0))}. \label{eqn:proof:thm:OneStage}
\end{align}

Using \eqref{eqn:ClairvoyantOptimal}, the above expression can be re-written as
\begin{align*}
\frac{V(T(1), T(0))}{V(T^*(1), T^*(0))} = \frac{\frac{1}{T(1)} \sigma^2(1) + \frac{1}{T(0)} \sigma^2(0)}{\frac{1}{T} \cdot (\sigma(1) + \sigma(0))^2}.
\end{align*}

When $\sigma(0) \ne 0$, denote $\rho = \sigma(1) /  \sigma(0) \in [0, +\infty)$.
Further denote 
\begin{align*}
g(\rho) = & \ \frac{\frac{1}{T(1)} \sigma^2(1) + \frac{1}{T(0)} \sigma^2(0)}{\frac{1}{T} \cdot (\sigma(1) + \sigma(0))^2} \\
= & \ \frac{\frac{1}{T(1)} \rho^2 + \frac{1}{T(0)}}{\frac{1}{T} \cdot (\rho + 1)^2} \\
= & \ \frac{T}{T(1)} \cdot \frac{\rho^2}{(\rho+1)^2} + \frac{T}{T(0)} \cdot \frac{1}{(\rho+1)^2}
\end{align*}

Taking first order derivative,
\begin{align*}
g'(\rho) = & \frac{2T}{(\rho+1)^3} \cdot \left( \frac{\rho}{T(1)} - \frac{1}{T(0)} \right).
\end{align*}

So $g(\rho)$ is an increasing function when $\rho > T(1) / T(0)$, and an decreasing function when $\rho < T(1) / T(0)$.
The maximum value of $g(\rho)$ is taken when either $\rho = 0$ or $\rho \to +\infty$.
Denote $g(+\infty) = \lim_{\rho \to +\infty} g(\rho)$.

Putting the above back to \eqref{eqn:proof:thm:OneStage}, we have, for any $(T(1), T(0))$ such that $T(1) > T(0) > 0$,
\begin{align*}
\sup_{\cF \in \sP} \ \frac{V(T(1), T(0))}{V(T^*(1), T^*(0))} = \max\{g(0), g(+\infty)\} = \max\left\{ \frac{T}{T(1)}, \frac{T}{T(0)} \right\} > 2,
\end{align*}
where the last inequality holds because $T(0) < T/2$.
This suggests that, if $T(1) > T(0) > 0$, then 
\begin{align*}
\inf_{\pi \in \Pi_0} \ \sup_{\cF \in \sP} \ \frac{V(T(1), T(0))}{V(T^*(1), T^*(0))} > 2.
\end{align*}

On the other hand, when $T(1) = T(0) = T/2$.
For any $\sigma(1), \sigma(0)$,
\begin{align}
\frac{V(\frac{T}{2}, \frac{T}{2})}{V(T^*(1), T^*(0))} = \frac{\frac{2}{T} \cdot (\sigma^2(1) + \sigma^2(0))}{\frac{1}{T} \cdot (\sigma(1) + \sigma(0))^2} = 2 \frac{\sigma^2(1) + \sigma^2(0) \ }{(\sigma(1) + \sigma(0))^2} \leq 2. \label{eqn:proof:opt}
\end{align}
This suggests that
\begin{align*}
\sup_{\cF \in \sP} \ \frac{V(\frac{T}{2}, \frac{T}{2})}{V(T^*(1), T^*(0))} = 2.
\end{align*}

Combining both cases, the optimal solution must be $T(1) = T(0) = T/2$.

To prove the second part of the Theorem, we focus on the inequality in \eqref{eqn:proof:opt}.
The inequality holds when either $\sigma(1) = 0$ or $\sigma(0) = 0$.
\hfill \halmos
\endproof

\subsection{Proof of Theorem~\ref{thm:2StageANA}}
\label{sec:proof:thm:2StageANA}

\proof{Proof of Theorem~\ref{thm:2StageANA}.}
Without loss of generality, we assume $\sigma(1) \geq \sigma(0)$ throughout the proof.
Our analysis of the two-stage adaptive Neyman allocation (Algorithm~\ref{alg:2StageANA}) will be based on the following two events.
\begin{align*}
\cE_1(1) = & \ \bigg\{ \left| \widehat{\sigma}^2_1(1) - \sigma^2(1) \right| < 2^{\frac{1}{2}} T^{-\frac{1}{4} + \frac{\eps}{2}} \sigma^2(1)\bigg\}, \\
\cE_1(0) = & \ \bigg\{ \left| \widehat{\sigma}^2_1(0) - \sigma^2(0) \right| < 2^{\frac{1}{2}} T^{-\frac{1}{4} + \frac{\eps}{2}} \sigma^2(0)\bigg\}.
\end{align*}
Denote $\cE = \cE_1(1) \cap \cE_1(0)$. Then $\Pr(\cE) = \Pr(\cE_1(1) \cap \cE_1(0)) \geq 1 - \Pr(\overline{\cE}_1(1)) - \Pr(\overline{\cE}_1(0))$.
We further have
\begin{align*}
\Pr(\cE) = & \ 1 - \Pr\left( \vert \widehat{\sigma}^2_1(1) - \sigma^2(1) \vert \geq 2^{\frac{1}{2}} T^{-\frac{1}{4} + \frac{\eps}{2}} \sigma^2(1) \right) - \Pr\left( \vert \widehat{\sigma}^2_1(0) - \sigma^2(0) \vert \geq 2^{\frac{1}{2}} T^{-\frac{1}{4} + \frac{\eps}{2}} \sigma^2(0) \right) \\
\geq & \ 1 - \frac{\kappa(1) \sigma^4(1)}{2 T^{-\frac{1}{2} + \eps} \sigma^4(1) T_1(1)} - \frac{\kappa(0) \sigma^4(0)}{2 T^{-\frac{1}{2} + \eps} \sigma^4(0) T_1(0)} \\
= & \ 1 - \frac{\kappa(1) + \kappa(0)}{T^{\eps}},
\end{align*}
where the inequality is due to Lemma~\ref{lem:LightTail}.

Conditional on the event $\cE$, we have
\begin{subequations}
\begin{align}
\sigma^2(1) \left( 1 - 2^{\frac{1}{2}} T^{-\frac{1}{4} + \frac{\eps}{2}} \right) \ \leq \ \widehat{\sigma}^2_1(1) \ \leq \ \sigma^2(1) \left( 1 + 2^{\frac{1}{2}} T^{-\frac{1}{4} + \frac{\eps}{2}} \right), \label{eqn:2Stage:ConfidenceBound1} \\
\sigma^2(0) \left( 1 - 2^{\frac{1}{2}} T^{-\frac{1}{4} + \frac{\eps}{2}} \right) \ \leq \ \widehat{\sigma}^2_1(0) \ \leq \ \sigma^2(0) \left( 1 + 2^{\frac{1}{2}} T^{-\frac{1}{4} + \frac{\eps}{2}} \right). \label{eqn:2Stage:ConfidenceBound0} 
\end{align}
\end{subequations}
Due to \eqref{eqn:2Stage:ConfidenceBound1} and \eqref{eqn:2Stage:ConfidenceBound0}, and given that $\sigma(1), \sigma(0) > 0$, we have $\widehat{\sigma}^2_1(1), \widehat{\sigma}^2_1(0) > 0$.
Denote $\rho = \frac{\sigma(1)}{\sigma(0)}$ and $\widehat{\rho} = \frac{\widehat{\sigma}_1(1)}{\widehat{\sigma}_1(0)}$.

Now we distinguish two cases, and discuss these two cases separately.
\begin{enumerate}
\item \textbf{Case 1}: 
\begin{align*}
\frac{\frac{1}{2}T^{\frac{1}{2}}}{T - \frac{1}{2}T^{\frac{1}{2}}} \leq \rho = \frac{\sigma(1)}{\sigma(0)} \leq \frac{T - \frac{1}{2}T^{\frac{1}{2}}}{\frac{1}{2}T^{\frac{1}{2}}}.
\end{align*}
\item \textbf{Case 2}:
\begin{align*}
\rho = \frac{\sigma(1)}{\sigma(0)} > \frac{T - \frac{1}{2}T^{\frac{1}{2}}}{\frac{1}{2}T^{\frac{1}{2}}}.
\end{align*}
\end{enumerate}
Note that, for case 2, we do not discuss $\rho = \frac{\sigma(1)}{\sigma(0)} < \frac{\frac{1}{2}T^{\frac{1}{2}}}{T - \frac{1}{2}T^{\frac{1}{2}}}$, because we assume that $\sigma(1) \geq \sigma(0)$.
For each of the above two cases, we further discuss two sub-cases.
The remaining of the proof is structured as enumerating all four cases.
After enumerating all four sub-cases we finish the proof.

\noindent\textbf{Case 1.1}:
\begin{align*}
\frac{\frac{1}{2}T^{\frac{1}{2}}}{T - \frac{1}{2}T^{\frac{1}{2}}} \leq \rho \leq \frac{T - \frac{1}{2}T^{\frac{1}{2}}}{\frac{1}{2}T^{\frac{1}{2}}}, && \text{and} && \frac{\frac{1}{2}T^{\frac{1}{2}}}{T - \frac{1}{2}T^{\frac{1}{2}}} \leq \widehat{\rho} \leq \frac{T - \frac{1}{2}T^{\frac{1}{2}}}{\frac{1}{2}T^{\frac{1}{2}}}.
\end{align*}
Since $\frac{\frac{1}{2}T^{\frac{1}{2}}}{T - \frac{1}{2}T^{\frac{1}{2}}} \leq \widehat{\rho} \leq \frac{T - \frac{1}{2}T^{\frac{1}{2}}}{\frac{1}{2}T^{\frac{1}{2}}}$, we have
\begin{align*}
\frac{\widehat{\sigma}_1(1)}{\widehat{\sigma}_1(1) + \widehat{\sigma}_1(0)} T \geq \frac{1}{1+\frac{T - \frac{1}{2}T^{\frac{1}{2}}}{\frac{1}{2}T^{\frac{1}{2}}}} \ T= \frac{1}{2}T^{\frac{1}{2}}, \\
\frac{\widehat{\sigma}_1(0)}{\widehat{\sigma}_1(1) + \widehat{\sigma}_1(0)} T \geq \frac{1}{\frac{T - \frac{1}{2}T^{\frac{1}{2}}}{\frac{1}{2}T^{\frac{1}{2}}}+1} \ T= \frac{1}{2}T^{\frac{1}{2}}.
\end{align*}
Due to this, Algorithm~\ref{alg:2StageANA} goes to Line 3 instead of Line 5 or Line 7.
The total numbers of treated and control units are given by \eqref{eqn:EstimatedOPT}. 
We re-write \eqref{eqn:EstimatedOPT} again as follows,
\begin{align*}
(T(1), T(0)) = (\frac{\widehat{\sigma}_1(1)}{\widehat{\sigma}_1(1) + \widehat{\sigma}_1(0)} T, \frac{\widehat{\sigma}_1(0)}{\widehat{\sigma}_1(1) + \widehat{\sigma}_1(0)} T).
\end{align*}
With a little abuse of notation, we write $V(T(1), T(0)\vert \cE)$ to stand for $V(T(1), T(0))$, where we emphasize that this is a random quantity (as $T(1)$ and $T(0)$ are random) that is conditional on event $\cE$.
Putting $(T(1), T(0))$ into \eqref{eqn:Obj}, we have, for any $\sigma(1), \sigma(0)$,
\begin{align}
\frac{V(T(1), T(0) \vert \cE)}{V(T^*(1), T^*(0))} = & \ \frac{\frac{1}{T(1)} \sigma^2(1) + \frac{1}{T(0)} \sigma^2(0)}{\frac{1}{T} (\sigma(1) + \sigma(0))^2} \nonumber \\
= & \ \frac{\left(1+\frac{\widehat{\sigma}_1(0)}{\widehat{\sigma}_1(1)}\right) \sigma^2(1) + \left(1+\frac{\widehat{\sigma}_1(1)}{\widehat{\sigma}_1(0)}\right) \sigma^2(0)}{(\sigma(1) + \sigma(0))^2} \nonumber \\
= & \ \frac{\sigma^2(1) + \sigma^2(0) + \frac{1}{\widehat{\rho}} \ \sigma^2(1) + \widehat{\rho} \ \sigma^2(1)}{(\sigma(1) + \sigma(0))^2} \nonumber \\
= & \ 1 + \frac{1}{(\sigma(1) + \sigma(0))^2}\left( \frac{1}{\widehat{\rho}} \ \sigma^2(1) + \widehat{\rho} \ \sigma^2(1) - 2 \sigma(1) \sigma(0) \right) \label{eqn:proof:2Stage:1}
\end{align}
Due to Lemma~\ref{lem:h:rhohat}, and using \eqref{eqn:2Stage:ConfidenceBound1} and \eqref{eqn:2Stage:ConfidenceBound0},
\begin{multline}
\frac{1}{(\sigma(1) + \sigma(0))^2}\left( \frac{1}{\widehat{\rho}} \ \sigma^2(1) + \widehat{\rho} \ \sigma^2(1) - 2 \sigma(1) \sigma(0) \right) \\
\leq \ \frac{\sigma(1) \sigma(0)}{(\sigma(1) + \sigma(0))^2} \left(\sqrt{\frac{1-2^{\frac{1}{2}} T^{-\frac{1}{4} + \frac{\eps}{2}}}{1+2^{\frac{1}{2}} T^{-\frac{1}{4} + \frac{\eps}{2}}}} + \sqrt{\frac{1+2^{\frac{1}{2}} T^{-\frac{1}{4} + \frac{\eps}{2}}}{1-2^{\frac{1}{2}} T^{-\frac{1}{4} + \frac{\eps}{2}}}} - 2\right). \label{eqn:proof:2Stage:2}
\end{multline}
Note that 
\begin{align}
\frac{\sigma(1) \sigma(0)}{(\sigma(1) + \sigma(0))^2} \leq \frac{1}{4}. \label{eqn:proof:2Stage:3}
\end{align}
Note also that
\begin{align}
\sqrt{\frac{1-2^{\frac{1}{2}} T^{-\frac{1}{4} + \frac{\eps}{2}}}{1+2^{\frac{1}{2}} T^{-\frac{1}{4} + \frac{\eps}{2}}}} + \sqrt{\frac{1+2^{\frac{1}{2}} T^{-\frac{1}{4} + \frac{\eps}{2}}}{1-2^{\frac{1}{2}} T^{-\frac{1}{4} + \frac{\eps}{2}}}} - 2 = & \ \frac{2}{\sqrt{1 - 2 T^{-\frac{1}{2} + \eps}}} - 2 \nonumber \\
= & \ 2 \left(1 - 2 T^{-\frac{1}{2} + \eps}\right)^{-\frac{1}{2}} - 2 \nonumber \\
\leq & \ 2 \left( 1 + 2 T^{-\frac{1}{2} + \eps} \right) - 2 \nonumber \\
= & \ 4 T^{-\frac{1}{2} + \eps}, \label{eqn:proof:2Stage:4}
\end{align}
where the inequality holds when $T^{\frac{1}{2} - \eps} \geq 2$. 
This is because $T \geq 16$ and $\eps \in (0,\frac{1}{8})$, so we have $T^{\frac{1}{2} - \eps} \geq T^{\frac{1}{4}} \geq 2$.

Combining \eqref{eqn:proof:2Stage:1} --- \eqref{eqn:proof:2Stage:4}, we have 
\begin{align*}
\frac{V(T(1), T(0) \vert \cE)}{V(T^*(1), T^*(0))} \leq 1 + T^{-\frac{1}{2} + \eps}.
\end{align*}

\noindent\textbf{Case 1.2}:
\begin{align*}
\frac{\frac{1}{2}T^{\frac{1}{2}}}{T - \frac{1}{2}T^{\frac{1}{2}}} \leq \rho \leq \frac{T - \frac{1}{2}T^{\frac{1}{2}}}{\frac{1}{2}T^{\frac{1}{2}}}, && \text{but} && \widehat{\rho} > \frac{T - \frac{1}{2}T^{\frac{1}{2}}}{\frac{1}{2}T^{\frac{1}{2}}} \ \text{or} \ \widehat{\rho} < \frac{\frac{1}{2}T^{\frac{1}{2}}}{T - \frac{1}{2}T^{\frac{1}{2}}}.
\end{align*}
If $\widehat{\rho} > \frac{T - \frac{1}{2}T^{\frac{1}{2}}}{\frac{1}{2}T^{\frac{1}{2}}}$, then 
\begin{align*}
\frac{\widehat{\sigma}_1(0)}{\widehat{\sigma}_1(1) + \widehat{\sigma}_1(0)} T < \frac{1}{\frac{T - \frac{1}{2}T^{\frac{1}{2}}}{\frac{1}{2}T^{\frac{1}{2}}}+1} \ T= \frac{1}{2}T^{\frac{1}{2}}.
\end{align*}
Due to this, Algorithm~\ref{alg:2StageANA} goes to Line 7.
The total numbers of treated and control units are given by $(T(1), T(0)) = (T - \frac{1}{2}T^{\frac{1}{2}}, \frac{1}{2}T^{\frac{1}{2}})$.

Note that, conditional on event $\cE$,
\begin{align*}
\rho \ = \ \frac{\sigma(1)}{\sigma(0)} \ \leq \ \frac{T - \frac{1}{2}T^{\frac{1}{2}}}{\frac{1}{2}T^{\frac{1}{2}}} \ < \ \widehat{\rho} \ \leq \ \frac{\sigma(1)}{\sigma(0)} \sqrt{\frac{1+2^{\frac{1}{2}} T^{-\frac{1}{4} + \frac{\eps}{2}}}{1-2^{\frac{1}{2}} T^{-\frac{1}{4} + \frac{\eps}{2}}}}.
\end{align*}
Putting $(T(1), T(0))$ into \eqref{eqn:Obj}, we have, for any $\sigma(1), \sigma(0)$,
\begin{align*}
\frac{V(T(1), T(0) \vert \cE)}{V(T^*(1), T^*(0))} = & \ \frac{\frac{1}{T(1)} \sigma^2(1) + \frac{1}{T(0)} \sigma^2(0)}{\frac{1}{T} (\sigma(1) + \sigma(0))^2} \\
= & \ \frac{\sigma^2(1) + \sigma^2(0) + \frac{\frac{1}{2}T^{\frac{1}{2}}}{T - \frac{1}{2}T^{\frac{1}{2}}} \sigma^2(1) + \frac{T - \frac{1}{2}T^{\frac{1}{2}}}{\frac{1}{2}T^{\frac{1}{2}}} \sigma^2(0)}{(\sigma(1) + \sigma(0))^2} \\
< & \ \frac{\sigma^2(1) + \sigma^2(0) + \frac{\sigma(0)}{\sigma(1)} \sqrt{\frac{1-2^{\frac{1}{2}} T^{-\frac{1}{4} + \frac{\eps}{2}}}{1+2^{\frac{1}{2}} T^{-\frac{1}{4} + \frac{\eps}{2}}}} \sigma^2(1) + \frac{\sigma(1)}{\sigma(0)} \sqrt{\frac{1+2^{\frac{1}{2}} T^{-\frac{1}{4} + \frac{\eps}{2}}}{1-2^{\frac{1}{2}} T^{-\frac{1}{4} + \frac{\eps}{2}}}} \sigma^2(0)}{(\sigma(1) + \sigma(0))^2} \\
= & \ 1 + \frac{\sigma(1) \sigma(0)}{(\sigma(1) + \sigma(0))^2} \left(\sqrt{\frac{1-2^{\frac{1}{2}} T^{-\frac{1}{4} + \frac{\eps}{2}}}{1+2^{\frac{1}{2}} T^{-\frac{1}{4} + \frac{\eps}{2}}}} + \sqrt{\frac{1+2^{\frac{1}{2}} T^{-\frac{1}{4} + \frac{\eps}{2}}}{1-2^{\frac{1}{2}} T^{-\frac{1}{4} + \frac{\eps}{2}}}} - 2\right).
\end{align*}
where the inequality is due to Lemma~\ref{lem:h:rhohat}.
Combining this with \eqref{eqn:proof:2Stage:3} and \eqref{eqn:proof:2Stage:4} we have again
\begin{align*}
\frac{V(T(1), T(0) \vert \cE)}{V(T^*(1), T^*(0))} \leq 1 + T^{-\frac{1}{2} + \eps}.
\end{align*}

If $\widehat{\rho} < \frac{\frac{1}{2}T^{\frac{1}{2}}}{T - \frac{1}{2}T^{\frac{1}{2}}}$, then Algorithm~\ref{alg:2StageANA} goes to Line 5.
\begin{align*}
\frac{\sigma(1)}{\sigma(0)} \sqrt{\frac{1-2^{\frac{1}{2}} T^{-\frac{1}{4} + \frac{\eps}{2}}}{1+2^{\frac{1}{2}} T^{-\frac{1}{4} + \frac{\eps}{2}}}} \leq \widehat{\rho} \ < \ \frac{\frac{1}{2}T^{\frac{1}{2}}}{T - \frac{1}{2}T^{\frac{1}{2}}} \ \leq \ \rho \ = \ \frac{\sigma(1)}{\sigma(0)},
\end{align*}
and the same analysis follows similarly.

\noindent\textbf{Case 2.1}:
\begin{align*}
\rho > \frac{T - \frac{1}{2}T^{\frac{1}{2}}}{\frac{1}{2}T^{\frac{1}{2}}}, && \text{and} && \widehat{\rho} > \frac{T - \frac{1}{2}T^{\frac{1}{2}}}{\frac{1}{2}T^{\frac{1}{2}}}.
\end{align*}
Since $\widehat{\rho} > \frac{T - \frac{1}{2}T^{\frac{1}{2}}}{\frac{1}{2}T^{\frac{1}{2}}}$, we have
\begin{align*}
\frac{\widehat{\sigma}_1(0)}{\widehat{\sigma}_1(1) + \widehat{\sigma}_1(0)} T < \frac{1}{\frac{T - \frac{1}{2}T^{\frac{1}{2}}}{\frac{1}{2}T^{\frac{1}{2}}}+1} \ T= \frac{1}{2}T^{\frac{1}{2}}.
\end{align*}
Due to this, Algorithm~\ref{alg:2StageANA} goes to Line 7.
The total numbers of treated and control units are given by $(T(1), T(0)) = (T - \frac{1}{2}T^{\frac{1}{2}}, \frac{1}{2}T^{\frac{1}{2}})$.

Putting $(T(1), T(0))$ into \eqref{eqn:Obj}, we have, for any $\sigma(1), \sigma(0)$,
\begin{align}
\frac{V(T(1), T(0) \vert \cE)}{V(T^*(1), T^*(0))} = & \ \frac{\frac{1}{T(1)} \sigma^2(1) + \frac{1}{T(0)} \sigma^2(0)}{\frac{1}{T} (\sigma(1) + \sigma(0))^2} \nonumber \\
= & \ \frac{T}{T - \frac{1}{2}T^{\frac{1}{2}}} \cdot \frac{\sigma^2(1)}{(\sigma(1) + \sigma(0))^2} + \frac{T}{\frac{1}{2}T^{\frac{1}{2}}} \cdot \frac{\sigma^2(0)}{(\sigma(1) + \sigma(0))^2} \nonumber \\
= & \ \frac{T}{T - \frac{1}{2}T^{\frac{1}{2}}} \cdot \frac{\rho^2}{(\rho+1)^2} + \frac{T}{\frac{1}{2}T^{\frac{1}{2}}} \cdot \frac{1}{(\rho+1)^2}. \label{eqn:proof:2Stage:5}
\end{align}
Due to Lemma~\ref{lem:g:rho}, since $\rho = \frac{\sigma(1)}{\sigma(0)} > \frac{T - \frac{1}{2}T^{\frac{1}{2}}}{\frac{1}{2}T^{\frac{1}{2}}}$, we know that the expression in \eqref{eqn:proof:2Stage:5} is increasing with respect to $\rho$.
So we have
\begin{align*}
\frac{V(T(1), T(0) \vert \cE)}{V(T^*(1), T^*(0))} \leq \lim_{\rho \to +\infty} \left(\frac{T}{T - \frac{1}{2}T^{\frac{1}{2}}} \cdot \frac{\rho^2}{(\rho+1)^2} + \frac{T}{\frac{1}{2}T^{\frac{1}{2}}} \cdot \frac{1}{(\rho+1)^2}\right) = \frac{T}{T - \frac{1}{2}T^{\frac{1}{2}}} \leq 1 + T^{-\frac{1}{2}},
\end{align*}
where the last inequality holds because $T \geq 1$.

\noindent\textbf{Case 2.2}:
\begin{align*}
\rho > \frac{T - \frac{1}{2}T^{\frac{1}{2}}}{\frac{1}{2}T^{\frac{1}{2}}}, && \text{and} && \widehat{\rho} \leq \frac{T - \frac{1}{2}T^{\frac{1}{2}}}{\frac{1}{2}T^{\frac{1}{2}}}
\end{align*}
Note that,
\begin{align*}
\widehat{\sigma}_1(1) \geq & \ \sigma(1) \sqrt{1 - 2^{\frac{1}{2}} T^{-\frac{1}{4}+\frac{\eps}{2}}} \\
> & \ \sigma(0) \frac{T - \frac{1}{2}T^{\frac{1}{2}}}{\frac{1}{2}T^{\frac{1}{2}}} \sqrt{1 - 2^{\frac{1}{2}} T^{-\frac{1}{4}+\frac{\eps}{2}}} \\
\geq & \ \widehat{\sigma}_1(0) \frac{T - \frac{1}{2}T^{\frac{1}{2}}}{\frac{1}{2}T^{\frac{1}{2}}} \sqrt{\frac{1 - 2^{\frac{1}{2}} T^{-\frac{1}{4}+\frac{\eps}{2}}}{1 + 2^{\frac{1}{2}} T^{-\frac{1}{4}+\frac{\eps}{2}}}} \\
\geq & \ \widehat{\sigma}_1(0) \frac{\frac{1}{2}T^{\frac{1}{2}}}{T - \frac{1}{2}T^{\frac{1}{2}}}.
\end{align*}
where the first inequality is due to \eqref{eqn:2Stage:ConfidenceBound1};
the second inequality is due to $\rho > \frac{T - \frac{1}{2}T^{\frac{1}{2}}}{\frac{1}{2}T^{\frac{1}{2}}}$;
the third inequality is due to \eqref{eqn:2Stage:ConfidenceBound0};
the last inequality is due to Lemma~\ref{lem:AlgebraicTrick1}.

The above shows that, in this case (Case 2.2), 
\begin{align*}
\widehat{\rho} \geq \frac{\frac{1}{2}T^{\frac{1}{2}}}{T - \frac{1}{2}T^{\frac{1}{2}}}.
\end{align*}
Since $\frac{\frac{1}{2}T^{\frac{1}{2}}}{T - \frac{1}{2}T^{\frac{1}{2}}} \leq \widehat{\rho} \leq \frac{T - \frac{1}{2}T^{\frac{1}{2}}}{\frac{1}{2}T^{\frac{1}{2}}}$, we have
\begin{align*}
\frac{\widehat{\sigma}_1(1)}{\widehat{\sigma}_1(1) + \widehat{\sigma}_1(0)} T \geq \frac{1}{1+\frac{T - \frac{1}{2}T^{\frac{1}{2}}}{\frac{1}{2}T^{\frac{1}{2}}}} \ T= \frac{1}{2}T^{\frac{1}{2}}, \\
\frac{\widehat{\sigma}_1(0)}{\widehat{\sigma}_1(1) + \widehat{\sigma}_1(0)} T \geq \frac{1}{\frac{T - \frac{1}{2}T^{\frac{1}{2}}}{\frac{1}{2}T^{\frac{1}{2}}}+1} \ T= \frac{1}{2}T^{\frac{1}{2}}.
\end{align*}
Due to this, Algorithm~\ref{alg:2StageANA} goes to Line 3 instead of Line 5 or Line 7.
The total numbers of treated and control units are given by \eqref{eqn:EstimatedOPT}, which we write again as follows,
\begin{align*}
(T(1), T(0)) = (\frac{\widehat{\sigma}_1(1)}{\widehat{\sigma}_1(1) + \widehat{\sigma}_1(0)} T, \frac{\widehat{\sigma}_1(0)}{\widehat{\sigma}_1(1) + \widehat{\sigma}_1(0)} T).
\end{align*}

Similar to Case 1.1, combining \eqref{eqn:proof:2Stage:1} --- \eqref{eqn:proof:2Stage:4}, we have 
\begin{align*}
\frac{V(T(1), T(0) \vert \cE)}{V(T^*(1), T^*(0))} \leq 1 + T^{-\frac{1}{2} + \eps}.
\end{align*}

To conclude, in all four cases, 
\begin{align*}
\frac{V(T(1), T(0) \vert \cE)}{V(T^*(1), T^*(0))} \leq 1 + T^{-\frac{1}{2} + \eps}.
\end{align*}
\hfill \halmos
\endproof

\subsection{Proof of Theorem~\ref{thm:MStageANA}}
\label{sec:proof:thm:MStageANA}

\proof{Proof of Theorem~\ref{thm:MStageANA}.}
We first show Algorithm~\ref{alg:MStageANA} is feasible.
To start, it is easy to see $1 < \beta_1 T^{\frac{1}{M}}$.
Then for any $m \leq M-2$, 
\begin{align*}
\beta_{m} T^{\frac{m}{M}} = 6 \cdot 15^{-\frac{m}{M}} \cdot T^{\frac{m}{M}} < 6 \cdot 15^{-\frac{m+1}{M}} \cdot T^{\frac{m+1}{M}} = \beta_{m+1} T^{\frac{m+1}{M}},
\end{align*}
where the inequality is because $T>15$.
Finally, 
\begin{align*}
\beta_{M-1} T^{\frac{M-1}{M}} = 6 \cdot 15^{-\frac{M-1}{M}} \cdot T^{\frac{M-1}{M}} \leq 6 \cdot 15^{-\frac{2}{3}} \cdot T^{\frac{M-1}{M}} \approx 0.9866 \cdot T^{\frac{M-1}{M}} < T^{\frac{M-1}{M}} \leq T,
\end{align*}
where the first inequality is because $M \geq 3$.
Combining all above we know Algorithm~\ref{alg:MStageANA} is feasible, i.e., $1 < \beta_1 T^{\frac{1}{M}} < ... < \beta_{M-1} T^{\frac{M-1}{M}} < T$.

Then we analyze the performance of Algorithm~\ref{alg:MStageANA}.
Our analysis of Algorithm~\ref{alg:MStageANA} relies on a clean event analysis, which has been widely used in the online learning literature to prove upper bounds \citep{badanidiyuru2018bandits, lattimore2020bandit, slivkins2019introduction}, and has been recently used in the stochastic control literature to prove lower bounds \citep{arlotto2019uniformly}.

To proceed with the clean event analysis, suppose there are two length-$T$ arrays for the treated and the control, respectively, with each value being an independent and identically distributed copy of the representative random variables $Y(1)$ and $Y(0)$, respectively.
When Algorithm~\ref{alg:MStageANA} suggests to conduct an $m$-th stage experiment parameterized by $(T_m(1), T_m(0))$, the observations from the $m$-th stage experiment are generated by reading the next $T_m(1)$ values from the treated array, and the next $T_m(0)$ values from the control array.
See Figure~\ref{tbl:CleanEvent} for an illustration.

\begin{table}[htb]
\centering
\TABLE{Illustration of the clear event analysis
\label{tbl:CleanEvent}}
{\begin{tabular}{|p{15mm}|C{14mm}|C{14mm}|C{14mm}|C{14mm}|C{14mm}|C{14mm}|C{14mm}|C{14mm}|}
\multicolumn{1}{c}{} & \multicolumn{4}{l}{\begin{minipage}[c][8mm][c]{64mm}\begin{center} \small estimates $\widehat{\psi}^2_m(1)$ \\ $\overbrace{\hspace*{66mm}}$\end{center} \end{minipage}} & \multicolumn{4}{c}{} \\ \cline{1-9}
Treated & $Z_1(1)$ & $Z_2(1)$ & $\ldots$ & $Z_s(1)$ & $\ldots$ & $Z_{s'}(1)$ & $\ldots$ & $Z_T(1)$ \\ \cline{1-9}
Control & $Z_1(0)$ & $Z_2(0)$ & $\ldots$ & $Z_s(0)$ & $\ldots$ & $Z_{s'}(0)$ & $\ldots$ & $Z_T(0)$ \\ \cline{1-9}
\multicolumn{1}{c}{} & \multicolumn{6}{l}{\begin{minipage}[c][2mm][c]{96mm}\begin{center} \small $\underbrace{\hspace*{100mm}}$ \\  estimates $\widehat{\psi}^2_{m'}(0)$ \end{center} \end{minipage}} & \multicolumn{2}{c}{\vspace{5mm}}
\end{tabular}
}
{\textit{Note}: In this illustration, the treated array contains random values $Z_1(1)$, $Z_2(1)$, ..., $Z_T(1)$ and the control array contains random values $Z_1(0)$, $Z_2(0)$, ..., $Z_T(0)$. In this illustration, we use the first $s = \frac{\beta_m}{2}T^{\frac{m}{M}}$ values in the treated array to compute the sample variance estimator $\widehat{\psi}^2_m(1)$, and the first $s' = \frac{\beta_{m'}}{2}T^{\frac{m'}{M}}$ values in the control array to compute the sample variance estimator $\widehat{\psi}^2_{m'}(0)$. In this table, all the sample variance estimators such as $\widehat{\psi}^2_m(1)$ and $\widehat{\psi}^2_{m'}(0)$ are all well-defined under a \emph{fixed} number of values.}
\end{table}

Even though Algorithm~\ref{alg:MStageANA} adaptively determines the number of treated and control units, it is always the first few values of of the two arrays that are read.
For any $m \leq M-1$, let $\widehat{\psi}^2_m(1)$ and $\widehat{\psi}^2_{m}(0)$ be the sample variance estimators obtained from reading the first $\frac{\beta_m}{2}T^{\frac{m}{M}}$ values in the treated array and control array, respectively.
Depending on the execution of Algorithm~\ref{alg:MStageANA}, only a few of the sample variance estimators $\widehat{\sigma}^2_m(1)$ or $\widehat{\sigma}^2_m(0)$ are calculated. 
When one sample variance estimator $\widehat{\sigma}^2_m(1)$ or $\widehat{\sigma}^2_m(0)$ is calculated following Algorithm~\ref{alg:MStageANA}, it is equivalent to reading the corresponding $\widehat{\psi}^2_m(1)$ or $\widehat{\psi}^2_m(0)$ from Table~\ref{tbl:CleanEvent}.

Define the following events.
For any $m \leq M-1$, define
\begin{align*}
\cE_m(1) = & \ \bigg\{ \left| \widehat{\psi}^2_m(1) - \sigma^2(1) \right| < 2^{\frac{1}{2}} \beta_m^{-\frac{1}{2}} T^{-\frac{m}{2M} + \frac{\eps}{2}} \sigma^2(1)\bigg\}, \\
\cE_m(0) = & \ \bigg\{ \left| \widehat{\psi}^2_m(0) - \sigma^2(0) \right| < 2^{\frac{1}{2}} \beta_m^{-\frac{1}{2}} T^{-\frac{m}{2M} + \frac{\eps}{2}} \sigma^2(0)\bigg\}.
\end{align*}
Denote the intersect of all above events as $\cE$, i.e., 
\begin{align*}
\cE = \bigcap_{m=1}^{M-1} \left(\cE_m(1) \cap \cE_m(0)\right).
\end{align*}
Then due to union bound, 
\begin{align*}
\Pr(\cE) \geq 1 - \sum_{m=1}^{M-1} \Pr(\overline{\cE}_m(1)) - \sum_{m=1}^{M-1} \Pr(\overline{\cE}_m(0)).
\end{align*}
We further have
\begin{align*}
& \Pr(\cE) \\
= & \ 1 - \sum_{m=1}^{M-1} \Pr\left( \vert \widehat{\psi}^2_m(1) - \sigma^2(1) \vert \geq 2^{\frac{1}{2}} \beta_m^{-\frac{1}{2}} T^{-\frac{m}{2M} + \frac{\eps}{2}} \sigma^2(1) \right) - \sum_{m=1}^{M-1} \Pr\left( \vert \widehat{\psi}^2_m(0) - \sigma^2(0) \vert \geq 2^{\frac{1}{2}} \beta_m^{-\frac{1}{2}} T^{-\frac{m}{2M} + \frac{\eps}{2}} \sigma^2(0) \right) \\
\geq & \ 1 - \sum_{m=1}^{M-1} \frac{\kappa(1) \sigma^4(1)}{2 \beta_m^{-1} T^{-\frac{m}{M} + \eps} \sigma^4(1) \frac{1}{2} \beta_m T^{\frac{m}{M}}} - \sum_{m=1}^{M-1} \frac{\kappa(0) \sigma^4(0)}{2 \beta_m^{-1} T^{-\frac{m}{M} + \eps} \sigma^4(0) \frac{1}{2} \beta_m T^{\frac{m}{M}}} \\
= & \ 1 - \sum_{m=1}^{M-1} \frac{\kappa(1) + \kappa(0)}{T^{\eps}} \\
= & \ 1 - (M-1) \frac{\kappa(1) + \kappa(0)}{T^{\eps}},
\end{align*}
where the inequality is due to Lemma~\ref{lem:LightTail}.

Conditional on the event $\cE$, we have, for any $m \leq M-1$,
\begin{subequations}
\begin{align}
\sigma^2(1) \left( 1 - 2^{\frac{1}{2}} \beta_m^{-\frac{1}{2}} T^{-\frac{m}{2M} + \frac{\eps}{2}} \right) \ \leq \ \widehat{\psi}^2_m(1) \ \leq \ \sigma^2(1) \left( 1 + 2^{\frac{1}{2}} \beta_m^{-\frac{1}{2}} T^{-\frac{m}{2M} + \frac{\eps}{2}} \right), \label{eqn:MStage:ConfidenceBound1} \\
\sigma^2(0) \left( 1 - 2^{\frac{1}{2}} \beta_m^{-\frac{1}{2}} T^{-\frac{m}{2M} + \frac{\eps}{2}} \right) \ \leq \ \widehat{\psi}^2_m(0) \ \leq \ \sigma^2(0) \left( 1 + 2^{\frac{1}{2}} \beta_m^{-\frac{1}{2}} T^{-\frac{m}{2M} + \frac{\eps}{2}} \right). \label{eqn:MStage:ConfidenceBound0} 
\end{align}
\end{subequations}

Since $\sigma(1), \sigma(0) > 0$, we can denote $\rho = \frac{\sigma(1)}{\sigma(0)}$.
For any $m \leq M-1$, when $\widehat{\sigma}^2_m(1)$ and $\widehat{\sigma}^2_m(0)$ are calculated during Algorithm~\ref{alg:MStageANA}, $\widehat{\sigma}^2_m(1) = \widehat{\psi}^2_m(1)$ and $\widehat{\sigma}^2_m(0) = \widehat{\psi}^2_m(0)$.
Conditional on the event $\cE$, due to \eqref{eqn:MStage:ConfidenceBound1} and \eqref{eqn:MStage:ConfidenceBound0}, and given that $\sigma(1), \sigma(0) > 0$, we have $\widehat{\sigma}^2_m(1), \widehat{\sigma}^2_m(0) > 0$.
Then we can denote $\widehat{\rho}_m = \frac{\widehat{\sigma}_m(1)}{\widehat{\sigma}_m(0)}$.

In the remaining of the analysis, we distinguish several cases and discuss these cases separately.
Recall that $\widehat{\rho}_1 = \frac{\widehat{\sigma}_1(1)}{\widehat{\sigma}_1(0)}$.
Without loss of generality, assume 
\begin{align}
\widehat{\rho}_1 \geq 1. \label{eqn:WLOG}
\end{align}

\noindent\underline{\textbf{Case 1}}: 
\begin{align*}
\widehat{\rho}_1 > \frac{T - \frac{1}{2} \beta_2 T^{\frac{2}{M}}}{\frac{1}{2} \beta_2 T^{\frac{2}{M}}}.
\end{align*}
\noindent \textbf{Case 1.1}: 
\begin{align*}
\widehat{\rho}_1 > \frac{T - \frac{1}{2} \beta_1 T^{\frac{1}{M}}}{\frac{1}{2} \beta_1 T^{\frac{1}{M}}}.
\end{align*}
In this case, 
\begin{align*}
\frac{\widehat{\sigma}_1(0)}{\widehat{\sigma}_1(1) + \widehat{\sigma}_1(0)} T < \frac{1}{\frac{T - \frac{1}{2} \beta_1 T^{\frac{1}{M}}}{\frac{1}{2} \beta_1 T^{\frac{1}{M}}}+1} T = \frac{1}{2} \beta_1 T^{\frac{1}{M}}.
\end{align*}
So Algorithm~\ref{alg:MStageANA} goes to Line~\ref{mrk:Case1} in the $1$-st stage experiment.
Then we have
\begin{align*}
(T(1), T(0)) = \bigg(T - \frac{1}{2} \beta_1 T^{\frac{1}{M}}, \frac{1}{2} \beta_1 T^{\frac{1}{M}}\bigg).
\end{align*}
With a little abuse of notation, we write $V(T(1), T(0)\vert \cE)$ to stand for $V(T(1), T(0))$, where we emphasize that this is a random quantity (as $T(1)$ and $T(0)$ are random) that is conditional on event $\cE$.
We can then express
\begin{align}
\frac{V(T(1), T(0) \vert \cE)}{V(T^*(1), T^*(0))} = & \ \frac{\frac{1}{T - \frac{1}{2} \beta_1 T^{\frac{1}{M}}} \sigma^2(1) + \frac{1}{\frac{1}{2} \beta_1 T^{\frac{1}{M}}} \sigma^2(0)}{\frac{1}{T} (\sigma(1) + \sigma(0))^2}. \label{eqn:StartingPoint}
\end{align}

Recall that $\rho = \frac{\sigma(1)}{\sigma(0)}$.
We further distinguish two cases. 

\textbf{First}, if $\rho < \frac{T - \frac{1}{2} \beta_1 T^{\frac{1}{M}}}{\frac{1}{2} \beta_1 T^{\frac{1}{M}}}$, then we write \eqref{eqn:StartingPoint} as 
\begin{align*}
\frac{V(T(1), T(0) \vert \cE)}{V(T^*(1), T^*(0))} = & \ \frac{\sigma^2(1) + \sigma^2(0) + \frac{\frac{1}{2} \beta_1 T^{\frac{1}{M}}}{T - \frac{1}{2} \beta_1 T^{\frac{1}{M}}} \sigma^2(1) + \frac{T - \frac{1}{2} \beta_1 T^{\frac{1}{M}}}{\frac{1}{2} \beta_1 T^{\frac{1}{M}}} \sigma^2(0)}{(\sigma(1) + \sigma(0))^2}.
\end{align*}
Note that, 
\begin{align}
\rho < \frac{T - \frac{1}{2} \beta_1 T^{\frac{1}{M}}}{\frac{1}{2} \beta_1 T^{\frac{1}{M}}} < \widehat{\rho}_1 \leq \rho \cdot \sqrt{ \frac{1+2^{\frac{1}{2}} \beta_1^{-\frac{1}{2}} T^{-\frac{1}{2M} + \frac{\eps}{2}}}{1-2^{\frac{1}{2}} \beta_1^{-\frac{1}{2}} T^{-\frac{1}{2M} + \frac{\eps}{2}}} }. \label{eqn:rhoRelations:Case1-1}
\end{align}
So we have
\begin{align}
\frac{V(T(1), T(0) \vert \cE)}{V(T^*(1), T^*(0))} \leq & \ \frac{\sigma^2(1) + \sigma^2(0) + \sigma(1)\sigma(0) \bigg( \sqrt{ \frac{1+2^{\frac{1}{2}} \beta_1^{-\frac{1}{2}} T^{-\frac{1}{2M} + \frac{\eps}{2}}}{1-2^{\frac{1}{2}} \beta_1^{-\frac{1}{2}} T^{-\frac{1}{2M} + \frac{\eps}{2}}} } + \sqrt{ \frac{1-2^{\frac{1}{2}} \beta_1^{-\frac{1}{2}} T^{-\frac{1}{2M} + \frac{\eps}{2}}}{1+2^{\frac{1}{2}} \beta_1^{-\frac{1}{2}} T^{-\frac{1}{2M} + \frac{\eps}{2}}} } \bigg)}{(\sigma(1) + \sigma(0))^2} \nonumber \\
= & \ 1 + \frac{\sigma(1) \sigma(0)}{(\sigma(1) + \sigma(0))^2} \cdot \bigg( \frac{2}{\sqrt{1 - 2 \beta_1^{-1} T^{ -\frac{1}{M} + \eps }}} - 2 \bigg) \nonumber \\
\leq & \ 1 + \frac{\sigma(1) \sigma(0)}{(\sigma(1) + \sigma(0))^2} \cdot 4 \beta_1^{-1} T^{ -\frac{1}{M} + \eps }, \label{eqn:BreakIntoTwoParts1}
\end{align}
where the first inequality is due to Lemma~\ref{lem:h:rhohat} and \eqref{eqn:rhoRelations:Case1-1}; the last inequality is due to Lemma~\ref{lem:AlgebraicTrick2}.

Note that, $\frac{\sigma(1) \sigma(0)}{(\sigma(1) + \sigma(0))^2} = \frac{\rho}{(\rho+1)^2}$ is a decreasing function when $\rho>1$.
Note also that,
\begin{align*}
\rho > \frac{T - \frac{1}{2} \beta_1 T^{\frac{1}{M}}}{\frac{1}{2} \beta_1 T^{\frac{1}{M}}} \cdot \sqrt{\frac{1-2^{\frac{1}{2}} \beta_1^{-\frac{1}{2}} T^{-\frac{1}{2M} + \frac{\eps}{2}}}{1+2^{\frac{1}{2}} \beta_1^{-\frac{1}{2}} T^{-\frac{1}{2M} + \frac{\eps}{2}}}} > \frac{1}{2} \frac{T - \frac{1}{2} \beta_1 T^{\frac{1}{M}}}{\frac{1}{2} \beta_1 T^{\frac{1}{M}}} > 1,
\end{align*}
where the first inequality is due to \eqref{eqn:rhoRelations:Case1-1}; the second inequality is due to Lemma~\ref{lem:AlgebraicTrick3}; the last inequality is due to Lemma~\ref{lem:AlgebraicTrick4}.

Then we have
\begin{align*}
\frac{\sigma(1) \sigma(0)}{(\sigma(1) + \sigma(0))^2} 
< \ \frac{\frac{1}{2} \frac{T - \frac{1}{2} \beta_1 T^{\frac{1}{M}}}{\frac{1}{2} \beta_1 T^{\frac{1}{M}}}}{\bigg( 1 + \frac{1}{2} \frac{T - \frac{1}{2} \beta_1 T^{\frac{1}{M}}}{\frac{1}{2} \beta_1 T^{\frac{1}{M}}} \bigg)^2}
= \ \frac{\beta_1 T^{\frac{1}{M}} (T - \frac{1}{2} \beta_1 T^{\frac{1}{M}}) }{(T + \frac{1}{2} \beta_1 T^{\frac{1}{M}})^2} 
\leq \ \frac{\beta_1 T^{\frac{1}{M}}}{T}.
\end{align*}
Putting this into \eqref{eqn:BreakIntoTwoParts1} we have 
\begin{align*}
\frac{V(T(1), T(0) \vert \cE)}{V(T^*(1), T^*(0))} \leq \ 1 + 4 T^{-1+\eps} < 1 + 4 \cdot 15^{-\frac{1}{M}} \cdot T^{-\frac{M-1}{M}+\eps},
\end{align*}
where the last inequality is because $T > 15$ so $T^{-1+\eps} = T^{-\frac{1}{M}} \cdot T^{-\frac{M-1}{M}+\eps} < 15^{-\frac{1}{M}} \cdot T^{-\frac{M-1}{M}+\eps}$.

\textbf{Second}, if $\rho \geq \frac{T - \frac{1}{2} \beta_1 T^{\frac{1}{M}}}{\frac{1}{2} \beta_1 T^{\frac{1}{M}}}$, then we write \eqref{eqn:StartingPoint} as 
\begin{align*}
\frac{V(T(1), T(0) \vert \cE)}{V(T^*(1), T^*(0))} = & \ \frac{T}{T - \frac{1}{2} \beta_1 T^{\frac{1}{M}}} \cdot \frac{\sigma^2(1)}{(\sigma(1) + \sigma(0))^2} + \frac{T}{\frac{1}{2} \beta_1 T^{\frac{1}{M}}} \cdot \frac{\sigma^2(0)}{(\sigma(1) + \sigma(0))^2}.
\end{align*}
So we have
\begin{align*}
\frac{V(T(1), T(0) \vert \cE)}{V(T^*(1), T^*(0))} \leq & \frac{T}{T - \frac{1}{2} \beta_1 T^{\frac{1}{M}}} = 1 + \frac{\frac{1}{2} \beta_1 T^{\frac{1}{M}}}{T - \frac{1}{2} \beta_1 T^{\frac{1}{M}}} < 1 + 4 \cdot 15^{-\frac{1}{M}} \cdot T^{-\frac{M-1}{M}},
\end{align*}
where the first inequality is due to Lemma~\ref{lem:g:rho}; the last inequality is due to Lemma~\ref{lem:AlgebraicTrick5}.

Combining $\rho < \frac{T - \frac{1}{2} \beta_1 T^{\frac{1}{M}}}{\frac{1}{2} \beta_1 T^{\frac{1}{M}}}$ and $\rho \geq \frac{T - \frac{1}{2} \beta_1 T^{\frac{1}{M}}}{\frac{1}{2} \beta_1 T^{\frac{1}{M}}}$ we have that in Case 1.1, 
\begin{align*}
\frac{V(T(1), T(0) \vert \cE)}{V(T^*(1), T^*(0))} \leq & \ 1 + 4 \cdot 15^{-\frac{1}{M}} \cdot T^{-\frac{M-1}{M}+\eps}.
\end{align*}

\noindent \textbf{Case 1.2}:
\begin{align*}
\frac{T - \frac{1}{2} \beta_2 T^{\frac{2}{M}}}{\frac{1}{2} \beta_2 T^{\frac{2}{M}}} < \widehat{\rho}_1 \leq \frac{T - \frac{1}{2} \beta_1 T^{\frac{1}{M}}}{\frac{1}{2} \beta_1 T^{\frac{1}{M}}}.
\end{align*}
In this case, 
\begin{align*}
\frac{1}{2} \beta_1 T^{\frac{1}{M}}  = \frac{1}{\frac{T - \frac{1}{2} \beta_1 T^{\frac{1}{M}}}{\frac{1}{2} \beta_1 T^{\frac{1}{M}}}+1} T \leq \frac{\widehat{\sigma}_1(0)}{\widehat{\sigma}_1(1) + \widehat{\sigma}_1(0)} T < \frac{1}{\frac{T - \frac{1}{2} \beta_2 T^{\frac{2}{M}}}{\frac{1}{2} \beta_2 T^{\frac{2}{M}}}+1} T = \frac{1}{2} \beta_2 T^{\frac{2}{M}}.
\end{align*}
So Algorithm~\ref{alg:MStageANA} goes to Line~\ref{mrk:Case2} in the $1$-st stage experiment.
Then we have
\begin{align*}
(T(1), T(0)) = \bigg(\frac{\widehat{\sigma}_1(1)}{\widehat{\sigma}_1(1) + \widehat{\sigma}_1(0)} T, \frac{\widehat{\sigma}_1(0)}{\widehat{\sigma}_1(1) + \widehat{\sigma}_1(0)} T\bigg).
\end{align*}
We can then express
\begin{align}
\frac{V(T(1), T(0) \vert \cE)}{V(T^*(1), T^*(0))} = & \ \frac{\sigma^2(1) + \sigma^2(0) + \frac{1}{\widehat{\rho}_1} \sigma^2(1) + \widehat{\rho}_1 \sigma^2(0)}{(\sigma(1) + \sigma(0))^2}. \label{eqn:ImmediateStartigPoint}
\end{align}

Recall that, conditional on $\cE$, \eqref{eqn:MStage:ConfidenceBound1} and \eqref{eqn:MStage:ConfidenceBound0} lead to
\begin{align*}
\rho \cdot \sqrt{\frac{1-2^{\frac{1}{2}} \beta_1^{-\frac{1}{2}} T^{-\frac{1}{2M} + \frac{\eps}{2}}}{1+2^{\frac{1}{2}} \beta_1^{-\frac{1}{2}} T^{-\frac{1}{2M} + \frac{\eps}{2}}}} \leq \widehat{\rho}_1 \leq \rho \sqrt{\frac{1+2^{\frac{1}{2}} \beta_1^{-\frac{1}{2}} T^{-\frac{1}{2M} + \frac{\eps}{2}}}{1-2^{\frac{1}{2}} \beta_1^{-\frac{1}{2}} T^{-\frac{1}{2M} + \frac{\eps}{2}}}}.
\end{align*}
So we have
\begin{align}
\frac{V(T(1), T(0) \vert \cE)}{V(T^*(1), T^*(0))} \leq & \ \frac{\sigma^2(1) + \sigma^2(0) + \sigma(1)\sigma(0) \bigg( \sqrt{ \frac{1+2^{\frac{1}{2}} \beta_1^{-\frac{1}{2}} T^{-\frac{1}{2M} + \frac{\eps}{2}}}{1-2^{\frac{1}{2}} \beta_1^{-\frac{1}{2}} T^{-\frac{1}{2M} + \frac{\eps}{2}}} } + \sqrt{ \frac{1-2^{\frac{1}{2}} \beta_1^{-\frac{1}{2}} T^{-\frac{1}{2M} + \frac{\eps}{2}}}{1+2^{\frac{1}{2}} \beta_1^{-\frac{1}{2}} T^{-\frac{1}{2M} + \frac{\eps}{2}}} } \bigg)}{(\sigma(1) + \sigma(0))^2} \nonumber \\
= & \ 1 + \frac{\sigma(1) \sigma(0)}{(\sigma(1) + \sigma(0))^2} \cdot \bigg( \frac{2}{\sqrt{1 - 2 \beta_1^{-1} T^{ -\frac{1}{M} + \eps }}} - 2 \bigg) \nonumber \\
\leq & \ 1 + \frac{\sigma(1) \sigma(0)}{(\sigma(1) + \sigma(0))^2} \cdot 4 \beta_1^{-1} T^{ -\frac{1}{M} + \eps }, \label{eqn:BreakIntoTwoParts2}
\end{align}
where the first inequality is due to Lemma~\ref{lem:h:rhohat}; the last inequality is due to Lemma~\ref{lem:AlgebraicTrick2}.

Note that, $\frac{\sigma(1) \sigma(0)}{(\sigma(1) + \sigma(0))^2} = \frac{\rho}{(\rho+1)^2}$ is a decreasing function when $\rho>1$.
Note also that,
\begin{align*}
\rho \geq \widehat{\rho}_1 \cdot \sqrt{\frac{1-2^{\frac{1}{2}} \beta_1^{-\frac{1}{2}} T^{-\frac{1}{2M} + \frac{\eps}{2}}}{1+2^{\frac{1}{2}} \beta_1^{-\frac{1}{2}} T^{-\frac{1}{2M} + \frac{\eps}{2}}}} > \frac{T - \frac{1}{2} \beta_2 T^{\frac{2}{M}}}{\frac{1}{2} \beta_2 T^{\frac{2}{M}}} \cdot \sqrt{\frac{1-2^{\frac{1}{2}} \beta_1^{-\frac{1}{2}} T^{-\frac{1}{2M} + \frac{\eps}{2}}}{1+2^{\frac{1}{2}} \beta_1^{-\frac{1}{2}} T^{-\frac{1}{2M} + \frac{\eps}{2}}}} > \frac{1}{2} \frac{T - \frac{1}{2} \beta_2 T^{\frac{2}{M}}}{\frac{1}{2} \beta_2 T^{\frac{2}{M}}} > 1,
\end{align*}
where the first inequality is due to \eqref{eqn:MStage:ConfidenceBound1} and \eqref{eqn:MStage:ConfidenceBound0}; the second inequality is due to the condition of Case 1.2; the third inequality is due to Lemma~\ref{lem:AlgebraicTrick3}; the last inequality is due to Lemma~\ref{lem:AlgebraicTrick4}.
Then we have
\begin{align*}
\frac{\sigma(1) \sigma(0)}{(\sigma(1) + \sigma(0))^2} 
< \ \frac{\frac{1}{2} \frac{T - \frac{1}{2} \beta_2 T^{\frac{2}{M}}}{\frac{1}{2} \beta_2 T^{\frac{2}{M}}}}{\bigg( 1 + \frac{1}{2} \frac{T - \frac{1}{2} \beta_2 T^{\frac{2}{M}}}{\frac{1}{2} \beta_2 T^{\frac{2}{M}}} \bigg)^2}
= \ \frac{\beta_2 T^{\frac{2}{M}} (T - \frac{1}{2} \beta_2 T^{\frac{2}{M}}) }{(T + \frac{1}{2} \beta_2 T^{\frac{2}{M}})^2} 
\leq \ \frac{\beta_2 T^{\frac{2}{M}}}{T}.
\end{align*}
Putting this into \eqref{eqn:BreakIntoTwoParts2} we have that in Case 1.2,
\begin{align*}
\frac{V(T(1), T(0) \vert \cE)}{V(T^*(1), T^*(0))} \leq \ 1 + \frac{4 \beta_2}{\beta_1} \cdot T^{-\frac{M-1}{M}+\eps} = \ 1 + 4 \cdot 15^{-\frac{1}{M}} \cdot T^{-\frac{M-1}{M}+\eps}.
\end{align*}

\noindent\underline{\textbf{Case 2}}:
\begin{align*}
\widehat{\rho}_1 \leq \frac{T - \frac{1}{2} \beta_2 T^{\frac{2}{M}}}{\frac{1}{2} \beta_2 T^{\frac{2}{M}}}.
\end{align*}
Due to \eqref{eqn:WLOG} we know that $\widehat{\sigma}_1(1) \geq \widehat{\sigma}_1(0)$.
In Case 2 we immediately have
\begin{align*}
\frac{\widehat{\sigma}_1(1)}{\widehat{\sigma}_1(1) + \widehat{\sigma}_1(0)} T \geq \frac{\widehat{\sigma}_1(0)}{\widehat{\sigma}_1(1) + \widehat{\sigma}_1(0)} T  \geq \frac{1}{\frac{T - \frac{1}{2} \beta_2 T^{\frac{2}{M}}}{\frac{1}{2} \beta_2 T^{\frac{2}{M}}}+1} T = \frac{1}{2} \beta_2 T^{\frac{2}{M}}.
\end{align*}
So Algorithm~\ref{alg:MStageANA} goes to Line~\ref{mrk:Case3} in the 1-st stage experiment. 
We further distinguish two cases.

\noindent\textbf{Case 2.1}:
\begin{align*}
\widehat{\rho}_1 \leq \frac{T - \frac{1}{2} \beta_2 T^{\frac{2}{M}}}{\frac{1}{2} \beta_2 T^{\frac{2}{M}}}, && \widehat{\rho}_2 > \frac{T - \frac{1}{2} \beta_2 T^{\frac{2}{M}}}{\frac{1}{2} \beta_2 T^{\frac{2}{M}}}.
\end{align*}
In this case, 
\begin{align*}
\frac{\widehat{\sigma}_2(0)}{\widehat{\sigma}_2(1) + \widehat{\sigma}_2(0)} T < \frac{1}{\frac{T - \frac{1}{2} \beta_2 T^{\frac{2}{M}}}{\frac{1}{2} \beta_2 T^{\frac{2}{M}}}+1} T = \frac{1}{2} \beta_2 T^{\frac{2}{M}}.
\end{align*}
So Algorithm~\ref{alg:MStageANA} goes to Line~\ref{mrk:Case1} in the $2$-nd stage experiment.
Then we have
\begin{align*}
(T(1), T(0)) = \bigg(T - \frac{1}{2} \beta_2 T^{\frac{2}{M}}, \frac{1}{2} \beta_2 T^{\frac{2}{M}}\bigg).
\end{align*}
We can express 
\begin{align*}
\frac{V(T(1), T(0) \vert \cE)}{V(T^*(1), T^*(0))} = & \ \frac{\sigma^2(1) + \sigma^2(0) + \frac{\frac{1}{2} \beta_2 T^{\frac{2}{M}}}{T - \frac{1}{2} \beta_2 T^{\frac{2}{M}}} \sigma^2(1) + \frac{T - \frac{1}{2} \beta_2 T^{\frac{2}{M}}}{\frac{1}{2} \beta_2 T^{\frac{2}{M}}} \sigma^2(0)}{(\sigma(1) + \sigma(0))^2}.
\end{align*}
Note that, 
\begin{multline}
\rho \cdot \sqrt{\frac{1-2^{\frac{1}{2}} \beta_1^{-\frac{1}{2}} T^{-\frac{1}{2M}+\frac{\eps}{2}}}{1+2^{\frac{1}{2}} \beta_1^{-\frac{1}{2}} T^{-\frac{1}{2M}+\frac{\eps}{2}}}} \leq \widehat{\rho}_1 \leq \frac{T - \frac{1}{2} \beta_2 T^{\frac{2}{M}}}{\frac{1}{2} \beta_2 T^{\frac{2}{M}}} \\
< \widehat{\rho}_2 \leq \rho \cdot \sqrt{\frac{1+2^{\frac{1}{2}} \beta_2^{-\frac{1}{2}} T^{-\frac{2}{2M}+\frac{\eps}{2}}}{1-2^{\frac{1}{2}} \beta_2^{-\frac{1}{2}} T^{-\frac{2}{2M}+\frac{\eps}{2}}}} < \rho \cdot \sqrt{\frac{1+2^{\frac{1}{2}} \beta_1^{-\frac{1}{2}} T^{-\frac{1}{2M}+\frac{\eps}{2}}}{1-2^{\frac{1}{2}} \beta_1^{-\frac{1}{2}} T^{-\frac{1}{2M}+\frac{\eps}{2}}}}, \label{eqn:TwoSidedBounds2}
\end{multline}
where the first and the fourth inequalities are due to \eqref{eqn:MStage:ConfidenceBound1} and \eqref{eqn:MStage:ConfidenceBound0}; the second and the third inequalities are due to the condition of Case 2.1;
the last inequality is because $\beta_1 T^{\frac{1}{M}} < \beta_2 T^{\frac{2}{M}}$ so we have $2^{\frac{1}{2}} \beta_2^{-\frac{1}{2}} T^{-\frac{2}{2M}+\frac{\eps}{2}} < 2^{\frac{1}{2}} \beta_1^{-\frac{1}{2}} T^{-\frac{1}{2M}+\frac{\eps}{2}}$.

Then we have
\begin{align}
\frac{V(T(1), T(0) \vert \cE)}{V(T^*(1), T^*(0))} \leq & \ \frac{\sigma^2(1) + \sigma^2(0) + \sigma(1) \sigma(0) \left( \sqrt{\frac{1+2^{\frac{1}{2}} \beta_1^{-\frac{1}{2}} T^{-\frac{1}{2M}+\frac{\eps}{2}}}{1-2^{\frac{1}{2}} \beta_1^{-\frac{1}{2}} T^{-\frac{1}{2M}+\frac{\eps}{2}}}} + \sqrt{\frac{1-2^{\frac{1}{2}} \beta_1^{-\frac{1}{2}} T^{-\frac{1}{2M}+\frac{\eps}{2}}}{1+2^{\frac{1}{2}} \beta_1^{-\frac{1}{2}} T^{-\frac{1}{2M}+\frac{\eps}{2}}}} \right) }{(\sigma(1) + \sigma(0))^2} \nonumber \\
= & \ 1 + \frac{\sigma(1)\sigma(0)}{(\sigma(1) + \sigma(0))^2} \cdot \left(\frac{2}{\sqrt{1-2\beta_1^{-1}T^{-\frac{1}{M}+\eps}}} - 2\right) \nonumber \\
\leq & \ 1 + \frac{\sigma(1)\sigma(0)}{(\sigma(1) + \sigma(0))^2} \cdot \left(4 \beta_1^{-1} T^{-\frac{1}{M}+\eps}\right), \label{eqn:BreakIntoTwoParts3}
\end{align}
where the first inequality is due to Lemma~\ref{lem:h:rhohat}; the last inequality is due to Lemma~\ref{lem:AlgebraicTrick2}.

Note that, $\frac{\sigma(1) \sigma(0)}{(\sigma(1) + \sigma(0))^2} = \frac{\rho}{(\rho+1)^2}$ is a decreasing function when $\rho>1$.
Note also that,
\begin{align*}
\rho > \frac{T - \frac{1}{2} \beta_2 T^{\frac{2}{M}}}{\frac{1}{2} \beta_2 T^{\frac{2}{M}}} \cdot \sqrt{\frac{1-2^{\frac{1}{2}} \beta_2^{-\frac{1}{2}} T^{-\frac{2}{2M} + \frac{\eps}{2}}}{1+2^{\frac{1}{2}} \beta_2^{-\frac{1}{2}} T^{-\frac{2}{2M} + \frac{\eps}{2}}}} > \frac{1}{2} \frac{T - \frac{1}{2} \beta_2 T^{\frac{2}{M}}}{\frac{1}{2} \beta_2 T^{\frac{2}{M}}} > 1,
\end{align*}
where the first inequality is due to \eqref{eqn:TwoSidedBounds2}; the second inequality is due to Lemma~\ref{lem:AlgebraicTrick3}; the last inequality is due to Lemma~\ref{lem:AlgebraicTrick4}.

Then we have
\begin{align*}
\frac{\sigma(1) \sigma(0)}{(\sigma(1) + \sigma(0))^2} 
< \ \frac{\frac{1}{2} \frac{T - \frac{1}{2} \beta_2 T^{\frac{2}{M}}}{\frac{1}{2} \beta_2 T^{\frac{2}{M}}}}{\bigg( 1 + \frac{1}{2} \frac{T - \frac{1}{2} \beta_2 T^{\frac{2}{M}}}{\frac{1}{2} \beta_2 T^{\frac{2}{M}}} \bigg)^2} 
= \ \frac{\beta_2 T^{\frac{2}{M}} (T - \frac{1}{2} \beta_2 T^{\frac{2}{M}}) }{(T + \frac{1}{2} \beta_2 T^{\frac{2}{M}})^2} 
\leq \ \frac{\beta_2 T^{\frac{2}{M}}}{T}.
\end{align*}
Putting this into \eqref{eqn:BreakIntoTwoParts3} we have that in Case 2.1,
\begin{align*}
\frac{V(T(1), T(0) \vert \cE)}{V(T^*(1), T^*(0))} \leq \ 1 + \frac{4\beta_2}{\beta_1}T^{-\frac{M-1}{M}+\eps} = 1 + 4 \cdot 15^{-\frac{1}{M}} \cdot T^{-\frac{M-1}{M}+\eps}.
\end{align*}

\noindent\textbf{Case 2.2}:
\begin{align*}
\widehat{\rho}_1 \leq \frac{T - \frac{1}{2} \beta_2 T^{\frac{2}{M}}}{\frac{1}{2} \beta_2 T^{\frac{2}{M}}}, && \frac{T - \frac{1}{2} \beta_3 T^{\frac{3}{M}}}{\frac{1}{2} \beta_3 T^{\frac{3}{M}}} < \widehat{\rho}_2 \leq \frac{T - \frac{1}{2} \beta_2 T^{\frac{2}{M}}}{\frac{1}{2} \beta_2 T^{\frac{2}{M}}}.
\end{align*}
In this case, 
\begin{align*}
\frac{1}{2} \beta_2 T^{\frac{2}{M}} = \frac{1}{\frac{T - \frac{1}{2} \beta_2 T^{\frac{2}{M}}}{\frac{1}{2} \beta_2 T^{\frac{2}{M}}}+1} T \leq \frac{\widehat{\sigma}_2(0)}{\widehat{\sigma}_2(1) + \widehat{\sigma}_2(0)} T < \frac{1}{\frac{T - \frac{1}{2} \beta_3 T^{\frac{3}{M}}}{\frac{1}{2} \beta_3 T^{\frac{3}{M}}}+1} T < \frac{1}{2} \beta_3 T^{\frac{3}{M}}.
\end{align*}
So Algorithm~\ref{alg:MStageANA} goes to Line~\ref{mrk:Case2} in the $2$-nd stage experiment.
Then we have
\begin{align*}
(T(1), T(0)) = \bigg(\frac{\widehat{\sigma}_2(1)}{\widehat{\sigma}_2(1) + \widehat{\sigma}_2(0)} T, \frac{\widehat{\sigma}_2(0)}{\widehat{\sigma}_2(1) + \widehat{\sigma}_2(0)} T\bigg).
\end{align*}
We can then express
\begin{align}
\frac{V(T(1), T(0) \vert \cE)}{V(T^*(1), T^*(0))} = & \ \frac{\sigma^2(1) + \sigma^2(0) + \frac{1}{\widehat{\rho}_2} \sigma^2(1) + \widehat{\rho}_2 \sigma^2(0)}{(\sigma(1) + \sigma(0))^2}. \label{eqn:ImmediateStartigPoint2}
\end{align}

Recall that, conditional on $\cE$, \eqref{eqn:MStage:ConfidenceBound1} and \eqref{eqn:MStage:ConfidenceBound0} lead to
\begin{align*}
\rho \cdot \sqrt{\frac{1-2^{\frac{1}{2}} \beta_2^{-\frac{1}{2}} T^{-\frac{2}{2M} + \frac{\eps}{2}}}{1+2^{\frac{1}{2}} \beta_2^{-\frac{1}{2}} T^{-\frac{2}{2M} + \frac{\eps}{2}}}} \leq \widehat{\rho}_2 \leq \rho \sqrt{\frac{1+2^{\frac{1}{2}} \beta_2^{-\frac{1}{2}} T^{-\frac{2}{2M} + \frac{\eps}{2}}}{1-2^{\frac{1}{2}} \beta_2^{-\frac{1}{2}} T^{-\frac{2}{2M} + \frac{\eps}{2}}}}.
\end{align*}
So we have
\begin{align}
\frac{V(T(1), T(0) \vert \cE)}{V(T^*(1), T^*(0))} \leq & \ \frac{\sigma^2(1) + \sigma^2(0) + \sigma(1)\sigma(0) \bigg( \sqrt{\frac{1-2^{\frac{1}{2}} \beta_2^{-\frac{1}{2}} T^{-\frac{2}{2M} + \frac{\eps}{2}}}{1+2^{\frac{1}{2}} \beta_2^{-\frac{1}{2}} T^{-\frac{2}{2M} + \frac{\eps}{2}}}} + \sqrt{\frac{1+2^{\frac{1}{2}} \beta_2^{-\frac{1}{2}} T^{-\frac{2}{2M} + \frac{\eps}{2}}}{1-2^{\frac{1}{2}} \beta_2^{-\frac{1}{2}} T^{-\frac{2}{2M} + \frac{\eps}{2}}}} \bigg)}{(\sigma(1) + \sigma(0))^2} \nonumber \\
= & \ 1 + \frac{\sigma(1) \sigma(0)}{(\sigma(1) + \sigma(0))^2} \cdot \bigg( \frac{2}{\sqrt{1 - 2 \beta_2^{-1} T^{ -\frac{2}{M} + \eps }}} - 2 \bigg) \nonumber \\
\leq & \ 1 + \frac{\sigma(1) \sigma(0)}{(\sigma(1) + \sigma(0))^2} \cdot 4 \beta_2^{-1} T^{ -\frac{2}{M} + \eps }, \label{eqn:BreakIntoTwoParts4}
\end{align}
where the first inequality is due to Lemma~\ref{lem:h:rhohat}; the last inequality is due to Lemma~\ref{lem:AlgebraicTrick2}.

Note that, $\frac{\sigma(1) \sigma(0)}{(\sigma(1) + \sigma(0))^2} = \frac{\rho}{(\rho+1)^2}$ is a decreasing function when $\rho>1$.
Note also that,
\begin{align*}
\rho \geq \widehat{\rho}_2 \cdot \sqrt{\frac{1-2^{\frac{1}{2}} \beta_2^{-\frac{1}{2}} T^{-\frac{2}{2M} + \frac{\eps}{2}}}{1+2^{\frac{1}{2}} \beta_2^{-\frac{1}{2}} T^{-\frac{2}{2M} + \frac{\eps}{2}}}} > \frac{T - \frac{1}{2} \beta_3 T^{\frac{3}{M}}}{\frac{1}{2} \beta_3 T^{\frac{3}{M}}} \cdot \sqrt{\frac{1-2^{\frac{1}{2}} \beta_2^{-\frac{1}{2}} T^{-\frac{2}{2M} + \frac{\eps}{2}}}{1+2^{\frac{1}{2}} \beta_2^{-\frac{1}{2}} T^{-\frac{2}{2M} + \frac{\eps}{2}}}} > \frac{1}{2} \frac{T - \frac{1}{2} \beta_3 T^{\frac{3}{M}}}{\frac{1}{2} \beta_3 T^{\frac{3}{M}}} > 1,
\end{align*}
where the first inequality is due to \eqref{eqn:MStage:ConfidenceBound1} and \eqref{eqn:MStage:ConfidenceBound0}; the second inequality is due to the condition of Case 2.2; the third inequality is due to Lemma~\ref{lem:AlgebraicTrick3}; the last inequality is due to Lemma~\ref{lem:AlgebraicTrick4}.
Then we have
\begin{align*}
\frac{\sigma(1) \sigma(0)}{(\sigma(1) + \sigma(0))^2} 
< \ \frac{\frac{1}{2} \frac{T - \frac{1}{2} \beta_3 T^{\frac{3}{M}}}{\frac{1}{2} \beta_3 T^{\frac{3}{M}}}}{\bigg( 1 + \frac{1}{2} \frac{T - \frac{1}{2} \beta_3 T^{\frac{3}{M}}}{\frac{1}{2} \beta_3 T^{\frac{3}{M}}} \bigg)^2}
= \ \frac{\beta_3 T^{\frac{3}{M}} (T - \frac{1}{2} \beta_3 T^{\frac{3}{M}}) }{(T + \frac{1}{2} \beta_3 T^{\frac{3}{M}})^2} 
\leq \ \frac{\beta_3 T^{\frac{3}{M}}}{T}
\end{align*}
Putting this into \eqref{eqn:BreakIntoTwoParts4} we have that in Case 2.2,
\begin{align*}
\frac{V(T(1), T(0) \vert \cE)}{V(T^*(1), T^*(0))} \leq \ 1 + \frac{4 \beta_3}{\beta_2} \cdot T^{-\frac{M-1}{M}+\eps} = \ 1 + 4 \cdot 15^{-\frac{1}{M}} \cdot T^{-\frac{M-1}{M}+\eps}.
\end{align*}

\noindent\underline{\textbf{Case $\bm{m}$}} (when $m \leq M-2$):
\begin{align*}
\widehat{\rho}_l \leq \frac{T - \frac{1}{2} \beta_{l+1} T^{\frac{l+1}{M}}}{\frac{1}{2} \beta_{l+1} T^{\frac{l+1}{M}}}, \ \forall \ l \leq m-1.
\end{align*}
Due to the condition of Case $m$, we immediately have
\begin{align*}
\frac{\widehat{\sigma}_{m-1}(0)}{\widehat{\sigma}_{m-1}(1) + \widehat{\sigma}_{m-1}(0)} T \geq \frac{1}{\frac{T - \frac{1}{2} \beta_{m} T^{\frac{m}{M}}}{\frac{1}{2} \beta_{m} T^{\frac{m}{M}}}+1} T = \frac{1}{2} \beta_{m} T^{\frac{m}{M}}.
\end{align*}
On the other hand, since
\begin{multline*}
\widehat{\rho}_{m-1} \geq \rho \sqrt{\frac{1-2^{\frac{1}{2}} \beta_{m-1}^{-\frac{1}{2}} T^{-\frac{m-1}{2M}+\frac{\eps}{2}}}{1+2^{\frac{1}{2}} \beta_{m-1}^{-\frac{1}{2}} T^{-\frac{m-1}{2M}+\frac{\eps}{2}}}} \geq \widehat{\rho}_1 \sqrt{\frac{1-2^{\frac{1}{2}} \beta_{1}^{-\frac{1}{2}} T^{-\frac{1}{2M}+\frac{\eps}{2}}}{1+2^{\frac{1}{2}} \beta_{1}^{-\frac{1}{2}} T^{-\frac{1}{2M}+\frac{\eps}{2}}}} \sqrt{\frac{1-2^{\frac{1}{2}} \beta_{m-1}^{-\frac{1}{2}} T^{-\frac{m-1}{2M}+\frac{\eps}{2}}}{1+2^{\frac{1}{2}} \beta_{m-1}^{-\frac{1}{2}} T^{-\frac{m-1}{2M}+\frac{\eps}{2}}}} \\
\geq \sqrt{\frac{1-2^{\frac{1}{2}} \beta_{1}^{-\frac{1}{2}} T^{-\frac{1}{2M}+\frac{\eps}{2}}}{1+2^{\frac{1}{2}} \beta_{1}^{-\frac{1}{2}} T^{-\frac{1}{2M}+\frac{\eps}{2}}}} \sqrt{\frac{1-2^{\frac{1}{2}} \beta_{m-1}^{-\frac{1}{2}} T^{-\frac{m-1}{2M}+\frac{\eps}{2}}}{1+2^{\frac{1}{2}} \beta_{m-1}^{-\frac{1}{2}} T^{-\frac{m-1}{2M}+\frac{\eps}{2}}}} > \frac{1}{4} \geq \frac{\frac{1}{2}\beta_m T^{\frac{m}{M}}}{T - \frac{1}{2}\beta_m T^{\frac{m}{M}}},
\end{multline*}
where the first and second inequalities are due to \eqref{eqn:MStage:ConfidenceBound1} and \eqref{eqn:MStage:ConfidenceBound0}; the third inequality is due to \eqref{eqn:WLOG}; the fourth inequality is due to Lemma~\ref{lem:AlgebraicTrick3}; the last inequality is due to Lemma~\ref{lem:AlgebraicTrick4}.
Due to the above sequence of inequalities, we have $\frac{1}{\widehat{\rho}_{m-1}} \leq \frac{T - \frac{1}{2}\beta_m T^{\frac{m}{M}}}{\frac{1}{2}\beta_m T^{\frac{m}{M}}}$, which leads to
\begin{align*}
\frac{\widehat{\sigma}_{m-1}(1)}{\widehat{\sigma}_{m-1}(1) + \widehat{\sigma}_{m-1}(0)} T  \geq \frac{1}{1+\frac{T - \frac{1}{2} \beta_{m} T^{\frac{m}{M}}}{\frac{1}{2} \beta_{m} T^{\frac{m}{M}}}} T = \frac{1}{2} \beta_{m} T^{\frac{m}{M}}.
\end{align*}
So Algorithm~\ref{alg:MStageANA} goes to Line~\ref{mrk:Case3} in the (m-1)-th stage experiment. 
We further distinguish two cases.

\noindent\textbf{Case $\bm{m}$.1}:
In addition to the conditions in Case $m$ above, we also have
\begin{align*}
\widehat{\rho}_m > \frac{T - \frac{1}{2} \beta_{m} T^{\frac{m}{M}}}{\frac{1}{2} \beta_{m} T^{\frac{m}{M}}}.
\end{align*}
Similar to the analysis in Case 2.1, we proceed with the following analysis.
In Case $m$.1, 
\begin{align*}
\frac{\widehat{\sigma}_m(0)}{\widehat{\sigma}_m(1) + \widehat{\sigma}_m(0)} T < \frac{1}{\frac{T - \frac{1}{2} \beta_m T^{\frac{m}{M}}}{\frac{1}{2} \beta_m T^{\frac{m}{M}}}+1} T = \frac{1}{2} \beta_m T^{\frac{m}{M}}.
\end{align*}
So Algorithm~\ref{alg:MStageANA} goes to Line~\ref{mrk:Case1} in the $m$-th stage experiment.
Then we have
\begin{align*}
(T(1), T(0)) = \bigg(T - \frac{1}{2} \beta_m T^{\frac{m}{M}}, \frac{1}{2} \beta_m T^{\frac{m}{M}}\bigg).
\end{align*}
We can express 
\begin{align*}
\frac{V(T(1), T(0) \vert \cE)}{V(T^*(1), T^*(0))} = & \ \frac{\sigma^2(1) + \sigma^2(0) + \frac{\frac{1}{2} \beta_m T^{\frac{m}{M}}}{T - \frac{1}{2} \beta_m T^{\frac{m}{M}}} \sigma^2(1) + \frac{T - \frac{1}{2} \beta_m T^{\frac{m}{M}}}{\frac{1}{2} \beta_m T^{\frac{m}{M}}} \sigma^2(0)}{(\sigma(1) + \sigma(0))^2}.
\end{align*}
Note that, 
\begin{multline}
\rho \cdot \sqrt{\frac{1-2^{\frac{1}{2}} \beta_{m-1}^{-\frac{1}{2}} T^{-\frac{m-1}{2M}+\frac{\eps}{2}}}{1+2^{\frac{1}{2}} \beta_{m-1}^{-\frac{1}{2}} T^{-\frac{m-1}{2M}+\frac{\eps}{2}}}} \leq \widehat{\rho}_{m-1} \leq \frac{T - \frac{1}{2} \beta_m T^{\frac{m}{M}}}{\frac{1}{2} \beta_m T^{\frac{m}{M}}} \\
< \widehat{\rho}_m \leq \rho \cdot \sqrt{\frac{1+2^{\frac{1}{2}} \beta_m^{-\frac{1}{2}} T^{-\frac{m}{2M}+\frac{\eps}{2}}}{1-2^{\frac{1}{2}} \beta_m^{-\frac{1}{2}} T^{-\frac{m}{2M}+\frac{\eps}{2}}}} < \rho \cdot \sqrt{\frac{1+2^{\frac{1}{2}} \beta_{m-1}^{-\frac{1}{2}} T^{-\frac{m-1}{2M}+\frac{\eps}{2}}}{1-2^{\frac{1}{2}} \beta_{m-1}^{-\frac{1}{2}} T^{-\frac{m-1}{2M}+\frac{\eps}{2}}}}, \label{eqn:TwoSidedBoundsm}
\end{multline}
where the first and the fourth inequalities are due to \eqref{eqn:MStage:ConfidenceBound1} and \eqref{eqn:MStage:ConfidenceBound0};
the second and the third inequalities are due to the condition of Case $m$.1;
the last inequality is because $\beta_{m-1} T^{\frac{m-1}{M}} < \beta_m T^{\frac{m}{M}}$ so we have $2^{\frac{1}{2}} \beta_m^{-\frac{1}{2}} T^{-\frac{m}{2M}+\frac{\eps}{2}} < 2^{\frac{1}{2}} \beta_{m-1}^{-\frac{1}{2}} T^{-\frac{m-1}{2M}+\frac{\eps}{2}}$.

Then we have
\begin{align}
\frac{V(T(1), T(0) \vert \cE)}{V(T^*(1), T^*(0))} \leq & \ \frac{\sigma^2(1) + \sigma^2(0) + \sigma(1) \sigma(0) \left( \sqrt{\frac{1+2^{\frac{1}{2}} \beta_{m-1}^{-\frac{1}{2}} T^{-\frac{m-1}{2M}+\frac{\eps}{2}}}{1-2^{\frac{1}{2}} \beta_{m-1}^{-\frac{1}{2}} T^{-\frac{m-1}{2M}+\frac{\eps}{2}}}} + \sqrt{\frac{1-2^{\frac{1}{2}} \beta_{m-1}^{-\frac{1}{2}} T^{-\frac{m-1}{2M}+\frac{\eps}{2}}}{1+2^{\frac{1}{2}} \beta_{m-1}^{-\frac{1}{2}} T^{-\frac{m-1}{2M}+\frac{\eps}{2}}}} \right) }{(\sigma(1) + \sigma(0))^2} \nonumber \\
= & \ 1 + \frac{\sigma(1)\sigma(0)}{(\sigma(1) + \sigma(0))^2} \cdot \left(\frac{2}{\sqrt{1-2\beta_{m-1}^{-1}T^{-\frac{m-1}{M}+\eps}}} - 2\right) \nonumber \\
\leq & \ 1 + \frac{\sigma(1)\sigma(0)}{(\sigma(1) + \sigma(0))^2} \cdot \left(4 \beta_{m-1}^{-1} T^{-\frac{m-1}{M}+\eps}\right), \label{eqn:BreakIntoTwoPartsm1}
\end{align}
where the first inequality is due to Lemma~\ref{lem:h:rhohat}; the last inequality is due to Lemma~\ref{lem:AlgebraicTrick2}.

Note that, $\frac{\sigma(1) \sigma(0)}{(\sigma(1) + \sigma(0))^2} = \frac{\rho}{(\rho+1)^2}$ is a decreasing function when $\rho>1$.
Note also that,
\begin{align*}
\rho > \frac{T - \frac{1}{2} \beta_m T^{\frac{m}{M}}}{\frac{1}{2} \beta_m T^{\frac{m}{M}}} \cdot \sqrt{\frac{1-2^{\frac{1}{2}} \beta_m^{-\frac{1}{2}} T^{-\frac{m}{2M} + \frac{\eps}{2}}}{1+2^{\frac{1}{2}} \beta_m^{-\frac{1}{2}} T^{-\frac{m}{2M} + \frac{\eps}{2}}}} > \frac{1}{2} \frac{T - \frac{1}{2} \beta_m T^{\frac{m}{M}}}{\frac{1}{2} \beta_m T^{\frac{m}{M}}} > 1,
\end{align*}
where the first inequality is due to \eqref{eqn:TwoSidedBoundsm}; the second inequality is due to Lemma~\ref{lem:AlgebraicTrick3}; the last inequality is due to Lemma~\ref{lem:AlgebraicTrick4}.

Then we have
\begin{align*}
\frac{\sigma(1) \sigma(0)}{(\sigma(1) + \sigma(0))^2} 
< \frac{\frac{1}{2} \frac{T - \frac{1}{2} \beta_m T^{\frac{m}{M}}}{\frac{1}{2} \beta_m T^{\frac{m}{M}}}}{\bigg( 1 + \frac{1}{2} \frac{T - \frac{1}{2} \beta_m T^{\frac{m}{M}}}{\frac{1}{2} \beta_m T^{\frac{m}{M}}} \bigg)^2} 
= \frac{\beta_m T^{\frac{m}{M}} (T - \frac{1}{2} \beta_m T^{\frac{m}{M}}) }{(T + \frac{1}{2} \beta_m T^{\frac{m}{M}})^2} 
\leq \frac{\beta_m T^{\frac{m}{M}}}{T}.
\end{align*}
Putting this into \eqref{eqn:BreakIntoTwoPartsm1} we have that in Case $m$.1,
\begin{align*}
\frac{V(T(1), T(0) \vert \cE)}{V(T^*(1), T^*(0))} \leq \ 1 + \frac{4\beta_m}{\beta_{m-1}}T^{-\frac{M-1}{M}+\eps} = 1 + 4 \cdot 15^{-\frac{1}{M}} \cdot T^{-\frac{M-1}{M}+\eps}.
\end{align*}

\noindent\textbf{Case $\bm{m}$.2}:
In addition to the conditions in Case $m$ above, we also have
\begin{align*}
\frac{T - \frac{1}{2} \beta_{m+1} T^{\frac{m+1}{M}}}{\frac{1}{2} \beta_{m+1} T^{\frac{m+1}{M}}} < \widehat{\rho}_m \leq \frac{T - \frac{1}{2} \beta_{m} T^{\frac{m}{M}}}{\frac{1}{2} \beta_{m} T^{\frac{m}{M}}}.
\end{align*}
Similar to the analysis in Case 2.2, we proceed with the following analysis.
In Case $m$.2, 
\begin{align*}
\frac{1}{2} \beta_m T^{\frac{m}{M}} = \frac{1}{\frac{T - \frac{1}{2} \beta_m T^{\frac{m}{M}}}{\frac{1}{2} \beta_m T^{\frac{m}{M}}}+1} T \leq \frac{\widehat{\sigma}_m(0)}{\widehat{\sigma}_m(1) + \widehat{\sigma}_m(0)} T < \frac{1}{\frac{T - \frac{1}{2} \beta_{m+1} T^{\frac{m+1}{M}}}{\frac{1}{2} \beta_{m+1} T^{\frac{m+1}{M}}}+1} T < \frac{1}{2} \beta_{m+1} T^{\frac{m+1}{M}}.
\end{align*}
So Algorithm~\ref{alg:MStageANA} goes to Line~\ref{mrk:Case2} in the $m$-th stage experiment.
Then we have
\begin{align*}
(T(1), T(0)) = \bigg(\frac{\widehat{\sigma}_m(1)}{\widehat{\sigma}_m(1) + \widehat{\sigma}_m(0)} T, \frac{\widehat{\sigma}_m(0)}{\widehat{\sigma}_m(1) + \widehat{\sigma}_m(0)} T\bigg).
\end{align*}
We can then express
\begin{align*}
\frac{V(T(1), T(0) \vert \cE)}{V(T^*(1), T^*(0))} = & \ \frac{\sigma^2(1) + \sigma^2(0) + \frac{1}{\widehat{\rho}_m} \sigma^2(1) + \widehat{\rho}_m \sigma^2(0)}{(\sigma(1) + \sigma(0))^2}.
\end{align*}
Recall that, conditional on $\cE$, \eqref{eqn:MStage:ConfidenceBound1} and \eqref{eqn:MStage:ConfidenceBound0} lead to
\begin{align*}
\rho \cdot \sqrt{\frac{1-2^{\frac{1}{2}} \beta_m^{-\frac{1}{2}} T^{-\frac{m}{2M} + \frac{\eps}{2}}}{1+2^{\frac{1}{2}} \beta_m^{-\frac{1}{2}} T^{-\frac{m}{2M} + \frac{\eps}{2}}}} \leq \widehat{\rho}_m \leq \rho \sqrt{\frac{1+2^{\frac{1}{2}} \beta_m^{-\frac{1}{2}} T^{-\frac{m}{2M} + \frac{\eps}{2}}}{1-2^{\frac{1}{2}} \beta_m^{-\frac{1}{2}} T^{-\frac{m}{2M} + \frac{\eps}{2}}}}.
\end{align*}
So we have
\begin{align}
\frac{V(T(1), T(0) \vert \cE)}{V(T^*(1), T^*(0))} \leq & \ \frac{\sigma^2(1) + \sigma^2(0) + \sigma(1)\sigma(0) \bigg( \sqrt{\frac{1-2^{\frac{1}{2}} \beta_m^{-\frac{1}{2}} T^{-\frac{m}{2M} + \frac{\eps}{2}}}{1+2^{\frac{1}{2}} \beta_m^{-\frac{1}{2}} T^{-\frac{m}{2M} + \frac{\eps}{2}}}} + \sqrt{\frac{1+2^{\frac{1}{2}} \beta_m^{-\frac{1}{2}} T^{-\frac{m}{2M} + \frac{\eps}{2}}}{1-2^{\frac{1}{2}} \beta_m^{-\frac{1}{2}} T^{-\frac{m}{2M} + \frac{\eps}{2}}}} \bigg)}{(\sigma(1) + \sigma(0))^2} \nonumber \\
= & \ 1 + \frac{\sigma(1) \sigma(0)}{(\sigma(1) + \sigma(0))^2} \cdot \bigg( \frac{2}{\sqrt{1 - 2 \beta_m^{-1} T^{ -\frac{m}{M} + \eps }}} - 2 \bigg) \nonumber \\
\leq & \ 1 + \frac{\sigma(1) \sigma(0)}{(\sigma(1) + \sigma(0))^2} \cdot 4 \beta_m^{-1} T^{ -\frac{m}{M} + \eps }, \label{eqn:BreakIntoTwoPartsm2}
\end{align}
where the first inequality is due to Lemma~\ref{lem:h:rhohat}; the last inequality is due to Lemma~\ref{lem:AlgebraicTrick2}.

Note that, $\frac{\sigma(1) \sigma(0)}{(\sigma(1) + \sigma(0))^2} = \frac{\rho}{(\rho+1)^2}$ is a decreasing function when $\rho>1$.
Note also that,
\begin{align*}
\rho \geq \widehat{\rho}_m \cdot \sqrt{\frac{1-2^{\frac{1}{2}} \beta_m^{-\frac{1}{2}} T^{-\frac{m}{2M} + \frac{\eps}{2}}}{1+2^{\frac{1}{2}} \beta_m^{-\frac{1}{2}} T^{-\frac{m}{2M} + \frac{\eps}{2}}}} > \frac{T - \frac{1}{2} \beta_{m+1} T^{\frac{m+1}{M}}}{\frac{1}{2} \beta_{m+1} T^{\frac{m+1}{M}}} \cdot \sqrt{\frac{1-2^{\frac{1}{2}} \beta_m^{-\frac{1}{2}} T^{-\frac{m}{2M} + \frac{\eps}{2}}}{1+2^{\frac{1}{2}} \beta_m^{-\frac{1}{2}} T^{-\frac{m}{2M} + \frac{\eps}{2}}}} > \frac{1}{2} \frac{T - \frac{1}{2} \beta_{m+1} T^{\frac{m+1}{M}}}{\frac{1}{2} \beta_{m+1} T^{\frac{m+1}{M}}} > 1,
\end{align*}
where the first inequality is due to \eqref{eqn:MStage:ConfidenceBound1} and \eqref{eqn:MStage:ConfidenceBound0}; the second inequality is due to the condition of Case 2.2; the third inequality is due to Lemma~\ref{lem:AlgebraicTrick3}; the last inequality is due to Lemma~\ref{lem:AlgebraicTrick4}.
Then we have
\begin{align*}
\frac{\sigma(1) \sigma(0)}{(\sigma(1) + \sigma(0))^2}
< \frac{\frac{1}{2} \frac{T - \frac{1}{2} \beta_{m+1} T^{\frac{m+1}{M}}}{\frac{1}{2} \beta_{m+1} T^{\frac{m+1}{M}}}}{\bigg( 1 + \frac{1}{2} \frac{T - \frac{1}{2} \beta_{m+1} T^{\frac{m+1}{M}}}{\frac{1}{2} \beta_{m+1} T^{\frac{m+1}{M}}} \bigg)^2}
= \frac{\beta_{m+1} T^{\frac{m+1}{M}} (T - \frac{1}{2} \beta_{m+1} T^{\frac{m+1}{M}}) }{(T + \frac{1}{2} \beta_{m+1} T^{\frac{m+1}{M}})^2} 
\leq \frac{\beta_{m+1} T^{\frac{m+1}{M}}}{T}
\end{align*}
Putting this into \eqref{eqn:BreakIntoTwoPartsm2} we have that in Case $m$.2,
\begin{align*}
\frac{V(T(1), T(0) \vert \cE)}{V(T^*(1), T^*(0))} \leq \ 1 + \frac{4 \beta_{m+1}}{\beta_m} \cdot T^{-\frac{M-1}{M}+\eps} = \ 1 + 4 \cdot 15^{-\frac{1}{M}} \cdot T^{-\frac{M-1}{M}+\eps}.
\end{align*}

\noindent\underline{\textbf{Case ($\bm{M-1}$)}}:
\begin{align*}
\widehat{\rho}_l \leq \frac{T - \frac{1}{2} \beta_{l+1} T^{\frac{l+1}{M}}}{\frac{1}{2} \beta_{l+1} T^{\frac{l+1}{M}}}, \ \forall \ l \leq M-2.
\end{align*}
Due to the condition of Case ($M-1$), we immediately have
\begin{align*}
\frac{\widehat{\sigma}_{M-2}(0)}{\widehat{\sigma}_{M-2}(1) + \widehat{\sigma}_{M-2}(0)} T \geq \frac{1}{\frac{T - \frac{1}{2} \beta_{M-1} T^{\frac{M-1}{M}}}{\frac{1}{2} \beta_{M-1} T^{\frac{M-1}{M}}}+1} T = \frac{1}{2} \beta_{M-1} T^{\frac{M-1}{M}}.
\end{align*}
On the other hand, since
\begin{multline*}
\widehat{\rho}_{M-2} \geq \rho \sqrt{\frac{1-2^{\frac{1}{2}} \beta_{M-2}^{-\frac{1}{2}} T^{-\frac{M-2}{2M}+\frac{\eps}{2}}}{1+2^{\frac{1}{2}} \beta_{M-2}^{-\frac{1}{2}} T^{-\frac{M-2}{2M}+\frac{\eps}{2}}}} \geq \widehat{\rho}_1 \sqrt{\frac{1-2^{\frac{1}{2}} \beta_{1}^{-\frac{1}{2}} T^{-\frac{1}{2M}+\frac{\eps}{2}}}{1+2^{\frac{1}{2}} \beta_{1}^{-\frac{1}{2}} T^{-\frac{1}{2M}+\frac{\eps}{2}}}} \sqrt{\frac{1-2^{\frac{1}{2}} \beta_{M-2}^{-\frac{1}{2}} T^{-\frac{M-2}{2M}+\frac{\eps}{2}}}{1+2^{\frac{1}{2}} \beta_{M-2}^{-\frac{1}{2}} T^{-\frac{M-2}{2M}+\frac{\eps}{2}}}} \\
\geq \sqrt{\frac{1-2^{\frac{1}{2}} \beta_{1}^{-\frac{1}{2}} T^{-\frac{1}{2M}+\frac{\eps}{2}}}{1+2^{\frac{1}{2}} \beta_{1}^{-\frac{1}{2}} T^{-\frac{1}{2M}+\frac{\eps}{2}}}} \sqrt{\frac{1-2^{\frac{1}{2}} \beta_{M-2}^{-\frac{1}{2}} T^{-\frac{M-2}{2M}+\frac{\eps}{2}}}{1+2^{\frac{1}{2}} \beta_{M-2}^{-\frac{1}{2}} T^{-\frac{M-2}{2M}+\frac{\eps}{2}}}} > \frac{1}{4} \geq \frac{\frac{1}{2}\beta_{M-1} T^{\frac{M-1}{M}}}{T - \frac{1}{2}\beta_{M-1} T^{\frac{M-1}{M}}},
\end{multline*}
where the first and second inequalities are due to \eqref{eqn:MStage:ConfidenceBound1} and \eqref{eqn:MStage:ConfidenceBound0}; the third inequality is due to \eqref{eqn:WLOG}; the fourth inequality is due to Lemma~\ref{lem:AlgebraicTrick3}; the last inequality is due to Lemma~\ref{lem:AlgebraicTrick4}.
Due to the above sequence of inequalities, we have $\frac{1}{\widehat{\rho}_{M-2}} \leq \frac{T - \frac{1}{2}\beta_{M-1} T^{\frac{M-1}{M}}}{\frac{1}{2}\beta_{M-1} T^{\frac{M-1}{M}}}$, which leads to
\begin{align*}
\frac{\widehat{\sigma}_{M-2}(1)}{\widehat{\sigma}_{M-2}(1) + \widehat{\sigma}_{M-2}(0)} T  \geq \frac{1}{1+\frac{T - \frac{1}{2} \beta_{M-1} T^{\frac{M-1}{M}}}{\frac{1}{2} \beta_{M-1} T^{\frac{M-1}{M}}}} T = \frac{1}{2} \beta_{M-1} T^{\frac{M-1}{M}}.
\end{align*}
So Algorithm~\ref{alg:MStageANA} goes to Line~\ref{mrk:Case3} in the $(M-2)$-th stage experiment. 
Then Algorithm~\ref{alg:MStageANA} goes to Line~\ref{mrk:LastStage} in the last stage.
We further distinguish two cases.

\noindent\textbf{Case ($\bm{M-1}$).1}:
In addition to the conditions in Case $(M-1)$ above, we also have
\begin{align*}
\widehat{\rho}_{M-1} > \frac{T - \frac{1}{2} \beta_{M-1} T^{\frac{M-1}{M}}}{\frac{1}{2} \beta_{M-1} T^{\frac{M-1}{M}}}.
\end{align*}
Similar to the analysis in Case $m$.1, we proceed with the following analysis.
In Case $(M-1)$.1, 
\begin{align*}
\frac{\widehat{\sigma}_{M-1}(0)}{\widehat{\sigma}_{M-1}(1) + \widehat{\sigma}_{M-1}(0)} T < \frac{1}{\frac{T - \frac{1}{2} \beta_{M-1} T^{\frac{M-1}{M}}}{\frac{1}{2} \beta_{M-1} T^{\frac{M-1}{M}}}+1} T = \frac{1}{2} \beta_{M-1} T^{\frac{M-1}{M}}.
\end{align*}
So Algorithm~\ref{alg:MStageANA} goes to Line~\ref{mrk:LastStage:Case1} in the $(M-1)$-th stage experiment, and we have
\begin{align*}
(T(1), T(0)) = \bigg(T - \frac{1}{2} \beta_{M-1} T^{\frac{M-1}{M}}, \frac{1}{2} \beta_{M-1} T^{\frac{M-1}{M}}\bigg).
\end{align*}
We can express 
\begin{align*}
\frac{V(T(1), T(0) \vert \cE)}{V(T^*(1), T^*(0))} = & \ \frac{\sigma^2(1) + \sigma^2(0) + \frac{\frac{1}{2} \beta_{M-1} T^{\frac{M-1}{M}}}{T - \frac{1}{2} \beta_{M-1} T^{\frac{M-1}{M}}} \sigma^2(1) + \frac{T - \frac{1}{2} \beta_{M-1} T^{\frac{M-1}{M}}}{\frac{1}{2} \beta_{M-1} T^{\frac{M-1}{M}}} \sigma^2(0)}{(\sigma(1) + \sigma(0))^2}.
\end{align*}
Note that, 
\begin{multline}
\rho \cdot \sqrt{\frac{1-2^{\frac{1}{2}} \beta_{M-2}^{-\frac{1}{2}} T^{-\frac{M-2}{2M}+\frac{\eps}{2}}}{1+2^{\frac{1}{2}} \beta_{M-2}^{-\frac{1}{2}} T^{-\frac{M-2}{2M}+\frac{\eps}{2}}}} \leq \widehat{\rho}_{M-2} \leq \frac{T - \frac{1}{2} \beta_{M-1} T^{\frac{M-1}{M}}}{\frac{1}{2} \beta_{M-1} T^{\frac{M-1}{M}}} \\
< \widehat{\rho}_{M-1} \leq \rho \cdot \sqrt{\frac{1+2^{\frac{1}{2}} \beta_{M-1}^{-\frac{1}{2}} T^{-\frac{M-1}{2M}+\frac{\eps}{2}}}{1-2^{\frac{1}{2}} \beta_{M-1}^{-\frac{1}{2}} T^{-\frac{M-1}{2M}+\frac{\eps}{2}}}} < \rho \cdot \sqrt{\frac{1+2^{\frac{1}{2}} \beta_{M-2}^{-\frac{1}{2}} T^{-\frac{M-2}{2M}+\frac{\eps}{2}}}{1-2^{\frac{1}{2}} \beta_{M-2}^{-\frac{1}{2}} T^{-\frac{M-2}{2M}+\frac{\eps}{2}}}}, \label{eqn:TwoSidedBoundsM-1}
\end{multline}
where the first and the fourth inequalities are due to \eqref{eqn:MStage:ConfidenceBound1} and \eqref{eqn:MStage:ConfidenceBound0};
the second and the third inequalities are due to the conditions of Case $(M-1)$.1;
the last inequality is because $\beta_{M-2} T^{\frac{M-2}{M}} < \beta_{M-1} T^{\frac{M-1}{M}}$ so we have $2^{\frac{1}{2}} \beta_{M-1}^{-\frac{1}{2}} T^{-\frac{M-1}{2M}+\frac{\eps}{2}} < 2^{\frac{1}{2}} \beta_{M-2}^{-\frac{1}{2}} T^{-\frac{M-2}{2M}+\frac{\eps}{2}}$.

Then we have
\begin{align}
\frac{V(T(1), T(0) \vert \cE)}{V(T^*(1), T^*(0))} \leq & \ \frac{\sigma^2(1) + \sigma^2(0) + \sigma(1) \sigma(0) \left( \sqrt{\frac{1+2^{\frac{1}{2}} \beta_{M-2}^{-\frac{1}{2}} T^{-\frac{M-2}{2M}+\frac{\eps}{2}}}{1-2^{\frac{1}{2}} \beta_{M-2}^{-\frac{1}{2}} T^{-\frac{M-2}{2M}+\frac{\eps}{2}}}} + \sqrt{\frac{1-2^{\frac{1}{2}} \beta_{M-2}^{-\frac{1}{2}} T^{-\frac{M-2}{2M}+\frac{\eps}{2}}}{1+2^{\frac{1}{2}} \beta_{M-2}^{-\frac{1}{2}} T^{-\frac{M-2}{2M}+\frac{\eps}{2}}}} \right) }{(\sigma(1) + \sigma(0))^2} \nonumber \\
= & \ 1 + \frac{\sigma(1)\sigma(0)}{(\sigma(1) + \sigma(0))^2} \cdot \left(\frac{2}{\sqrt{1-2\beta_{M-2}^{-1}T^{-\frac{M-2}{M}+\eps}}} - 2\right) \nonumber \\
\leq & \ 1 + \frac{\sigma(1)\sigma(0)}{(\sigma(1) + \sigma(0))^2} \cdot \left(4 \beta_{M-2}^{-1} T^{-\frac{M-2}{M}+\eps}\right), \label{eqn:BreakIntoTwoPartsM-11}
\end{align}
where the first inequality is due to Lemma~\ref{lem:h:rhohat}; the last inequality is due to Lemma~\ref{lem:AlgebraicTrick2}.

Note that, $\frac{\sigma(1) \sigma(0)}{(\sigma(1) + \sigma(0))^2} = \frac{\rho}{(\rho+1)^2}$ is a decreasing function when $\rho>1$.
Note also that,
\begin{align*}
\rho > \frac{T - \frac{1}{2} \beta_{M-1} T^{\frac{M-1}{M}}}{\frac{1}{2} \beta_{M-1} T^{\frac{M-1}{M}}} \cdot \sqrt{\frac{1-2^{\frac{1}{2}} \beta_{M-1}^{-\frac{1}{2}} T^{-\frac{M-1}{2M} + \frac{\eps}{2}}}{1+2^{\frac{1}{2}} \beta_{M-1}^{-\frac{1}{2}} T^{-\frac{M-1}{2M} + \frac{\eps}{2}}}} > \frac{1}{2} \frac{T - \frac{1}{2} \beta_{M-1} T^{\frac{M-1}{M}}}{\frac{1}{2} \beta_{M-1} T^{\frac{M-1}{M}}} > 1,
\end{align*}
where the first inequality is due to \eqref{eqn:TwoSidedBoundsM-1}; the second inequality is due to Lemma~\ref{lem:AlgebraicTrick3}; the last inequality is due to Lemma~\ref{lem:AlgebraicTrick4}.

Then we have
\begin{align*}
\frac{\sigma(1) \sigma(0)}{(\sigma(1) + \sigma(0))^2} 
< \frac{\frac{1}{2} \frac{T - \frac{1}{2} \beta_{M-1} T^{\frac{M-1}{M}}}{\frac{1}{2} \beta_{M-1} T^{\frac{M-1}{M}}}}{\bigg( 1 + \frac{1}{2} \frac{T - \frac{1}{2} \beta_{M-1} T^{\frac{M-1}{M}}}{\frac{1}{2} \beta_{M-1} T^{\frac{M-1}{M}}} \bigg)^2} 
= \frac{\beta_{M-1} T^{\frac{M-1}{M}} (T - \frac{1}{2} \beta_{M-1} T^{\frac{M-1}{M}}) }{(T + \frac{1}{2} \beta_{M-1} T^{\frac{M-1}{M}})^2} 
\leq \frac{\beta_{M-1} T^{\frac{M-1}{M}}}{T}.
\end{align*}
Putting this into \eqref{eqn:BreakIntoTwoPartsM-11} we have that in Case $(M-1)$.1,
\begin{align*}
\frac{V(T(1), T(0) \vert \cE)}{V(T^*(1), T^*(0))} \leq \ 1 + \frac{4\beta_{M-1}}{\beta_{M-2}}T^{-\frac{M-1}{M}+\eps} = 1 + 4 \cdot 15^{-\frac{1}{M}} \cdot T^{-\frac{M-1}{M}+\eps}.
\end{align*}

\noindent\textbf{Case ($\bm{M-1}$).2}:
In addition to the conditions in Case $(M-1)$ above, we also have
\begin{align*}
\widehat{\rho}_{M-1} \leq \frac{T - \frac{1}{2} \beta_{M-1} T^{\frac{M-1}{M}}}{\frac{1}{2} \beta_{M-1} T^{\frac{M-1}{M}}}.
\end{align*}
Due to the condition of Case ($M-1$).2, we immediately have
\begin{align*}
\frac{\widehat{\sigma}_{M-1}(0)}{\widehat{\sigma}_{M-1}(1) + \widehat{\sigma}_{M-1}(0)} T \geq \frac{1}{\frac{T - \frac{1}{2} \beta_{M-1} T^{\frac{M-1}{M}}}{\frac{1}{2} \beta_{M-1} T^{\frac{M-1}{M}}}+1} T = \frac{1}{2} \beta_{M-1} T^{\frac{M-1}{M}}.
\end{align*}
On the other hand, since
\begin{multline*}
\widehat{\rho}_{M-1} \geq \rho \sqrt{\frac{1-2^{\frac{1}{2}} \beta_{M-1}^{-\frac{1}{2}} T^{-\frac{M-1}{2M}+\frac{\eps}{2}}}{1+2^{\frac{1}{2}} \beta_{M-1}^{-\frac{1}{2}} T^{-\frac{M-1}{2M}+\frac{\eps}{2}}}} \geq \widehat{\rho}_1 \sqrt{\frac{1-2^{\frac{1}{2}} \beta_{1}^{-\frac{1}{2}} T^{-\frac{1}{2M}+\frac{\eps}{2}}}{1+2^{\frac{1}{2}} \beta_{1}^{-\frac{1}{2}} T^{-\frac{1}{2M}+\frac{\eps}{2}}}} \sqrt{\frac{1-2^{\frac{1}{2}} \beta_{M-1}^{-\frac{1}{2}} T^{-\frac{M-1}{2M}+\frac{\eps}{2}}}{1+2^{\frac{1}{2}} \beta_{M-1}^{-\frac{1}{2}} T^{-\frac{M-1}{2M}+\frac{\eps}{2}}}} \\
\geq \sqrt{\frac{1-2^{\frac{1}{2}} \beta_{1}^{-\frac{1}{2}} T^{-\frac{1}{2M}+\frac{\eps}{2}}}{1+2^{\frac{1}{2}} \beta_{1}^{-\frac{1}{2}} T^{-\frac{1}{2M}+\frac{\eps}{2}}}} \sqrt{\frac{1-2^{\frac{1}{2}} \beta_{M-1}^{-\frac{1}{2}} T^{-\frac{M-1}{2M}+\frac{\eps}{2}}}{1+2^{\frac{1}{2}} \beta_{M-1}^{-\frac{1}{2}} T^{-\frac{M-1}{2M}+\frac{\eps}{2}}}} > \frac{1}{4} \geq \frac{\frac{1}{2}\beta_{M-1} T^{\frac{M-1}{M}}}{T - \frac{1}{2}\beta_{M-1} T^{\frac{M-1}{M}}},
\end{multline*}
where the first and second inequalities are due to \eqref{eqn:MStage:ConfidenceBound1} and \eqref{eqn:MStage:ConfidenceBound0}; the third inequality is due to \eqref{eqn:WLOG}; the fourth inequality is due to Lemma~\ref{lem:AlgebraicTrick3}; the last inequality is due to Lemma~\ref{lem:AlgebraicTrick4}.
Due to the above sequence of inequalities, we have $\frac{1}{\widehat{\rho}_{M-1}} \leq \frac{T - \frac{1}{2}\beta_{M-1} T^{\frac{M-1}{M}}}{\frac{1}{2}\beta_{M-1} T^{\frac{M-1}{M}}}$, which leads to
\begin{align*}
\frac{\widehat{\sigma}_{M-1}(1)}{\widehat{\sigma}_{M-1}(1) + \widehat{\sigma}_{M-1}(0)} T  \geq \frac{1}{1+\frac{T - \frac{1}{2} \beta_{M-1} T^{\frac{M-1}{M}}}{\frac{1}{2} \beta_{M-1} T^{\frac{M-1}{M}}}} T = \frac{1}{2} \beta_{M-1} T^{\frac{M-1}{M}}.
\end{align*}
So Algorithm~\ref{alg:MStageANA} goes to Line~\ref{mrk:LastStage:Case2} in the $(M-1)$-th stage experiment. 
Then we have
\begin{align*}
(T(1), T(0)) = \bigg(\frac{\widehat{\sigma}_{M-1}(1)}{\widehat{\sigma}_{M-1}(1) + \widehat{\sigma}_{M-1}(0)} T, \frac{\widehat{\sigma}_{M-1}(0)}{\widehat{\sigma}_{M-1}(1) + \widehat{\sigma}_{M-1}(0)} T\bigg).
\end{align*}
We can then express
\begin{align*}
\frac{V(T(1), T(0) \vert \cE)}{V(T^*(1), T^*(0))} = & \ \frac{\sigma^2(1) + \sigma^2(0) + \frac{1}{\widehat{\rho}_{M-1}} \sigma^2(1) + \widehat{\rho}_{M-1} \sigma^2(0)}{(\sigma(1) + \sigma(0))^2}.
\end{align*}
Recall that, conditional on $\cE$, \eqref{eqn:MStage:ConfidenceBound1} and \eqref{eqn:MStage:ConfidenceBound0} lead to
\begin{align*}
\rho \cdot \sqrt{\frac{1-2^{\frac{1}{2}} \beta_{M-1}^{-\frac{1}{2}} T^{-\frac{M-1}{2M} + \frac{\eps}{2}}}{1+2^{\frac{1}{2}} \beta_{M-1}^{-\frac{1}{2}} T^{-\frac{M-1}{2M} + \frac{\eps}{2}}}} \leq \widehat{\rho}_{M-1} \leq \rho \sqrt{\frac{1+2^{\frac{1}{2}} \beta_{M-1}^{-\frac{1}{2}} T^{-\frac{M-1}{2M} + \frac{\eps}{2}}}{1-2^{\frac{1}{2}} \beta_{M-1}^{-\frac{1}{2}} T^{-\frac{M-1}{2M} + \frac{\eps}{2}}}}.
\end{align*}
So we have
\begin{align*}
\frac{V(T(1), T(0) \vert \cE)}{V(T^*(1), T^*(0))} \leq & \ \frac{\sigma^2(1) + \sigma^2(0) + \sigma(1)\sigma(0) \bigg( \sqrt{\frac{1-2^{\frac{1}{2}} \beta_{M-1}^{-\frac{1}{2}} T^{-\frac{M-1}{2M} + \frac{\eps}{2}}}{1+2^{\frac{1}{2}} \beta_{M-1}^{-\frac{1}{2}} T^{-\frac{M-1}{2M} + \frac{\eps}{2}}}} + \sqrt{\frac{1+2^{\frac{1}{2}} \beta_{M-1}^{-\frac{1}{2}} T^{-\frac{M-1}{2M} + \frac{\eps}{2}}}{1-2^{\frac{1}{2}} \beta_{M-1}^{-\frac{1}{2}} T^{-\frac{M-1}{2M} + \frac{\eps}{2}}}} \bigg)}{(\sigma(1) + \sigma(0))^2} \\
= & \ 1 + \frac{\sigma(1) \sigma(0)}{(\sigma(1) + \sigma(0))^2} \cdot \bigg( \frac{2}{\sqrt{1 - 2 \beta_{M-1}^{-1} T^{ -\frac{M-1}{M} + \eps }}} - 2 \bigg) \\
\leq & \ 1 + \frac{\sigma(1) \sigma(0)}{(\sigma(1) + \sigma(0))^2} \cdot 4 \beta_{M-1}^{-1} T^{ -\frac{M-1}{M} + \eps } \\
\leq & \ 1 + \beta_{M-1}^{-1} T^{ -\frac{M-1}{M} + \eps },
\end{align*}
where the first inequality is due to Lemma~\ref{lem:h:rhohat}; the second inequality is due to Lemma~\ref{lem:AlgebraicTrick2}; the last inequality is because $\frac{\sigma(1) \sigma(0)}{(\sigma(1) + \sigma(0))^2} \leq \frac{1}{4}$.

Finally, using the definition of $\beta_{M-1} = 6 \cdot 15^{-\frac{M-1}{M}}$,
\begin{align*}
\beta_{M-1}^{-1} = \frac{1}{6} \cdot 15^{\frac{M-1}{M}} = \frac{15}{6} \cdot 15^{-\frac{1}{15}} < 4 \cdot 15^{-\frac{1}{15}}.
\end{align*}
So we have
\begin{align*}
\frac{V(T(1), T(0) \vert \cE)}{V(T^*(1), T^*(0))} \leq 1 + \beta_{M-1}^{-1} T^{ -\frac{M-1}{M} + \eps } < 1 + 4 \cdot 15^{-\frac{1}{15}} T^{ -\frac{M-1}{M} + \eps }.
\end{align*}

To conclude, in all cases, we have shown that
\begin{align*}
\frac{V(T(1), T(0) \vert \cE)}{V(T^*(1), T^*(0))} \leq 1 + 4 \cdot 15^{-\frac{1}{15}} T^{ -\frac{M-1}{M} + \eps }.
\end{align*}
\hfill \halmos
\endproof

\subsection{Proof of Theorem~\ref{thm:ANA:LB}}
\label{sec:proof:thm:LB}

\proof{Proof of Theorem~\ref{thm:ANA:LB}.}
Fix any adaptive design of experiment $\pi$.
Let $T \geq 4$ and define
\begin{align*}
\eps = \frac{1}{3T^{\frac{1}{2}}} \leq \frac{1}{6}.
\end{align*}
Let there be two discrete probability distributions $\nu$ and $\nu'$, defined as follows.
Both distributions have three discrete supports $\{-1,0,1\}$.
The probability mass for distribution $\nu$ is given by 
\begin{align*}
p_{-1} = \frac{1}{3}, && p_{0} = \frac{1}{3}, && p_{1} = \frac{1}{3}.
\end{align*}
The probability mass for distribution $\nu'$ is given by 
\begin{align*}
p'_{-1} = \frac{1}{3}+\frac{\eps}{2}, && p'_{0} = \frac{1}{3}-\eps, && p'_{1} = \frac{1}{3}+\frac{\eps}{2}.
\end{align*}
Then we immediately have
\begin{align*}
\sigma^2(\nu) = \frac{2}{3}, && \sigma^2(\nu') = \frac{2}{3} + \eps.
\end{align*}

Moreover, we upper bound the KL-divergences of these two probability distributions as follows.
\begin{align}
D_{KL}(\nu \vert\vert \nu') = & \ \frac{1}{3} \log\left(\frac{1}{1+\frac{3}{2}\eps}\right) + \frac{1}{3} \log\left(\frac{1}{1-3\eps}\right) + \frac{1}{3} \log\left(\frac{1}{1+\frac{3}{2}\eps}\right) \nonumber \\
= & \ \frac{1}{3} \log\left( \frac{1}{1-\frac{27}{4}\eps^2-\frac{27}{4}\eps^3} \right) \nonumber \\
\leq & \ \frac{1}{3} \log\left( 1 + \frac{27}{2} \eps^2\right) \nonumber \\
\leq & \ \frac{1}{3} \cdot \frac{27}{2} \eps^2 \nonumber \\
= & \ \frac{9}{2} \eps^2, \label{eqn:KL-UB1}
\end{align}
where the first inequality is due to Lemma~\ref{lem:AlgebraicTrick7};
the second inequality is because for any $x>0$, $\log{(1+x)} \leq x$.

On the other hand, the KL-divergence calculated in the other way is upper bounded by.
\begin{align}
D_{KL}(\nu' \vert\vert \nu) = & \ \left(\frac{1}{3}+\frac{1}{2}\eps\right) \log\left(1+\frac{3}{2}\eps\right) + \left(\frac{1}{3} - \eps\right) \log\left(1-3\eps\right) + \left(\frac{1}{3}+\frac{1}{2}\eps\right) \log\left(1+\frac{3}{2}\eps\right) \nonumber \\
= & \ \left(\frac{2}{3}+\eps\right) \log\left(1+\frac{3}{2}\eps\right) + \left(\frac{1}{3}-\eps\right) \log\left(1-3\eps\right) \nonumber \\
\leq & \ \left(\frac{2}{3}+\eps\right) \left(\frac{3}{2}\eps\right) + \left(\frac{1}{3}-\eps\right) \left(-3\eps\right) \nonumber \\
\leq & \ \eps + \frac{3}{2} \eps^2 - \eps + 3 \eps^2 \nonumber \\
= & \ \frac{9}{2} \eps^2, \label{eqn:KL-UB2}
\end{align}
where the first inequality is because first, for any $x>0$, $\log{(1+x)} \leq x$, and second, for any $0<x<1$, $\log{(1-x)} \leq -x$.

We will use these two probability distributions to construct two problem instances.
Consider the first problem instance where $Y(1) \sim \nu', Y(0) \sim \nu$.
Denote $\Pr_{\nu',\nu}$ as the probability distribution induced by this problem instance and by the design of experiment $\pi$, where we drop the dependence on $\pi$ as it is clear from the context.
Denote $\bE_{\nu',\nu}$ as the expectation taken under $\Pr_{\nu',\nu}$.

Similarly, consider the second problem instance where $Y(1) \sim \nu, Y(0) \sim \nu'$.
Denote $\Pr_{\nu,\nu'}$ as the probability distribution induced by this problem instance and by the design of experiment $\pi$, where we drop the dependence on $\pi$.
Denote $\bE_{\nu,\nu'}$ as the expectation taken under $\Pr_{\nu,\nu'}$.

Now we focus on the first instance $(Y(1), Y(0)) \sim (\nu', \nu)$.
Note that $\sigma^2(1) = \sigma^2(\nu') > \sigma^2(\nu) = \sigma^2(0)$.
When $T^\pi(1) \leq \frac{T}{2}$, we have
\begin{align*}
\frac{T^\pi(1)}{T^\pi(0)} \leq 1 < \frac{\sigma(1)}{\sigma(0)}.
\end{align*}
Due to Lemma~\ref{lem:h:rhohat},
\begin{align*}
\frac{\bE_{\nu',\nu}\left[ V(T^\pi(1), T^\pi(0)) \right]}{V\left( T^*(1), T^*(0) \right)} = \frac{\sigma^2(1) + \sigma^2(0) + \frac{T^\pi(0)}{T^\pi(1)}\sigma^2(1)+\frac{T^\pi(1)}{T^\pi(0)}\sigma^2(0)}{( \sigma(1)+\sigma(0) )^2} \geq \frac{2\sigma^2(1)+2\sigma^2(0)}{( \sigma(1)+\sigma(0) )^2}.
\end{align*}
On the other hand, when $T^\pi(1) > \frac{T}{2}$, the ratio is greater or equal to $1$, i.e., 
\begin{align*}
\frac{\bE_{\nu',\nu}\left[ V(T^\pi(1), T^\pi(0)) \right]}{V\left( T^*(1), T^*(0) \right)} \geq 1.
\end{align*}
Putting the above two cases together we have
\begin{align}
\frac{\bE_{\nu',\nu}\left[ V(T^\pi(1), T^\pi(0)) \right]}{V\left( T^*(1), T^*(0) \right)} 
\geq & \ \mathrm{Pr}_{\nu',\nu}\left(T^\pi(1) \leq \frac{T}{2}\right) \cdot \frac{2 \cdot (\frac{2}{3} + \eps) + 2 \cdot (\frac{2}{3}) }{\left(\sqrt{\frac{2}{3}+\eps} + \sqrt{\frac{2}{3}}\right)^2} + \mathrm{Pr}_{\nu',\nu}\left(T^\pi(1) > \frac{T}{2}\right) \cdot 1. \nonumber \\
= & \ 1 + \mathrm{Pr}_{\nu',\nu}\left(T^\pi(1) \leq \frac{T}{2}\right) \cdot \frac{\frac{4}{3} + \eps - \frac{4}{3}\sqrt{1+\frac{3}{2}\eps}}{\frac{4}{3} + \eps + \frac{4}{3}\sqrt{1+\frac{3}{2}\eps}}. \label{eqn:LB:BreakIntoFewParts}
\end{align}
We further have
\begin{align*}
\frac{\frac{4}{3} + \eps - \frac{4}{3}\sqrt{1+\frac{3}{2}\eps}}{\frac{4}{3} + \eps + \frac{4}{3}\sqrt{1+\frac{3}{2}\eps}} \geq \frac{\frac{4}{3} + \eps - \frac{4}{3}(1 + \frac{3}{4}\eps - \frac{9}{64}\eps^2)}{\frac{4}{3} + \eps + \frac{4}{3}(1 + \frac{3}{4}\eps)} = \frac{\frac{3}{16}\eps^2}{\frac{8}{3}+2\eps} \geq \frac{\eps^2}{16},
\end{align*}
where the first inequality is due to Lemma~\ref{lem:AlgebraicTrick6}; the last inequality is due to $\eps \leq \frac{1}{6}$.
Putting the above inequality into \eqref{eqn:LB:BreakIntoFewParts} we have
\begin{align}
\frac{\bE_{\nu',\nu}\left[ V(T^\pi(1), T^\pi(0)) \right]}{V\left( T^*(1), T^*(0) \right)} 
> 1 + \mathrm{Pr}_{\nu',\nu}\left(T^\pi(1) \leq \frac{T}{2}\right) \cdot \frac{\eps^2}{16}. \label{eqn:LB:FirstInstance}
\end{align}

Next we focus on the second instance $(Y(1), Y(0)) \sim (\nu, \nu')$.
Similar to the above analysis, we have
\begin{align}
\frac{\bE\left[ V(T^\pi(1), T^\pi(0)) \right]}{V\left( T^*(1), T^*(0) \right)} 
\geq & \ \mathrm{Pr}_{\nu,\nu'}\left(T^\pi(1) \leq \frac{T}{2}\right) \cdot 1 + \mathrm{Pr}_{\nu,\nu'}\left(T^\pi(1) > \frac{T}{2}\right) \cdot \frac{2 \cdot (\frac{2}{3}) + 2 \cdot (\frac{2}{3} + \eps) }{\left(\sqrt{\frac{2}{3}} + \sqrt{\frac{2}{3}+\eps}\right)^2}. \nonumber \\
= & \ 1 + \mathrm{Pr}_{\nu,\nu'}\left(T^\pi(1) > \frac{T}{2}\right) \cdot \frac{\frac{4}{3} + \eps - \frac{4}{3}\sqrt{1+\frac{3}{2}\eps}}{\frac{4}{3} + \eps + \frac{4}{3}\sqrt{1+\frac{3}{2}\eps}}. \nonumber \\
> & \ 1 + \mathrm{Pr}_{\nu,\nu'}\left(T^\pi(1) > \frac{T}{2}\right) \cdot \frac{\eps^2}{16}. \label{eqn:LB:SecondInstance}
\end{align}

Combining \eqref{eqn:LB:FirstInstance} and \eqref{eqn:LB:SecondInstance} we have
\begin{align}
\frac{\bE_{\nu',\nu}\left[ V(T^\pi(1), T^\pi(0)) \right]}{V\left( T^*(1), T^*(0) \right)}  + & \frac{\bE\left[ V(T^\pi(1), T^\pi(0)) \right]}{V\left( T^*(1), T^*(0) \right)} \nonumber \\
\geq & \ 2 + \frac{\eps^2}{16} \cdot \left( \mathrm{Pr}_{\nu',\nu}\left(T^\pi(1) \leq \frac{T}{2}\right) + \mathrm{Pr}_{\nu,\nu'}\left(T^\pi(1) > \frac{T}{2}\right) \right) \nonumber \\
\geq & \ 2 + \frac{\eps^2}{16} \cdot \exp\left\{- D_{KL}\left( \mathrm{Pr}_{\nu',\nu}, \mathrm{Pr}_{\nu,\nu'} \right)\right\}, \label{eqn:LB:Bretagnolle-Huber}
\end{align}
where the second inequality is due to Bretagnolle-Huber inequality \citep{bretagnolle1979estimation, lattimore2020bandit}.

Next we upper bound $D_{KL}\left( \mathrm{Pr}_{\nu',\nu}, \mathrm{Pr}_{\nu,\nu'} \right)$.
\begin{align*}
D_{KL}(\mathrm{Pr}_{\nu',\nu} \vert \vert \mathrm{Pr}_{\nu,\nu'}) = & \ \bE_{\nu',\nu}\left[T^\pi(1)\right] D_{KL}(\nu' \vert\vert \nu) + \bE_{\nu',\nu}\left[T^\pi(0)\right] D_{KL}(\nu \vert\vert \nu') \\
\leq & \ \bE_{\nu',\nu}\left[T^\pi(1)\right] \frac{9}{2}\eps^2 + \bE_{\nu',\nu}\left[T^\pi(0)\right] \frac{9}{2}\eps^2 \\
= & T \cdot \frac{9}{2}\eps^2,
\end{align*}
where the inequality is due to \eqref{eqn:KL-UB1} and \eqref{eqn:KL-UB2}.
Putting this into \eqref{eqn:LB:Bretagnolle-Huber} we have
\begin{multline*}
\frac{\bE_{\nu',\nu}\left[ V(T^\pi(1), T^\pi(0)) \right]}{V\left( T^*(1), T^*(0) \right)} + \frac{\bE\left[ V(T^\pi(1), T^\pi(0)) \right]}{V\left( T^*(1), T^*(0) \right)} \geq 2 + \frac{\eps^2}{16} \exp\left\{- T \cdot \frac{9}{2}\eps^2\right\} \\
= 2+ \frac{1}{144 \exp{\left\{\frac{1}{2}\right\}}} \cdot T^{-1} \geq 2 + \frac{1}{240} T^{-1},
\end{multline*}
where the equality is using $\eps = \frac{1}{3T^{\frac{1}{2}}}$.
Using the above inequality we have
\begin{align*}
\max\left\{\frac{\bE_{\nu',\nu}\left[ V(T^\pi(1), T^\pi(0)) \right]}{V\left( T^*(1), T^*(0) \right)},  \frac{\bE\left[ V(T^\pi(1), T^\pi(0)) \right]}{V\left( T^*(1), T^*(0) \right)}\right\} \geq 1 + \frac{1}{480} \cdot T^{-1}.
\end{align*}
\hfill \halmos
\endproof

\subsection{Proof of Theorem~\ref{thm:estimation}}
\label{sec:proof:thm:estimation}

\proof{Proof of Theorem~\ref{thm:estimation}.}
Our proof proceeds by identifying the following sequence of realizations of the sample variances,
\begin{align*}
\widehat{\sigma}^2_1(1) & = a_1(1), & \widehat{\sigma}^2_2(1) & = a_2(1), & ..., & & \widehat{\sigma}^2_{M-1}(1) & = a_{M-1}(1), \\
\widehat{\sigma}^2_1(0) & = a_1(0), & \widehat{\sigma}^2_2(0) & = a_2(0), & ..., & & \widehat{\sigma}^2_{M-1}(0) & = a_{M-1}(0). 
\end{align*}
We denote the above event using $\cE(a_1(1), ..., a_{M-1}(1), a_1(0), ..., a_{M-1}(0)) := \cE(\bm{a})$.

Below we show that, for any $\bm{a} \in \bR^{2(M-1)}$, and conditional on event $\cE(\bm{a})$, the difference-in-means estimator as defined in \eqref{eqn:Estimator} is unbiased, that is,
\begin{align*}
\bE\Big[ \widehat{\tau} \Big\vert \cE(\bm{a}) \Big] = \tau.
\end{align*}

Note that, conditional on $\cE(\bm{a})$, our adaptive Neyman allocation algorithm (including both Algorithm~\ref{alg:2StageANA} and Algorithm~\ref{alg:MStageANA}) will uniquely determine the number of treated and control units assigned in each stage, which are $T_1(1), T_1(0), ..., T_M(1), T_M(1).$
In other words, conditional on $\cE(\bm{a})$, we can think of $T_1(1), T_1(0), ..., T_M(1), T_M(1)$ as constants.
Consequently, conditional on $\cE(\bm{a})$, we can also think of $T(1) = \sum_{m=1}^M T_m(1)$ and $T(0) = \sum_{m=1}^M T_m(0)$ as constants.

Now we focus on the difference-in-means estimator.
\begin{align}
\bE\bigg[ \frac{1}{T(1)} \sum_{t=1}^T Y_t \bI\{W_t=1\} \Big\vert \cE(\bm{a}) \bigg] 
= & \bE\bigg[ \frac{1}{T(1)} \sum_{m=1}^M \sum_{t=\sum_{l=1}^{m-1} T_l + 1}^{\sum_{l=1}^m T_l} Y_t \bI\{W_t=1\} \Big\vert \cE(\bm{a}) \bigg] \nonumber \\
= & \frac{1}{T(1)} \sum_{m=1}^M \sum_{t=\sum_{l=1}^{m-1} T_l + 1}^{\sum_{l=1}^m T_l} \bE\Big[ Y_t(1) \Big\vert \cE(\bm{a}) \Big] \cdot \bE\Big[ \bI\{W_t=1\} \Big\vert \cE(\bm{a}) \Big] \nonumber \\
= & \frac{1}{T(1)} \sum_{m=1}^M \sum_{t=\sum_{l=1}^{m-1} T_l + 1}^{\sum_{l=1}^m T_l} \frac{T_m(1)}{T_m} \ \bE\Big[ Y_t(1) \Big\vert \cE(\bm{a}) \Big] \label{eqn:CountingConditionalExpectation}
\end{align}
where the first equality is counting by each stage;
the second equality is because conditional on $\cE(\bm{a})$, $T_1(1), T_1(0), ..., T_M(1), T_M(1)$ are all fixed, so that $\bI\{W_t=1\}$ only depends on the randomization and thus is independent of $Y_t$.

Next, we focus on $\bE\big[ Y_t(1) \Big\vert \cE(\bm{a}) \big]$.
Because of Lemma~\ref{lem:CommonKnowledge:SumOfSquares}, the event $\widehat{\sigma}^2_m(1) = a_m(1)$ can be written as
\begin{align}
\frac{\sum_{i=1}^{\sum_{l=1}^m T_l} \sum_{j=1}^{\sum_{l=1}^m T_l} \big(Y_i(1) - Y_j(1)\big)^2 \bI\{W_i=W_j=1\}}{2 \sum_{l=1}^m T_l(1) \big(\sum_{l=1}^m T_l(1) - 1\big)} \ = \ a_m(1). \label{eqn:OneEvent}
\end{align}
Because of Assumption~\ref{asp:symmetric} and because the potential outcomes $(Y_1(1), Y_1(0))$, $(Y_2(1), Y_2(0))$, ..., $(Y_T(1), Y_T(0))$ are mutually independent, the joint distribution of 
\begin{align*}
\big(Y_1(1), Y_2(1), ..., Y_T(1)\big)
\end{align*}
and the joint distribution of
\begin{align*}
\big(2\bE[Y(1)] - Y_1(1), 2\bE[Y(1)] - Y_2(1), ..., 2\bE[Y(1)] - Y_T(1)\big)
\end{align*}
are identical.
Replacing all the random variables $\big(Y_1(1), Y_2(1), ..., Y_T(1)\big)$ by the random variables $\big(2\bE[Y(1)] - Y_1(1), 2\bE[Y(1)] - Y_2(1), ..., 2\bE[Y(1)] - Y_T(1)\big)$, the event $\widehat{\sigma}^2_m(1) = a_m(1)$ is written as
\begin{align*}
& \frac{\sum_{i=1}^{\sum_{l=1}^m T_l} \sum_{j=1}^{\sum_{l=1}^m T_l} \Big( \big(2\bE[Y(1)] - Y_i(1) \big) - \big(2\bE[Y(1)] - Y_j(1)\big) \Big)^2 \bI\{W_i=W_j=1\}}{2 \sum_{l=1}^m T_l(1) \big(\sum_{l=1}^m T_l(1) - 1\big)} \\
= \ & \frac{\sum_{i=1}^{\sum_{l=1}^m T_l} \sum_{j=1}^{\sum_{l=1}^m T_l} \big(Y_i(1) - Y_j(1)\big)^2 \bI\{W_i=W_j=1\}}{2 \sum_{l=1}^m T_l(1) \big(\sum_{l=1}^m T_l(1) - 1\big)} \\
= \ & a_m(1).
\end{align*}
This above expression coincides with \eqref{eqn:OneEvent}.

Similarly, we can replace all the random variables $\big(Y_1(0), Y_2(0), ..., Y_T(0)\big)$ by the random variables $\big(2\bE[Y(0)] - Y_1(0), 2\bE[Y(0)] - Y_2(0), ..., 2\bE[Y(0)] - Y_T(0)\big)$, and the event $\widehat{\sigma}^2_m(0) = a_m(0)$ has the same expression. 
Consequently, 
\begin{align*}
\bE\big[ Y_t(1) \Big\vert \cE(\bm{a}) \big] = \bE\big[ 2\bE[Y(1)] - Y_t(1) \Big\vert \cE(\bm{a}) \big] = 2\bE[Y(1)] - \bE\big[ Y_t(1) \Big\vert \cE(\bm{a}) \big],
\end{align*}
which yields
\begin{align}
\bE\big[ Y_t(1) \Big\vert \cE(\bm{a}) \big] = \bE[Y(1)]. \label{eqn:Unpredictability}
\end{align}

Putting \eqref{eqn:Unpredictability} into \eqref{eqn:CountingConditionalExpectation}, we have
\begin{align*}
\bE\bigg[ \frac{1}{T(1)} \sum_{t=1}^T Y_t \bI\{W_t=1\} \Big\vert \cE(\bm{a}) \bigg] = \frac{1}{T(1)} \sum_{m=1}^M \sum_{t=\sum_{l=1}^{m-1} T_l + 1}^{\sum_{l=1}^m T_l} \frac{T_m(1)}{T_m} \ \bE\big[ Y(1) \big] = \bE\big[ Y(1) \big].
\end{align*}

Similarly,
\begin{align*}
\bE\bigg[ \frac{1}{T(0)} \sum_{t=1}^T Y_t \bI\{W_t=0\} \Big\vert \cE(\bm{a}) \bigg] = \bE\big[ Y(0) \big].
\end{align*}
Combining both equalities we finish the proof.
\hfill \halmos
\endproof

\subsection{Proof of Theorem~\ref{thm:inference}}
\label{sec:proof:thm:inference}

As suggested in \citet{chen2025characterization, khamaru2024inference}, as long as some notion of stability condition holds, one can establish central limit theorems for the sample means, despite that the data is collected in an adaptive fashion.
In Section~\ref{sec:StabilityCondition}, we state the stability condition of our adaptive Neyman allocation algorithms.
Then in Section~\ref{sec:martingaleCLT}, we use the martingale central limit theorem from \citep{brown1971martingale} to prove Theorem~\ref{thm:inference} by checking the Lindeberg condition.

\subsubsection{Stability condition.}
\label{sec:StabilityCondition}

To check the stability condition, we need to construct a pair of deterministic quantities $(T^*(1), T^*(0))$ to compare with the pair of random variables $(T(1), T(0))$.

When $M=2$, define $(T^*(1), T^*(0))$ as follows,
\begin{align}
(T^*(1), T^*(0)) = \left\{
\begin{aligned}
& \left(T-\frac{1}{2} T^{\frac{1}{2}}, \frac{1}{2} T^{\frac{1}{2}}\right), & & \quad \text{if\ } \frac{\sigma(1)}{\sigma(0)} > \frac{T-\frac{1}{2} T^{\frac{1}{2}}}{\frac{1}{2} T^{\frac{1}{2}}},\\
& \left(\frac{\sigma(1)}{\sigma(1) + \sigma(0)} T, \frac{\sigma(0)}{\sigma(1) + \sigma(0)} T\right), & & \quad \text{if\ } \frac{\frac{1}{2} T^{\frac{1}{2}}}{T-\frac{1}{2} T^{\frac{1}{2}}} \leq \frac{\sigma(1)}{\sigma(0)} \leq \frac{T-\frac{1}{2} T^{\frac{1}{2}}}{\frac{1}{2} T^{\frac{1}{2}}},\\
& \left(\frac{1}{2} T^{\frac{1}{2}}, T-\frac{1}{2} T^{\frac{1}{2}}\right), & & \quad \text{if\ } \frac{\sigma(1)}{\sigma(0)} < \frac{\frac{1}{2} T^{\frac{1}{2}}}{T-\frac{1}{2} T^{\frac{1}{2}}}.\\
\end{aligned}
\right. \label{eqn:defn:Stability:2}
\end{align}
When $M \geq 3$, define $(T^*(1), T^*(0))$ as follows,
\begin{align}
(T^*(1), T^*(0)) = \left\{
\begin{aligned}
& \left(T-\frac{1}{2} \beta_1 T^{\frac{1}{M}}, \frac{1}{2} \beta_1 T^{\frac{1}{M}}\right), & & \quad \text{if\ } \frac{\sigma(1)}{\sigma(0)} > \frac{T-\frac{1}{2} \beta_1 T^{\frac{1}{M}}}{\frac{1}{2} \beta_1 T^{\frac{1}{M}}},\\
& \left(\frac{\sigma(1)}{\sigma(1) + \sigma(0)} T, \frac{\sigma(0)}{\sigma(1) + \sigma(0)} T\right), & & \quad \text{if\ } \frac{\frac{1}{2} \beta_1 T^{\frac{1}{M}}}{T-\frac{1}{2} \beta_1 T^{\frac{1}{M}}} \leq \frac{\sigma(1)}{\sigma(0)} \leq \frac{T-\frac{1}{2} \beta_1 T^{\frac{1}{M}}}{\frac{1}{2} \beta_1 T^{\frac{1}{M}}},\\
& \left(\frac{1}{2} \beta_1 T^{\frac{1}{M}}, T-\frac{1}{2} \beta_1 T^{\frac{1}{M}}\right), & & \quad \text{if\ } \frac{\sigma(1)}{\sigma(0)} < \frac{\frac{1}{2} \beta_1 T^{\frac{1}{M}}}{T-\frac{1}{2} \beta_1 T^{\frac{1}{M}}}.\\
\end{aligned}
\right. \label{eqn:defn:Stability:3}
\end{align}

Using the above definitions, we define the stability condition of our adaptive Neyman allocation algorithms below.

\begin{lemma}[Stability Condition]
\label{lem:Stability}
When $M=2$, use Algorithm~\ref{alg:2StageANA} under $\beta = 1$, and set $0 < \eps \leq \frac{1}{8}$.
When $M \geq 3$, use Algorithm~\ref{alg:MStageANA} under $\beta_m = 6 \cdot 15^{-\frac{m}{M}}$, and set $0 < \eps \leq \min\{\frac{1}{M}, \frac{1}{100}\}$.
Let $(T(1), T(0))$ be the number of total treated and control units from the algorithm, respectively. 
Under Assumption~\ref{asp:kurtosis}, there exists $(T^*(1), T^*(0))$ which depends on $\sigma(1), \sigma(0)$, and $T$, such that 
\begin{enumerate}[label=(\roman*)]
\item $T^*(1), T^*(0) \to +\infty$ as $T \to +\infty$; \label{enum:StabilityCondition1}
\item $\frac{T(1)}{T^*(1)} \xrightarrow{p} 1$ and $\frac{T(0)}{T^*(0)} \xrightarrow{p} 1$ as $T \to +\infty$, where $\xrightarrow{p}$ stands for convergence in probability. \label{enum:StabilityCondition2}
\end{enumerate}
\end{lemma}

Below we prove Lemma~\ref{lem:Stability} separately when $M=2$ and when $M \geq 3$ because the algorithms that we use are different.

\proof{Proof of Lemma~\ref{lem:Stability} when $M=2$.}
We can explicitly verify that Condition~\ref{enum:StabilityCondition1} in Lemma~\ref{lem:Stability} is satisfied.
Below we prove Condition~\ref{enum:StabilityCondition2}.

Without loss of generality, we assume $\sigma(1) \geq \sigma(0)$ throughout the proof.
Our analysis of the two-stage adaptive Neyman allocation (Algorithm~\ref{alg:2StageANA}) will be based on the following two events.
\begin{align*}
\cE_1(1) = & \ \bigg\{ \left| \widehat{\sigma}^2_1(1) - \sigma^2(1) \right| < 2^{\frac{1}{2}} T^{-\frac{1}{4} + \frac{\eps}{2}} \sigma^2(1)\bigg\}, \\
\cE_1(0) = & \ \bigg\{ \left| \widehat{\sigma}^2_1(0) - \sigma^2(0) \right| < 2^{\frac{1}{2}} T^{-\frac{1}{4} + \frac{\eps}{2}} \sigma^2(0)\bigg\}.
\end{align*}
Denote $\cE = \cE_1(1) \cap \cE_1(0)$. Then $\Pr(\cE) = \Pr(\cE_1(1) \cap \cE_1(0)) \geq 1 - \Pr(\overline{\cE}_1(1)) - \Pr(\overline{\cE}_1(0))$.
We further have
\begin{align*}
\Pr(\cE) = & \ 1 - \Pr\left( \vert \widehat{\sigma}^2_1(1) - \sigma^2(1) \vert \geq 2^{\frac{1}{2}} T^{-\frac{1}{4} + \frac{\eps}{2}} \sigma^2(1) \right) - \Pr\left( \vert \widehat{\sigma}^2_1(0) - \sigma^2(0) \vert \geq 2^{\frac{1}{2}} T^{-\frac{1}{4} + \frac{\eps}{2}} \sigma^2(0) \right) \\
\geq & \ 1 - \frac{\kappa(1) \sigma^4(1)}{2 T^{-\frac{1}{2} + \eps} \sigma^4(1) T_1(1)} - \frac{\kappa(0) \sigma^4(0)}{2 T^{-\frac{1}{2} + \eps} \sigma^4(0) T_1(0)} \\
= & \ 1 - \frac{\kappa(1) + \kappa(0)}{T^{\eps}},
\end{align*}
where the inequality is due to Lemma~\ref{lem:LightTail}.

Conditional on the event $\cE$, we have
\begin{subequations}
\begin{align}
\sigma^2(1) \left( 1 - 2^{\frac{1}{2}} T^{-\frac{1}{4} + \frac{\eps}{2}} \right) \ \leq \ \widehat{\sigma}^2_1(1) \ \leq \ \sigma^2(1) \left( 1 + 2^{\frac{1}{2}} T^{-\frac{1}{4} + \frac{\eps}{2}} \right), \label{eqn:2Stage:ConfidenceBound1rename} \\
\sigma^2(0) \left( 1 - 2^{\frac{1}{2}} T^{-\frac{1}{4} + \frac{\eps}{2}} \right) \ \leq \ \widehat{\sigma}^2_1(0) \ \leq \ \sigma^2(0) \left( 1 + 2^{\frac{1}{2}} T^{-\frac{1}{4} + \frac{\eps}{2}} \right). \label{eqn:2Stage:ConfidenceBound0rename} 
\end{align}
\end{subequations}
Due to \eqref{eqn:2Stage:ConfidenceBound1rename} and \eqref{eqn:2Stage:ConfidenceBound0rename}, and given that $\sigma(1), \sigma(0) > 0$, we have $\widehat{\sigma}^2_1(1), \widehat{\sigma}^2_1(0) > 0$.
Denote $\rho = \frac{\sigma(1)}{\sigma(0)}$ and $\widehat{\rho} = \frac{\widehat{\sigma}_1(1)}{\widehat{\sigma}_1(0)}$.

Now we distinguish two cases, and discuss these two cases separately.
\begin{enumerate}
\item \textbf{Case 1}: 
\begin{align*}
\frac{\frac{1}{2}T^{\frac{1}{2}}}{T - \frac{1}{2}T^{\frac{1}{2}}} \leq \rho = \frac{\sigma(1)}{\sigma(0)} \leq \frac{T - \frac{1}{2}T^{\frac{1}{2}}}{\frac{1}{2}T^{\frac{1}{2}}}.
\end{align*}
\item \textbf{Case 2}:
\begin{align*}
\rho = \frac{\sigma(1)}{\sigma(0)} > \frac{T - \frac{1}{2}T^{\frac{1}{2}}}{\frac{1}{2}T^{\frac{1}{2}}}.
\end{align*}
\end{enumerate}
Note that, for case 2, we do not discuss $\rho = \frac{\sigma(1)}{\sigma(0)} < \frac{\frac{1}{2}T^{\frac{1}{2}}}{T - \frac{1}{2}T^{\frac{1}{2}}}$, because we assume that $\sigma(1) \geq \sigma(0)$.
For each of the above two cases, we further discuss two sub-cases.
The remaining of the proof is structured as enumerating all four cases.
After enumerating all four sub-cases we finish the proof.

\noindent\textbf{Case 1.1}:
\begin{align*}
\frac{\frac{1}{2}T^{\frac{1}{2}}}{T - \frac{1}{2}T^{\frac{1}{2}}} \leq \rho \leq \frac{T - \frac{1}{2}T^{\frac{1}{2}}}{\frac{1}{2}T^{\frac{1}{2}}}, && \text{and} && \frac{\frac{1}{2}T^{\frac{1}{2}}}{T - \frac{1}{2}T^{\frac{1}{2}}} \leq \widehat{\rho} \leq \frac{T - \frac{1}{2}T^{\frac{1}{2}}}{\frac{1}{2}T^{\frac{1}{2}}}.
\end{align*}

Since $\frac{\frac{1}{2}T^{\frac{1}{2}}}{T - \frac{1}{2}T^{\frac{1}{2}}} \leq \widehat{\rho} \leq \frac{T - \frac{1}{2}T^{\frac{1}{2}}}{\frac{1}{2}T^{\frac{1}{2}}}$, we have
\begin{align*}
\frac{\widehat{\sigma}_1(1)}{\widehat{\sigma}_1(1) + \widehat{\sigma}_1(0)} T \geq \frac{1}{1+\frac{T - \frac{1}{2}T^{\frac{1}{2}}}{\frac{1}{2}T^{\frac{1}{2}}}} \ T= \frac{1}{2}T^{\frac{1}{2}}, \\
\frac{\widehat{\sigma}_1(0)}{\widehat{\sigma}_1(1) + \widehat{\sigma}_1(0)} T \geq \frac{1}{\frac{T - \frac{1}{2}T^{\frac{1}{2}}}{\frac{1}{2}T^{\frac{1}{2}}}+1} \ T= \frac{1}{2}T^{\frac{1}{2}}.
\end{align*}
As a result, Algorithm~\ref{alg:2StageANA} goes to Line 3 instead of Line 5 or Line 7.
The total numbers of treated and control units are given by \eqref{eqn:EstimatedOPT}. 
We re-write \eqref{eqn:EstimatedOPT} again as follows,
\begin{align*}
(T(1), T(0)) = (\frac{\widehat{\sigma}_1(1)}{\widehat{\sigma}_1(1) + \widehat{\sigma}_1(0)} T, \frac{\widehat{\sigma}_1(0)}{\widehat{\sigma}_1(1) + \widehat{\sigma}_1(0)} T).
\end{align*}

On the other hand, since $\frac{\frac{1}{2}T^{\frac{1}{2}}}{T - \frac{1}{2}T^{\frac{1}{2}}} \leq \rho \leq \frac{T - \frac{1}{2}T^{\frac{1}{2}}}{\frac{1}{2}T^{\frac{1}{2}}}$, following \eqref{eqn:defn:Stability:2} we have
\begin{align*}
(T^*(1), T^*(0)) = (\frac{\sigma(1)}{\sigma(1) + \sigma(0)} T, \frac{\sigma(0)}{\sigma(1) + \sigma(0)} T).
\end{align*}

Conditional on event $\cE$, we have
\begin{align*}
\left\vert \frac{T(1)}{T^*(1)} - 1 \right\vert = \left\vert \frac{\frac{\widehat{\sigma}_1(1)}{\widehat{\sigma}_1(1) + \widehat{\sigma}_1(0)} T}{\frac{\sigma(1)}{\sigma(1) + \sigma(0)} T} - 1 \right\vert \leq \max \left\{ \frac{1 + 2^{\frac{1}{2}} T^{-\frac{1}{4} + \frac{\eps}{2}}}{1 - 2^{\frac{1}{2}} T^{-\frac{1}{4} + \frac{\eps}{2}}} - 1, 1 - \frac{1 - 2^{\frac{1}{2}} T^{-\frac{1}{4} + \frac{\eps}{2}}}{1 + 2^{\frac{1}{2}} T^{-\frac{1}{4} + \frac{\eps}{2}}} \right\}
\end{align*}
where the inequality is due to \eqref{eqn:2Stage:ConfidenceBound1rename} and \eqref{eqn:2Stage:ConfidenceBound0rename}.

So conditional on event $\cE$, we have $\left\vert \frac{T(1)}{T^*(1)} - 1 \right\vert \to 0$ as $T \to +\infty$.
In addition, $1 - \Pr(\cE) = \frac{\kappa(1) + \kappa(0)}{T^{\eps}} \to 0$ as $T \to +\infty$.
Combining these two, we have $\frac{T(1)}{T^*(1)} \xrightarrow{p} 1$ as $T \to +\infty$.
Similarly, $\frac{T(0)}{T^*(0)} \xrightarrow{p} 1$ as $T \to +\infty$.

\noindent\textbf{Case 1.2}:
\begin{align*}
\frac{\frac{1}{2}T^{\frac{1}{2}}}{T - \frac{1}{2}T^{\frac{1}{2}}} \leq \rho \leq \frac{T - \frac{1}{2}T^{\frac{1}{2}}}{\frac{1}{2}T^{\frac{1}{2}}}, && \text{but} && \widehat{\rho} > \frac{T - \frac{1}{2}T^{\frac{1}{2}}}{\frac{1}{2}T^{\frac{1}{2}}} \ \text{or} \ \widehat{\rho} < \frac{\frac{1}{2}T^{\frac{1}{2}}}{T - \frac{1}{2}T^{\frac{1}{2}}}.
\end{align*}

If $\widehat{\rho} > \frac{T - \frac{1}{2}T^{\frac{1}{2}}}{\frac{1}{2}T^{\frac{1}{2}}}$, then 
\begin{align*}
\frac{\widehat{\sigma}_1(0)}{\widehat{\sigma}_1(1) + \widehat{\sigma}_1(0)} T < \frac{1}{\frac{T - \frac{1}{2}T^{\frac{1}{2}}}{\frac{1}{2}T^{\frac{1}{2}}}+1} \ T= \frac{1}{2}T^{\frac{1}{2}}.
\end{align*}
Due to this, Algorithm~\ref{alg:2StageANA} goes to Line 7.
The total numbers of treated and control units are given by
\begin{align*}
(T(1), T(0)) = (T - \frac{1}{2}T^{\frac{1}{2}}, \frac{1}{2}T^{\frac{1}{2}}).
\end{align*}

On the other hand, since $\frac{\frac{1}{2}T^{\frac{1}{2}}}{T - \frac{1}{2}T^{\frac{1}{2}}} \leq \rho \leq \frac{T - \frac{1}{2}T^{\frac{1}{2}}}{\frac{1}{2}T^{\frac{1}{2}}}$, following \eqref{eqn:defn:Stability:2} we have
\begin{align*}
(T^*(1), T^*(0)) = (\frac{\sigma(1)}{\sigma(1) + \sigma(0)} T, \frac{\sigma(0)}{\sigma(1) + \sigma(0)} T).
\end{align*}

Note that,
\begin{align*}
\rho \ \leq \ \frac{T - \frac{1}{2}T^{\frac{1}{2}}}{\frac{1}{2}T^{\frac{1}{2}}} \ < \ \widehat{\rho} \ \leq \ \rho \sqrt{\frac{1+2^{\frac{1}{2}} T^{-\frac{1}{4} + \frac{\eps}{2}}}{1-2^{\frac{1}{2}} T^{-\frac{1}{4} + \frac{\eps}{2}}}}.
\end{align*}

Conditional on event $\cE$, we have
\begin{multline*}
\left\vert \frac{T(1)}{T^*(1)} - 1 \right\vert = \left\vert \frac{T - \frac{1}{2}T^{\frac{1}{2}}}{\frac{\sigma(1)}{\sigma(1) + \sigma(0)} T} - 1 \right\vert = \left\vert \bigg(1+\frac{1}{\rho}\bigg) \frac{T-\frac{1}{2}T^{\frac{1}{2}}}{T} - 1 \right\vert \\
\leq \bigg(1+\frac{\frac{1}{2}T^{\frac{1}{2}}}{T-\frac{1}{2}T^{\frac{1}{2}}} \sqrt{\frac{1+2^{\frac{1}{2}} T^{-\frac{1}{4} + \frac{\eps}{2}}}{1-2^{\frac{1}{2}} T^{-\frac{1}{4} + \frac{\eps}{2}}}} \bigg) \frac{T-\frac{1}{2}T^{\frac{1}{2}}}{T} - 1 
= \frac{1}{2T^{\frac{1}{2}}} \bigg(\sqrt{\frac{1+2^{\frac{1}{2}} T^{-\frac{1}{4} + \frac{\eps}{2}}}{1-2^{\frac{1}{2}} T^{-\frac{1}{4} + \frac{\eps}{2}}}}-1\bigg),
\end{multline*}
where the inequality is because $\big(1+\frac{1}{\rho}\big) \cdot \frac{T-\frac{1}{2}T^{\frac{1}{2}}}{T} - 1$ is decreasing in $\rho$ and equals $0$ when $\rho = \frac{T - \frac{1}{2}T^{\frac{1}{2}}}{\frac{1}{2}T^{\frac{1}{2}}}$.

So conditional on event $\cE$, we have $\left\vert \frac{T(1)}{T^*(1)} - 1 \right\vert \to 0$ as $T \to +\infty$.
In addition, $1 - \Pr(\cE) = \frac{\kappa(1) + \kappa(0)}{T^{\eps}} \to 0$ as $T \to +\infty$.
Combining these two, we have $\frac{T(1)}{T^*(1)} \xrightarrow{p} 1$ as $T \to +\infty$.

Conditional on event $\cE$, we have
\begin{multline*}
\left\vert \frac{T(0)}{T^*(0)} - 1 \right\vert = \left\vert \frac{\frac{1}{2}T^{\frac{1}{2}}}{\frac{\sigma(0)}{\sigma(1) + \sigma(0)} T} - 1 \right\vert = \left\vert \big(\rho + 1\big) \frac{1}{2T^{\frac{1}{2}}} - 1 \right\vert \\
\leq 1 - \bigg( \frac{\frac{1}{2}T^{\frac{1}{2}}}{T - \frac{1}{2}T^{\frac{1}{2}}} \sqrt{\frac{1-2^{\frac{1}{2}} T^{-\frac{1}{4} + \frac{\eps}{2}}}{1+2^{\frac{1}{2}} T^{-\frac{1}{4} + \frac{\eps}{2}}}} + 1 \bigg) \frac{1}{2T^{\frac{1}{2}}} 
= \frac{T - \frac{1}{2}T^{\frac{1}{2}}}{T} \bigg( 1 - \sqrt{\frac{1-2^{\frac{1}{2}} T^{-\frac{1}{4} + \frac{\eps}{2}}}{1+2^{\frac{1}{2}} T^{-\frac{1}{4} + \frac{\eps}{2}}}} \bigg),
\end{multline*}
where the inequality is because $\big(\rho + 1\big) \frac{1}{2T^{\frac{1}{2}}} - 1$ is increasing in $\rho$ and equals $0$ when $\rho = \frac{T - \frac{1}{2}T^{\frac{1}{2}}}{\frac{1}{2}T^{\frac{1}{2}}}$.

So conditional on event $\cE$, we have $\left\vert \frac{T(0)}{T^*(0)} - 1 \right\vert \to 0$ as $T \to +\infty$.
In addition, $1 - \Pr(\cE) = \frac{\kappa(1) + \kappa(0)}{T^{\eps}} \to 0$ as $T \to +\infty$.
Combining these two, we have $\frac{T(0)}{T^*(0)} \xrightarrow{p} 1$ as $T \to +\infty$.

If $\widehat{\rho} < \frac{\frac{1}{2}T^{\frac{1}{2}}}{T - \frac{1}{2}T^{\frac{1}{2}}}$, then Algorithm~\ref{alg:2StageANA} goes to Line 5 and the same analysis follows similarly.

\noindent\textbf{Case 2.1}:
\begin{align*}
\rho > \frac{T - \frac{1}{2}T^{\frac{1}{2}}}{\frac{1}{2}T^{\frac{1}{2}}}, && \text{and} && \widehat{\rho} > \frac{T - \frac{1}{2}T^{\frac{1}{2}}}{\frac{1}{2}T^{\frac{1}{2}}}.
\end{align*}

Since $\widehat{\rho} > \frac{T - \frac{1}{2}T^{\frac{1}{2}}}{\frac{1}{2}T^{\frac{1}{2}}}$, we have
\begin{align*}
\frac{\widehat{\sigma}_1(0)}{\widehat{\sigma}_1(1) + \widehat{\sigma}_1(0)} T < \frac{1}{\frac{T - \frac{1}{2}T^{\frac{1}{2}}}{\frac{1}{2}T^{\frac{1}{2}}}+1} \ T= \frac{1}{2}T^{\frac{1}{2}}.
\end{align*}
As a result, Algorithm~\ref{alg:2StageANA} goes to Line 7.
The total numbers of treated and control units are given by
\begin{align*}
(T(1), T(0)) = (T - \frac{1}{2}T^{\frac{1}{2}}, \frac{1}{2}T^{\frac{1}{2}}).
\end{align*}

On the other hand, since $\rho > \frac{T - \frac{1}{2}T^{\frac{1}{2}}}{\frac{1}{2}T^{\frac{1}{2}}}$, following \eqref{eqn:defn:Stability:2} we have
\begin{align*}
(T^*(1), T^*(0)) = (T - \frac{1}{2}T^{\frac{1}{2}}, \frac{1}{2}T^{\frac{1}{2}}).
\end{align*}

So conditional on event $\cE$, we have $\frac{T(1)}{T^*(1)} = 1$.
In addition, $1 - \Pr(\cE) = \frac{\kappa(1) + \kappa(0)}{T^{\eps}} \to 0$ as $T \to +\infty$.
Combining these two, we have $\frac{T(1)}{T^*(1)} \xrightarrow{p} 1$ as $T \to +\infty$.
Similarly, $\frac{T(0)}{T^*(0)} \xrightarrow{p} 1$ as $T \to +\infty$.

\noindent\textbf{Case 2.2}:
\begin{align*}
\rho > \frac{T - \frac{1}{2}T^{\frac{1}{2}}}{\frac{1}{2}T^{\frac{1}{2}}}, && \text{and} && \widehat{\rho} \leq \frac{T - \frac{1}{2}T^{\frac{1}{2}}}{\frac{1}{2}T^{\frac{1}{2}}}
\end{align*}
Note that,
\begin{align*}
\widehat{\sigma}_1(1) \geq & \ \sigma(1) \sqrt{1 - 2^{\frac{1}{2}} T^{-\frac{1}{4}+\frac{\eps}{2}}} \\
> & \ \sigma(0) \frac{T - \frac{1}{2}T^{\frac{1}{2}}}{\frac{1}{2}T^{\frac{1}{2}}} \sqrt{1 - 2^{\frac{1}{2}} T^{-\frac{1}{4}+\frac{\eps}{2}}} \\
\geq & \ \widehat{\sigma}_1(0) \frac{T - \frac{1}{2}T^{\frac{1}{2}}}{\frac{1}{2}T^{\frac{1}{2}}} \sqrt{\frac{1 - 2^{\frac{1}{2}} T^{-\frac{1}{4}+\frac{\eps}{2}}}{1 + 2^{\frac{1}{2}} T^{-\frac{1}{4}+\frac{\eps}{2}}}} \\
\geq & \ \widehat{\sigma}_1(0) \frac{\frac{1}{2}T^{\frac{1}{2}}}{T - \frac{1}{2}T^{\frac{1}{2}}}.
\end{align*}
where the first inequality is due to \eqref{eqn:2Stage:ConfidenceBound1rename};
the second inequality is due to $\rho > \frac{T - \frac{1}{2}T^{\frac{1}{2}}}{\frac{1}{2}T^{\frac{1}{2}}}$;
the third inequality is due to \eqref{eqn:2Stage:ConfidenceBound0rename};
the last inequality is due to Lemma~\ref{lem:AlgebraicTrick1}.

The above shows that, in this case (Case 2.2), 
\begin{align*}
\widehat{\rho} \geq \frac{\frac{1}{2}T^{\frac{1}{2}}}{T - \frac{1}{2}T^{\frac{1}{2}}}.
\end{align*}
Since $\frac{\frac{1}{2}T^{\frac{1}{2}}}{T - \frac{1}{2}T^{\frac{1}{2}}} \leq \widehat{\rho} \leq \frac{T - \frac{1}{2}T^{\frac{1}{2}}}{\frac{1}{2}T^{\frac{1}{2}}}$, we have
\begin{align*}
\frac{\widehat{\sigma}_1(1)}{\widehat{\sigma}_1(1) + \widehat{\sigma}_1(0)} T \geq \frac{1}{1+\frac{T - \frac{1}{2}T^{\frac{1}{2}}}{\frac{1}{2}T^{\frac{1}{2}}}} \ T= \frac{1}{2}T^{\frac{1}{2}}, \\
\frac{\widehat{\sigma}_1(0)}{\widehat{\sigma}_1(1) + \widehat{\sigma}_1(0)} T \geq \frac{1}{\frac{T - \frac{1}{2}T^{\frac{1}{2}}}{\frac{1}{2}T^{\frac{1}{2}}}+1} \ T= \frac{1}{2}T^{\frac{1}{2}}.
\end{align*}
Due to this, Algorithm~\ref{alg:2StageANA} goes to Line 3 instead of Line 5 or Line 7.
The total numbers of treated and control units are given by \eqref{eqn:EstimatedOPT}, which we write again as follows,
\begin{align*}
(T(1), T(0)) = (\frac{\widehat{\sigma}_1(1)}{\widehat{\sigma}_1(1) + \widehat{\sigma}_1(0)} T, \frac{\widehat{\sigma}_1(0)}{\widehat{\sigma}_1(1) + \widehat{\sigma}_1(0)} T).
\end{align*}

On the other hand, since $\rho > \frac{T - \frac{1}{2}T^{\frac{1}{2}}}{\frac{1}{2}T^{\frac{1}{2}}}$, following \eqref{eqn:defn:Stability:2} we have
\begin{align*}
(T^*(1), T^*(0)) = (T - \frac{1}{2}T^{\frac{1}{2}}, \frac{1}{2}T^{\frac{1}{2}}).
\end{align*}

Note that,
\begin{align*}
\widehat{\rho} \ \leq \ \frac{T - \frac{1}{2}T^{\frac{1}{2}}}{\frac{1}{2}T^{\frac{1}{2}}} \ < \ \rho \ \leq \ \widehat{\rho} \sqrt{\frac{1+2^{\frac{1}{2}} T^{-\frac{1}{4} + \frac{\eps}{2}}}{1-2^{\frac{1}{2}} T^{-\frac{1}{4} + \frac{\eps}{2}}}}.
\end{align*}

Conditional on event $\cE$, we have
\begin{multline*}
\left\vert \frac{T(1)}{T^*(1)} - 1 \right\vert = \left\vert \frac{\frac{\widehat{\sigma}_1(1)}{\widehat{\sigma}_1(1) + \widehat{\sigma}_1(0)} T}{T - \frac{1}{2}T^{\frac{1}{2}}} - 1 \right\vert = \left\vert \frac{\widehat{\rho}}{\widehat{\rho}+1} \frac{T}{T-\frac{1}{2}T^{\frac{1}{2}}} - 1 \right\vert \\
\leq 1 - \frac{T \sqrt{\frac{1 - 2^{\frac{1}{2}} T^{-\frac{1}{4}+\frac{\eps}{2}}}{1 + 2^{\frac{1}{2}} T^{-\frac{1}{4}+\frac{\eps}{2}}}} }{\frac{1}{2}T^{\frac{1}{2}} + \big(T-\frac{1}{2}T^{\frac{1}{2}}\big) \sqrt{\frac{1 - 2^{\frac{1}{2}} T^{-\frac{1}{4}+\frac{\eps}{2}}}{1 + 2^{\frac{1}{2}} T^{-\frac{1}{4}+\frac{\eps}{2}}}} }
= \frac{\frac{1}{2}T^{\frac{1}{2}} - \frac{1}{2}T^{\frac{1}{2}}\sqrt{\frac{1 - 2^{\frac{1}{2}} T^{-\frac{1}{4}+\frac{\eps}{2}}}{1 + 2^{\frac{1}{2}} T^{-\frac{1}{4}+\frac{\eps}{2}}}} }{\frac{1}{2}T^{\frac{1}{2}} + \big(T-\frac{1}{2}T^{\frac{1}{2}}\big) \sqrt{\frac{1 - 2^{\frac{1}{2}} T^{-\frac{1}{4}+\frac{\eps}{2}}}{1 + 2^{\frac{1}{2}} T^{-\frac{1}{4}+\frac{\eps}{2}}}} }
\end{multline*}
where the inequality is because $\frac{\widehat{\rho}}{\widehat{\rho}+1} \frac{T}{T-\frac{1}{2}T^{\frac{1}{2}}} - 1$ is increasing in $\widehat{\rho}$ and equals $0$ when $\widehat{\rho} = \frac{T - \frac{1}{2}T^{\frac{1}{2}}}{\frac{1}{2}T^{\frac{1}{2}}}$.

So conditional on event $\cE$, we have $\left\vert \frac{T(1)}{T^*(1)} - 1 \right\vert \to 0$ as $T \to +\infty$.
In addition, $1 - \Pr(\cE) = \frac{\kappa(1) + \kappa(0)}{T^{\eps}} \to 0$ as $T \to +\infty$.
Combining these two, we have $\frac{T(1)}{T^*(1)} \xrightarrow{p} 1$ as $T \to +\infty$.

Conditional on event $\cE$, we have
\begin{multline*}
\left\vert \frac{T(0)}{T^*(0)} - 1 \right\vert = \left\vert \frac{\frac{\widehat{\sigma}_1(0)}{\widehat{\sigma}_1(1) + \widehat{\sigma}_1(0)} T}{\frac{1}{2}T^{\frac{1}{2}}} - 1 \right\vert = \left\vert \frac{1}{\widehat{\rho}+1} \frac{T}{\frac{1}{2}T^{\frac{1}{2}}} - 1 \right\vert 
\leq \frac{T}{\frac{1}{2}T^{\frac{1}{2}} + \big(T-\frac{1}{2}T^{\frac{1}{2}}\big) \sqrt{\frac{1 - 2^{\frac{1}{2}} T^{-\frac{1}{4}+\frac{\eps}{2}}}{1 + 2^{\frac{1}{2}} T^{-\frac{1}{4}+\frac{\eps}{2}}}} } - 1 \\
=  \frac{\big(T - \frac{1}{2}T^{\frac{1}{2}}\big) \bigg( 1 - \sqrt{\frac{1 - 2^{\frac{1}{2}} T^{-\frac{1}{4}+\frac{\eps}{2}}}{1 + 2^{\frac{1}{2}} T^{-\frac{1}{4}+\frac{\eps}{2}}}} \bigg)}{T \sqrt{\frac{1 - 2^{\frac{1}{2}} T^{-\frac{1}{4}+\frac{\eps}{2}}}{1 + 2^{\frac{1}{2}} T^{-\frac{1}{4}+\frac{\eps}{2}}}} + \frac{1}{2}T^{\frac{1}{2}} \bigg( 1 - \sqrt{\frac{1 - 2^{\frac{1}{2}} T^{-\frac{1}{4}+\frac{\eps}{2}}}{1 + 2^{\frac{1}{2}} T^{-\frac{1}{4}+\frac{\eps}{2}}}} \bigg) }
\leq \frac{T \bigg( 1 - \sqrt{\frac{1 - 2^{\frac{1}{2}} T^{-\frac{1}{4}+\frac{\eps}{2}}}{1 + 2^{\frac{1}{2}} T^{-\frac{1}{4}+\frac{\eps}{2}}}} \bigg)}{T \sqrt{\frac{1 - 2^{\frac{1}{2}} T^{-\frac{1}{4}+\frac{\eps}{2}}}{1 + 2^{\frac{1}{2}} T^{-\frac{1}{4}+\frac{\eps}{2}}}} }
= \sqrt{\frac{1 + 2^{\frac{1}{2}} T^{-\frac{1}{4}+\frac{\eps}{2}}}{1 - 2^{\frac{1}{2}} T^{-\frac{1}{4}+\frac{\eps}{2}}}} - 1
\end{multline*}
where the first inequality is because $\frac{1}{\widehat{\rho}+1} \frac{T}{\frac{1}{2}T^{\frac{1}{2}}} - 1$ is decreasing in $\widehat{\rho}$ and equals $0$ when $\widehat{\rho} = \frac{T - \frac{1}{2}T^{\frac{1}{2}}}{\frac{1}{2}T^{\frac{1}{2}}}$.

So conditional on event $\cE$, we have $\left\vert \frac{T(0)}{T^*(0)} - 1 \right\vert \to 0$ as $T \to +\infty$.
In addition, $1 - \Pr(\cE) = \frac{\kappa(1) + \kappa(0)}{T^{\eps}} \to 0$ as $T \to +\infty$.
Combining these two, we have $\frac{T(0)}{T^*(0)} \xrightarrow{p} 1$ as $T \to +\infty$.

To conclude, in all cases, as $T \to +\infty$,
\begin{align*}
\frac{T(1)}{T^*(1)} \xrightarrow{p} 1, && \frac{T(0)}{T^*(0)} \xrightarrow{p} 1.
\end{align*}
This proves Condition~\ref{enum:StabilityCondition2}, and completes the proof of Lemma~\ref{lem:Stability} when $M=2$.
\hfill \halmos
\endproof

\proof{Proof of Lemma~\ref{lem:Stability} when $M \geq 3$.}
We can explicitly verify that Condition~\ref{enum:StabilityCondition1} in Lemma~\ref{lem:Stability} is satisfied.
Below we prove Condition~\ref{enum:StabilityCondition2}.

We borrow the same clean event analysis as in the proof of Theorem~\ref{thm:MStageANA}.
To proceed with the clean event analysis, suppose there are two length-$T$ arrays for the treated and the control, respectively, with each value being an independent and identically distributed copy of the representative random variables $Y(1)$ and $Y(0)$, respectively.
When Algorithm~\ref{alg:MStageANA} suggests to conduct an $m$-th stage experiment parameterized by $(T_m(1), T_m(0))$, the observations from the $m$-th stage experiment are generated by reading the next $T_m(1)$ values from the treated array, and the next $T_m(0)$ values from the control array.

Even though Algorithm~\ref{alg:MStageANA} adaptively determines the number of treated and control units, it is always the first few values of of the two arrays that are read.
For any $m \leq M-1$, let $\widehat{\psi}^2_m(1)$ and $\widehat{\psi}^2_{m}(0)$ be the sample variance estimators obtained from reading the first $\frac{\beta_m}{2}T^{\frac{m}{M}}$ values in the treated array and control array, respectively.
Depending on the execution of Algorithm~\ref{alg:MStageANA}, only a few of the sample variance estimators $\widehat{\sigma}^2_m(1)$ or $\widehat{\sigma}^2_m(0)$ are calculated. 
When one sample variance estimator $\widehat{\sigma}^2_m(1)$ or $\widehat{\sigma}^2_m(0)$ is calculated following Algorithm~\ref{alg:MStageANA}, it is equivalent to reading the corresponding $\widehat{\psi}^2_m(1)$ or $\widehat{\psi}^2_m(0)$ from the array.

Define the following events.
For any $m \leq M-1$, define
\begin{align*}
\cE_m(1) = & \ \bigg\{ \left| \widehat{\psi}^2_m(1) - \sigma^2(1) \right| < 2^{\frac{1}{2}} \beta_m^{-\frac{1}{2}} T^{-\frac{m}{2M} + \frac{\eps}{2}} \sigma^2(1)\bigg\}, \\
\cE_m(0) = & \ \bigg\{ \left| \widehat{\psi}^2_m(0) - \sigma^2(0) \right| < 2^{\frac{1}{2}} \beta_m^{-\frac{1}{2}} T^{-\frac{m}{2M} + \frac{\eps}{2}} \sigma^2(0)\bigg\}.
\end{align*}
Denote the intersect of all above events as $\cE$, i.e., 
\begin{align*}
\cE = \bigcap_{m=1}^{M-1} \left(\cE_m(1) \cap \cE_m(0)\right).
\end{align*}
Then due to union bound, 
\begin{align*}
\Pr(\cE) \geq 1 - \sum_{m=1}^{M-1} \Pr(\overline{\cE}_m(1)) - \sum_{m=1}^{M-1} \Pr(\overline{\cE}_m(0)).
\end{align*}
We further have
\begin{align*}
& \Pr(\cE) \\
= & \ 1 - \sum_{m=1}^{M-1} \Pr\left( \vert \widehat{\psi}^2_m(1) - \sigma^2(1) \vert \geq 2^{\frac{1}{2}} \beta_m^{-\frac{1}{2}} T^{-\frac{m}{2M} + \frac{\eps}{2}} \sigma^2(1) \right) - \sum_{m=1}^{M-1} \Pr\left( \vert \widehat{\psi}^2_m(0) - \sigma^2(0) \vert \geq 2^{\frac{1}{2}} \beta_m^{-\frac{1}{2}} T^{-\frac{m}{2M} + \frac{\eps}{2}} \sigma^2(0) \right) \\
\geq & \ 1 - \sum_{m=1}^{M-1} \frac{\kappa(1) \sigma^4(1)}{2 \beta_m^{-1} T^{-\frac{m}{M} + \eps} \sigma^4(1) \frac{1}{2} \beta_m T^{\frac{m}{M}}} - \sum_{m=1}^{M-1} \frac{\kappa(0) \sigma^4(0)}{2 \beta_m^{-1} T^{-\frac{m}{M} + \eps} \sigma^4(0) \frac{1}{2} \beta_m T^{\frac{m}{M}}} \\
= & \ 1 - \sum_{m=1}^{M-1} \frac{\kappa(1) + \kappa(0)}{T^{\eps}} \\
= & \ 1 - (M-1) \frac{\kappa(1) + \kappa(0)}{T^{\eps}},
\end{align*}
where the inequality is due to Lemma~\ref{lem:LightTail}.

Conditional on the event $\cE$, we have, for any $m \leq M-1$,
\begin{subequations}
\begin{align}
\sigma^2(1) \left( 1 - 2^{\frac{1}{2}} \beta_m^{-\frac{1}{2}} T^{-\frac{m}{2M} + \frac{\eps}{2}} \right) \ \leq \ \widehat{\psi}^2_m(1) \ \leq \ \sigma^2(1) \left( 1 + 2^{\frac{1}{2}} \beta_m^{-\frac{1}{2}} T^{-\frac{m}{2M} + \frac{\eps}{2}} \right), \label{eqn:MStage:ConfidenceBound1rename} \\
\sigma^2(0) \left( 1 - 2^{\frac{1}{2}} \beta_m^{-\frac{1}{2}} T^{-\frac{m}{2M} + \frac{\eps}{2}} \right) \ \leq \ \widehat{\psi}^2_m(0) \ \leq \ \sigma^2(0) \left( 1 + 2^{\frac{1}{2}} \beta_m^{-\frac{1}{2}} T^{-\frac{m}{2M} + \frac{\eps}{2}} \right). \label{eqn:MStage:ConfidenceBound0rename} 
\end{align}
\end{subequations}

Since $\sigma(1), \sigma(0) > 0$, we can denote $\rho = \frac{\sigma(1)}{\sigma(0)}$.
For any $m \leq M-1$, when $\widehat{\sigma}^2_m(1)$ and $\widehat{\sigma}^2_m(0)$ are calculated during Algorithm~\ref{alg:MStageANA}, $\widehat{\sigma}^2_m(1) = \widehat{\psi}^2_m(1)$ and $\widehat{\sigma}^2_m(0) = \widehat{\psi}^2_m(0)$.
Conditional on the event $\cE$, due to \eqref{eqn:MStage:ConfidenceBound1rename} and \eqref{eqn:MStage:ConfidenceBound0rename}, and given that $\sigma(1), \sigma(0) > 0$, we have $\widehat{\sigma}^2_m(1), \widehat{\sigma}^2_m(0) > 0$.
Then we can denote $\widehat{\rho}_m = \frac{\widehat{\sigma}_m(1)}{\widehat{\sigma}_m(0)}$.

In the remaining of the analysis, we distinguish several cases and discuss these cases separately.
Recall that $\widehat{\rho}_m = \frac{\widehat{\sigma}_m(1)}{\widehat{\sigma}_m(0)}$, and that $\rho = \frac{\sigma(1)}{\sigma(0)}$.
Without loss of generality, assume 
\begin{align}
\widehat{\rho}_1 \geq 1. \label{eqn:WLOG:rename}
\end{align}

\noindent\underline{\textbf{Case 1}}: 
\begin{align*}
\widehat{\rho}_1 > \frac{T - \frac{1}{2} \beta_2 T^{\frac{2}{M}}}{\frac{1}{2} \beta_2 T^{\frac{2}{M}}}.
\end{align*}
\noindent \textbf{Case 1.1}: 
\begin{align*}
\widehat{\rho}_1 > \frac{T - \frac{1}{2} \beta_1 T^{\frac{1}{M}}}{\frac{1}{2} \beta_1 T^{\frac{1}{M}}}.
\end{align*}
In this case, 
\begin{align*}
\frac{\widehat{\sigma}_1(0)}{\widehat{\sigma}_1(1) + \widehat{\sigma}_1(0)} T < \frac{1}{\frac{T - \frac{1}{2} \beta_1 T^{\frac{1}{M}}}{\frac{1}{2} \beta_1 T^{\frac{1}{M}}}+1} T = \frac{1}{2} \beta_1 T^{\frac{1}{M}}.
\end{align*}
So Algorithm~\ref{alg:MStageANA} goes to Line~\ref{mrk:Case1} in the $1$-st stage experiment.
Then we have
\begin{align*}
(T(1), T(0)) = \bigg(T - \frac{1}{2} \beta_1 T^{\frac{1}{M}}, \frac{1}{2} \beta_1 T^{\frac{1}{M}}\bigg).
\end{align*}

We further distinguish two cases. 

\textbf{First}, $\rho < \frac{T - \frac{1}{2} \beta_1 T^{\frac{1}{M}}}{\frac{1}{2} \beta_1 T^{\frac{1}{M}}}$. 
Note that, conditional on $\cE$,
\begin{align}
\rho < \frac{T - \frac{1}{2} \beta_1 T^{\frac{1}{M}}}{\frac{1}{2} \beta_1 T^{\frac{1}{M}}} < \widehat{\rho}_1 \leq \rho \sqrt{ \frac{1+2^{\frac{1}{2}} \beta_1^{-\frac{1}{2}} T^{-\frac{1}{2M} + \frac{\eps}{2}}}{1-2^{\frac{1}{2}} \beta_1^{-\frac{1}{2}} T^{-\frac{1}{2M} + \frac{\eps}{2}}} } \label{eqn:rhoRelations:Case1-1rename}
\end{align}

Note also that,
\begin{align*}
\rho > \frac{T - \frac{1}{2} \beta_1 T^{\frac{1}{M}}}{\frac{1}{2} \beta_1 T^{\frac{1}{M}}} \cdot \sqrt{\frac{1-2^{\frac{1}{2}} \beta_1^{-\frac{1}{2}} T^{-\frac{1}{2M} + \frac{\eps}{2}}}{1+2^{\frac{1}{2}} \beta_1^{-\frac{1}{2}} T^{-\frac{1}{2M} + \frac{\eps}{2}}}} > \frac{1}{2} \frac{T - \frac{1}{2} \beta_1 T^{\frac{1}{M}}}{\frac{1}{2} \beta_1 T^{\frac{1}{M}}} > 1,
\end{align*}
where the first inequality is due to \eqref{eqn:rhoRelations:Case1-1rename}; the second inequality is due to Lemma~\ref{lem:AlgebraicTrick3}; the last inequality is due to Lemma~\ref{lem:AlgebraicTrick4}.

As a result,
\begin{align*}
(T^*(1), T^*(0)) = \bigg(\frac{\sigma(1)}{\sigma(1)+\sigma(0)}T, \frac{\sigma(0)}{\sigma(1)+\sigma(0)}T\bigg).
\end{align*}

Conditional on event $\cE$, we have
\begin{multline*}
\left\vert \frac{T(1)}{T^*(1)} - 1 \right\vert = \left\vert \frac{T - \frac{1}{2} \beta_1 T^{\frac{1}{M}}}{\frac{\sigma(1)}{\sigma(1) + \sigma(0)} T} - 1 \right\vert = \left\vert \big( \frac{1}{\rho}+1 \big) \frac{T-\frac{1}{2} \beta_1 T^{\frac{1}{M}}}{T} - 1 \right\vert \\
\leq \bigg( \frac{\frac{1}{2} \beta_1 T^{\frac{1}{M}}}{T-\frac{1}{2} \beta_1 T^{\frac{1}{M}}} \sqrt{\frac{1+2^{\frac{1}{2}} \beta_1^{-\frac{1}{2}} T^{-\frac{1}{2M} + \frac{\eps}{2}}}{1-2^{\frac{1}{2}} \beta_1^{-\frac{1}{2}} T^{-\frac{1}{2M} + \frac{\eps}{2}}}} + 1\bigg) \frac{T-\frac{1}{2} \beta_1 T^{\frac{1}{M}}}{T} - 1,
\end{multline*}
where the inequality is because $\big( \frac{1}{\rho}+1 \big) \frac{T-\frac{1}{2} \beta_1 T^{\frac{1}{M}}}{T} - 1$ is decreasing in $\rho$ and equals $0$ when $\rho = \frac{T - \frac{1}{2}\beta_1 T^{\frac{1}{M}}}{\frac{1}{2}\beta_1 T^{\frac{1}{M}}}$; and due to \eqref{eqn:rhoRelations:Case1-1rename} we have a lower bound for $\rho$.

So conditional on event $\cE$, we have $\left\vert \frac{T(1)}{T^*(1)} - 1 \right\vert \to 0$ as $T \to +\infty$.
In addition, $1 - \Pr(\cE) = (M-1) \frac{\kappa(1) + \kappa(0)}{T^{\eps}} \to 0$ as $T \to +\infty$.
Combining these two, we have $\frac{T(1)}{T^*(1)} \xrightarrow{p} 1$ as $T \to +\infty$.

Conditional on event $\cE$, we have
\begin{multline*}
\left\vert \frac{T(0)}{T^*(0)} - 1 \right\vert = \left\vert \frac{\frac{1}{2} \beta_1 T^{\frac{1}{M}}}{\frac{\sigma(0)}{\sigma(1) + \sigma(0)} T} - 1 \right\vert = \left\vert \big( \rho+1 \big) \frac{\frac{1}{2} \beta_1 T^{\frac{1}{M}}}{T} - 1 \right\vert \\
\leq 1 - \bigg( 1 + \frac{T - \frac{1}{2} \beta_1 T^{\frac{1}{M}}}{\frac{1}{2} \beta_1 T^{\frac{1}{M}}} \sqrt{\frac{1-2^{\frac{1}{2}} \beta_1^{-\frac{1}{2}} T^{-\frac{1}{2M} + \frac{\eps}{2}}}{1+2^{\frac{1}{2}} \beta_1^{-\frac{1}{2}} T^{-\frac{1}{2M} + \frac{\eps}{2}}}}\bigg) \frac{\frac{1}{2} \beta_1 T^{\frac{1}{M}}}{T} 
\leq \frac{T - \frac{1}{2} \beta_1 T^{\frac{1}{M}}}{T} 2^{\frac{1}{2}} \beta_1^{-\frac{1}{2}} T^{-\frac{1}{2M} + \frac{\eps}{2}}
\end{multline*}
where the first inequality is because $\big( \rho+1 \big) \frac{\frac{1}{2} \beta_1 T^{\frac{1}{M}}}{T} - 1$ is increasing in $\rho$ and equals $0$ when $\rho = \frac{T - \frac{1}{2}\beta_1 T^{\frac{1}{M}}}{\frac{1}{2}\beta_1 T^{\frac{1}{M}}}$,
the second inequality is because for any $\delta \in [0,1), 1- \delta \leq \sqrt{\frac{1-\delta}{1+\delta}}$.

So conditional on event $\cE$, we have $\left\vert \frac{T(0)}{T^*(0)} - 1 \right\vert \to 0$ as $T \to +\infty$.
In addition, $1 - \Pr(\cE) = (M-1) \frac{\kappa(1) + \kappa(0)}{T^{\eps}} \to 0$ as $T \to +\infty$.
Combining these two, we have $\frac{T(0)}{T^*(0)} \xrightarrow{p} 1$ as $T \to +\infty$.

\textbf{Second}, if $\rho \geq \frac{T - \frac{1}{2} \beta_1 T^{\frac{1}{M}}}{\frac{1}{2} \beta_1 T^{\frac{1}{M}}}$, then we have
\begin{align*}
(T^*(1), T^*(0)) = \bigg(T - \frac{1}{2} \beta_1 T^{\frac{1}{M}}, \frac{1}{2} \beta_1 T^{\frac{1}{M}}\bigg).
\end{align*}

So conditional on event $\cE$, we have $\frac{T(1)}{T^*(1)} = 1$.
In addition, $1 - \Pr(\cE) = (M-1) \frac{\kappa(1) + \kappa(0)}{T^{\eps}} \to 0$ as $T \to +\infty$.
Combining these two, we have $\frac{T(1)}{T^*(1)} \xrightarrow{p} 1$ as $T \to +\infty$.
Similarly, $\frac{T(0)}{T^*(0)} \xrightarrow{p} 1$ as $T \to +\infty$.

\noindent \textbf{Case 1.2}:
\begin{align*}
\frac{T - \frac{1}{2} \beta_2 T^{\frac{2}{M}}}{\frac{1}{2} \beta_2 T^{\frac{2}{M}}} < \widehat{\rho}_1 \leq \frac{T - \frac{1}{2} \beta_1 T^{\frac{1}{M}}}{\frac{1}{2} \beta_1 T^{\frac{1}{M}}}.
\end{align*}
In this case, 
\begin{align*}
\frac{1}{2} \beta_1 T^{\frac{1}{M}}  = \frac{1}{\frac{T - \frac{1}{2} \beta_1 T^{\frac{1}{M}}}{\frac{1}{2} \beta_1 T^{\frac{1}{M}}}+1} T \leq \frac{\widehat{\sigma}_1(0)}{\widehat{\sigma}_1(1) + \widehat{\sigma}_1(0)} T < \frac{1}{\frac{T - \frac{1}{2} \beta_2 T^{\frac{2}{M}}}{\frac{1}{2} \beta_2 T^{\frac{2}{M}}}+1} T = \frac{1}{2} \beta_2 T^{\frac{2}{M}}.
\end{align*}
So Algorithm~\ref{alg:MStageANA} goes to Line~\ref{mrk:Case2} in the $1$-st stage experiment.
Then we have
\begin{align*}
(T(1), T(0)) = \bigg(\frac{\widehat{\sigma}_1(1)}{\widehat{\sigma}_1(1) + \widehat{\sigma}_1(0)} T, \frac{\widehat{\sigma}_1(0)}{\widehat{\sigma}_1(1) + \widehat{\sigma}_1(0)} T\bigg).
\end{align*}

We further distinguish two cases.

\textbf{First}, $\rho < \frac{T - \frac{1}{2} \beta_1 T^{\frac{1}{M}}}{\frac{1}{2} \beta_1 T^{\frac{1}{M}}}$. 
Note that, conditional on $\cE$,
\begin{align*}
\rho \geq \widehat{\rho}_1 \cdot \sqrt{\frac{1-2^{\frac{1}{2}} \beta_1^{-\frac{1}{2}} T^{-\frac{1}{2M} + \frac{\eps}{2}}}{1+2^{\frac{1}{2}} \beta_1^{-\frac{1}{2}} T^{-\frac{1}{2M} + \frac{\eps}{2}}}} > \frac{T - \frac{1}{2} \beta_2 T^{\frac{2}{M}}}{\frac{1}{2} \beta_2 T^{\frac{2}{M}}} \cdot \sqrt{\frac{1-2^{\frac{1}{2}} \beta_1^{-\frac{1}{2}} T^{-\frac{1}{2M} + \frac{\eps}{2}}}{1+2^{\frac{1}{2}} \beta_1^{-\frac{1}{2}} T^{-\frac{1}{2M} + \frac{\eps}{2}}}} > \frac{1}{2} \frac{T - \frac{1}{2} \beta_2 T^{\frac{2}{M}}}{\frac{1}{2} \beta_2 T^{\frac{2}{M}}} > 1,
\end{align*}
where the first inequality is due to \eqref{eqn:MStage:ConfidenceBound1rename} and \eqref{eqn:MStage:ConfidenceBound0rename}; the second inequality is due to the condition of Case 1.2; the third inequality is due to Lemma~\ref{lem:AlgebraicTrick3}; the last inequality is due to Lemma~\ref{lem:AlgebraicTrick4}.

As a result,
\begin{align*}
(T^*(1), T^*(0)) = \bigg(\frac{\sigma(1)}{\sigma(1)+\sigma(0)}T, \frac{\sigma(0)}{\sigma(1)+\sigma(0)}T\bigg).
\end{align*}

Conditional on event $\cE$, we have
\begin{align*}
\left\vert \frac{T(1)}{T^*(1)} - 1 \right\vert = \left\vert \frac{\frac{\widehat{\sigma}_1(1)}{\widehat{\sigma}_1(1) + \widehat{\sigma}_1(0)} T}{\frac{\sigma(1)}{\sigma(1) + \sigma(0)} T} - 1 \right\vert \leq \max \left\{ \frac{1 + 2^{\frac{1}{2}} \beta_1^{-\frac{1}{2}} T^{-\frac{1}{2M} + \frac{\eps}{2}}}{1 - 2^{\frac{1}{2}} \beta_1^{-\frac{1}{2}} T^{-\frac{1}{2M} + \frac{\eps}{2}}} - 1, 1 - \frac{1 - 2^{\frac{1}{2}} \beta_1^{-\frac{1}{2}} T^{-\frac{1}{2M} + \frac{\eps}{2}}}{1 + 2^{\frac{1}{2}} \beta_1^{-\frac{1}{2}} T^{-\frac{1}{2M} + \frac{\eps}{2}}} \right\},
\end{align*}
where the inequality is due to \eqref{eqn:MStage:ConfidenceBound1rename} and \eqref{eqn:MStage:ConfidenceBound0rename}. 

So conditional on event $\cE$, we have $\left\vert \frac{T(1)}{T^*(1)} - 1 \right\vert \to 0$ as $T \to +\infty$.
In addition, $1 - \Pr(\cE) = (M-1) \frac{\kappa(1) + \kappa(0)}{T^{\eps}} \to 0$ as $T \to +\infty$.
Combining these two, we have $\frac{T(1)}{T^*(1)} \xrightarrow{p} 1$ as $T \to +\infty$.
Similarly, we have $\frac{T(0)}{T^*(0)} \xrightarrow{p} 1$ as $T \to +\infty$.

\textbf{Second}, if $\rho \geq \frac{T - \frac{1}{2} \beta_1 T^{\frac{1}{M}}}{\frac{1}{2} \beta_1 T^{\frac{1}{M}}}$, then we have
\begin{align*}
(T^*(1), T^*(0)) = \bigg(T - \frac{1}{2} \beta_1 T^{\frac{1}{M}}, \frac{1}{2} \beta_1 T^{\frac{1}{M}}\bigg).
\end{align*}

Note that, conditional on $\cE$, 
\begin{align*}
\widehat{\rho}_1 \leq \frac{T - \frac{1}{2} \beta_1 T^{\frac{1}{M}}}{\frac{1}{2} \beta_1 T^{\frac{1}{M}}} \leq \rho \leq \widehat{\rho}_1 \sqrt{ \frac{1+2^{\frac{1}{2}} \beta_1^{-\frac{1}{2}} T^{-\frac{1}{2M} + \frac{\eps}{2}}}{1-2^{\frac{1}{2}} \beta_1^{-\frac{1}{2}} T^{-\frac{1}{2M} + \frac{\eps}{2}}} } 
\end{align*}

Conditional on event $\cE$, we have
\begin{multline*}
\left\vert \frac{T(1)}{T^*(1)} - 1 \right\vert = \left\vert \frac{\frac{\widehat{\sigma}_1(0)}{\widehat{\sigma}_1(1) + \widehat{\sigma}_1(0)} T}{T - \frac{1}{2} \beta_1 T^{\frac{1}{M}}} - 1 \right\vert = \left\vert \frac{\widehat{\rho}_1}{\widehat{\rho}_1+1} \cdot \frac{T}{T - \frac{1}{2} \beta_1 T^{\frac{1}{M}}} - 1 \right\vert \\
\leq 1 - \frac{T}{T - \frac{1}{2} \beta_1 T^{\frac{1}{M}}} \cdot \frac{\frac{T - \frac{1}{2} \beta_1 T^{\frac{1}{M}}}{\frac{1}{2} \beta_1 T^{\frac{1}{M}}} ({1-2^{\frac{1}{2}} \beta_1^{-\frac{1}{2}} T^{-\frac{1}{2M} + \frac{\eps}{2}}})}{\frac{T - \frac{1}{2} \beta_1 T^{\frac{1}{M}}}{\frac{1}{2} \beta_1 T^{\frac{1}{M}}} ({1-2^{\frac{1}{2}} \beta_1^{-\frac{1}{2}} T^{-\frac{1}{2M} + \frac{\eps}{2}}})+1} \\
= \frac{2^{-\frac{1}{2}} \beta_1^{\frac{1}{2}} T^{\frac{1}{2M}+\frac{\eps}{2}}}{T(1-2^{\frac{1}{2}} \beta_1^{-\frac{1}{2}} T^{-\frac{1}{2M} + \frac{\eps}{2}}) + 2^{-\frac{1}{2}} \beta_1^{\frac{1}{2}} T^{\frac{1}{2M}+\frac{\eps}{2}}},
\end{multline*}
where the first inequality is because $\frac{\widehat{\rho}_1}{\widehat{\rho}_1+1} \frac{T}{T - \frac{1}{2} \beta_1 T^{\frac{1}{M}}} - 1$ is increasing in $\widehat{\rho}_1$ and equals $0$ when $\widehat{\rho}_1 = \frac{T - \frac{1}{2}\beta_1 T^{\frac{1}{M}}}{\frac{1}{2}\beta_1 T^{\frac{1}{M}}}$,
the second inequality is because for any $\delta \in [0,1), 1- \delta \leq \sqrt{\frac{1-\delta}{1+\delta}}$ and we lower bound $\widehat{\rho}_1$ with $\frac{T - \frac{1}{2} \beta_1 T^{\frac{1}{M}}}{\frac{1}{2} \beta_1 T^{\frac{1}{M}}} ({1-2^{\frac{1}{2}} \beta_1^{-\frac{1}{2}} T^{-\frac{1}{2M} + \frac{\eps}{2}}})$.

So conditional on event $\cE$, we have $\left\vert \frac{T(1)}{T^*(1)} - 1 \right\vert \to 0$ as $T \to +\infty$.
In addition, $1 - \Pr(\cE) = (M-1) \frac{\kappa(1) + \kappa(0)}{T^{\eps}} \to 0$ as $T \to +\infty$.
Combining these two, we have $\frac{T(1)}{T^*(1)} \xrightarrow{p} 1$ as $T \to +\infty$.

Conditional on event $\cE$, we have
\begin{multline*}
\left\vert \frac{T(0)}{T^*(0)} - 1 \right\vert = \left\vert \frac{\frac{\widehat{\sigma}_1(1)}{\widehat{\sigma}_1(1) + \widehat{\sigma}_1(0)} T}{\frac{1}{2} \beta_1 T^{\frac{1}{M}}} - 1 \right\vert = \left\vert \frac{1}{\widehat{\rho}_1+1} \cdot \frac{T}{\frac{1}{2} \beta_1 T^{\frac{1}{M}}} - 1 \right\vert \\
\leq \frac{T}{\frac{1}{2} \beta_1 T^{\frac{1}{M}} + (T-\frac{1}{2} \beta_1 T^{\frac{1}{M}})(1-2^{\frac{1}{2}} \beta_1^{-\frac{1}{2}} T^{-\frac{1}{2M} + \frac{\eps}{2}})} - 1 \\
\leq \frac{T}{T(1-2^{\frac{1}{2}} \beta_1^{-\frac{1}{2}} T^{-\frac{1}{2M} + \frac{\eps}{2}})} - 1 
= \frac{2^{\frac{1}{2}} \beta_1^{-\frac{1}{2}} T^{-\frac{1}{2M} + \frac{\eps}{2}}}{1-2^{\frac{1}{2}} \beta_1^{-\frac{1}{2}} T^{-\frac{1}{2M} + \frac{\eps}{2}}},
\end{multline*}
where the first inequality is because $\frac{1}{\widehat{\rho}_1+1} \frac{T}{\frac{1}{2} \beta_1 T^{\frac{1}{M}}} - 1$ is decreasing in $\widehat{\rho}_1$ and equals $0$ when $\widehat{\rho}_1 = \frac{T - \frac{1}{2}\beta_1 T^{\frac{1}{M}}}{\frac{1}{2}\beta_1 T^{\frac{1}{M}}}$,
the second inequality is because for any $\delta \in [0,1), 1- \delta \leq \sqrt{\frac{1-\delta}{1+\delta}}$ and we lower bound $\widehat{\rho}_1$ with $\frac{T - \frac{1}{2} \beta_1 T^{\frac{1}{M}}}{\frac{1}{2} \beta_1 T^{\frac{1}{M}}} ({1-2^{\frac{1}{2}} \beta_1^{-\frac{1}{2}} T^{-\frac{1}{2M} + \frac{\eps}{2}}})$.

So conditional on event $\cE$, we have $\left\vert \frac{T(0)}{T^*(0)} - 1 \right\vert \to 0$ as $T \to +\infty$.
In addition, $1 - \Pr(\cE) = (M-1) \frac{\kappa(1) + \kappa(0)}{T^{\eps}} \to 0$ as $T \to +\infty$.
Combining these two, we have $\frac{T(0)}{T^*(0)} \xrightarrow{p} 1$ as $T \to +\infty$.

\noindent\underline{\textbf{Case 2}}:
\begin{align*}
\widehat{\rho}_1 \leq \frac{T - \frac{1}{2} \beta_2 T^{\frac{2}{M}}}{\frac{1}{2} \beta_2 T^{\frac{2}{M}}}.
\end{align*}
Due to \eqref{eqn:WLOG:rename} we know that $\widehat{\sigma}_1(1) \geq \widehat{\sigma}_1(0)$.
In Case 2 we immediately have
\begin{align*}
\frac{\widehat{\sigma}_1(1)}{\widehat{\sigma}_1(1) + \widehat{\sigma}_1(0)} T \geq \frac{\widehat{\sigma}_1(0)}{\widehat{\sigma}_1(1) + \widehat{\sigma}_1(0)} T  \geq \frac{1}{\frac{T - \frac{1}{2} \beta_2 T^{\frac{2}{M}}}{\frac{1}{2} \beta_2 T^{\frac{2}{M}}}+1} T = \frac{1}{2} \beta_2 T^{\frac{2}{M}}.
\end{align*}
So Algorithm~\ref{alg:MStageANA} goes to Line~\ref{mrk:Case3} in the 1-st stage experiment. 
We further distinguish two cases.

\noindent\textbf{Case 2.1}:
\begin{align*}
\widehat{\rho}_1 \leq \frac{T - \frac{1}{2} \beta_2 T^{\frac{2}{M}}}{\frac{1}{2} \beta_2 T^{\frac{2}{M}}}, && \widehat{\rho}_2 > \frac{T - \frac{1}{2} \beta_2 T^{\frac{2}{M}}}{\frac{1}{2} \beta_2 T^{\frac{2}{M}}}.
\end{align*}
In this case, 
\begin{align*}
\frac{\widehat{\sigma}_2(0)}{\widehat{\sigma}_2(1) + \widehat{\sigma}_2(0)} T < \frac{1}{\frac{T - \frac{1}{2} \beta_2 T^{\frac{2}{M}}}{\frac{1}{2} \beta_2 T^{\frac{2}{M}}}+1} T = \frac{1}{2} \beta_2 T^{\frac{2}{M}}.
\end{align*}
So Algorithm~\ref{alg:MStageANA} goes to Line~\ref{mrk:Case1} in the $2$-nd stage experiment.
Then we have
\begin{align*}
(T(1), T(0)) = \bigg(T - \frac{1}{2} \beta_2 T^{\frac{2}{M}}, \frac{1}{2} \beta_2 T^{\frac{2}{M}}\bigg).
\end{align*}

Note that, as $T \to +\infty$,
\begin{align*}
\rho \leq \widehat{\rho}_1 \sqrt{\frac{1+2^{\frac{1}{2}} \beta_1^{-\frac{1}{2}} T^{-\frac{1}{2M}+\frac{\eps}{2}}}{1-2^{\frac{1}{2}} \beta_1^{-\frac{1}{2}} T^{-\frac{1}{2M}+\frac{\eps}{2}}}} 
\leq \frac{T - \frac{1}{2} \beta_2 T^{\frac{2}{M}}}{\frac{1}{2} \beta_2 T^{\frac{2}{M}}} \sqrt{\frac{1+2^{\frac{1}{2}} \beta_1^{-\frac{1}{2}} T^{-\frac{1}{2M}+\frac{\eps}{2}}}{1-2^{\frac{1}{2}} \beta_1^{-\frac{1}{2}} T^{-\frac{1}{2M}+\frac{\eps}{2}}}} 
= \frac{T - \frac{1}{2} \beta_2 T^{\frac{2}{M}}}{\frac{1}{2} \beta_2 T^{\frac{2}{M}}}
< \frac{T - \frac{1}{2} \beta_1 T^{\frac{1}{M}}}{\frac{1}{2} \beta_1 T^{\frac{1}{M}}}
\end{align*}
where the first inequality is due to \eqref{eqn:MStage:ConfidenceBound1rename} and \eqref{eqn:MStage:ConfidenceBound0rename}; the second inequality is due to the condition of Case~2;
the equality holds when $T \to +\infty$;
the last inequality is because $\beta_1 T^{\frac{1}{M}} < \beta_2 T^{\frac{2}{M}}$.

Note also that,
\begin{align*}
\rho \geq \widehat{\rho}_2 \sqrt{\frac{1-2^{\frac{1}{2}} \beta_2^{-\frac{1}{2}} T^{-\frac{2}{2M} + \frac{\eps}{2}}}{1+2^{\frac{1}{2}} \beta_2^{-\frac{1}{2}} T^{-\frac{2}{2M} + \frac{\eps}{2}}}} \geq \frac{T - \frac{1}{2} \beta_2 T^{\frac{2}{M}}}{\frac{1}{2} \beta_2 T^{\frac{2}{M}}} \cdot \sqrt{\frac{1-2^{\frac{1}{2}} \beta_2^{-\frac{1}{2}} T^{-\frac{2}{2M} + \frac{\eps}{2}}}{1+2^{\frac{1}{2}} \beta_2^{-\frac{1}{2}} T^{-\frac{2}{2M} + \frac{\eps}{2}}}} > \frac{1}{2} \frac{T - \frac{1}{2} \beta_2 T^{\frac{2}{M}}}{\frac{1}{2} \beta_2 T^{\frac{2}{M}}} > 1,
\end{align*}
where the first inequality is due to \eqref{eqn:MStage:ConfidenceBound1rename} and \eqref{eqn:MStage:ConfidenceBound0rename}; the second inequality is due to the condition of Case 2.1; the third inequality is due to Lemma~\ref{lem:AlgebraicTrick3}; the last inequality is due to Lemma~\ref{lem:AlgebraicTrick4}.

As a result, as $T \to +\infty$, we have $\frac{\frac{1}{2} \beta_1 T^{\frac{1}{M}}}{T - \frac{1}{2} \beta_1 T^{\frac{1}{M}}} < \rho < \frac{T - \frac{1}{2} \beta_1 T^{\frac{1}{M}}}{\frac{1}{2} \beta_1 T^{\frac{1}{M}}}$, so
\begin{align*}
(T^*(1), T^*(0)) = \bigg(\frac{\sigma(1)}{\sigma(1)+\sigma(0)}T, \frac{\sigma(0)}{\sigma(1)+\sigma(0)}T\bigg).
\end{align*}

Conditional on event $\cE$, we have
\begin{multline*}
\left\vert \frac{T(1)}{T^*(1)} - 1 \right\vert = \left\vert \frac{T - \frac{1}{2} \beta_2 T^{\frac{2}{M}}}{\frac{\sigma(1)}{\sigma(1) + \sigma(0)} T} - 1 \right\vert = \left\vert \big( \frac{1}{\rho}+1 \big) \frac{T-\frac{1}{2} \beta_2 T^{\frac{2}{M}}}{T} - 1 \right\vert \\
\leq \bigg( \frac{\frac{1}{2} \beta_2 T^{\frac{2}{M}}}{T-\frac{1}{2} \beta_2 T^{\frac{2}{M}}} \sqrt{\frac{1+2^{\frac{1}{2}} \beta_2^{-\frac{1}{2}} T^{-\frac{2}{2M} + \frac{\eps}{2}}}{1-2^{\frac{1}{2}} \beta_2^{-\frac{1}{2}} T^{-\frac{2}{2M} + \frac{\eps}{2}}}} + 1\bigg) \frac{T-\frac{1}{2} \beta_2 T^{\frac{2}{M}}}{T} - 1,
\end{multline*}
where the inequality is because $\big( \frac{1}{\rho}+1 \big) \frac{T-\frac{1}{2} \beta_2 T^{\frac{2}{M}}}{T} - 1$ is decreasing in $\rho$ and equals $0$ when $\rho = \frac{T - \frac{1}{2}\beta_2 T^{\frac{2}{M}}}{\frac{1}{2}\beta_2 T^{\frac{2}{M}}}$; and we lower bound $\rho$ by $\frac{T - \frac{1}{2} \beta_2 T^{\frac{2}{M}}}{\frac{1}{2} \beta_2 T^{\frac{2}{M}}} \cdot \sqrt{\frac{1-2^{\frac{1}{2}} \beta_2^{-\frac{1}{2}} T^{-\frac{2}{2M} + \frac{\eps}{2}}}{1+2^{\frac{1}{2}} \beta_2^{-\frac{1}{2}} T^{-\frac{2}{2M} + \frac{\eps}{2}}}}$.

So conditional on event $\cE$, we have $\left\vert \frac{T(1)}{T^*(1)} - 1 \right\vert \to 0$ as $T \to +\infty$.
In addition, $1 - \Pr(\cE) = (M-1) \frac{\kappa(1) + \kappa(0)}{T^{\eps}} \to 0$ as $T \to +\infty$.
Combining these two, we have $\frac{T(1)}{T^*(1)} \xrightarrow{p} 1$ as $T \to +\infty$.

Conditional on event $\cE$, we have
\begin{multline*}
\left\vert \frac{T(0)}{T^*(0)} - 1 \right\vert = \left\vert \frac{\frac{1}{2} \beta_2 T^{\frac{2}{M}}}{\frac{\sigma(0)}{\sigma(1) + \sigma(0)} T} - 1 \right\vert = \left\vert \big( \rho+1 \big) \frac{\frac{1}{2} \beta_2 T^{\frac{2}{M}}}{T} - 1 \right\vert \\
\leq 1 - \bigg( 1 + \frac{T - \frac{1}{2} \beta_2 T^{\frac{2}{M}}}{\frac{1}{2} \beta_2 T^{\frac{2}{M}}} \sqrt{\frac{1-2^{\frac{1}{2}} \beta_2^{-\frac{1}{2}} T^{-\frac{2}{2M} + \frac{\eps}{2}}}{1+2^{\frac{1}{2}} \beta_2^{-\frac{1}{2}} T^{-\frac{2}{2M} + \frac{\eps}{2}}}}\bigg) \frac{\frac{1}{2} \beta_2 T^{\frac{2}{M}}}{T} 
\leq \frac{T - \frac{1}{2} \beta_2 T^{\frac{2}{M}}}{T} 2^{\frac{1}{2}} \beta_2^{-\frac{1}{2}} T^{-\frac{2}{2M} + \frac{\eps}{2}}
\end{multline*}
where the first inequality is because $\big( \rho+1 \big) \frac{\frac{1}{2} \beta_2 T^{\frac{2}{M}}}{T} - 1$ is increasing in $\rho$ and equals $0$ when $\rho = \frac{T - \frac{1}{2}\beta_2 T^{\frac{2}{M}}}{\frac{1}{2}\beta_2 T^{\frac{2}{M}}}$,
the second inequality is because for any $\delta \in [0,1), 1- \delta \leq \sqrt{\frac{1-\delta}{1+\delta}}$.

So conditional on event $\cE$, we have $\left\vert \frac{T(0)}{T^*(0)} - 1 \right\vert \to 0$ as $T \to +\infty$.
In addition, $1 - \Pr(\cE) = (M-1) \frac{\kappa(1) + \kappa(0)}{T^{\eps}} \to 0$ as $T \to +\infty$.
Combining these two, we have $\frac{T(0)}{T^*(0)} \xrightarrow{p} 1$ as $T \to +\infty$.

\noindent\textbf{Case 2.2}:
\begin{align*}
\widehat{\rho}_1 \leq \frac{T - \frac{1}{2} \beta_2 T^{\frac{2}{M}}}{\frac{1}{2} \beta_2 T^{\frac{2}{M}}}, && \frac{T - \frac{1}{2} \beta_3 T^{\frac{3}{M}}}{\frac{1}{2} \beta_3 T^{\frac{3}{M}}} < \widehat{\rho}_2 \leq \frac{T - \frac{1}{2} \beta_2 T^{\frac{2}{M}}}{\frac{1}{2} \beta_2 T^{\frac{2}{M}}}.
\end{align*}
In this case, 
\begin{align*}
\frac{1}{2} \beta_2 T^{\frac{2}{M}} = \frac{1}{\frac{T - \frac{1}{2} \beta_2 T^{\frac{2}{M}}}{\frac{1}{2} \beta_2 T^{\frac{2}{M}}}+1} T \leq \frac{\widehat{\sigma}_2(0)}{\widehat{\sigma}_2(1) + \widehat{\sigma}_2(0)} T < \frac{1}{\frac{T - \frac{1}{2} \beta_3 T^{\frac{3}{M}}}{\frac{1}{2} \beta_3 T^{\frac{3}{M}}}+1} T < \frac{1}{2} \beta_3 T^{\frac{3}{M}}.
\end{align*}
So Algorithm~\ref{alg:MStageANA} goes to Line~\ref{mrk:Case2} in the $2$-nd stage experiment.
Then we have
\begin{align*}
(T(1), T(0)) = \bigg(\frac{\widehat{\sigma}_2(1)}{\widehat{\sigma}_2(1) + \widehat{\sigma}_2(0)} T, \frac{\widehat{\sigma}_2(0)}{\widehat{\sigma}_2(1) + \widehat{\sigma}_2(0)} T\bigg).
\end{align*}

Note that, as $T \to +\infty$,
\begin{align*}
\rho \leq \widehat{\rho}_2 \sqrt{\frac{1+2^{\frac{1}{2}} \beta_2^{-\frac{1}{2}} T^{-\frac{2}{2M}+\frac{\eps}{2}}}{1-2^{\frac{1}{2}} \beta_2^{-\frac{1}{2}} T^{-\frac{2}{2M}+\frac{\eps}{2}}}} 
\leq \frac{T - \frac{1}{2} \beta_2 T^{\frac{2}{M}}}{\frac{1}{2} \beta_2 T^{\frac{2}{M}}} \sqrt{\frac{1+2^{\frac{1}{2}} \beta_1^{-\frac{1}{2}} T^{-\frac{1}{2M}+\frac{\eps}{2}}}{1-2^{\frac{1}{2}} \beta_1^{-\frac{1}{2}} T^{-\frac{1}{2M}+\frac{\eps}{2}}}} 
\leq \frac{T - \frac{1}{2} \beta_1 T^{\frac{1}{M}}}{\frac{1}{2} \beta_1 T^{\frac{1}{M}}}
\end{align*}
where the first inequality is due to \eqref{eqn:MStage:ConfidenceBound1rename} and \eqref{eqn:MStage:ConfidenceBound0rename}; the second inequality is due to the condition of Case~2.2;
the last inequality is because $\beta_1 T^{\frac{1}{M}} < \beta_2 T^{\frac{2}{M}}$ and holds as $T \to +\infty$.

Note also that,
\begin{align*}
\rho \geq \widehat{\rho}_2 \cdot \sqrt{\frac{1-2^{\frac{1}{2}} \beta_2^{-\frac{1}{2}} T^{-\frac{2}{2M} + \frac{\eps}{2}}}{1+2^{\frac{1}{2}} \beta_2^{-\frac{1}{2}} T^{-\frac{2}{2M} + \frac{\eps}{2}}}} > \frac{T - \frac{1}{2} \beta_3 T^{\frac{3}{M}}}{\frac{1}{2} \beta_3 T^{\frac{3}{M}}} \cdot \sqrt{\frac{1-2^{\frac{1}{2}} \beta_2^{-\frac{1}{2}} T^{-\frac{2}{2M} + \frac{\eps}{2}}}{1+2^{\frac{1}{2}} \beta_2^{-\frac{1}{2}} T^{-\frac{2}{2M} + \frac{\eps}{2}}}} > \frac{1}{2} \frac{T - \frac{1}{2} \beta_3 T^{\frac{3}{M}}}{\frac{1}{2} \beta_3 T^{\frac{3}{M}}} > 1,
\end{align*}
where the first inequality is due to \eqref{eqn:MStage:ConfidenceBound1rename} and \eqref{eqn:MStage:ConfidenceBound0rename}; the second inequality is due to the condition of Case 2.2; the third inequality is due to Lemma~\ref{lem:AlgebraicTrick3}; the last inequality is due to Lemma~\ref{lem:AlgebraicTrick4}.

As a result, as $T \to +\infty$,
\begin{align*}
(T^*(1), T^*(0)) = \bigg(\frac{\sigma(1)}{\sigma(1)+\sigma(0)}T, \frac{\sigma(0)}{\sigma(1)+\sigma(0)}T\bigg).
\end{align*}

Conditional on event $\cE$, we have
\begin{align*}
\left\vert \frac{T(1)}{T^*(1)} - 1 \right\vert = \left\vert \frac{\frac{\widehat{\sigma}_2(1)}{\widehat{\sigma}_2(1) + \widehat{\sigma}_2(0)} T}{\frac{\sigma(1)}{\sigma(1) + \sigma(0)} T} - 1 \right\vert 
\leq \max \left\{ \frac{1 + 2^{\frac{1}{2}} \beta_2^{-\frac{1}{2}} T^{-\frac{2}{2M} + \frac{\eps}{2}}}{1 - 2^{\frac{1}{2}} \beta_2^{-\frac{1}{2}} T^{-\frac{2}{2M} + \frac{\eps}{2}}} - 1, 1 - \frac{1 - 2^{\frac{1}{2}} \beta_2^{-\frac{1}{2}} T^{-\frac{2}{2M} + \frac{\eps}{2}}}{1 + 2^{\frac{1}{2}} \beta_2^{-\frac{1}{2}} T^{-\frac{2}{2M} + \frac{\eps}{2}}} \right\},
\end{align*}
where the inequality is due to \eqref{eqn:MStage:ConfidenceBound1rename} and \eqref{eqn:MStage:ConfidenceBound0rename}. 

So conditional on event $\cE$, we have $\left\vert \frac{T(1)}{T^*(1)} - 1 \right\vert \to 0$ as $T \to +\infty$.
In addition, $1 - \Pr(\cE) = (M-1) \frac{\kappa(1) + \kappa(0)}{T^{\eps}} \to 0$ as $T \to +\infty$.
Combining these two, we have $\frac{T(1)}{T^*(1)} \xrightarrow{p} 1$ as $T \to +\infty$.
Similarly, we have $\frac{T(0)}{T^*(0)} \xrightarrow{p} 1$ as $T \to +\infty$.

\noindent\underline{\textbf{Case $\bm{m}$}} (when $m \leq M-2$):
\begin{align*}
\widehat{\rho}_l \leq \frac{T - \frac{1}{2} \beta_{l+1} T^{\frac{l+1}{M}}}{\frac{1}{2} \beta_{l+1} T^{\frac{l+1}{M}}}, \ \forall \ l \leq m-1.
\end{align*}
Due to the condition of Case $m$, we immediately have
\begin{align*}
\frac{\widehat{\sigma}_{m-1}(0)}{\widehat{\sigma}_{m-1}(1) + \widehat{\sigma}_{m-1}(0)} T \geq \frac{1}{\frac{T - \frac{1}{2} \beta_{m} T^{\frac{m}{M}}}{\frac{1}{2} \beta_{m} T^{\frac{m}{M}}}+1} T = \frac{1}{2} \beta_{m} T^{\frac{m}{M}}.
\end{align*}
On the other hand, since
\begin{multline*}
\widehat{\rho}_{m-1} \geq \rho \sqrt{\frac{1-2^{\frac{1}{2}} \beta_{m-1}^{-\frac{1}{2}} T^{-\frac{m-1}{2M}+\frac{\eps}{2}}}{1+2^{\frac{1}{2}} \beta_{m-1}^{-\frac{1}{2}} T^{-\frac{m-1}{2M}+\frac{\eps}{2}}}} \geq \widehat{\rho}_1 \sqrt{\frac{1-2^{\frac{1}{2}} \beta_{1}^{-\frac{1}{2}} T^{-\frac{1}{2M}+\frac{\eps}{2}}}{1+2^{\frac{1}{2}} \beta_{1}^{-\frac{1}{2}} T^{-\frac{1}{2M}+\frac{\eps}{2}}}} \sqrt{\frac{1-2^{\frac{1}{2}} \beta_{m-1}^{-\frac{1}{2}} T^{-\frac{m-1}{2M}+\frac{\eps}{2}}}{1+2^{\frac{1}{2}} \beta_{m-1}^{-\frac{1}{2}} T^{-\frac{m-1}{2M}+\frac{\eps}{2}}}} \\
\geq \sqrt{\frac{1-2^{\frac{1}{2}} \beta_{1}^{-\frac{1}{2}} T^{-\frac{1}{2M}+\frac{\eps}{2}}}{1+2^{\frac{1}{2}} \beta_{1}^{-\frac{1}{2}} T^{-\frac{1}{2M}+\frac{\eps}{2}}}} \sqrt{\frac{1-2^{\frac{1}{2}} \beta_{m-1}^{-\frac{1}{2}} T^{-\frac{m-1}{2M}+\frac{\eps}{2}}}{1+2^{\frac{1}{2}} \beta_{m-1}^{-\frac{1}{2}} T^{-\frac{m-1}{2M}+\frac{\eps}{2}}}} > \frac{1}{4} \geq \frac{\frac{1}{2}\beta_m T^{\frac{m}{M}}}{T - \frac{1}{2}\beta_m T^{\frac{m}{M}}},
\end{multline*}
where the first and second inequalities are due to \eqref{eqn:MStage:ConfidenceBound1rename} and \eqref{eqn:MStage:ConfidenceBound0rename}; the third inequality is due to \eqref{eqn:WLOG:rename}; the fourth inequality is due to Lemma~\ref{lem:AlgebraicTrick3}; the last inequality is due to Lemma~\ref{lem:AlgebraicTrick4}.
Due to the above sequence of inequalities, we have $\frac{1}{\widehat{\rho}_{m-1}} \leq \frac{T - \frac{1}{2}\beta_m T^{\frac{m}{M}}}{\frac{1}{2}\beta_m T^{\frac{m}{M}}}$, which leads to
\begin{align*}
\frac{\widehat{\sigma}_{m-1}(1)}{\widehat{\sigma}_{m-1}(1) + \widehat{\sigma}_{m-1}(0)} T  \geq \frac{1}{1+\frac{T - \frac{1}{2} \beta_{m} T^{\frac{m}{M}}}{\frac{1}{2} \beta_{m} T^{\frac{m}{M}}}} T = \frac{1}{2} \beta_{m} T^{\frac{m}{M}}.
\end{align*}
So Algorithm~\ref{alg:MStageANA} goes to Line~\ref{mrk:Case3} in the (m-1)-th stage experiment. 
We further distinguish two cases.

\noindent\textbf{Case $\bm{m}$.1}:
In addition to the conditions in Case $m$ above, we also have
\begin{align*}
\widehat{\rho}_m > \frac{T - \frac{1}{2} \beta_{m} T^{\frac{m}{M}}}{\frac{1}{2} \beta_{m} T^{\frac{m}{M}}}.
\end{align*}
Similar to the analysis in Case 2.1, we proceed with the following analysis.
In Case $m$.1, 
\begin{align*}
\frac{\widehat{\sigma}_m(0)}{\widehat{\sigma}_m(1) + \widehat{\sigma}_m(0)} T < \frac{1}{\frac{T - \frac{1}{2} \beta_m T^{\frac{m}{M}}}{\frac{1}{2} \beta_m T^{\frac{m}{M}}}+1} T = \frac{1}{2} \beta_m T^{\frac{m}{M}}.
\end{align*}
So Algorithm~\ref{alg:MStageANA} goes to Line~\ref{mrk:Case1} in the $m$-th stage experiment.
Then we have
\begin{align*}
(T(1), T(0)) = \bigg(T - \frac{1}{2} \beta_m T^{\frac{m}{M}}, \frac{1}{2} \beta_m T^{\frac{m}{M}}\bigg).
\end{align*}

Note that, as $T \to +\infty$,
\begin{multline*}
\rho \leq \widehat{\rho}_{m-1} \sqrt{\frac{1+2^{\frac{1}{2}} \beta_{m-1}^{-\frac{1}{2}} T^{-\frac{m-1}{2M}+\frac{\eps}{2}}}{1-2^{\frac{1}{2}} \beta_{m-1}^{-\frac{1}{2}} T^{-\frac{m-1}{2M}+\frac{\eps}{2}}}} 
\leq \frac{T - \frac{1}{2} \beta_m T^{\frac{m}{M}}}{\frac{1}{2} \beta_m T^{\frac{m}{M}}} \sqrt{\frac{1+2^{\frac{1}{2}} \beta_{m-1}^{-\frac{1}{2}} T^{-\frac{m-1}{2M}+\frac{\eps}{2}}}{1-2^{\frac{1}{2}} \beta_{m-1}^{-\frac{1}{2}} T^{-\frac{m-1}{2M}+\frac{\eps}{2}}}} \\
= \frac{T - \frac{1}{2} \beta_m T^{\frac{m}{M}}}{\frac{1}{2} \beta_m T^{\frac{m}{M}}}
\leq \frac{T - \frac{1}{2} \beta_1 T^{\frac{1}{M}}}{\frac{1}{2} \beta_1 T^{\frac{1}{M}}}
\end{multline*}
where the first inequality is due to \eqref{eqn:MStage:ConfidenceBound1rename} and \eqref{eqn:MStage:ConfidenceBound0rename}; the second inequality is due to the condition of Case~m;
the equality holds when $T \to +\infty$;
the last inequality is because $\beta_1 T^{\frac{1}{M}} < \beta_m T^{\frac{m}{M}}$.

Note also that,
\begin{align*}
\rho \geq \widehat{\rho}_m \sqrt{\frac{1-2^{\frac{1}{2}} \beta_m^{-\frac{1}{2}} T^{-\frac{m}{2M} + \frac{\eps}{2}}}{1+2^{\frac{1}{2}} \beta_m^{-\frac{1}{2}} T^{-\frac{m}{2M} + \frac{\eps}{2}}}} 
\geq \frac{T - \frac{1}{2} \beta_m T^{\frac{m}{M}}}{\frac{1}{2} \beta_m T^{\frac{m}{M}}} \cdot \sqrt{\frac{1-2^{\frac{1}{2}} \beta_m^{-\frac{1}{2}} T^{-\frac{m}{2M} + \frac{\eps}{2}}}{1+2^{\frac{1}{2}} \beta_m^{-\frac{1}{2}} T^{-\frac{m}{2M} + \frac{\eps}{2}}}} > \frac{1}{2} \frac{T - \frac{1}{2} \beta_m T^{\frac{m}{M}}}{\frac{1}{2} \beta_m T^{\frac{m}{M}}} > 1,
\end{align*}
where the first inequality is due to \eqref{eqn:MStage:ConfidenceBound1rename} and \eqref{eqn:MStage:ConfidenceBound0rename}; the second inequality is due to the condition of Case~m.1; the third inequality is due to Lemma~\ref{lem:AlgebraicTrick3}; the last inequality is due to Lemma~\ref{lem:AlgebraicTrick4}.

As a result, as $T \to +\infty$, we have $\frac{\frac{1}{2} \beta_1 T^{\frac{1}{M}}}{T - \frac{1}{2} \beta_1 T^{\frac{1}{M}}} < \rho < \frac{T - \frac{1}{2} \beta_1 T^{\frac{1}{M}}}{\frac{1}{2} \beta_1 T^{\frac{1}{M}}}$, so
\begin{align*}
(T^*(1), T^*(0)) = \bigg(\frac{\sigma(1)}{\sigma(1)+\sigma(0)}T, \frac{\sigma(0)}{\sigma(1)+\sigma(0)}T\bigg).
\end{align*}

Conditional on event $\cE$, we have
\begin{multline*}
\left\vert \frac{T(1)}{T^*(1)} - 1 \right\vert = \left\vert \frac{T - \frac{1}{2} \beta_m T^{\frac{m}{M}}}{\frac{\sigma(1)}{\sigma(1) + \sigma(0)} T} - 1 \right\vert = \left\vert \big( \frac{1}{\rho}+1 \big) \frac{T-\frac{1}{2} \beta_m T^{\frac{m}{M}}}{T} - 1 \right\vert \\
\leq \bigg( \frac{\frac{1}{2} \beta_m T^{\frac{m}{M}}}{T-\frac{1}{2} \beta_m T^{\frac{m}{M}}} \sqrt{\frac{1+2^{\frac{1}{2}} \beta_m^{-\frac{1}{2}} T^{-\frac{m}{2M} + \frac{\eps}{2}}}{1-2^{\frac{1}{2}} \beta_m^{-\frac{1}{2}} T^{-\frac{m}{2M} + \frac{\eps}{2}}}} + 1\bigg) \frac{T-\frac{1}{2} \beta_m T^{\frac{m}{M}}}{T} - 1,
\end{multline*}
where the inequality is because $\big( \frac{1}{\rho}+1 \big) \frac{T-\frac{1}{2} \beta_m T^{\frac{m}{M}}}{T} - 1$ is decreasing in $\rho$ and equals $0$ when $\rho = \frac{T - \frac{1}{2}\beta_m T^{\frac{m}{M}}}{\frac{1}{2}\beta_m T^{\frac{m}{M}}}$; and we lower bound $\rho$ by $\frac{T - \frac{1}{2} \beta_m T^{\frac{m}{M}}}{\frac{1}{2} \beta_m T^{\frac{m}{M}}} \cdot \sqrt{\frac{1-2^{\frac{1}{2}} \beta_m^{-\frac{1}{2}} T^{-\frac{m}{2M} + \frac{\eps}{2}}}{1+2^{\frac{1}{2}} \beta_m^{-\frac{1}{2}} T^{-\frac{m}{2M} + \frac{\eps}{2}}}}$.

So conditional on event $\cE$, we have $\left\vert \frac{T(1)}{T^*(1)} - 1 \right\vert \to 0$ as $T \to +\infty$.
In addition, $1 - \Pr(\cE) = (M-1) \frac{\kappa(1) + \kappa(0)}{T^{\eps}} \to 0$ as $T \to +\infty$.
Combining these two, we have $\frac{T(1)}{T^*(1)} \xrightarrow{p} 1$ as $T \to +\infty$.

Conditional on event $\cE$, we have
\begin{multline*}
\left\vert \frac{T(0)}{T^*(0)} - 1 \right\vert = \left\vert \frac{\frac{1}{2} \beta_m T^{\frac{m}{M}}}{\frac{\sigma(0)}{\sigma(1) + \sigma(0)} T} - 1 \right\vert = \left\vert \big( \rho+1 \big) \frac{\frac{1}{2} \beta_m T^{\frac{m}{M}}}{T} - 1 \right\vert \\
\leq 1 - \bigg( 1 + \frac{T - \frac{1}{2} \beta_m T^{\frac{m}{M}}}{\frac{1}{2} \beta_m T^{\frac{m}{M}}} \sqrt{\frac{1-2^{\frac{1}{2}} \beta_m^{-\frac{1}{2}} T^{-\frac{m}{2M} + \frac{\eps}{2}}}{1+2^{\frac{1}{2}} \beta_m^{-\frac{1}{2}} T^{-\frac{m}{2M} + \frac{\eps}{2}}}}\bigg) \frac{\frac{1}{2} \beta_m T^{\frac{m}{M}}}{T} 
\leq \frac{T - \frac{1}{2} \beta_m T^{\frac{m}{M}}}{T} 2^{\frac{1}{2}} \beta_m^{-\frac{1}{2}} T^{-\frac{m}{2M} + \frac{\eps}{2}}
\end{multline*}
where the first inequality is because $\big( \rho+1 \big) \frac{\frac{1}{2} \beta_m T^{\frac{m}{M}}}{T} - 1$ is increasing in $\rho$ and equals $0$ when $\rho = \frac{T - \frac{1}{2}\beta_m T^{\frac{m}{M}}}{\frac{1}{2}\beta_m T^{\frac{m}{M}}}$,
the second inequality is because for any $\delta \in [0,1), 1- \delta \leq \sqrt{\frac{1-\delta}{1+\delta}}$.

So conditional on event $\cE$, we have $\left\vert \frac{T(0)}{T^*(0)} - 1 \right\vert \to 0$ as $T \to +\infty$.
In addition, $1 - \Pr(\cE) = (M-1) \frac{\kappa(1) + \kappa(0)}{T^{\eps}} \to 0$ as $T \to +\infty$.
Combining these two, we have $\frac{T(0)}{T^*(0)} \xrightarrow{p} 1$ as $T \to +\infty$.

\noindent\textbf{Case $\bm{m}$.2}:
In addition to the conditions in Case $m$ above, we also have
\begin{align*}
\frac{T - \frac{1}{2} \beta_{m+1} T^{\frac{m+1}{M}}}{\frac{1}{2} \beta_{m+1} T^{\frac{m+1}{M}}} < \widehat{\rho}_m \leq \frac{T - \frac{1}{2} \beta_{m} T^{\frac{m}{M}}}{\frac{1}{2} \beta_{m} T^{\frac{m}{M}}}.
\end{align*}
Similar to the analysis in Case 2.2, we proceed with the following analysis.
In Case $m$.2, 
\begin{align*}
\frac{1}{2} \beta_m T^{\frac{m}{M}} = \frac{1}{\frac{T - \frac{1}{2} \beta_m T^{\frac{m}{M}}}{\frac{1}{2} \beta_m T^{\frac{m}{M}}}+1} T \leq \frac{\widehat{\sigma}_m(0)}{\widehat{\sigma}_m(1) + \widehat{\sigma}_m(0)} T < \frac{1}{\frac{T - \frac{1}{2} \beta_{m+1} T^{\frac{m+1}{M}}}{\frac{1}{2} \beta_{m+1} T^{\frac{m+1}{M}}}+1} T < \frac{1}{2} \beta_{m+1} T^{\frac{m+1}{M}}.
\end{align*}
So Algorithm~\ref{alg:MStageANA} goes to Line~\ref{mrk:Case2} in the $m$-th stage experiment.
Then we have
\begin{align*}
(T(1), T(0)) = \bigg(\frac{\widehat{\sigma}_m(1)}{\widehat{\sigma}_m(1) + \widehat{\sigma}_m(0)} T, \frac{\widehat{\sigma}_m(0)}{\widehat{\sigma}_m(1) + \widehat{\sigma}_m(0)} T\bigg).
\end{align*}

Note that, as $T \to +\infty$,
\begin{align*}
\rho \leq \widehat{\rho}_m \sqrt{\frac{1+2^{\frac{1}{2}} \beta_m^{-\frac{1}{2}} T^{-\frac{m}{2M}+\frac{\eps}{2}}}{1-2^{\frac{1}{2}} \beta_m^{-\frac{1}{2}} T^{-\frac{m}{2M}+\frac{\eps}{2}}}} 
\leq \frac{T - \frac{1}{2} \beta_m T^{\frac{m}{M}}}{\frac{1}{2} \beta_m T^{\frac{m}{M}}} \sqrt{\frac{1+2^{\frac{1}{2}} \beta_m^{-\frac{1}{2}} T^{-\frac{m}{2M}+\frac{\eps}{2}}}{1-2^{\frac{1}{2}} \beta_m^{-\frac{1}{2}} T^{-\frac{m}{2M}+\frac{\eps}{2}}}} 
\leq \frac{T - \frac{1}{2} \beta_1 T^{\frac{1}{M}}}{\frac{1}{2} \beta_1 T^{\frac{1}{M}}}
\end{align*}
where the first inequality is due to \eqref{eqn:MStage:ConfidenceBound1rename} and \eqref{eqn:MStage:ConfidenceBound0rename}; the second inequality is due to the condition of Case~m.2;
the last inequality is because $\beta_1 T^{\frac{1}{M}} < \beta_m T^{\frac{m}{M}}$ and holds as $T \to +\infty$. 

Note also that,
\begin{align*}
\rho \geq \widehat{\rho}_m \cdot \sqrt{\frac{1-2^{\frac{1}{2}} \beta_m^{-\frac{1}{2}} T^{-\frac{m}{2M} + \frac{\eps}{2}}}{1+2^{\frac{1}{2}} \beta_m^{-\frac{1}{2}} T^{-\frac{m}{2M} + \frac{\eps}{2}}}} > \frac{T - \frac{1}{2} \beta_{m+1} T^{\frac{m+1}{M}}}{\frac{1}{2} \beta_{m+1} T^{\frac{m+1}{M}}} \cdot \sqrt{\frac{1-2^{\frac{1}{2}} \beta_m^{-\frac{1}{2}} T^{-\frac{m}{2M} + \frac{\eps}{2}}}{1+2^{\frac{1}{2}} \beta_m^{-\frac{1}{2}} T^{-\frac{m}{2M} + \frac{\eps}{2}}}} > \frac{1}{2} \frac{T - \frac{1}{2} \beta_{m+1} T^{\frac{m+1}{M}}}{\frac{1}{2} \beta_{m+1} T^{\frac{m+1}{M}}} > 1,
\end{align*}
where the first inequality is due to \eqref{eqn:MStage:ConfidenceBound1rename} and \eqref{eqn:MStage:ConfidenceBound0rename}; the second inequality is due to the condition of Case m.2; the third inequality is due to Lemma~\ref{lem:AlgebraicTrick3}; the last inequality is due to Lemma~\ref{lem:AlgebraicTrick4}.

As a result, as $T \to +\infty$,
\begin{align*}
(T^*(1), T^*(0)) = \bigg(\frac{\sigma(1)}{\sigma(1)+\sigma(0)}T, \frac{\sigma(0)}{\sigma(1)+\sigma(0)}T\bigg).
\end{align*}

Conditional on event $\cE$, we have
\begin{align*}
\left\vert \frac{T(1)}{T^*(1)} - 1 \right\vert = \left\vert \frac{\frac{\widehat{\sigma}_m(1)}{\widehat{\sigma}_m(1) + \widehat{\sigma}_m(0)} T}{\frac{\sigma(1)}{\sigma(1) + \sigma(0)} T} - 1 \right\vert 
\leq \max \left\{ \frac{1 + 2^{\frac{1}{2}} \beta_m^{-\frac{1}{2}} T^{-\frac{m}{2M} + \frac{\eps}{2}}}{1 - 2^{\frac{1}{2}} \beta_m^{-\frac{1}{2}} T^{-\frac{m}{2M} + \frac{\eps}{2}}} - 1, 1 - \frac{1 - 2^{\frac{1}{2}} \beta_m^{-\frac{1}{2}} T^{-\frac{m}{2M} + \frac{\eps}{2}}}{1 + 2^{\frac{1}{2}} \beta_m^{-\frac{1}{2}} T^{-\frac{m}{2M} + \frac{\eps}{2}}} \right\},
\end{align*}
where the inequality is due to \eqref{eqn:MStage:ConfidenceBound1rename} and \eqref{eqn:MStage:ConfidenceBound0rename}. 

So conditional on event $\cE$, we have $\left\vert \frac{T(1)}{T^*(1)} - 1 \right\vert \to 0$ as $T \to +\infty$.
In addition, $1 - \Pr(\cE) = (M-1) \frac{\kappa(1) + \kappa(0)}{T^{\eps}} \to 0$ as $T \to +\infty$.
Combining these two, we have $\frac{T(1)}{T^*(1)} \xrightarrow{p} 1$ as $T \to +\infty$.
Similarly, we have $\frac{T(0)}{T^*(0)} \xrightarrow{p} 1$ as $T \to +\infty$.

\noindent\underline{\textbf{Case ($\bm{M-1}$)}}:
\begin{align*}
\widehat{\rho}_l \leq \frac{T - \frac{1}{2} \beta_{l+1} T^{\frac{l+1}{M}}}{\frac{1}{2} \beta_{l+1} T^{\frac{l+1}{M}}}, \ \forall \ l \leq M-2.
\end{align*}
Due to the condition of Case ($M-1$), we immediately have
\begin{align*}
\frac{\widehat{\sigma}_{M-2}(0)}{\widehat{\sigma}_{M-2}(1) + \widehat{\sigma}_{M-2}(0)} T \geq \frac{1}{\frac{T - \frac{1}{2} \beta_{M-1} T^{\frac{M-1}{M}}}{\frac{1}{2} \beta_{M-1} T^{\frac{M-1}{M}}}+1} T = \frac{1}{2} \beta_{M-1} T^{\frac{M-1}{M}}.
\end{align*}
On the other hand, since
\begin{multline*}
\widehat{\rho}_{M-2} \geq \rho \sqrt{\frac{1-2^{\frac{1}{2}} \beta_{M-2}^{-\frac{1}{2}} T^{-\frac{M-2}{2M}+\frac{\eps}{2}}}{1+2^{\frac{1}{2}} \beta_{M-2}^{-\frac{1}{2}} T^{-\frac{M-2}{2M}+\frac{\eps}{2}}}} \geq \widehat{\rho}_1 \sqrt{\frac{1-2^{\frac{1}{2}} \beta_{1}^{-\frac{1}{2}} T^{-\frac{1}{2M}+\frac{\eps}{2}}}{1+2^{\frac{1}{2}} \beta_{1}^{-\frac{1}{2}} T^{-\frac{1}{2M}+\frac{\eps}{2}}}} \sqrt{\frac{1-2^{\frac{1}{2}} \beta_{M-2}^{-\frac{1}{2}} T^{-\frac{M-2}{2M}+\frac{\eps}{2}}}{1+2^{\frac{1}{2}} \beta_{M-2}^{-\frac{1}{2}} T^{-\frac{M-2}{2M}+\frac{\eps}{2}}}} \\
\geq \sqrt{\frac{1-2^{\frac{1}{2}} \beta_{1}^{-\frac{1}{2}} T^{-\frac{1}{2M}+\frac{\eps}{2}}}{1+2^{\frac{1}{2}} \beta_{1}^{-\frac{1}{2}} T^{-\frac{1}{2M}+\frac{\eps}{2}}}} \sqrt{\frac{1-2^{\frac{1}{2}} \beta_{M-2}^{-\frac{1}{2}} T^{-\frac{M-2}{2M}+\frac{\eps}{2}}}{1+2^{\frac{1}{2}} \beta_{M-2}^{-\frac{1}{2}} T^{-\frac{M-2}{2M}+\frac{\eps}{2}}}} > \frac{1}{4} \geq \frac{\frac{1}{2}\beta_{M-1} T^{\frac{M-1}{M}}}{T - \frac{1}{2}\beta_{M-1} T^{\frac{M-1}{M}}},
\end{multline*}
where the first and second inequalities are due to \eqref{eqn:MStage:ConfidenceBound1rename} and \eqref{eqn:MStage:ConfidenceBound0rename}; the third inequality is due to \eqref{eqn:WLOG:rename}; the fourth inequality is due to Lemma~\ref{lem:AlgebraicTrick3}; the last inequality is due to Lemma~\ref{lem:AlgebraicTrick4}.
Due to the above sequence of inequalities, we have $\frac{1}{\widehat{\rho}_{M-2}} \leq \frac{T - \frac{1}{2}\beta_{M-1} T^{\frac{M-1}{M}}}{\frac{1}{2}\beta_{M-1} T^{\frac{M-1}{M}}}$, which leads to
\begin{align*}
\frac{\widehat{\sigma}_{M-2}(1)}{\widehat{\sigma}_{M-2}(1) + \widehat{\sigma}_{M-2}(0)} T  \geq \frac{1}{1+\frac{T - \frac{1}{2} \beta_{M-1} T^{\frac{M-1}{M}}}{\frac{1}{2} \beta_{M-1} T^{\frac{M-1}{M}}}} T = \frac{1}{2} \beta_{M-1} T^{\frac{M-1}{M}}.
\end{align*}
So Algorithm~\ref{alg:MStageANA} goes to Line~\ref{mrk:Case3} in the $(M-2)$-th stage experiment. 
Then Algorithm~\ref{alg:MStageANA} goes to Line~\ref{mrk:LastStage} in the last stage.
We further distinguish two cases.

\noindent\textbf{Case ($\bm{M-1}$).1}:
In addition to the conditions in Case $(M-1)$ above, we also have
\begin{align*}
\widehat{\rho}_{M-1} > \frac{T - \frac{1}{2} \beta_{M-1} T^{\frac{M-1}{M}}}{\frac{1}{2} \beta_{M-1} T^{\frac{M-1}{M}}}.
\end{align*}
Similar to the analysis in Case $m$.1, we proceed with the following analysis.
In Case $(M-1)$.1, 
\begin{align*}
\frac{\widehat{\sigma}_{M-1}(0)}{\widehat{\sigma}_{M-1}(1) + \widehat{\sigma}_{M-1}(0)} T < \frac{1}{\frac{T - \frac{1}{2} \beta_{M-1} T^{\frac{M-1}{M}}}{\frac{1}{2} \beta_{M-1} T^{\frac{M-1}{M}}}+1} T = \frac{1}{2} \beta_{M-1} T^{\frac{M-1}{M}}.
\end{align*}
So Algorithm~\ref{alg:MStageANA} goes to Line~\ref{mrk:LastStage:Case1} in the $(M-1)$-th stage experiment, and we have
\begin{align*}
(T(1), T(0)) = \bigg(T - \frac{1}{2} \beta_{M-1} T^{\frac{M-1}{M}}, \frac{1}{2} \beta_{M-1} T^{\frac{M-1}{M}}\bigg).
\end{align*}

Note that, as $T \to +\infty$,
\begin{multline*}
\rho \leq \widehat{\rho}_{M-2} \sqrt{\frac{1+2^{\frac{1}{2}} \beta_{M-2}^{-\frac{1}{2}} T^{-\frac{M-2}{2M}+\frac{\eps}{2}}}{1-2^{\frac{1}{2}} \beta_{M-2}^{-\frac{1}{2}} T^{-\frac{M-2}{2M}+\frac{\eps}{2}}}} 
\leq \frac{T - \frac{1}{2} \beta_{M-1} T^{\frac{M-1}{M}}}{\frac{1}{2} \beta_{M-1} T^{\frac{M-1}{M}}} \sqrt{\frac{1+2^{\frac{1}{2}} \beta_{M-2}^{-\frac{1}{2}} T^{-\frac{M-2}{2M}+\frac{\eps}{2}}}{1-2^{\frac{1}{2}} \beta_{M-2}^{-\frac{1}{2}} T^{-\frac{M-2}{2M}+\frac{\eps}{2}}}} \\
= \frac{T - \frac{1}{2} \beta_{M-1} T^{\frac{M-1}{M}}}{\frac{1}{2} \beta_{M-1} T^{\frac{M-1}{M}}}
\leq \frac{T - \frac{1}{2} \beta_1 T^{\frac{1}{M}}}{\frac{1}{2} \beta_1 T^{\frac{1}{M}}}
\end{multline*}
where the first inequality is due to \eqref{eqn:MStage:ConfidenceBound1rename} and \eqref{eqn:MStage:ConfidenceBound0rename}; the second inequality is due to the condition of Case~(M-1);
the equality holds when $T \to +\infty$;
the last inequality is because $\beta_1 T^{\frac{1}{M}} < \beta_{M-1} T^{\frac{M-1}{M}}$.

Note also that,
\begin{multline*}
\rho \geq \widehat{\rho}_{M-1} \sqrt{\frac{1-2^{\frac{1}{2}} \beta_{M-1}^{-\frac{1}{2}} T^{-\frac{M-1}{2M} + \frac{\eps}{2}}}{1+2^{\frac{1}{2}} \beta_{M-1}^{-\frac{1}{2}} T^{-\frac{M-1}{2M} + \frac{\eps}{2}}}} 
\geq \frac{T - \frac{1}{2} \beta_{M-1} T^{\frac{M-1}{M}}}{\frac{1}{2} \beta_{M-1} T^{\frac{M-1}{M}}} \cdot \sqrt{\frac{1-2^{\frac{1}{2}} \beta_{M-1}^{-\frac{1}{2}} T^{-\frac{M-1}{2M} + \frac{\eps}{2}}}{1+2^{\frac{1}{2}} \beta_{M-1}^{-\frac{1}{2}} T^{-\frac{M-1}{2M} + \frac{\eps}{2}}}} \\
> \frac{1}{2} \frac{T - \frac{1}{2} \beta_{M-1} T^{\frac{M-1}{M}}}{\frac{1}{2} \beta_{M-1} T^{\frac{M-1}{M}}} > 1,
\end{multline*}
where the first inequality is due to \eqref{eqn:MStage:ConfidenceBound1rename} and \eqref{eqn:MStage:ConfidenceBound0rename}; the second inequality is due to the condition of Case~(M-1).1; the third inequality is due to Lemma~\ref{lem:AlgebraicTrick3}; the last inequality is due to Lemma~\ref{lem:AlgebraicTrick4}.

As a result, as $T \to +\infty$, we have $\frac{\frac{1}{2} \beta_1 T^{\frac{1}{M}}}{T - \frac{1}{2} \beta_1 T^{\frac{1}{M}}} < \rho < \frac{T - \frac{1}{2} \beta_1 T^{\frac{1}{M}}}{\frac{1}{2} \beta_1 T^{\frac{1}{M}}}$, so
\begin{align*}
(T^*(1), T^*(0)) = \bigg(\frac{\sigma(1)}{\sigma(1)+\sigma(0)}T, \frac{\sigma(0)}{\sigma(1)+\sigma(0)}T\bigg).
\end{align*}

Conditional on event $\cE$, we have
\begin{multline*}
\left\vert \frac{T(1)}{T^*(1)} - 1 \right\vert = \left\vert \frac{T - \frac{1}{2} \beta_{M-1} T^{\frac{M-1}{M}}}{\frac{\sigma(1)}{\sigma(1) + \sigma(0)} T} - 1 \right\vert = \left\vert \big( \frac{1}{\rho}+1 \big) \frac{T-\frac{1}{2} \beta_{M-1} T^{\frac{M-1}{M}}}{T} - 1 \right\vert \\
\leq \bigg( \frac{\frac{1}{2} \beta_{M-1} T^{\frac{M-1}{M}}}{T-\frac{1}{2} \beta_{M-1} T^{\frac{M-1}{M}}} \sqrt{\frac{1+2^{\frac{1}{2}} \beta_{M-1}^{-\frac{1}{2}} T^{-\frac{M-1}{2M} + \frac{\eps}{2}}}{1-2^{\frac{1}{2}} \beta_{M-1}^{-\frac{1}{2}} T^{-\frac{M-1}{2M} + \frac{\eps}{2}}}} + 1\bigg) \frac{T-\frac{1}{2} \beta_{M-1} T^{\frac{M-1}{M}}}{T} - 1,
\end{multline*}
where the inequality is because $\big( \frac{1}{\rho}+1 \big) \frac{T-\frac{1}{2} \beta_{M-1} T^{\frac{M-1}{M}}}{T} - 1$ is decreasing in $\rho$ and equals $0$ when $\rho = \frac{T - \frac{1}{2}\beta_{M-1} T^{\frac{M-1}{M}}}{\frac{1}{2}\beta_{M-1} T^{\frac{M-1}{M}}}$; and we lower bound $\rho$ by $\frac{T - \frac{1}{2} \beta_{M-1} T^{\frac{M-1}{M}}}{\frac{1}{2} \beta_{M-1} T^{\frac{M-1}{M}}} \cdot \sqrt{\frac{1-2^{\frac{1}{2}} \beta_{M-1}^{-\frac{1}{2}} T^{-\frac{M-1}{2M} + \frac{\eps}{2}}}{1+2^{\frac{1}{2}} \beta_{M-1}^{-\frac{1}{2}} T^{-\frac{M-1}{2M} + \frac{\eps}{2}}}}$.

So conditional on event $\cE$, we have $\left\vert \frac{T(1)}{T^*(1)} - 1 \right\vert \to 0$ as $T \to +\infty$.
In addition, $1 - \Pr(\cE) = (M-1) \frac{\kappa(1) + \kappa(0)}{T^{\eps}} \to 0$ as $T \to +\infty$.
Combining these two, we have $\frac{T(1)}{T^*(1)} \xrightarrow{p} 1$ as $T \to +\infty$.

Conditional on event $\cE$, we have
\begin{multline*}
\left\vert \frac{T(0)}{T^*(0)} - 1 \right\vert = \left\vert \frac{\frac{1}{2} \beta_{M-1} T^{\frac{M-1}{M}}}{\frac{\sigma(0)}{\sigma(1) + \sigma(0)} T} - 1 \right\vert = \left\vert \big( \rho+1 \big) \frac{\frac{1}{2} \beta_{M-1} T^{\frac{M-1}{M}}}{T} - 1 \right\vert \\
\leq 1 - \bigg( 1 + \frac{T - \frac{1}{2} \beta_{M-1} T^{\frac{M-1}{M}}}{\frac{1}{2} \beta_{M-1} T^{\frac{M-1}{M}}} \sqrt{\frac{1-2^{\frac{1}{2}} \beta_{M-1}^{-\frac{1}{2}} T^{-\frac{M-1}{2M} + \frac{\eps}{2}}}{1+2^{\frac{1}{2}} \beta_{M-1}^{-\frac{1}{2}} T^{-\frac{M-1}{2M} + \frac{\eps}{2}}}}\bigg) \frac{\frac{1}{2} \beta_{M-1} T^{\frac{M-1}{M}}}{T} \\
\leq \frac{T - \frac{1}{2} \beta_{M-1} T^{\frac{M-1}{M}}}{T} 2^{\frac{1}{2}} \beta_{M-1}^{-\frac{1}{2}} T^{-\frac{M-1}{2M} + \frac{\eps}{2}}
\end{multline*}
where the first inequality is because $\big( \rho+1 \big) \frac{\frac{1}{2} \beta_{M-1} T^{\frac{M-1}{M}}}{T} - 1$ is increasing in $\rho$ and equals $0$ when $\rho = \frac{T - \frac{1}{2}\beta_{M-1} T^{\frac{M-1}{M}}}{\frac{1}{2}\beta_{M-1} T^{\frac{M-1}{M}}}$,
the second inequality is because for any $\delta \in [0,1), 1- \delta \leq \sqrt{\frac{1-\delta}{1+\delta}}$.

So conditional on event $\cE$, we have $\left\vert \frac{T(0)}{T^*(0)} - 1 \right\vert \to 0$ as $T \to +\infty$.
In addition, $1 - \Pr(\cE) = (M-1) \frac{\kappa(1) + \kappa(0)}{T^{\eps}} \to 0$ as $T \to +\infty$.
Combining these two, we have $\frac{T(0)}{T^*(0)} \xrightarrow{p} 1$ as $T \to +\infty$.

\noindent\textbf{Case ($\bm{M-1}$).2}:
In addition to the conditions in Case $(M-1)$ above, we also have
\begin{align*}
\widehat{\rho}_{M-1} \leq \frac{T - \frac{1}{2} \beta_{M-1} T^{\frac{M-1}{M}}}{\frac{1}{2} \beta_{M-1} T^{\frac{M-1}{M}}}.
\end{align*}
Due to the condition of Case ($M-1$).2, we immediately have
\begin{align*}
\frac{\widehat{\sigma}_{M-1}(0)}{\widehat{\sigma}_{M-1}(1) + \widehat{\sigma}_{M-1}(0)} T \geq \frac{1}{\frac{T - \frac{1}{2} \beta_{M-1} T^{\frac{M-1}{M}}}{\frac{1}{2} \beta_{M-1} T^{\frac{M-1}{M}}}+1} T = \frac{1}{2} \beta_{M-1} T^{\frac{M-1}{M}}.
\end{align*}
On the other hand, since
\begin{multline*}
\widehat{\rho}_{M-1} \geq \rho \sqrt{\frac{1-2^{\frac{1}{2}} \beta_{M-1}^{-\frac{1}{2}} T^{-\frac{M-1}{2M}+\frac{\eps}{2}}}{1+2^{\frac{1}{2}} \beta_{M-1}^{-\frac{1}{2}} T^{-\frac{M-1}{2M}+\frac{\eps}{2}}}} \geq \widehat{\rho}_1 \sqrt{\frac{1-2^{\frac{1}{2}} \beta_{1}^{-\frac{1}{2}} T^{-\frac{1}{2M}+\frac{\eps}{2}}}{1+2^{\frac{1}{2}} \beta_{1}^{-\frac{1}{2}} T^{-\frac{1}{2M}+\frac{\eps}{2}}}} \sqrt{\frac{1-2^{\frac{1}{2}} \beta_{M-1}^{-\frac{1}{2}} T^{-\frac{M-1}{2M}+\frac{\eps}{2}}}{1+2^{\frac{1}{2}} \beta_{M-1}^{-\frac{1}{2}} T^{-\frac{M-1}{2M}+\frac{\eps}{2}}}} \\
\geq \sqrt{\frac{1-2^{\frac{1}{2}} \beta_{1}^{-\frac{1}{2}} T^{-\frac{1}{2M}+\frac{\eps}{2}}}{1+2^{\frac{1}{2}} \beta_{1}^{-\frac{1}{2}} T^{-\frac{1}{2M}+\frac{\eps}{2}}}} \sqrt{\frac{1-2^{\frac{1}{2}} \beta_{M-1}^{-\frac{1}{2}} T^{-\frac{M-1}{2M}+\frac{\eps}{2}}}{1+2^{\frac{1}{2}} \beta_{M-1}^{-\frac{1}{2}} T^{-\frac{M-1}{2M}+\frac{\eps}{2}}}} > \frac{1}{4} \geq \frac{\frac{1}{2}\beta_{M-1} T^{\frac{M-1}{M}}}{T - \frac{1}{2}\beta_{M-1} T^{\frac{M-1}{M}}},
\end{multline*}
where the first and second inequalities are due to \eqref{eqn:MStage:ConfidenceBound1rename} and \eqref{eqn:MStage:ConfidenceBound0rename}; the third inequality is due to \eqref{eqn:WLOG:rename}; the fourth inequality is due to Lemma~\ref{lem:AlgebraicTrick3}; the last inequality is due to Lemma~\ref{lem:AlgebraicTrick4}.
Due to the above sequence of inequalities, we have $\frac{1}{\widehat{\rho}_{M-1}} \leq \frac{T - \frac{1}{2}\beta_{M-1} T^{\frac{M-1}{M}}}{\frac{1}{2}\beta_{M-1} T^{\frac{M-1}{M}}}$, which leads to
\begin{align*}
\frac{\widehat{\sigma}_{M-1}(1)}{\widehat{\sigma}_{M-1}(1) + \widehat{\sigma}_{M-1}(0)} T  \geq \frac{1}{1+\frac{T - \frac{1}{2} \beta_{M-1} T^{\frac{M-1}{M}}}{\frac{1}{2} \beta_{M-1} T^{\frac{M-1}{M}}}} T = \frac{1}{2} \beta_{M-1} T^{\frac{M-1}{M}}.
\end{align*}
So Algorithm~\ref{alg:MStageANA} goes to Line~\ref{mrk:LastStage:Case2} in the $(M-1)$-th stage experiment. 
Then we have
\begin{align*}
(T(1), T(0)) = \bigg(\frac{\widehat{\sigma}_{M-1}(1)}{\widehat{\sigma}_{M-1}(1) + \widehat{\sigma}_{M-1}(0)} T, \frac{\widehat{\sigma}_{M-1}(0)}{\widehat{\sigma}_{M-1}(1) + \widehat{\sigma}_{M-1}(0)} T\bigg).
\end{align*}

Note that, as $T \to +\infty$,
\begin{align*}
\rho \leq \widehat{\rho}_{M-1} \sqrt{\frac{1+2^{\frac{1}{2}} \beta_{M-1}^{-\frac{1}{2}} T^{-\frac{M-1}{2M}+\frac{\eps}{2}}}{1-2^{\frac{1}{2}} \beta_{M-1}^{-\frac{1}{2}} T^{-\frac{M-1}{2M}+\frac{\eps}{2}}}} 
\leq \frac{T - \frac{1}{2} \beta_{M-1} T^{\frac{M-1}{M}}}{\frac{1}{2} \beta_{M-1} T^{\frac{M-1}{M}}} \sqrt{\frac{1+2^{\frac{1}{2}} \beta_{M-1}^{-\frac{1}{2}} T^{-\frac{M-1}{2M}+\frac{\eps}{2}}}{1-2^{\frac{1}{2}} \beta_{M-1}^{-\frac{1}{2}} T^{-\frac{M-1}{2M}+\frac{\eps}{2}}}} 
\leq \frac{T - \frac{1}{2} \beta_1 T^{\frac{1}{M}}}{\frac{1}{2} \beta_1 T^{\frac{1}{M}}}
\end{align*}
where the first inequality is due to \eqref{eqn:MStage:ConfidenceBound1rename} and \eqref{eqn:MStage:ConfidenceBound0rename}; the second inequality is due to the condition of Case~2.2;
the last inequality is because $\beta_1 T^{\frac{1}{M}} < \beta_m T^{\frac{m}{M}}$ and holds as $T \to +\infty$. 

Note also that,
\begin{align*}
\rho \geq \widehat{\rho}_1 \sqrt{\frac{1-2^{\frac{1}{2}} \beta_1^{-\frac{1}{2}} T^{-\frac{1}{2M} + \frac{\eps}{2}}}{1+2^{\frac{1}{2}} \beta_1^{-\frac{1}{2}} T^{-\frac{1}{2M} + \frac{\eps}{2}}}} \geq \sqrt{\frac{1-2^{\frac{1}{2}} \beta_1^{-\frac{1}{2}} T^{-\frac{1}{2M} + \frac{\eps}{2}}}{1+2^{\frac{1}{2}} \beta_1^{-\frac{1}{2}} T^{-\frac{1}{2M} + \frac{\eps}{2}}}} \geq \frac{1}{2} \geq \frac{\frac{1}{2} \beta_1 T^{\frac{1}{M}}}{T - \frac{1}{2} \beta_1 T^{\frac{1}{M}}}
\end{align*}
where the first inequality is due to \eqref{eqn:MStage:ConfidenceBound1rename} and \eqref{eqn:MStage:ConfidenceBound0rename}; the second inequality is due to \eqref{eqn:WLOG:rename}; the third inequality is due to Lemma~\ref{lem:AlgebraicTrick3}; the last inequality is due to Lemma~\ref{lem:AlgebraicTrick4}.

As a result, as $T \to +\infty$,
\begin{align*}
(T^*(1), T^*(0)) = \bigg(\frac{\sigma(1)}{\sigma(1)+\sigma(0)}T, \frac{\sigma(0)}{\sigma(1)+\sigma(0)}T\bigg).
\end{align*}

Conditional on event $\cE$, we have
\begin{align*}
\left\vert \frac{T(1)}{T^*(1)} - 1 \right\vert = \left\vert \frac{\frac{\widehat{\sigma}_{M-1}(1)}{\widehat{\sigma}_{M-1}(1) + \widehat{\sigma}_{M-1}(0)} T}{\frac{\sigma(1)}{\sigma(1) + \sigma(0)} T} - 1 \right\vert 
\leq \max \left\{ \frac{1 + 2^{\frac{1}{2}} \beta_{M-1}^{-\frac{1}{2}} T^{-\frac{M-1}{2M} + \frac{\eps}{2}}}{1 - 2^{\frac{1}{2}} \beta_{M-1}^{-\frac{1}{2}} T^{-\frac{M-1}{2M} + \frac{\eps}{2}}} - 1, 1 - \frac{1 - 2^{\frac{1}{2}} \beta_{M-1}^{-\frac{1}{2}} T^{-\frac{M-1}{2M} + \frac{\eps}{2}}}{1 + 2^{\frac{1}{2}} \beta_{M-1}^{-\frac{1}{2}} T^{-\frac{M-1}{2M} + \frac{\eps}{2}}} \right\},
\end{align*}
where the inequality is due to \eqref{eqn:MStage:ConfidenceBound1rename} and \eqref{eqn:MStage:ConfidenceBound0rename}. 

So conditional on event $\cE$, we have $\left\vert \frac{T(1)}{T^*(1)} - 1 \right\vert \to 0$ as $T \to +\infty$.
In addition, $1 - \Pr(\cE) = (M-1) \frac{\kappa(1) + \kappa(0)}{T^{\eps}} \to 0$ as $T \to +\infty$.
Combining these two, we have $\frac{T(1)}{T^*(1)} \xrightarrow{p} 1$ as $T \to +\infty$.
Similarly, we have $\frac{T(0)}{T^*(0)} \xrightarrow{p} 1$ as $T \to +\infty$.

To conclude, in all cases, we have shown that $\frac{T(1)}{T^*(1)} \xrightarrow{p} 1$ and $\frac{T(0)}{T^*(0)} \xrightarrow{p} 1$ as $T \to +\infty$.
\hfill \halmos
\endproof

\subsubsection{Martingale central limit theorem.}
\label{sec:martingaleCLT}

\proof{Proof of Theorem~\ref{thm:inference}.}
To use the martingale central limit theorem to prove Theorem~\ref{thm:inference}, we adopt the following ``tape'' view of each run of the adaptive Neyman allocation algorithm.
For any fixed $T$, suppose there are two length-$T$ arrays for the treated and the control, respectively, with each value being an independent and identically distributed copy of the representative random variables $Y(1)$ and $Y(0)$, respectively.
When Algorithm~\ref{alg:2StageANA} or Algorithm~\ref{alg:MStageANA} assigns treatment and control to unit $t$, it reads the corresponding $Y_t(1)$ or $Y_t(0)$ from one of the two arrays.
See Table~\ref{tbl:TapeView} for an illustration.

\begin{table}[htb]
\centering
\TABLE{Illustration of the tape view
\label{tbl:TapeView}}
{\begin{tabular}{|p{15mm}|C{14mm}|C{14mm}|C{14mm}|C{14mm}|C{14mm}|C{14mm}|}
\cline{1-7}
Treated & \cellcolor{gray!20} $Z_1(1)$ & $Z_2(1)$ & $Z_3(1)$ & \cellcolor{gray!20} $Z_4(1)$ & $\ldots$ & \cellcolor{gray!20} $Z_T(1)$ \\ \cline{1-7}
Control & $Z_1(0)$ & \cellcolor{gray!20} $Z_2(0)$ & \cellcolor{gray!20} $Z_3(0)$ & $Z_4(0)$ & $\ldots$ & $Z_T(0)$ \\ \cline{1-7}
\end{tabular}
}
{\textit{Note}: In this illustration, the treated array contains random values $Y_1(1)$, $Y_2(1)$, ..., $Y_T(1)$ and the control array contains random values $Y_1(0)$, $Y_2(0)$, ..., $Y_T(0)$. The gray background stands for a trajectory of the random variables that we read.}
\end{table}

We can also take a ``sequential'' view of the completely randomized design.
For a completely randomized experiment involving $T_m = T_m(1)+T_m(0)$ units, with $T_m(1)$ and $T_m(0)$ units in the treatment and control groups, respectively, we conduct the sequential experiment as follows.
The first unit is randomly assigned into the treatment group with probability $\frac{T_m(1)}{T_m}$ and control group with probability $\frac{T_m(0)}{T_m}$.
When there are already $N(1) \leq T_m(1)$ and $N(0) \leq T_m(0)$ units in the treated and control groups, the next unit is randomly assigned into the treatment group with probability $\frac{T_m(1)-N(1)}{T_m-N(1)-N(0)}$ and control group with probability $\frac{T_m(0)-N(0)}{T_m-N(1)-N(0)}$.

We now define $\sF_t = \sigma(W_1, Y_1(W_1), ..., W_t, Y_t(W_t))$ to be a filtration defined on the first $t$ treatment assignments and observed outcomes.
Denote the following random variables
\begin{align*}
X_t(1) & = \frac{1}{\sigma(1) \sqrt{T^*(1)}} \Big( Y_t(1) - \bE[Y(1)] \Big) \bI\{W_t=1\}, \\
X_t(0) & = \frac{1}{\sigma(0) \sqrt{T^*(0)}} \Big( Y_t(0) - \bE[Y(0)] \Big) \bI\{W_t=0\}.
\end{align*}
Note that $X_t(1)$ and $X_t(0)$ are not the sample means.
On the denominator, $T^*(1)$ and $T^*(0)$ are deterministic quantities as defined in \eqref{eqn:defn:Stability:2} when $M=2$ or \eqref{eqn:defn:Stability:3} when $M \geq 3$.

We first show that $\{X_t(1)\}_{t=1,2,...}$ (and $\{X_t(0)\}_{t=1,2,...}$) is a martingale difference sequence.
To see this, note that
\begin{align*}
\big[X_t(1) \big\vert \sF_{t-1}\big] & = \frac{1}{\sigma(1) \sqrt{T^*(1)}} \bE\Big[ Y_t(1) - \bE[Y(1)] \Big\vert \sF_{t-1} \Big] \bE\Big[\bI\{W_t=1\} \Big\vert \sF_{t-1} \Big] \\
& = \frac{1}{\sigma(1) \sqrt{T^*(1)}} \cdot 0 \cdot \bE\Big[\bI\{W_t=1\} \Big\vert \sF_{t-1} \Big] \\
& = 0,
\end{align*}
where the second equality holds because $Y_t(1)$ is independent of the filtration $\sF_{t-1}$.

Then, for any constants $\alpha(1), \alpha(0) > 0$, denote $X_t = \alpha(1) X_t(1) + \alpha(0) X_t(0)$.
We see that $\bE\big[X_t \big\vert \sF_{t-1}\big] = 0$.
So $\{X_t\}_{t=1,2,...}$ is a martingale difference sequence.

Now we check the first condition in Lemma~\ref{lem:martingaleCLT}.
\begin{align*}
& \sum_{t=1}^T \bE\Big[ X_t^2 \Big\vert \sF_{t-1} \Big] \\
= & \sum_{t=1}^T \bigg\{ \frac{\alpha^2(1)}{\sigma^2(1) T^*(1)} \bE\Big[ \big(Y_t(1) - \bE[Y(1)]\big)^2 \bI\{W_t=1\}^2 \Big\vert \sF_{t-1} \Big] \\
& \quad + \frac{2\alpha(1)\alpha(0)}{\sigma(1)\sigma(0) \sqrt{T^*(1)T^*(0)}} \bE\Big[ \big(Y_t(1) - \bE[Y(1)]\big) \big(Y_t(0) - \bE[Y(0)]\big) \bI\{W_t=1\} \bI\{W_t=0\} \Big\vert \sF_{t-1} \Big] \\
& \qquad + \frac{\alpha^2(0)}{\sigma^2(0) T^*(0)} \bE\Big[ \big(Y_t(0) - \bE[Y(0)]\big)^2 \bI\{W_t=0\}^2 \Big\vert \sF_{t-1} \Big] \bigg\} \\
= & \sum_{t=1}^T \bigg\{ \frac{\alpha^2(1)}{\sigma^2(1) T^*(1)} \bE\Big[ \big(Y_t(1) - \bE[Y(1)]\big)^2 \bI\{W_t=1\} \Big\vert \sF_{t-1} \Big] \\
& \qquad + \frac{\alpha^2(0)}{\sigma^2(0) T^*(0)} \bE\Big[ \big(Y_t(0) - \bE[Y(0)]\big)^2 \bI\{W_t=0\} \Big\vert \sF_{t-1} \Big] \bigg\} \\
= & \sum_{t=1}^T \bigg\{ \frac{\alpha^2(1)}{\sigma^2(1) T^*(1)} \bE\Big[ \big(Y_t(1) - \bE[Y(1)]\big)^2 \Big\vert \sF_{t-1} \Big] \bE\Big[ \bI\{W_t=1\} \Big\vert \sF_{t-1} \Big] \\
& \qquad + \frac{\alpha^2(0)}{\sigma^2(0) T^*(0)} \bE\Big[ \big(Y_t(0) - \bE[Y(0)]\big)^2 \Big\vert \sF_{t-1} \Big] \bE\Big[\bI\{W_t=0\} \Big\vert \sF_{t-1} \Big] \bigg\} \\
= & \frac{\alpha^2(1) T(1)}{T^*(1)} + \frac{\alpha^2(0) T(0)}{T^*(0)} 
\end{align*}
where the second equality is because $\bI\{W_t=1\} \bI\{W_t=0\} = 0$;
the last equality is because $Y_t(1)$ and $Y_t(0)$ are independent of the filtration $\sF_{t-1}$ so $\bE\big[ \big(Y_t(1) - \bE[Y(1)]\big)^2 \big\vert \sF_{t-1} \big] = \sigma^2(1)$ and $\bE\big[ \big(Y_t(0) - \bE[Y(0)]\big)^2 \big\vert \sF_{t-1} \big] = \sigma^2(0)$.

Using Lemma~\ref{lem:Stability}, as $T \to +\infty$, we have
\begin{align*}
\sum_{t=1}^T \bE\Big[ X_t^2 \Big\vert \sF_{t-1} \Big] \xrightarrow{p} \alpha^2(1) + \alpha^2(0).
\end{align*}
So this satisfies the first condition in Lemma~\ref{lem:martingaleCLT}.

Now we check the second condition in Lemma~\ref{lem:martingaleCLT}.
Denote $\alpha = \sqrt{\alpha^2(1) + \alpha^2(0)}$.
Note that, for any $\eps > 0$ and any $t \in [T]$,
\begin{align}
\bE\Big[ X_t^2 \bI\{ \vert X_t \vert \geq \eps \alpha \} \Big\vert \sF_{t-1} \Big]
\leq \bE\Big[ X_t^2 \cdot \frac{\vert X_t \vert^2}{\eps^2 \alpha^2} \Big\vert \sF_{t-1} \Big]
= \frac{1}{\eps^2 \alpha^2} \bE\Big[ X_t^4 \Big\vert \sF_{t-1} \Big], \label{eqn:LindebergtoLyapunov}
\end{align}
where the first inequality is because either $\vert X_t \vert \geq \eps \alpha$, in which case $\bI\{ \vert X_t \vert \geq \eps \alpha \} = 1 \leq \frac{\vert X_t \vert^2}{\eps^2 \alpha^2}$, or $\vert X_t \vert < \eps \alpha$, in which case $\bI\{ \vert X_t \vert \geq \eps \alpha \} = 0 \leq \frac{\vert X_t \vert^2}{\eps^2 \alpha^2}$.
Note that,
\begin{align*}
\sum_{t=1}^T \bE\Big[ X_t^4 \Big\vert \sF_{t-1} \Big]
= & \sum_{t=1}^T \bigg\{ \frac{\alpha^4(1)}{\sigma^4(1) \big(T^*(1)\big)^2} \bE\Big[ \big(Y_t(1) - \bE[Y(1)]\big)^4 \bI\{W_t=1\}^4 \Big\vert \sF_{t-1} \Big] \\
& \qquad + \frac{\alpha^4(0)}{\sigma^4(0) \big(T^*(0)\big)^2} \bE\Big[ \big(Y_t(0) - \bE[Y(0)]\big)^4 \bI\{W_t=0\}^4 \Big\vert \sF_{t-1} \Big] \bigg\} \\
= & \sum_{t=1}^T \bigg\{ \frac{\alpha^4(1)}{\sigma^4(1) \big(T^*(1)\big)^2} \bE\Big[ \big(Y_t(1) - \bE[Y(1)]\big)^4 \Big\vert \sF_{t-1} \Big] \Big[ \bI\{W_t=1\} \Big\vert \sF_{t-1} \Big] \\
& \qquad + \frac{\alpha^4(0)}{\sigma^4(0) \big(T^*(0)\big)^2} \bE\Big[ \big(Y_t(0) - \bE[Y(0)]\big)^4 \Big\vert \sF_{t-1} \Big] \Big[ \bI\{W_t=0\} \Big\vert \sF_{t-1} \Big] \bigg\} \\
= & \ \frac{\alpha^4(1) T(1)}{\big(T^*(1)\big)^2} + \frac{\alpha^4(0) T(0)}{\big(T^*(0)\big)^2},
\end{align*}
where the first equality is because all the cross terms containing $\bI\{W_t=1\} \bI\{W_t=0\}$ are equal to $0$;
the last equality is because $Y_t(1)$ and $Y_t(0)$ are independent of the filtration $\sF_{t-1}$ so $\bE\big[ \big(Y_t(1) - \bE[Y(1)]\big)^4 \big\vert \sF_{t-1} \big] = \sigma^4(1)$ and $\bE\big[ \big(Y_t(0) - \bE[Y(0)]\big)^4 \big\vert \sF_{t-1} \big] = \sigma^4(0)$.

Using Lemma~\ref{lem:Stability}, as $T \to +\infty$, we have
\begin{align*}
\sum_{t=1}^T \bE\Big[ X_t^4 \Big\vert \sF_{t-1} \Big] \xrightarrow{p} \frac{\alpha^4(1)}{T^*(1)} + \frac{\alpha^4(0)}{T^*(0)} \to 0.
\end{align*}
Due to \eqref{eqn:LindebergtoLyapunov}, as $T \to +\infty$, we have
\begin{align*}
\bE\Big[ X_t^2 \bI\{ \vert X_t \vert \geq \eps \alpha \} \Big\vert \sF_{t-1} \Big] \xrightarrow{p} 0.
\end{align*}
So this satisfies the second condition in Lemma~\ref{lem:martingaleCLT}.

Following Lemma~\ref{lem:martingaleCLT}, we have that for any $\alpha(1), \alpha(0)$,
\begin{align}
\lim_{T \to +\infty} \sum_{t=1}^T X_t \xrightarrow{d} \cN\big(0, \alpha^2(1) + \alpha^2(0) \big). \label{eqn:NormalDist}
\end{align}

Now we would like to apply the Cramer-Wold Theorem to show a joint normal distribution.
Let there be a two-dimensional multivariate normal distribution denoted as
\begin{align*}
(X(1), X(0)) \sim \cN\big(0, \mathbb{I}_2\big),
\end{align*}
where $\mathbb{I}_2 = \begin{bmatrix}
1 & 0 \\
0 & 1
\end{bmatrix}$
stands for the $2 \times 2$ identity matrix.
For any $\alpha(1), \alpha(0)$, we know that
\begin{align*}
\alpha(1) X(1) + \alpha(0) X(0) \sim \cN\big(0, \alpha^2(1) + \alpha^2(0)\big)
\end{align*}
follows a normal distribution, which is the same distribution as \eqref{eqn:NormalDist}.
Following the Cramer-Wold Theorem, 
\begin{align*}
\lim_{T \to +\infty} \begin{pmatrix}
\sum_{t=1}^T X_t(1) \\
\sum_{t=1}^T X_t(0)
\end{pmatrix}
\xrightarrow{d} \cN\big(0, \mathbb{I}_2\big).
\end{align*}

Finally, note that from Lemma~\ref{lem:Stability} we have 
\begin{align*}
\lim_{T \to +\infty} \frac{T(1)}{T^*(1)} \xrightarrow{p} 1, \lim_{T \to +\infty} \frac{T(0)}{T^*(0)} \xrightarrow{p} 1.
\end{align*}
Following the Slutsky Theorem, we have
\begin{align*}
\lim_{T \to +\infty} \begin{pmatrix}
\frac{1}{\sqrt{T(1)}} \sum_{t=1}^T (Y_t - \bE[Y(1)]) \bI\{W_t=1\}\\
\frac{1}{\sqrt{T(0)}} \sum_{t=1}^T (Y_t - \bE[Y(0)]) \bI\{W_t=0\}
\end{pmatrix} 
\xrightarrow{d} \cN\left( 
\begin{pmatrix}
0 \\
0
\end{pmatrix},
\begin{bmatrix}
\sigma^2(1) & 0 \\
0 & \sigma^2(0)
\end{bmatrix}
\right).
\end{align*}
\hfill \halmos
\endproof

\subsection{Proof of Proposition~\ref{prop:SampleVarianceConsistent}}
\label{sec:proof:prop:SampleVarianceConsistent}

\proof{Proof of Proposition~\ref{prop:SampleVarianceConsistent}.}
For any fixed $T$, suppose there are two length-$T$ arrays for the treated and the control, respectively, with each value being an independent and identically distributed copy of the representative random variables $Y(1)$ and $Y(0)$, respectively.
When Algorithm~\ref{alg:2StageANA} or Algorithm~\ref{alg:MStageANA} assigns treatment or control to unit $t$, we read the \textit{next} value from the treated or control array.
Note that, we read the next value, instead of the $t$-th value, from the corresponding array.
In other words, even though Algorithm~\ref{alg:2StageANA} or Algorithm~\ref{alg:MStageANA} adaptively determines the number of treated and control units, it is always the first few values of of the two arrays that are read.
See Table~\ref{tbl:FirstFewValues} for an illustration.

\begin{table}[htb]
\centering
\TABLE{Illustration of the reading first few values from an array
\label{tbl:FirstFewValues}}
{\begin{tabular}{|p{15mm}|C{14mm}|C{14mm}|C{14mm}|C{14mm}|C{14mm}|C{14mm}|C{14mm}|C{14mm}|}
\multicolumn{1}{c}{} & \multicolumn{4}{l}{\begin{minipage}[c][8mm][c]{64mm}\begin{center} \small estimates $\widehat{\sigma}^2(1)$ \\ $\overbrace{\hspace*{66mm}}$\end{center} \end{minipage}} & \multicolumn{4}{c}{} \\ \cline{1-9}
Treated & $Z_1(1)$ & $Z_2(1)$ & $\ldots$ & $Z_s(1)$ & $\ldots$ & $Z_{s'}(1)$ & $\ldots$ & $Z_T(1)$ \\ \cline{1-9}
Control & $Z_1(0)$ & $Z_2(0)$ & $\ldots$ & $Z_s(0)$ & $\ldots$ & $Z_{s'}(0)$ & $\ldots$ & $Z_T(0)$ \\ \cline{1-9}
\multicolumn{1}{c}{} & \multicolumn{6}{l}{\begin{minipage}[c][2mm][c]{96mm}\begin{center} \small $\underbrace{\hspace*{100mm}}$ \\  estimates $\widehat{\sigma}^2(0)$ \end{center} \end{minipage}} & \multicolumn{2}{c}{\vspace{5mm}}
\end{tabular}
}
{\textit{Note}: In this illustration, the treated array contains random values $Z_1(1)$, $Z_2(1)$, ..., $Z_T(1)$ and the control array contains random values $Z_1(0)$, $Z_2(0)$, ..., $Z_T(0)$. In this illustration, we use the first $s = T(1)$ values in the treated array to compute the sample variance estimator $\widehat{\sigma}^2(1)$, and the first $s' = T(0)$ values in the control array to compute the sample variance estimator $\widehat{\sigma}^2(0)$.}
\end{table}

Note that the sample variance estimators can be expressed as
\begin{align*}
\widehat{\sigma}^2(1) = & \ \frac{T(1)}{T(1)-1} \Bigg( \frac{1}{T(1)} \sum_{t=1}^T Y_t^2 \bI\{W_t=1\} - \bigg( \frac{1}{T(1)} \sum_{t=1}^T Y_t \bI\{W_t=1\} \bigg)^2 \Bigg), \\
\widehat{\sigma}^2(0) = & \ \frac{T(0)}{T(0)-1} \Bigg( \frac{1}{T(0)} \sum_{t=1}^T Y_t^2 \bI\{W_t=0\} - \bigg( \frac{1}{T(0)} \sum_{t=1}^T Y_t \bI\{W_t=0\} \bigg)^2 \Bigg).
\end{align*}

Now we focus on $\widehat{\sigma}^2(1)$.
Define
\begin{align*}
Z_{1,T} = & \ \frac{1}{T(1)} \sum_{t=1}^T Y_t \bI\{W_t=1\}, & Z_{2,T} = & \ \frac{1}{T(1)} \sum_{t=1}^T Y_t^2 \bI\{W_t=1\}.
\end{align*}
Because $\bE\big[Y(1)\big] < +\infty$ and $\bE\big[Y^2(1)\big] < +\infty$, due to the strong law of large numbers, as $T \to +\infty$,
\begin{align}
\frac{1}{T} \sum_{t=1}^T Y_t(1) \xrightarrow{a.s.} \bE\big[Y(1)\big], && \frac{1}{T} \sum_{t=1}^T Y_t^2(1) \xrightarrow{a.s.} \bE\big[Y^2(1)\big]. \label{eqn:YConverge}
\end{align}
Then, following Lemma~\ref{lem:Stability}, as $T \to +\infty$,
\begin{align}
T(1) \xrightarrow{p} +\infty. \label{eqn:NConverge}
\end{align}

Since $Z_{1,T}$ (and $Z_{2,T}$) can be interpreted as taking the average of the first $T(1)$ random variables (and their squares), following Lemma~\ref{lem:LLNRandomIndices} and combining \eqref{eqn:YConverge} and \eqref{eqn:NConverge}, we have that, as $T \to +\infty$,
\begin{align*}
Z_{1,T} \xrightarrow{p} \bE\big[Y(1)\big], && Z_{2,T} \xrightarrow{p} \bE\big[Y^2(1)\big].
\end{align*}
So we have, as $T \to +\infty$,
\begin{align*}
\widehat{\sigma}^2(1) \xrightarrow{p} \sigma^2(1).
\end{align*}
Similarly, the other part $\widehat{\sigma}^2(0) \xrightarrow{p} \sigma^2(0)$ follows.
\hfill \halmos
\endproof

\subsection{Proof of Corollary~\ref{coro:2StageANA}}
\label{sec:proof:coro:2StageANA}

\subsubsection{Establishing a high probability bound.}
We first establish a high probability bound for the proof of Corollary~\ref{coro:2StageANA}.

\begin{lemma}
\label{lem:2Stage:NewHighProbBound}
Let $T \geq 320^\frac{5}{4} C^5$.
Let $\beta = 4C^2 (\log{T})^{\frac{1}{2}}$ in Algorithm~\ref{alg:2StageANA}.
Let $(T(1), T(0))$ be the number of total treated and control units from Algorithm~\ref{alg:2StageANA}, respectively. Under Assumption~\ref{asp:bounded}, there exists an event that happens with probability at least $1 - \frac{4}{T^2}$, conditional on which
\begin{align*}
\sup_{\cF \in \sP^{[C]}} \ \frac{\bE[V(T(1), T(0))]}{V(T^*(1), T^*(0))} \leq 1 + 4 C^2 T^{-\frac{1}{2}} (\log{T})^{\frac{1}{2}}.
\end{align*}
\end{lemma}

\proof{Proof of Lemma~\ref{lem:2Stage:NewHighProbBound}.}
Without loss of generality, we assume $\sigma(1) \geq \sigma(0)$ throughout the proof.
We consider the following two events.
\begin{align*}
\cE_1(1) = & \ \bigg\{ \left| \widehat{\sigma}^2_1(1) - \sigma^2(1) \right| < 2 C T^{-\frac{1}{4}} (\log{T})^{\frac{1}{4}} \sigma^2(1)\bigg\}, \\
\cE_1(0) = & \ \bigg\{ \left| \widehat{\sigma}^2_1(0) - \sigma^2(0) \right| < 2 C T^{-\frac{1}{4}} (\log{T})^{\frac{1}{4}} \sigma^2(0)\bigg\}.
\end{align*}
Denote $\cE = \cE_1(1) \cap \cE_1(0)$. Then $\Pr(\cE) = \Pr(\cE_1(1) \cap \cE_1(0)) \geq 1 - \Pr(\overline{\cE}_1(1)) - \Pr(\overline{\cE}_1(0))$.
We further have
\begin{align*}
\Pr(\cE) = & \ 1 - \Pr\left( \vert \widehat{\sigma}^2_1(1) - \sigma^2(1) \vert \geq 2 C T^{-\frac{1}{4}} (\log{T})^{\frac{1}{4}} \sigma^2(1) \right) - \Pr\left( \vert \widehat{\sigma}^2_1(0) - \sigma^2(0) \vert \geq 2 C T^{-\frac{1}{4}} (\log{T})^{\frac{1}{4}} \sigma^2(0) \right) \\
\geq & \ 1 - 2 \exp\left\{-\frac{4 C^2 T^{-\frac{1}{2}} (\log{T})^{\frac{1}{2}} \sigma^4(1) \cdot T_1(1)}{8 C^4 \sigma^4(1)}\right\} - 2 \exp\left\{-\frac{4 C^2 T^{-\frac{1}{2}} (\log{T})^{\frac{1}{2}} \sigma^4(0) \cdot T_1(0)}{8 C^4 \sigma^4(0)}\right\} \\
= & \ 1 - 4 T^{-2},
\end{align*}
where the inequality is due to Lemma~\ref{lem:ExponentialTail}; 
and the last equality is using $T_1(1) = T_1(0) = 4 C^2 T^{\frac{1}{2}} (\log{T})^{\frac{1}{2}}$

Conditional on the event $\cE$, we have
\begin{subequations}
\begin{align}
\sigma^2(1) \left( 1 - 2 C T^{-\frac{1}{4}} (\log{T})^{\frac{1}{4}} \right) \ \leq \ \widehat{\sigma}^2_1(1) \ \leq \ \sigma^2(1) \left( 1 + 2 C T^{-\frac{1}{4}} (\log{T})^{\frac{1}{4}} \right), \label{eqn:2Stage:NewHighProbBound:ConfidenceBound1} \\
\sigma^2(0) \left( 1 - 2 C T^{-\frac{1}{4}} (\log{T})^{\frac{1}{4}} \right) \ \leq \ \widehat{\sigma}^2_1(0) \ \leq \ \sigma^2(0) \left( 1 + 2 C T^{-\frac{1}{4}} (\log{T})^{\frac{1}{4}} \right). \label{eqn:2Stage:NewHighProbBound:ConfidenceBound0} 
\end{align}
\end{subequations}
Due to \eqref{eqn:2Stage:NewHighProbBound:ConfidenceBound1} and \eqref{eqn:2Stage:NewHighProbBound:ConfidenceBound0}, and given that $\sigma(1), \sigma(0) > 0$, we have $\widehat{\sigma}^2_1(1), \widehat{\sigma}^2_1(0) > 0$.
Denote $\rho = \frac{\sigma(1)}{\sigma(0)}$ and $\widehat{\rho} = \frac{\widehat{\sigma}_1(1)}{\widehat{\sigma}_1(0)}$.

Now we distinguish two cases, and discuss these two cases separately.
\begin{enumerate}
\item \textbf{Case 1}: 
\begin{align*}
\frac{\frac{1}{2}\beta T^{\frac{1}{2}}}{T - \frac{1}{2}\beta T^{\frac{1}{2}}} \leq \rho = \frac{\sigma(1)}{\sigma(0)} \leq \frac{T - \frac{1}{2}\beta T^{\frac{1}{2}}}{\frac{1}{2}\beta T^{\frac{1}{2}}}.
\end{align*}
\item \textbf{Case 2}:
\begin{align*}
\rho = \frac{\sigma(1)}{\sigma(0)} > \frac{T - \frac{1}{2}\beta T^{\frac{1}{2}}}{\frac{1}{2}\beta T^{\frac{1}{2}}}.
\end{align*}
\end{enumerate}
Note that, for case 2, we do not discuss $\rho = \frac{\sigma(1)}{\sigma(0)} < \frac{\frac{1}{2}\beta T^{\frac{1}{2}}}{T - \frac{1}{2}\beta T^{\frac{1}{2}}}$, because we assume that $\sigma(1) \geq \sigma(0)$.
For each of the above two cases, we further discuss two sub-cases.
The remaining of the proof is structured as enumerating all four cases.
After enumerating all four sub-cases we finish the proof.

\noindent\textbf{Case 1.1}:
\begin{align*}
\frac{\frac{1}{2}\beta T^{\frac{1}{2}}}{T - \frac{1}{2}\beta T^{\frac{1}{2}}} \leq \rho \leq \frac{T - \frac{1}{2}\beta T^{\frac{1}{2}}}{\frac{1}{2}\beta T^{\frac{1}{2}}}, && \text{and} && \frac{\frac{1}{2}\beta T^{\frac{1}{2}}}{T - \frac{1}{2}\beta T^{\frac{1}{2}}} \leq \widehat{\rho} \leq \frac{T - \frac{1}{2}\beta T^{\frac{1}{2}}}{\frac{1}{2}\beta T^{\frac{1}{2}}}.
\end{align*}
Since $\frac{\frac{1}{2}\beta T^{\frac{1}{2}}}{T - \frac{1}{2}\beta T^{\frac{1}{2}}} \leq \widehat{\rho} \leq \frac{T - \frac{1}{2}\beta T^{\frac{1}{2}}}{\frac{1}{2}\beta T^{\frac{1}{2}}}$, we have
\begin{align*}
\frac{\widehat{\sigma}_1(1)}{\widehat{\sigma}_1(1) + \widehat{\sigma}_1(0)} T \geq \frac{1}{1+\frac{T - \frac{1}{2}\beta T^{\frac{1}{2}}}{\frac{1}{2}\beta T^{\frac{1}{2}}}} \ T= \frac{1}{2}\beta T^{\frac{1}{2}}, \\
\frac{\widehat{\sigma}_1(0)}{\widehat{\sigma}_1(1) + \widehat{\sigma}_1(0)} T \geq \frac{1}{\frac{T - \frac{1}{2}\beta T^{\frac{1}{2}}}{\frac{1}{2}\beta T^{\frac{1}{2}}}+1} \ T= \frac{1}{2}\beta T^{\frac{1}{2}}.
\end{align*}
Due to this, Algorithm~\ref{alg:2StageANA} goes to Line 3 instead of Line 5 or Line 7.
The total numbers of treated and control units are given by \eqref{eqn:EstimatedOPT}. 
We re-write \eqref{eqn:EstimatedOPT} again as follows,
\begin{align*}
(T(1), T(0)) = (\frac{\widehat{\sigma}_1(1)}{\widehat{\sigma}_1(1) + \widehat{\sigma}_1(0)} T, \frac{\widehat{\sigma}_1(0)}{\widehat{\sigma}_1(1) + \widehat{\sigma}_1(0)} T).
\end{align*}
Putting $(T(1), T(0))$ into \eqref{eqn:Obj}, we have, for any $\sigma(1), \sigma(0)$,
\begin{align}
\frac{V(T(1), T(0) \vert \cE)}{V(T^*(1), T^*(0))} = & \ \frac{\frac{1}{T(1)} \sigma^2(1) + \frac{1}{T(0)} \sigma^2(0)}{\frac{1}{T} (\sigma(1) + \sigma(0))^2} \nonumber \\
= & \ \frac{\left(1+\frac{\widehat{\sigma}_1(0)}{\widehat{\sigma}_1(1)}\right) \sigma^2(1) + \left(1+\frac{\widehat{\sigma}_1(1)}{\widehat{\sigma}_1(0)}\right) \sigma^2(0)}{(\sigma(1) + \sigma(0))^2} \nonumber \\
= & \ \frac{\sigma^2(1) + \sigma^2(0) + \frac{1}{\widehat{\rho}} \ \sigma^2(1) + \widehat{\rho} \ \sigma^2(1)}{(\sigma(1) + \sigma(0))^2} \nonumber \\
= & \ 1 + \frac{1}{(\sigma(1) + \sigma(0))^2}\left( \frac{1}{\widehat{\rho}} \ \sigma^2(1) + \widehat{\rho} \ \sigma^2(1) - 2 \sigma(1) \sigma(0) \right) \label{eqn:proof:2Stage:1:NewHighProbBound}
\end{align}
Due to Lemma~\ref{lem:h:rhohat}, and using \eqref{eqn:2Stage:NewHighProbBound:ConfidenceBound1} and \eqref{eqn:2Stage:NewHighProbBound:ConfidenceBound0},
\begin{multline}
\frac{1}{(\sigma(1) + \sigma(0))^2}\left( \frac{1}{\widehat{\rho}} \ \sigma^2(1) + \widehat{\rho} \ \sigma^2(1) - 2 \sigma(1) \sigma(0) \right) \\
\leq \ \frac{\sigma(1) \sigma(0)}{(\sigma(1) + \sigma(0))^2} \left(\sqrt{\frac{1-2 C T^{-\frac{1}{4}} (\log{T})^{\frac{1}{4}}}{1+2 C T^{-\frac{1}{4}} (\log{T})^{\frac{1}{4}}}} + \sqrt{\frac{1+2 C T^{-\frac{1}{4}} (\log{T})^{\frac{1}{4}}}{1-2 C T^{-\frac{1}{4}} (\log{T})^{\frac{1}{4}}}} - 2\right). \label{eqn:proof:2Stage:2:NewHighProbBound}
\end{multline}
Note that 
\begin{align}
\frac{\sigma(1) \sigma(0)}{(\sigma(1) + \sigma(0))^2} \leq \frac{1}{4}. \label{eqn:proof:2Stage:3:NewHighProbBound}
\end{align}
Note also that
\begin{align}
\sqrt{\frac{1-2 C T^{-\frac{1}{4}} (\log{T})^{\frac{1}{4}}}{1+2 C T^{-\frac{1}{4}} (\log{T})^{\frac{1}{4}}}} + \sqrt{\frac{1+2 C T^{-\frac{1}{4}} (\log{T})^{\frac{1}{4}}}{1-2 C T^{-\frac{1}{4}} (\log{T})^{\frac{1}{4}}}} - 2 = & \ \frac{2}{\sqrt{1 - 4 C^2 T^{-\frac{1}{2}} (\log{T})^{\frac{1}{2}}}} - 2 \nonumber \\
= & \ 2 \left(1 - 4 C^2 T^{-\frac{1}{2}} (\log{T})^{\frac{1}{2}}\right)^{-\frac{1}{2}} - 2 \nonumber \\
\leq & \ 2 \left( 1 + 4 C^2 T^{-\frac{1}{2}} (\log{T})^{\frac{1}{2}} \right) - 2 \nonumber \\
= & \ 8 C^2 T^{-\frac{1}{2}} (\log{T})^{\frac{1}{2}}, \label{eqn:proof:2Stage:4:NewHighProbBound}
\end{align}
where the inequality is due to Lemma~\ref{lem:AlgebraicTrick1:Refined}-(iii).

Combining \eqref{eqn:proof:2Stage:1:NewHighProbBound} --- \eqref{eqn:proof:2Stage:4:NewHighProbBound}, we have 
\begin{align*}
\frac{V(T(1), T(0) \vert \cE)}{V(T^*(1), T^*(0))} \leq 1 + 2 C^2 T^{-\frac{1}{2}} (\log{T})^{\frac{1}{2}}.
\end{align*}

\noindent\textbf{Case 1.2}:
\begin{align*}
\frac{\frac{1}{2}\beta T^{\frac{1}{2}}}{T - \frac{1}{2}\beta T^{\frac{1}{2}}} \leq \rho \leq \frac{T - \frac{1}{2}\beta T^{\frac{1}{2}}}{\frac{1}{2}\beta T^{\frac{1}{2}}}, && \text{but} && \widehat{\rho} > \frac{T - \frac{1}{2}\beta T^{\frac{1}{2}}}{\frac{1}{2}\beta T^{\frac{1}{2}}} \ \text{or} \ \widehat{\rho} < \frac{\frac{1}{2}\beta T^{\frac{1}{2}}}{T - \frac{1}{2}\beta T^{\frac{1}{2}}}.
\end{align*}
If $\widehat{\rho} > \frac{T - \frac{1}{2}\beta T^{\frac{1}{2}}}{\frac{1}{2}\beta T^{\frac{1}{2}}}$, then 
\begin{align*}
\frac{\widehat{\sigma}_1(0)}{\widehat{\sigma}_1(1) + \widehat{\sigma}_1(0)} T < \frac{1}{\frac{T - \frac{1}{2}\beta T^{\frac{1}{2}}}{\frac{1}{2}\beta T^{\frac{1}{2}}}+1} \ T= \frac{1}{2}\beta T^{\frac{1}{2}}.
\end{align*}
Due to this, Algorithm~\ref{alg:2StageANA} goes to Line 7.
The total numbers of treated and control units are given by $(T(1), T(0)) = (T - \frac{1}{2}\beta T^{\frac{1}{2}}, \frac{1}{2}\beta T^{\frac{1}{2}})$.

Note that,
\begin{align*}
\rho \ = \ \frac{\sigma(1)}{\sigma(0)} \ \leq \ \frac{T - \frac{1}{2}\beta T^{\frac{1}{2}}}{\frac{1}{2}\beta T^{\frac{1}{2}}} \ < \ \widehat{\rho} \ \leq \ \frac{\sigma(1)}{\sigma(0)} \sqrt{\frac{1+2 C T^{-\frac{1}{4}} (\log{T})^{\frac{1}{4}}}{1-2 C T^{-\frac{1}{4}} (\log{T})^{\frac{1}{4}}}}.
\end{align*}
Then we have, for any $\sigma(1), \sigma(0)$,
\begin{align*}
\frac{V(T(1), T(0) \vert \cE)}{V(T^*(1), T^*(0))} = & \ \frac{\frac{1}{T(1)} \sigma^2(1) + \frac{1}{T(0)} \sigma^2(0)}{\frac{1}{T} (\sigma(1) + \sigma(0))^2} \\
= & \ \frac{\sigma^2(1) + \sigma^2(0) + \frac{\frac{1}{2}\beta T^{\frac{1}{2}}}{T - \frac{1}{2}\beta T^{\frac{1}{2}}} \sigma^2(1) + \frac{T - \frac{1}{2}\beta T^{\frac{1}{2}}}{\frac{1}{2}\beta T^{\frac{1}{2}}} \sigma^2(0)}{(\sigma(1) + \sigma(0))^2} \\
< & \ \frac{\sigma^2(1) + \sigma^2(0) + \frac{\sigma(0)}{\sigma(1)} \sqrt{\frac{1-2 C T^{-\frac{1}{4}} (\log{T})^{\frac{1}{4}}}{1+2 C T^{-\frac{1}{4}} (\log{T})^{\frac{1}{4}}}} \sigma^2(1) + \frac{\sigma(1)}{\sigma(0)} \sqrt{\frac{1+2 C T^{-\frac{1}{4}} (\log{T})^{\frac{1}{4}}}{1-2 C T^{-\frac{1}{4}} (\log{T})^{\frac{1}{4}}}} \sigma^2(0)}{(\sigma(1) + \sigma(0))^2} \\
= & \ 1 + \frac{\sigma(1) \sigma(0)}{(\sigma(1) + \sigma(0))^2} \left(\sqrt{\frac{1-2 C T^{-\frac{1}{4}} (\log{T})^{\frac{1}{4}}}{1+2 C T^{-\frac{1}{4}} (\log{T})^{\frac{1}{4}}}} + \sqrt{\frac{1+2 C T^{-\frac{1}{4}} (\log{T})^{\frac{1}{4}}}{1-2 C T^{-\frac{1}{4}} (\log{T})^{\frac{1}{4}}}} - 2\right).
\end{align*}
where the inequality is due to Lemma~\ref{lem:h:rhohat}.
Combining this with \eqref{eqn:proof:2Stage:3:NewHighProbBound} and \eqref{eqn:proof:2Stage:4:NewHighProbBound} we have again
\begin{align*}
\frac{V(T(1), T(0) \vert \cE)}{V(T^*(1), T^*(0))} \leq 1 + 2 C^2 T^{-\frac{1}{2}} (\log{T})^{\frac{1}{2}}.
\end{align*}

If $\widehat{\rho} < \frac{\frac{1}{2}\beta T^{\frac{1}{2}}}{T - \frac{1}{2}\beta T^{\frac{1}{2}}}$, then Algorithm~\ref{alg:2StageANA} goes to Line 5.
\begin{align*}
\frac{\sigma(1)}{\sigma(0)} \sqrt{\frac{1-2 C T^{-\frac{1}{4}} (\log{T})^{\frac{1}{4}}}{1+2 C T^{-\frac{1}{4}} (\log{T})^{\frac{1}{4}}}} \leq \widehat{\rho} \ < \ \frac{\frac{1}{2}\beta T^{\frac{1}{2}}}{T - \frac{1}{2}\beta T^{\frac{1}{2}}} \ \leq \ \rho \ = \ \frac{\sigma(1)}{\sigma(0)},
\end{align*}
and the same analysis follows similarly.

\noindent\textbf{Case 2.1}:
\begin{align*}
\rho > \frac{T - \frac{1}{2}\beta T^{\frac{1}{2}}}{\frac{1}{2}\beta T^{\frac{1}{2}}}, && \text{and} && \widehat{\rho} > \frac{T - \frac{1}{2}\beta T^{\frac{1}{2}}}{\frac{1}{2}\beta T^{\frac{1}{2}}}.
\end{align*}
Since $\widehat{\rho} > \frac{T - \frac{1}{2}\beta T^{\frac{1}{2}}}{\frac{1}{2}\beta T^{\frac{1}{2}}}$, we have
\begin{align*}
\frac{\widehat{\sigma}_1(0)}{\widehat{\sigma}_1(1) + \widehat{\sigma}_1(0)} T < \frac{1}{\frac{T - \frac{1}{2}\beta T^{\frac{1}{2}}}{\frac{1}{2}\beta T^{\frac{1}{2}}}+1} \ T= \frac{1}{2}\beta T^{\frac{1}{2}}.
\end{align*}
Due to this, Algorithm~\ref{alg:2StageANA} goes to Line 7.
The total numbers of treated and control units are given by $(T(1), T(0)) = (T - \frac{1}{2}\beta T^{\frac{1}{2}}, \frac{1}{2}\beta T^{\frac{1}{2}})$.

Putting $(T(1), T(0))$ into \eqref{eqn:Obj}, we have, for any $\sigma(1), \sigma(0)$,
\begin{align}
\frac{V(T(1), T(0) \vert \cE)}{V(T^*(1), T^*(0))} = & \ \frac{\frac{1}{T(1)} \sigma^2(1) + \frac{1}{T(0)} \sigma^2(0)}{\frac{1}{T} (\sigma(1) + \sigma(0))^2} \nonumber \\
= & \ \frac{T}{T - \frac{1}{2}\beta T^{\frac{1}{2}}} \cdot \frac{\sigma^2(1)}{(\sigma(1) + \sigma(0))^2} + \frac{T}{\frac{1}{2}\beta T^{\frac{1}{2}}} \cdot \frac{\sigma^2(0)}{(\sigma(1) + \sigma(0))^2} \nonumber \\
= & \ \frac{T}{T - \frac{1}{2}\beta T^{\frac{1}{2}}} \cdot \frac{\rho^2}{(\rho+1)^2} + \frac{T}{\frac{1}{2}\beta T^{\frac{1}{2}}} \cdot \frac{1}{(\rho+1)^2}. \label{eqn:proof:2Stage:5:NewHighProbBound}
\end{align}
Due to Lemma~\ref{lem:g:rho}, since $\rho = \frac{\sigma(1)}{\sigma(0)} > \frac{T - \frac{1}{2}\beta T^{\frac{1}{2}}}{\frac{1}{2}\beta T^{\frac{1}{2}}}$, we know that the expression in \eqref{eqn:proof:2Stage:5:NewHighProbBound} is increasing with respect to $\rho$.
So we have
\begin{multline*}
\frac{V(T(1), T(0) \vert \cE)}{V(T^*(1), T^*(0))} \leq \lim_{\rho \to +\infty} \left(\frac{T}{T - \frac{1}{2}\beta T^{\frac{1}{2}}} \cdot \frac{\rho^2}{(\rho+1)^2} + \frac{T}{\frac{1}{2}\beta T^{\frac{1}{2}}} \cdot \frac{1}{(\rho+1)^2}\right) \\
= \frac{T}{T - \frac{1}{2}\beta T^{\frac{1}{2}}} = \frac{T}{T - 2 C^2 T^{\frac{1}{2}} (\log{T})^{\frac{1}{2}}} \leq 1 + 4 C^2 T^{-\frac{1}{2}} (\log{T})^{\frac{1}{2}},
\end{multline*}
where the last inequality holds because $T \geq 64 C^4 \log{T} > 16 C^4 \log{T}$.

\noindent\textbf{Case 2.2}:
\begin{align*}
\rho > \frac{T - \frac{1}{2}\beta T^{\frac{1}{2}}}{\frac{1}{2}\beta T^{\frac{1}{2}}}, && \text{and} && \widehat{\rho} \leq \frac{T - \frac{1}{2}\beta T^{\frac{1}{2}}}{\frac{1}{2}\beta T^{\frac{1}{2}}}
\end{align*}
Note that,
\begin{align*}
\widehat{\sigma}_1(1) \geq & \ \sigma(1) \sqrt{1 - 2 C T^{-\frac{1}{4}} (\log{T})^{\frac{1}{4}}} \\
> & \ \sigma(0) \frac{T - \frac{1}{2}\beta T^{\frac{1}{2}}}{\frac{1}{2}\beta T^{\frac{1}{2}}} \sqrt{1 - 2 C T^{-\frac{1}{4}} (\log{T})^{\frac{1}{4}}} \\
\geq & \ \widehat{\sigma}_1(0) \frac{T - \frac{1}{2}\beta T^{\frac{1}{2}}}{\frac{1}{2}\beta T^{\frac{1}{2}}} \sqrt{\frac{1 - 2 C T^{-\frac{1}{4}} (\log{T})^{\frac{1}{4}}}{1 + 2 C T^{-\frac{1}{4}} (\log{T})^{\frac{1}{4}}}} \\
\geq & \ \widehat{\sigma}_1(0) \frac{\frac{1}{2}\beta T^{\frac{1}{2}}}{T - \frac{1}{2}\beta T^{\frac{1}{2}}}.
\end{align*}
where the first inequality is due to \eqref{eqn:2Stage:NewHighProbBound:ConfidenceBound1};
the second inequality is due to $\rho > \frac{T - \frac{1}{2}\beta T^{\frac{1}{2}}}{\frac{1}{2}\beta T^{\frac{1}{2}}}$;
the third inequality is due to \eqref{eqn:2Stage:NewHighProbBound:ConfidenceBound0};
the last inequality is due to Lemma~\ref{lem:AlgebraicTrick1:Refined}-(ii).

The above shows that, in this case (Case 2.2), 
\begin{align*}
\widehat{\rho} \geq \frac{\frac{1}{2}\beta T^{\frac{1}{2}}}{T - \frac{1}{2}\beta T^{\frac{1}{2}}}.
\end{align*}
Since $\frac{\frac{1}{2}\beta T^{\frac{1}{2}}}{T - \frac{1}{2}\beta T^{\frac{1}{2}}} \leq \widehat{\rho} \leq \frac{T - \frac{1}{2}\beta T^{\frac{1}{2}}}{\frac{1}{2}\beta T^{\frac{1}{2}}}$, we have
\begin{align*}
\frac{\widehat{\sigma}_1(1)}{\widehat{\sigma}_1(1) + \widehat{\sigma}_1(0)} T \geq \frac{1}{1+\frac{T - \frac{1}{2}\beta T^{\frac{1}{2}}}{\frac{1}{2}\beta T^{\frac{1}{2}}}} \ T= \frac{1}{2}\beta T^{\frac{1}{2}}, \\
\frac{\widehat{\sigma}_1(0)}{\widehat{\sigma}_1(1) + \widehat{\sigma}_1(0)} T \geq \frac{1}{\frac{T - \frac{1}{2}\beta T^{\frac{1}{2}}}{\frac{1}{2}\beta T^{\frac{1}{2}}}+1} \ T= \frac{1}{2}\beta T^{\frac{1}{2}}.
\end{align*}
Due to this, Algorithm~\ref{alg:2StageANA} goes to Line 3 instead of Line 5 or Line 7.
The total numbers of treated and control units are given by \eqref{eqn:EstimatedOPT}, which we write again as follows,
\begin{align*}
(T(1), T(0)) = (\frac{\widehat{\sigma}_1(1)}{\widehat{\sigma}_1(1) + \widehat{\sigma}_1(0)} T, \frac{\widehat{\sigma}_1(0)}{\widehat{\sigma}_1(1) + \widehat{\sigma}_1(0)} T).
\end{align*}

Similar to Case 1.1, combining \eqref{eqn:proof:2Stage:1:NewHighProbBound} --- \eqref{eqn:proof:2Stage:4:NewHighProbBound}, we have 
\begin{align*}
\frac{V(T(1), T(0) \vert \cE)}{V(T^*(1), T^*(0))} \leq 1 + 2 C^2 T^{-\frac{1}{2}} (\log{T})^{\frac{1}{2}}.
\end{align*}

To conclude, in all four cases, 
\begin{align*}
\frac{V(T(1), T(0) \vert \cE)}{V(T^*(1), T^*(0))} \leq 1 + 4 C^2 T^{-\frac{1}{2}} (\log{T})^{\frac{1}{2}}.
\end{align*}
\hfill \halmos
\endproof

\subsubsection{Completing the proof of Corollary~\ref{coro:2StageANA}.}
\proof{Proof of Corollary~\ref{coro:2StageANA}.}
We first show Algorithm~\ref{alg:2StageANA} is feasible under $\beta = 4 C^2 (\log{T})^{\frac{1}{2}}$.
This is because
\begin{align*}
\beta T^{\frac{1}{2}} = \frac{1}{2} \cdot 8 C^2 (\log{T})^{\frac{1}{2}} \cdot T^{\frac{1}{2}} \leq \frac{1}{2} \cdot T^{\frac{1}{2}} \cdot T^{\frac{1}{2}} = \frac{1}{2} T,
\end{align*}
where the inequality is due to Lemma~\ref{lem:AlgebraicTrick:Basic:1}.

Next, due to Lemma~\ref{lem:2Stage:NewHighProbBound}, conditional on $\cE$ that happens with probability at least $1 - \frac{4}{T^2}$, 
\begin{align}
\frac{V(T(1), T(0)\vert \cE)}{V(T^*(1), T^*(0))} \leq 1 + 4 C^2 T^{-\frac{1}{2}} (\log{T})^{\frac{1}{2}}. \label{eqn:coro:2StageANA:1}
\end{align}

On the other hand, on the low probability event $\overline{\cE}$ that happens with probability at most $\frac{4}{T^2}$,
\begin{align}
\frac{V(T(1), T(0) \vert \overline{\cE})}{V(T^*(1), T^*(0))} = & \ \frac{T}{T - \frac{1}{2} \beta T^{\frac{1}{2}}} \cdot \frac{\sigma^2(1)}{(\sigma(1) + \sigma(0))^2} + \frac{T}{\frac{1}{2} \beta T^{\frac{1}{2}}} \cdot \frac{\sigma^2(0)}{(\sigma(1) + \sigma(0))^2} \nonumber \\
\leq & \ \max\left\{\frac{T}{T - \frac{1}{2} \beta T^{\frac{1}{2}}}, \frac{T}{\frac{1}{2} \beta T^{\frac{1}{2}}}\right\} \nonumber \\
= & \ 2 \beta^{-1} T^{\frac{1}{2}}, \nonumber \\
= & \ 2^{-1} C^{-2} (\log{T})^{-\frac{1}{2}} T^{\frac{1}{2}}, \label{eqn:coro:2StageANA:2}
\end{align}
where the inequality is due to Lemma~\ref{lem:g:rho}.

So overall we have
\begin{align*}
\sup_{\cF \in \sP^{[C]}} \ \frac{\bE[V(T(1), T(0))]}{V(T^*(1), T^*(0))} \leq & \ \left(1 - \frac{4}{T^2}\right) \left( 1 + 4 C^2 T^{-\frac{1}{2}} (\log{T})^{\frac{1}{2}} \right) + \frac{4}{T^2} \cdot 2^{-1} C^{-2} (\log{T})^{-\frac{1}{2}} T^{\frac{1}{2}} \\
\leq & \ 1 + 4 C^2 T^{-\frac{1}{2}} (\log{T})^{\frac{1}{2}} + \frac{2}{T} \cdot C^{-2} (\log{T})^{-\frac{1}{2}} \cdot T^{-\frac{1}{2}} \\
\leq & \ 1 + 4 C^2 T^{-\frac{1}{2}} (\log{T})^{\frac{1}{2}} + 1 \cdot C^2 (\log{T})^{\frac{1}{2}} \cdot T^{-\frac{1}{2}} \\
= & \ 1 + 5 C^2 T^{-\frac{1}{2}} (\log{T})^{\frac{1}{2}},
\end{align*}
where the first inequality is using the total law of probability, and upper bounding the two parts using \eqref{eqn:coro:2StageANA:1} and \eqref{eqn:coro:2StageANA:2};
the second probability is upper bounding $1 - \frac{4}{T^2}$ by 1;
the third inequality is because $T \geq 2$ and $C^4 \log{T} \geq 1$.
\hfill \halmos
\endproof

\subsection{Proof of Corollary~\ref{coro:MStageANA}}
\label{sec:proof:coro:MStageANA}
\subsubsection{Establishing a high probability bound.}
We first establish a high probability bound for the proof of Corollary~\ref{coro:MStageANA}.

\begin{lemma}
\label{lem:NewHighProbBound}
Let $M \geq 3$ and $T \geq (\frac{5000}{3})^{\frac{5}{4}} C^5$.
Let the tuning parameters from Algorithm~\ref{alg:MStageANA} be defined as $\beta_m = \frac{400}{3} C^4 \log{T} \cdot (\frac{1000}{3} C^4 \log{T})^{-\frac{m}{M}}$.
Let $(T(1), T(0))$ be the total number of treated and control units from Algorithm~\ref{alg:MStageANA}, respectively. 
Under Assumption~\ref{asp:kurtosis}, there exists an event that happens with probability at least $1 - \frac{4}{T^2}$, conditional on which
\begin{align*}
\sup_{\cF \in \sP^{[C]}} \ \frac{\bE[V(T(1), T(0))]}{V(T^*(1), T^*(0))} \leq 1 + 96 \cdot \left(\frac{1000}{3}\right)^{-\frac{1}{M}} C^{\frac{4(M-1)}{M}} T^{-\frac{M-1}{M}} (\log{T})^{\frac{M-1}{M}}.
\end{align*}
\end{lemma}

\proof{Proof of Lemma~\ref{lem:NewHighProbBound}.}
We proceed with the similar clean event analysis as in Theorem~\ref{thm:MStageANA}.
Suppose there are two length-$T$ arrays for the treated and the control, respectively, with each value being an independent and identically distributed copy of the representative random variables $Y(1)$ and $Y(0)$, respectively.
When Algorithm~\ref{alg:MStageANA} suggests to conduct an $m$-th stage experiment parameterized by $(T_m(1), T_m(0))$, the observations from the $m$-th stage experiment are generated by reading the next $T_m(1)$ values from the treated array, and the next $T_m(0)$ values from the control array.

Even though Algorithm~\ref{alg:MStageANA} adaptively determines the number of treated and control units, it is always the first few values of of the two arrays that are read.
For any $m \leq M-1$, let $\widehat{\psi}^2_m(1)$ and $\widehat{\psi}^2_{m}(0)$ be the sample variance estimators obtained from reading the first $\frac{\beta_m}{2}T^{\frac{m}{M}}$ values in the treated array and control array, respectively.
Depending on the execution of Algorithm~\ref{alg:MStageANA}, only a few of the sample variance estimators $\widehat{\sigma}^2_m(1)$ or $\widehat{\sigma}^2_m(0)$ are calculated. 
When one sample variance estimator $\widehat{\sigma}^2_m(1)$ or $\widehat{\sigma}^2_m(0)$ is calculated following Algorithm~\ref{alg:MStageANA}, it is equivalent to reading the corresponding $\widehat{\psi}^2_m(1)$ or $\widehat{\psi}^2_m(0)$ from the array.

Define the following events.
For any $m \leq M-1$, define
\begin{align*}
\cE_m(1) = & \ \bigg\{ \left| \widehat{\psi}^2_m(1) - \sigma^2(1) \right| < 48^{\frac{1}{2}} C^2 \beta_m^{-\frac{1}{2}} T^{-\frac{m}{2M}} (\log{T})^{\frac{1}{2}} \sigma^2(1)\bigg\}, \\
\cE_m(0) = & \ \bigg\{ \left| \widehat{\psi}^2_m(0) - \sigma^2(0) \right| < 48^{\frac{1}{2}} C^2 \beta_m^{-\frac{1}{2}} T^{-\frac{m}{2M}} (\log{T})^{\frac{1}{2}} \sigma^2(0)\bigg\}.
\end{align*}
Denote the intersect of all above events as $\cE$, i.e., 
\begin{align*}
\cE = \bigcap_{m=1}^{M-1} \left(\cE_m(1) \cap \cE_m(0)\right).
\end{align*}
Then due to union bound, 
\begin{align*}
\Pr(\cE) \geq 1 - \sum_{m=1}^{M-1} \Pr(\overline{\cE}_m(1)) - \sum_{m=1}^{M-1} \Pr(\overline{\cE}_m(0)).
\end{align*}
We further have
\begin{align*}
\Pr(\cE) = & \ 1 - \sum_{m=1}^{M-1} \Pr\left( \vert \widehat{\psi}^2_m(1) - \sigma^2(1) \vert \geq 48^{\frac{1}{2}} C^2 \beta_m^{-\frac{1}{2}} T^{-\frac{m}{2M}} (\log{T})^{\frac{1}{2}} \sigma^2(1) \right) \\
& \qquad - \sum_{m=1}^{M-1} \Pr\left( \vert \widehat{\psi}^2_m(0) - \sigma^2(0) \vert \geq 48^{\frac{1}{2}} C^2 \beta_m^{-\frac{1}{2}} T^{-\frac{m}{2M}} (\log{T})^{\frac{1}{2}} \sigma^2(0) \right) \\
\geq & \ 1 - \sum_{m=1}^{M-1} 2 \exp\left\{ -\frac{48 C^4 \beta_m^{-1} T^{-\frac{m}{M}} \log{T} \sigma^4(1) \cdot \frac{1}{2}\beta_m T^{\frac{m}{M}}}{8 C^4 \sigma^4(1)} \right\} \\
& \qquad - \sum_{m=1}^{M-1} 2 \exp\left\{ -\frac{48 C^4 \beta_m^{-1} T^{-\frac{m}{M}} \log{T} \sigma^4(0) \cdot \frac{1}{2}\beta_m T^{\frac{m}{M}}}{8 C^4 \sigma^4(0)} \right\} \\
= & \ 1 - \sum_{m=1}^{M-1} 4\exp\{-3 \log{T}\} \\
= & \ 1 - (M-1) \frac{4}{T^3} \\
\geq & \ 1 - \frac{4}{T^2},
\end{align*}
where the first inequality is due to Lemma~\ref{lem:ExponentialTail}.

Conditional on the event $\cE$, we have, for any $m \leq M-1$,
\begin{subequations}
\begin{align}
\sigma^2(1) \left( 1 - 48^{\frac{1}{2}} C^2 \beta_m^{-\frac{1}{2}} T^{-\frac{m}{2M}} (\log{T})^{\frac{1}{2}} \right) \ \leq \ \widehat{\psi}^2_m(1) \ \leq \ \sigma^2(1) \left( 1 + 48^{\frac{1}{2}} C^2 \beta_m^{-\frac{1}{2}} T^{-\frac{m}{2M}} (\log{T})^{\frac{1}{2}} \right), \label{eqn:MStage:NewHighProbBound:ConfidenceBound1} \\
\sigma^2(0) \left( 1 - 48^{\frac{1}{2}} C^2 \beta_m^{-\frac{1}{2}} T^{-\frac{m}{2M}} (\log{T})^{\frac{1}{2}} \right) \ \leq \ \widehat{\psi}^2_m(0) \ \leq \ \sigma^2(0) \left( 1 + 48^{\frac{1}{2}} C^2 \beta_m^{-\frac{1}{2}} T^{-\frac{m}{2M}} (\log{T})^{\frac{1}{2}} \right). \label{eqn:MStage:NewHighProbBound:ConfidenceBound0} 
\end{align}
\end{subequations}

Since $\sigma(1), \sigma(0) > 0$, we can denote $\rho = \frac{\sigma(1)}{\sigma(0)}$.
For any $m \leq M-1$, when $\widehat{\sigma}^2_m(1)$ and $\widehat{\sigma}^2_m(0)$ are calculated during Algorithm~\ref{alg:MStageANA}, $\widehat{\sigma}^2_m(1) = \widehat{\psi}^2_m(1)$ and $\widehat{\sigma}^2_m(0) = \widehat{\psi}^2_m(0)$.
Conditional on the event $\cE$, due to \eqref{eqn:MStage:NewHighProbBound:ConfidenceBound1} and \eqref{eqn:MStage:NewHighProbBound:ConfidenceBound0}, and given that $\sigma(1), \sigma(0) > 0$, we have $\widehat{\sigma}^2_m(1), \widehat{\sigma}^2_m(0) > 0$.
Then we can denote $\widehat{\rho}_m = \frac{\widehat{\sigma}_m(1)}{\widehat{\sigma}_m(0)}$.

In the remaining of the analysis, we distinguish several cases and discuss these cases separately.
Recall that $\widehat{\rho}_1 = \frac{\widehat{\sigma}_1(1)}{\widehat{\sigma}_1(0)}$.
Without loss of generality, assume 
\begin{align}
\widehat{\rho}_1 \geq 1. \label{eqn:WLOG:NewHighProbBound}
\end{align}

\noindent\underline{\textbf{Case 1}}: 
\begin{align*}
\widehat{\rho}_1 > \frac{T - \frac{1}{2} \beta_2 T^{\frac{2}{M}}}{\frac{1}{2} \beta_2 T^{\frac{2}{M}}}.
\end{align*}
\noindent \textbf{Case 1.1}: 
\begin{align*}
\widehat{\rho}_1 > \frac{T - \frac{1}{2} \beta_1 T^{\frac{1}{M}}}{\frac{1}{2} \beta_1 T^{\frac{1}{M}}}.
\end{align*}
In this case, 
\begin{align*}
\frac{\widehat{\sigma}_1(0)}{\widehat{\sigma}_1(1) + \widehat{\sigma}_1(0)} T < \frac{1}{\frac{T - \frac{1}{2} \beta_1 T^{\frac{1}{M}}}{\frac{1}{2} \beta_1 T^{\frac{1}{M}}}+1} T = \frac{1}{2} \beta_1 T^{\frac{1}{M}}.
\end{align*}
So Algorithm~\ref{alg:MStageANA} goes to Line~\ref{mrk:Case1} in the $1$-st stage experiment.
Then we have
\begin{align*}
(T(1), T(0)) = \bigg(T - \frac{1}{2} \beta_1 T^{\frac{1}{M}}, \frac{1}{2} \beta_1 T^{\frac{1}{M}}\bigg).
\end{align*}
We can then express
\begin{align}
\frac{V(T(1), T(0) \vert \cE)}{V(T^*(1), T^*(0))} = & \ \frac{\frac{1}{T - \frac{1}{2} \beta_1 T^{\frac{1}{M}}} \sigma^2(1) + \frac{1}{\frac{1}{2} \beta_1 T^{\frac{1}{M}}} \sigma^2(0)}{\frac{1}{T} (\sigma(1) + \sigma(0))^2}. \label{eqn:StartingPoint:NewHighProbBound}
\end{align}

Recall that $\rho = \frac{\sigma(1)}{\sigma(0)}$.
We further distinguish two cases. 

\textbf{First}, if $\rho < \frac{T - \frac{1}{2} \beta_1 T^{\frac{1}{M}}}{\frac{1}{2} \beta_1 T^{\frac{1}{M}}}$, then we write \eqref{eqn:StartingPoint:NewHighProbBound} as 
\begin{align*}
\frac{V(T(1), T(0) \vert \cE)}{V(T^*(1), T^*(0))} = & \ \frac{\sigma^2(1) + \sigma^2(0) + \frac{\frac{1}{2} \beta_1 T^{\frac{1}{M}}}{T - \frac{1}{2} \beta_1 T^{\frac{1}{M}}} \sigma^2(1) + \frac{T - \frac{1}{2} \beta_1 T^{\frac{1}{M}}}{\frac{1}{2} \beta_1 T^{\frac{1}{M}}} \sigma^2(0)}{(\sigma(1) + \sigma(0))^2}.
\end{align*}
Note that, 
\begin{align}
\rho < \frac{T - \frac{1}{2} \beta_1 T^{\frac{1}{M}}}{\frac{1}{2} \beta_1 T^{\frac{1}{M}}} < \widehat{\rho}_1 \leq \rho \cdot \sqrt{ \frac{1+48^{\frac{1}{2}} C^2 \beta_1^{-\frac{1}{2}} T^{-\frac{1}{2M}} (\log{T})^{\frac{1}{2}}}{1-48^{\frac{1}{2}} C^2 \beta_1^{-\frac{1}{2}} T^{-\frac{1}{2M}} (\log{T})^{\frac{1}{2}}} }. \label{eqn:rhoRelations:Case1-1:NewHighProbBound}
\end{align}
So we have
\begin{align}
\frac{V(T(1), T(0) \vert \cE)}{V(T^*(1), T^*(0))} \leq & \ \frac{\sigma^2(1) + \sigma^2(0) + \sigma(1)\sigma(0) \bigg( \sqrt{ \frac{1+48^{\frac{1}{2}} C^2 \beta_1^{-\frac{1}{2}} T^{-\frac{1}{2M}} (\log{T})^{\frac{1}{2}}}{1-48^{\frac{1}{2}} C^2 \beta_1^{-\frac{1}{2}} T^{-\frac{1}{2M}} (\log{T})^{\frac{1}{2}}} } + \sqrt{ \frac{1-48^{\frac{1}{2}} C^2 \beta_1^{-\frac{1}{2}} T^{-\frac{1}{2M}} (\log{T})^{\frac{1}{2}}}{1+48^{\frac{1}{2}} C^2 \beta_1^{-\frac{1}{2}} T^{-\frac{1}{2M}} (\log{T})^{\frac{1}{2}}} } \bigg)}{(\sigma(1) + \sigma(0))^2} \nonumber \\
= & \ 1 + \frac{\sigma(1) \sigma(0)}{(\sigma(1) + \sigma(0))^2} \cdot \bigg( \frac{2}{\sqrt{1 - 48 C^4 \beta_1^{-1} T^{-\frac{1}{M}} \log{T}}} - 2 \bigg) \nonumber \\
\leq & \ 1 + \frac{\sigma(1) \sigma(0)}{(\sigma(1) + \sigma(0))^2} \cdot \left(96 C^4 \beta_1^{-1} T^{-\frac{1}{M}} \log{T}\right), \label{eqn:BreakIntoTwoParts1:NewHighProbBound}
\end{align}
where the first inequality is due to Lemma~\ref{lem:h:rhohat} and \eqref{eqn:rhoRelations:Case1-1:NewHighProbBound}; the last inequality is due to Lemma~\ref{lem:AlgebraicTrick2:Refined}.

Note that, $\frac{\sigma(1) \sigma(0)}{(\sigma(1) + \sigma(0))^2} = \frac{\rho}{(\rho+1)^2}$ is a decreasing function when $\rho>1$.
Note also that,
\begin{align*}
\rho > \frac{T - \frac{1}{2} \beta_1 T^{\frac{1}{M}}}{\frac{1}{2} \beta_1 T^{\frac{1}{M}}} \cdot \sqrt{\frac{1-48^{\frac{1}{2}} C^2 \beta_1^{-\frac{1}{2}} T^{-\frac{1}{2M}} (\log{T})^{\frac{1}{2}}}{1+48^{\frac{1}{2}} C^2 \beta_1^{-\frac{1}{2}} T^{-\frac{1}{2M}} (\log{T})^{\frac{1}{2}}}} > \frac{1}{2} \frac{T - \frac{1}{2} \beta_1 T^{\frac{1}{M}}}{\frac{1}{2} \beta_1 T^{\frac{1}{M}}} > 1,
\end{align*}
where the first inequality is due to \eqref{eqn:rhoRelations:Case1-1:NewHighProbBound}; the second inequality is due to Lemma~\ref{lem:AlgebraicTrick3:Refined}; the last inequality is due to Lemma~\ref{lem:AlgebraicTrick4:Refined}.

Then we have
\begin{align*}
\frac{\sigma(1) \sigma(0)}{(\sigma(1) + \sigma(0))^2} 
< \ \frac{\frac{1}{2} \frac{T - \frac{1}{2} \beta_1 T^{\frac{1}{M}}}{\frac{1}{2} \beta_1 T^{\frac{1}{M}}}}{\bigg( 1 + \frac{1}{2} \frac{T - \frac{1}{2} \beta_1 T^{\frac{1}{M}}}{\frac{1}{2} \beta_1 T^{\frac{1}{M}}} \bigg)^2}
= \ \frac{\beta_1 T^{\frac{1}{M}} (T - \frac{1}{2} \beta_1 T^{\frac{1}{M}}) }{(T + \frac{1}{2} \beta_1 T^{\frac{1}{M}})^2} 
\leq \ \frac{\beta_1 T^{\frac{1}{M}}}{T}.
\end{align*}
Putting this into \eqref{eqn:BreakIntoTwoParts1:NewHighProbBound} we have 
\begin{align*}
\frac{V(T(1), T(0) \vert \cE)}{V(T^*(1), T^*(0))} \leq \ 1 + 96 C^4 T^{-1} \log{T} < 1 + 96 \cdot \left(\frac{1000}{3}\right)^{-\frac{1}{M}} C^{\frac{4(M-1)}{M}} T^{-\frac{M-1}{M}} (\log{T})^{\frac{M-1}{M}},
\end{align*}
where the last inequality is due to Lemma~\ref{lem:AlgebraicTrick:Basic}.

\textbf{Second}, if $\rho \geq \frac{T - \frac{1}{2} \beta_1 T^{\frac{1}{M}}}{\frac{1}{2} \beta_1 T^{\frac{1}{M}}}$, then we write \eqref{eqn:StartingPoint:NewHighProbBound} as 
\begin{align*}
\frac{V(T(1), T(0) \vert \cE)}{V(T^*(1), T^*(0))} = & \ \frac{T}{T - \frac{1}{2} \beta_1 T^{\frac{1}{M}}} \cdot \frac{\sigma^2(1)}{(\sigma(1) + \sigma(0))^2} + \frac{T}{\frac{1}{2} \beta_1 T^{\frac{1}{M}}} \cdot \frac{\sigma^2(0)}{(\sigma(1) + \sigma(0))^2}.
\end{align*}
So we have
\begin{multline*}
\frac{V(T(1), T(0) \vert \cE)}{V(T^*(1), T^*(0))} \leq \frac{T}{T - \frac{1}{2} \beta_1 T^{\frac{1}{M}}} = 1 + \frac{\frac{1}{2} \beta_1 T^{\frac{1}{M}}}{T - \frac{1}{2} \beta_1 T^{\frac{1}{M}}} \\
< 1 + 96 \cdot \left(\frac{1000}{3}\right)^{-\frac{1}{M}} C^{\frac{4(M-1)}{M}} T^{-\frac{M-1}{M}} (\log{T})^{\frac{M-1}{M}},
\end{multline*}
where the first inequality is due to Lemma~\ref{lem:g:rho}; the last inequality is due to Lemma~\ref{lem:AlgebraicTrick5:Refined}.

Combining $\rho < \frac{T - \frac{1}{2} \beta_1 T^{\frac{1}{M}}}{\frac{1}{2} \beta_1 T^{\frac{1}{M}}}$ and $\rho \geq \frac{T - \frac{1}{2} \beta_1 T^{\frac{1}{M}}}{\frac{1}{2} \beta_1 T^{\frac{1}{M}}}$, we have that in Case 1.1, 
\begin{align*}
\frac{V(T(1), T(0) \vert \cE)}{V(T^*(1), T^*(0))} \leq 1 + 96 \cdot \left(\frac{1000}{3}\right)^{-\frac{1}{M}} C^{\frac{4(M-1)}{M}} T^{-\frac{M-1}{M}} (\log{T})^{\frac{M-1}{M}}.
\end{align*}

\noindent \textbf{Case 1.2}:
\begin{align*}
\frac{T - \frac{1}{2} \beta_2 T^{\frac{2}{M}}}{\frac{1}{2} \beta_2 T^{\frac{2}{M}}} < \widehat{\rho}_1 \leq \frac{T - \frac{1}{2} \beta_1 T^{\frac{1}{M}}}{\frac{1}{2} \beta_1 T^{\frac{1}{M}}}.
\end{align*}
In this case, 
\begin{align*}
\frac{1}{2} \beta_1 T^{\frac{1}{M}}  = \frac{1}{\frac{T - \frac{1}{2} \beta_1 T^{\frac{1}{M}}}{\frac{1}{2} \beta_1 T^{\frac{1}{M}}}+1} T \leq \frac{\widehat{\sigma}_1(0)}{\widehat{\sigma}_1(1) + \widehat{\sigma}_1(0)} T < \frac{1}{\frac{T - \frac{1}{2} \beta_2 T^{\frac{2}{M}}}{\frac{1}{2} \beta_2 T^{\frac{2}{M}}}+1} T = \frac{1}{2} \beta_2 T^{\frac{2}{M}}.
\end{align*}
So Algorithm~\ref{alg:MStageANA} goes to Line~\ref{mrk:Case2} in the $1$-st stage experiment.
Then we have
\begin{align*}
(T(1), T(0)) = \bigg(\frac{\widehat{\sigma}_1(1)}{\widehat{\sigma}_1(1) + \widehat{\sigma}_1(0)} T, \frac{\widehat{\sigma}_1(0)}{\widehat{\sigma}_1(1) + \widehat{\sigma}_1(0)} T\bigg).
\end{align*}
We can then express
\begin{align}
\frac{V(T(1), T(0) \vert \cE)}{V(T^*(1), T^*(0))} = & \ \frac{\sigma^2(1) + \sigma^2(0) + \frac{1}{\widehat{\rho}_1} \sigma^2(1) + \widehat{\rho}_1 \sigma^2(0)}{(\sigma(1) + \sigma(0))^2}. \label{eqn:ImmediateStartigPoint:NewHighProbBound}
\end{align}

Recall that, conditional on $\cE$, \eqref{eqn:MStage:NewHighProbBound:ConfidenceBound1} and \eqref{eqn:MStage:NewHighProbBound:ConfidenceBound0} lead to
\begin{align*}
\rho \cdot \sqrt{\frac{1-48^{\frac{1}{2}} C^2 \beta_1^{-\frac{1}{2}} T^{-\frac{1}{2M}} (\log{T})^{\frac{1}{2}}}{1+48^{\frac{1}{2}} C^2 \beta_1^{-\frac{1}{2}} T^{-\frac{1}{2M}} (\log{T})^{\frac{1}{2}}}} \leq \widehat{\rho}_1 \leq \rho \sqrt{\frac{1+48^{\frac{1}{2}} C^2 \beta_1^{-\frac{1}{2}} T^{-\frac{1}{2M}} (\log{T})^{\frac{1}{2}}}{1-48^{\frac{1}{2}} C^2 \beta_1^{-\frac{1}{2}} T^{-\frac{1}{2M}} (\log{T})^{\frac{1}{2}}}}.
\end{align*}
So we have
\begin{align}
\frac{V(T(1), T(0) \vert \cE)}{V(T^*(1), T^*(0))} \leq & \ \frac{\sigma^2(1) + \sigma^2(0) + \sigma(1)\sigma(0) \bigg( \sqrt{ \frac{1+48^{\frac{1}{2}} C^2 \beta_1^{-\frac{1}{2}} T^{-\frac{1}{2M}} (\log{T})^{\frac{1}{2}}}{1-48^{\frac{1}{2}} C^2 \beta_1^{-\frac{1}{2}} T^{-\frac{1}{2M}} (\log{T})^{\frac{1}{2}}} } + \sqrt{ \frac{1-48^{\frac{1}{2}} C^2 \beta_1^{-\frac{1}{2}} T^{-\frac{1}{2M}} (\log{T})^{\frac{1}{2}}}{1+48^{\frac{1}{2}} C^2 \beta_1^{-\frac{1}{2}} T^{-\frac{1}{2M}} (\log{T})^{\frac{1}{2}}} } \bigg)}{(\sigma(1) + \sigma(0))^2} \nonumber \\
= & \ 1 + \frac{\sigma(1) \sigma(0)}{(\sigma(1) + \sigma(0))^2} \cdot \bigg( \frac{2}{\sqrt{1 - 48 C^4 \beta_1^{-1} T^{-\frac{1}{M}} \log{T}}} - 2 \bigg) \nonumber \\
\leq & \ 1 + \frac{\sigma(1) \sigma(0)}{(\sigma(1) + \sigma(0))^2} \cdot \left( 96 C^4 \beta_1^{-1} T^{-\frac{1}{M}} \log{T} \right), \label{eqn:BreakIntoTwoParts2:NewHighProbBound}
\end{align}
where the first inequality is due to Lemma~\ref{lem:h:rhohat}; the last inequality is due to Lemma~\ref{lem:AlgebraicTrick2:Refined}.

Note that, $\frac{\sigma(1) \sigma(0)}{(\sigma(1) + \sigma(0))^2} = \frac{\rho}{(\rho+1)^2}$ is a decreasing function when $\rho>1$.
Note also that,
\begin{multline*}
\rho \geq \widehat{\rho}_1 \cdot \sqrt{\frac{1-48^{\frac{1}{2}} C^2 \beta_1^{-\frac{1}{2}} T^{-\frac{1}{2M}} (\log{T})^{\frac{1}{2}}}{1+48^{\frac{1}{2}} C^2 \beta_1^{-\frac{1}{2}} T^{-\frac{1}{2M}} (\log{T})^{\frac{1}{2}}}} \\
> \frac{T - \frac{1}{2} \beta_2 T^{\frac{2}{M}}}{\frac{1}{2} \beta_2 T^{\frac{2}{M}}} \cdot \sqrt{\frac{1-48^{\frac{1}{2}} C^2 \beta_1^{-\frac{1}{2}} T^{-\frac{1}{2M}} (\log{T})^{\frac{1}{2}}}{1+48^{\frac{1}{2}} C^2 \beta_1^{-\frac{1}{2}} T^{-\frac{1}{2M}} (\log{T})^{\frac{1}{2}}}} > \frac{1}{2} \frac{T - \frac{1}{2} \beta_2 T^{\frac{2}{M}}}{\frac{1}{2} \beta_2 T^{\frac{2}{M}}} > 1,
\end{multline*}
where the first inequality is due to \eqref{eqn:MStage:NewHighProbBound:ConfidenceBound1} and \eqref{eqn:MStage:NewHighProbBound:ConfidenceBound0}; the second inequality is due to the condition of Case 1.2; the third inequality is due to Lemma~\ref{lem:AlgebraicTrick3:Refined}; the last inequality is due to Lemma~\ref{lem:AlgebraicTrick4:Refined}.
Then we have
\begin{align*}
\frac{\sigma(1) \sigma(0)}{(\sigma(1) + \sigma(0))^2}
< \ \frac{\frac{1}{2} \frac{T - \frac{1}{2} \beta_2 T^{\frac{2}{M}}}{\frac{1}{2} \beta_2 T^{\frac{2}{M}}}}{\bigg( 1 + \frac{1}{2} \frac{T - \frac{1}{2} \beta_2 T^{\frac{2}{M}}}{\frac{1}{2} \beta_2 T^{\frac{2}{M}}} \bigg)^2}
= \ \frac{\beta_2 T^{\frac{2}{M}} (T - \frac{1}{2} \beta_2 T^{\frac{2}{M}}) }{(T + \frac{1}{2} \beta_2 T^{\frac{2}{M}})^2} 
\leq \ \frac{\beta_2 T^{\frac{2}{M}}}{T}.
\end{align*}
Putting this into \eqref{eqn:BreakIntoTwoParts2:NewHighProbBound} we have that in Case 1.2,
\begin{align*}
\frac{V(T(1), T(0) \vert \cE)}{V(T^*(1), T^*(0))} \leq 1 + \frac{96 C^4 \beta_2}{\beta_1} \cdot T^{-\frac{M-1}{M}} \log{T} = 1 + 96 \cdot \left(\frac{1000}{3}\right)^{-\frac{1}{M}} C^{\frac{4(M-1)}{M}} T^{-\frac{M-1}{M}} (\log{T})^{\frac{M-1}{M}}.
\end{align*}

\noindent\underline{\textbf{Case 2}}:
\begin{align*}
\widehat{\rho}_1 \leq \frac{T - \frac{1}{2} \beta_2 T^{\frac{2}{M}}}{\frac{1}{2} \beta_2 T^{\frac{2}{M}}}.
\end{align*}
Due to \eqref{eqn:WLOG:rename} we know that $\widehat{\sigma}_1(1) \geq \widehat{\sigma}_1(0)$.
In Case 2 we immediately have
\begin{align*}
\frac{\widehat{\sigma}_1(1)}{\widehat{\sigma}_1(1) + \widehat{\sigma}_1(0)} T \geq \frac{\widehat{\sigma}_1(0)}{\widehat{\sigma}_1(1) + \widehat{\sigma}_1(0)} T  \geq \frac{1}{\frac{T - \frac{1}{2} \beta_2 T^{\frac{2}{M}}}{\frac{1}{2} \beta_2 T^{\frac{2}{M}}}+1} T = \frac{1}{2} \beta_2 T^{\frac{2}{M}}.
\end{align*}
So Algorithm~\ref{alg:MStageANA} goes to Line~\ref{mrk:Case3} in the 1-st stage experiment. 
We further distinguish two cases.

\noindent\textbf{Case 2.1}:
\begin{align*}
\widehat{\rho}_1 \leq \frac{T - \frac{1}{2} \beta_2 T^{\frac{2}{M}}}{\frac{1}{2} \beta_2 T^{\frac{2}{M}}}, && \widehat{\rho}_2 > \frac{T - \frac{1}{2} \beta_2 T^{\frac{2}{M}}}{\frac{1}{2} \beta_2 T^{\frac{2}{M}}}.
\end{align*}
In this case, 
\begin{align*}
\frac{\widehat{\sigma}_2(0)}{\widehat{\sigma}_2(1) + \widehat{\sigma}_2(0)} T < \frac{1}{\frac{T - \frac{1}{2} \beta_2 T^{\frac{2}{M}}}{\frac{1}{2} \beta_2 T^{\frac{2}{M}}}+1} T = \frac{1}{2} \beta_2 T^{\frac{2}{M}}.
\end{align*}
So Algorithm~\ref{alg:MStageANA} goes to Line~\ref{mrk:Case1} in the $2$-nd stage experiment.
Then we have
\begin{align*}
(T(1), T(0)) = \bigg(T - \frac{1}{2} \beta_2 T^{\frac{2}{M}}, \frac{1}{2} \beta_2 T^{\frac{2}{M}}\bigg).
\end{align*}
We can express 
\begin{align*}
\frac{V(T(1), T(0) \vert \cE)}{V(T^*(1), T^*(0))} = & \ \frac{\sigma^2(1) + \sigma^2(0) + \frac{\frac{1}{2} \beta_2 T^{\frac{2}{M}}}{T - \frac{1}{2} \beta_2 T^{\frac{2}{M}}} \sigma^2(1) + \frac{T - \frac{1}{2} \beta_2 T^{\frac{2}{M}}}{\frac{1}{2} \beta_2 T^{\frac{2}{M}}} \sigma^2(0)}{(\sigma(1) + \sigma(0))^2}.
\end{align*}
Note that, 
\begin{multline}
\rho \cdot \sqrt{\frac{1-48^{\frac{1}{2}} C^2 \beta_1^{-\frac{1}{2}} T^{-\frac{1}{2M}} (\log{T})^{\frac{1}{2}}}{1+48^{\frac{1}{2}} C^2 \beta_1^{-\frac{1}{2}} T^{-\frac{1}{2M}} (\log{T})^{\frac{1}{2}}}} \leq \widehat{\rho}_1 \leq \frac{T - \frac{1}{2} \beta_2 T^{\frac{2}{M}}}{\frac{1}{2} \beta_2 T^{\frac{2}{M}}} < \widehat{\rho}_2 \\
\leq \rho \cdot \sqrt{\frac{1+48^{\frac{1}{2}} C^2 \beta_2^{-\frac{1}{2}} T^{-\frac{2}{2M}} (\log{T})^{\frac{1}{2}}}{1-48^{\frac{1}{2}} C^2 \beta_2^{-\frac{1}{2}} T^{-\frac{2}{2M}} (\log{T})^{\frac{1}{2}}}} < \rho \cdot \sqrt{\frac{1+48^{\frac{1}{2}} C^2 \beta_1^{-\frac{1}{2}} T^{-\frac{1}{2M}} (\log{T})^{\frac{1}{2}}}{1-48^{\frac{1}{2}} C^2 \beta_1^{-\frac{1}{2}} T^{-\frac{1}{2M}} (\log{T})^{\frac{1}{2}}}}, \label{eqn:TwoSidedBounds2:NewHighProbBound}
\end{multline}
where the first and the fourth inequalities are due to \eqref{eqn:MStage:NewHighProbBound:ConfidenceBound1} and \eqref{eqn:MStage:NewHighProbBound:ConfidenceBound0}; the second and the third inequalities are due to the condition of Case 2.1;
the last inequality is because $\beta_1 T^{\frac{1}{M}} < \beta_2 T^{\frac{2}{M}}$ so we have $48^{\frac{1}{2}} C^2 \beta_2^{-\frac{1}{2}} T^{-\frac{2}{2M}} (\log{T})^{\frac{1}{2}} < 48^{\frac{1}{2}} C^2 \beta_1^{-\frac{1}{2}} T^{-\frac{1}{2M}} (\log{T})^{\frac{1}{2}}$.

Then we have
\begin{align}
\frac{V(T(1), T(0) \vert \cE)}{V(T^*(1), T^*(0))} \leq & \ \frac{\sigma^2(1) + \sigma^2(0) + \sigma(1) \sigma(0) \left( \sqrt{\frac{1+48^{\frac{1}{2}} C^2 \beta_1^{-\frac{1}{2}} T^{-\frac{1}{2M}} (\log{T})^{\frac{1}{2}}}{1-48^{\frac{1}{2}} C^2 \beta_1^{-\frac{1}{2}} T^{-\frac{1}{2M}} (\log{T})^{\frac{1}{2}}}} + \sqrt{\frac{1-48^{\frac{1}{2}} C^2 \beta_1^{-\frac{1}{2}} T^{-\frac{1}{2M}} (\log{T})^{\frac{1}{2}}}{1+48^{\frac{1}{2}} C^2 \beta_1^{-\frac{1}{2}} T^{-\frac{1}{2M}} (\log{T})^{\frac{1}{2}}}} \right) }{(\sigma(1) + \sigma(0))^2} \nonumber \\
= & \ 1 + \frac{\sigma(1)\sigma(0)}{(\sigma(1) + \sigma(0))^2} \cdot \left(\frac{2}{\sqrt{1-48 C^4 \beta_1^{-1} T^{-\frac{1}{M}} \log{T}}} - 2\right) \nonumber \\
\leq & \ 1 + \frac{\sigma(1)\sigma(0)}{(\sigma(1) + \sigma(0))^2} \cdot \left(96 C^4 \beta_1^{-1} T^{-\frac{1}{M}} \log{T}\right), \label{eqn:BreakIntoTwoParts3:NewHighProbBound}
\end{align}
where the first inequality is due to Lemma~\ref{lem:h:rhohat}; the last inequality is due to Lemma~\ref{lem:AlgebraicTrick2:Refined}.

Note that, $\frac{\sigma(1) \sigma(0)}{(\sigma(1) + \sigma(0))^2} = \frac{\rho}{(\rho+1)^2}$ is a decreasing function when $\rho>1$.
Note also that,
\begin{align*}
\rho > \frac{T - \frac{1}{2} \beta_2 T^{\frac{2}{M}}}{\frac{1}{2} \beta_2 T^{\frac{2}{M}}} \cdot \sqrt{\frac{1-48^{\frac{1}{2}} C^2 \beta_2^{-\frac{1}{2}} T^{-\frac{2}{2M}} (\log{T})^{\frac{1}{2}}}{1+48^{\frac{1}{2}} C^2 \beta_2^{-\frac{1}{2}} T^{-\frac{2}{2M}} (\log{T})^{\frac{1}{2}}}} > \frac{1}{2} \frac{T - \frac{1}{2} \beta_2 T^{\frac{2}{M}}}{\frac{1}{2} \beta_2 T^{\frac{2}{M}}} > 1,
\end{align*}
where the first inequality is due to \eqref{eqn:TwoSidedBounds2:NewHighProbBound}; the second inequality is due to Lemma~\ref{lem:AlgebraicTrick3:Refined}; the last inequality is due to Lemma~\ref{lem:AlgebraicTrick4:Refined}.

Then we have
\begin{align*}
\frac{\sigma(1) \sigma(0)}{(\sigma(1) + \sigma(0))^2}
< \ \frac{\frac{1}{2} \frac{T - \frac{1}{2} \beta_2 T^{\frac{2}{M}}}{\frac{1}{2} \beta_2 T^{\frac{2}{M}}}}{\bigg( 1 + \frac{1}{2} \frac{T - \frac{1}{2} \beta_2 T^{\frac{2}{M}}}{\frac{1}{2} \beta_2 T^{\frac{2}{M}}} \bigg)^2} 
= \ \frac{\beta_2 T^{\frac{2}{M}} (T - \frac{1}{2} \beta_2 T^{\frac{2}{M}}) }{(T + \frac{1}{2} \beta_2 T^{\frac{2}{M}})^2} 
\leq \ \frac{\beta_2 T^{\frac{2}{M}}}{T}.
\end{align*}
Putting this into \eqref{eqn:BreakIntoTwoParts3:NewHighProbBound} we have that in Case 2.1,
\begin{align*}
\frac{V(T(1), T(0) \vert \cE)}{V(T^*(1), T^*(0))} \leq 1 + \frac{96 C^4 \beta_2}{\beta_1} \cdot T^{-\frac{M-1}{M}} \log{T} = 1 + 96 \cdot \left(\frac{1000}{3}\right)^{-\frac{1}{M}} C^{\frac{4(M-1)}{M}} T^{-\frac{M-1}{M}} (\log{T})^{\frac{M-1}{M}}.
\end{align*}

\noindent\textbf{Case 2.2}:
\begin{align*}
\widehat{\rho}_1 \leq \frac{T - \frac{1}{2} \beta_2 T^{\frac{2}{M}}}{\frac{1}{2} \beta_2 T^{\frac{2}{M}}}, && \frac{T - \frac{1}{2} \beta_3 T^{\frac{3}{M}}}{\frac{1}{2} \beta_3 T^{\frac{3}{M}}} < \widehat{\rho}_2 \leq \frac{T - \frac{1}{2} \beta_2 T^{\frac{2}{M}}}{\frac{1}{2} \beta_2 T^{\frac{2}{M}}}.
\end{align*}
In this case, 
\begin{align*}
\frac{1}{2} \beta_2 T^{\frac{2}{M}} = \frac{1}{\frac{T - \frac{1}{2} \beta_2 T^{\frac{2}{M}}}{\frac{1}{2} \beta_2 T^{\frac{2}{M}}}+1} T \leq \frac{\widehat{\sigma}_2(0)}{\widehat{\sigma}_2(1) + \widehat{\sigma}_2(0)} T < \frac{1}{\frac{T - \frac{1}{2} \beta_3 T^{\frac{3}{M}}}{\frac{1}{2} \beta_3 T^{\frac{3}{M}}}+1} T < \frac{1}{2} \beta_3 T^{\frac{3}{M}}.
\end{align*}
So Algorithm~\ref{alg:MStageANA} goes to Line~\ref{mrk:Case2} in the $2$-nd stage experiment.
Then we have
\begin{align*}
(T(1), T(0)) = \bigg(\frac{\widehat{\sigma}_2(1)}{\widehat{\sigma}_2(1) + \widehat{\sigma}_2(0)} T, \frac{\widehat{\sigma}_2(0)}{\widehat{\sigma}_2(1) + \widehat{\sigma}_2(0)} T\bigg).
\end{align*}
We can then express
\begin{align}
\frac{V(T(1), T(0) \vert \cE)}{V(T^*(1), T^*(0))} = & \ \frac{\sigma^2(1) + \sigma^2(0) + \frac{1}{\widehat{\rho}_2} \sigma^2(1) + \widehat{\rho}_2 \sigma^2(0)}{(\sigma(1) + \sigma(0))^2}. \label{eqn:ImmediateStartigPoint2:NewHighProbBound}
\end{align}

Recall that, conditional on $\cE$, \eqref{eqn:MStage:NewHighProbBound:ConfidenceBound1} and \eqref{eqn:MStage:NewHighProbBound:ConfidenceBound0} lead to
\begin{align*}
\rho \cdot \sqrt{\frac{1-48^{\frac{1}{2}} C^2 \beta_2^{-\frac{1}{2}} T^{-\frac{2}{2M}} (\log{T})^{\frac{1}{2}}}{1+48^{\frac{1}{2}} C^2 \beta_2^{-\frac{1}{2}} T^{-\frac{2}{2M}} (\log{T})^{\frac{1}{2}}}} \leq \widehat{\rho}_2 \leq \rho \sqrt{\frac{1+48^{\frac{1}{2}} C^2 \beta_2^{-\frac{1}{2}} T^{-\frac{2}{2M}} (\log{T})^{\frac{1}{2}}}{1-48^{\frac{1}{2}} C^2 \beta_2^{-\frac{1}{2}} T^{-\frac{2}{2M}} (\log{T})^{\frac{1}{2}}}}.
\end{align*}
So we have
\begin{align}
\frac{V(T(1), T(0) \vert \cE)}{V(T^*(1), T^*(0))} \leq & \ \frac{\sigma^2(1) + \sigma^2(0) + \sigma(1)\sigma(0) \bigg( \sqrt{\frac{1-48^{\frac{1}{2}} C^2 \beta_2^{-\frac{1}{2}} T^{-\frac{2}{2M}} (\log{T})^{\frac{1}{2}}}{1+48^{\frac{1}{2}} C^2 \beta_2^{-\frac{1}{2}} T^{-\frac{2}{2M}} (\log{T})^{\frac{1}{2}}}} + \sqrt{\frac{1+48^{\frac{1}{2}} C^2 \beta_2^{-\frac{1}{2}} T^{-\frac{2}{2M}} (\log{T})^{\frac{1}{2}}}{1-48^{\frac{1}{2}} C^2 \beta_2^{-\frac{1}{2}} T^{-\frac{2}{2M}} (\log{T})^{\frac{1}{2}}}} \bigg)}{(\sigma(1) + \sigma(0))^2} \nonumber \\
= & \ 1 + \frac{\sigma(1) \sigma(0)}{(\sigma(1) + \sigma(0))^2} \cdot \bigg( \frac{2}{\sqrt{1 - 48 C^4 \beta_2^{-1} T^{-\frac{2}{M}} \log{T}}} - 2 \bigg) \nonumber \\
\leq & \ 1 + \frac{\sigma(1) \sigma(0)}{(\sigma(1) + \sigma(0))^2} \cdot \left(96 C^4 \beta_2^{-1} T^{-\frac{2}{M}} \log{T}\right), \label{eqn:BreakIntoTwoParts4:NewHighProbBound}
\end{align}
where the first inequality is due to Lemma~\ref{lem:h:rhohat}; the last inequality is due to Lemma~\ref{lem:AlgebraicTrick2:Refined}.

Note that, $\frac{\sigma(1) \sigma(0)}{(\sigma(1) + \sigma(0))^2} = \frac{\rho}{(\rho+1)^2}$ is a decreasing function when $\rho>1$.
Note also that,
\begin{multline*}
\rho \geq \widehat{\rho}_2 \cdot \sqrt{\frac{1-48^{\frac{1}{2}} C^2 \beta_2^{-\frac{1}{2}} T^{-\frac{2}{2M}} (\log{T})^{\frac{1}{2}}}{1+48^{\frac{1}{2}} C^2 \beta_2^{-\frac{1}{2}} T^{-\frac{2}{2M}} (\log{T})^{\frac{1}{2}}}} \\
> \frac{T - \frac{1}{2} \beta_3 T^{\frac{3}{M}}}{\frac{1}{2} \beta_3 T^{\frac{3}{M}}} \cdot \sqrt{\frac{1-48^{\frac{1}{2}} C^2 \beta_2^{-\frac{1}{2}} T^{-\frac{2}{2M}} (\log{T})^{\frac{1}{2}}}{1+48^{\frac{1}{2}} C^2 \beta_2^{-\frac{1}{2}} T^{-\frac{2}{2M}} (\log{T})^{\frac{1}{2}}}} > \frac{1}{2} \frac{T - \frac{1}{2} \beta_3 T^{\frac{3}{M}}}{\frac{1}{2} \beta_3 T^{\frac{3}{M}}} > 1,
\end{multline*}
where the first inequality is due to \eqref{eqn:MStage:NewHighProbBound:ConfidenceBound1} and \eqref{eqn:MStage:NewHighProbBound:ConfidenceBound0}; the second inequality is due to the condition of Case 2.2; the third inequality is due to Lemma~\ref{lem:AlgebraicTrick3:Refined}; the last inequality is due to Lemma~\ref{lem:AlgebraicTrick4:Refined}.
Then we have
\begin{align*}
\frac{\sigma(1) \sigma(0)}{(\sigma(1) + \sigma(0))^2} 
< \ \frac{\frac{1}{2} \frac{T - \frac{1}{2} \beta_3 T^{\frac{3}{M}}}{\frac{1}{2} \beta_3 T^{\frac{3}{M}}}}{\bigg( 1 + \frac{1}{2} \frac{T - \frac{1}{2} \beta_3 T^{\frac{3}{M}}}{\frac{1}{2} \beta_3 T^{\frac{3}{M}}} \bigg)^2}
= \ \frac{\beta_3 T^{\frac{3}{M}} (T - \frac{1}{2} \beta_3 T^{\frac{3}{M}}) }{(T + \frac{1}{2} \beta_3 T^{\frac{3}{M}})^2} 
\leq \ \frac{\beta_3 T^{\frac{3}{M}}}{T}
\end{align*}
Putting this into \eqref{eqn:BreakIntoTwoParts4:NewHighProbBound} we have that in Case 2.2,
\begin{align*}
\frac{V(T(1), T(0) \vert \cE)}{V(T^*(1), T^*(0))} \leq 1 + \frac{96 C^4 \beta_3}{\beta_2} \cdot T^{-\frac{M-1}{M}} \log{T} = 1 + 96 \cdot \left(\frac{1000}{3}\right)^{-\frac{1}{M}} C^{\frac{4(M-1)}{M}} T^{-\frac{M-1}{M}} (\log{T})^{\frac{M-1}{M}}.
\end{align*}

\noindent\underline{\textbf{Case $\bm{m}$}} (when $m \leq M-2$):
\begin{align*}
\widehat{\rho}_l \leq \frac{T - \frac{1}{2} \beta_{l+1} T^{\frac{l+1}{M}}}{\frac{1}{2} \beta_{l+1} T^{\frac{l+1}{M}}}, \ \forall \ l \leq m-1.
\end{align*}
Due to the condition of Case $m$, we immediately have
\begin{align*}
\frac{\widehat{\sigma}_{m-1}(0)}{\widehat{\sigma}_{m-1}(1) + \widehat{\sigma}_{m-1}(0)} T \geq \frac{1}{\frac{T - \frac{1}{2} \beta_{m} T^{\frac{m}{M}}}{\frac{1}{2} \beta_{m} T^{\frac{m}{M}}}+1} T = \frac{1}{2} \beta_{m} T^{\frac{m}{M}}.
\end{align*}
On the other hand, since
\begin{align*}
\widehat{\rho}_{m-1} \geq & \ \rho \sqrt{\frac{1-48^{\frac{1}{2}} C^2 \beta_{m-1}^{-\frac{1}{2}} T^{-\frac{m-1}{2M}} (\log{T})^{\frac{1}{2}}}{1+48^{\frac{1}{2}} C^2 \beta_{m-1}^{-\frac{1}{2}} T^{-\frac{m-1}{2M}} (\log{T})^{\frac{1}{2}}}} \\
\geq & \ \widehat{\rho}_1 \sqrt{\frac{1-48^{\frac{1}{2}} C^2 \beta_{1}^{-\frac{1}{2}} T^{-\frac{1}{2M}} (\log{T})^{\frac{1}{2}}}{1+48^{\frac{1}{2}} C^2 \beta_{1}^{-\frac{1}{2}} T^{-\frac{1}{2M}} (\log{T})^{\frac{1}{2}}}} \sqrt{\frac{1-48^{\frac{1}{2}} C^2 \beta_{m-1}^{-\frac{1}{2}} T^{-\frac{m-1}{2M}} (\log{T})^{\frac{1}{2}}}{1+48^{\frac{1}{2}} C^2 \beta_{m-1}^{-\frac{1}{2}} T^{-\frac{m-1}{2M}} (\log{T})^{\frac{1}{2}}}} \\
\geq & \ \sqrt{\frac{1-48^{\frac{1}{2}} C^2 \beta_{1}^{-\frac{1}{2}} T^{-\frac{1}{2M}} (\log{T})^{\frac{1}{2}}}{1+48^{\frac{1}{2}} C^2 \beta_{1}^{-\frac{1}{2}} T^{-\frac{1}{2M}} (\log{T})^{\frac{1}{2}}}} \sqrt{\frac{1-48^{\frac{1}{2}} C^2 \beta_{m-1}^{-\frac{1}{2}} T^{-\frac{m-1}{2M}} (\log{T})^{\frac{1}{2}}}{1+48^{\frac{1}{2}} C^2 \beta_{m-1}^{-\frac{1}{2}} T^{-\frac{m-1}{2M}} (\log{T})^{\frac{1}{2}}}} \\
> & \ \frac{1}{4} \\
\geq & \ \frac{\frac{1}{2}\beta_m T^{\frac{m}{M}}}{T - \frac{1}{2}\beta_m T^{\frac{m}{M}}},
\end{align*}
where the first and second inequalities are due to \eqref{eqn:MStage:NewHighProbBound:ConfidenceBound1} and \eqref{eqn:MStage:NewHighProbBound:ConfidenceBound0}; the third inequality is due to \eqref{eqn:WLOG:NewHighProbBound}; the fourth inequality is due to Lemma~\ref{lem:AlgebraicTrick3:Refined}; the last inequality is due to Lemma~\ref{lem:AlgebraicTrick4:Refined}.
Due to the above sequence of inequalities, we have $\frac{1}{\widehat{\rho}_{m-1}} \leq \frac{T - \frac{1}{2}\beta_m T^{\frac{m}{M}}}{\frac{1}{2}\beta_m T^{\frac{m}{M}}}$, which leads to
\begin{align*}
\frac{\widehat{\sigma}_{m-1}(1)}{\widehat{\sigma}_{m-1}(1) + \widehat{\sigma}_{m-1}(0)} T  \geq \frac{1}{1+\frac{T - \frac{1}{2} \beta_{m} T^{\frac{m}{M}}}{\frac{1}{2} \beta_{m} T^{\frac{m}{M}}}} T = \frac{1}{2} \beta_{m} T^{\frac{m}{M}}.
\end{align*}
So Algorithm~\ref{alg:MStageANA} goes to Line~\ref{mrk:Case3} in the (m-1)-th stage experiment. 
We further distinguish two cases.

\noindent\textbf{Case $\bm{m}$.1}:
In addition to the conditions in Case $m$ above, we also have
\begin{align*}
\widehat{\rho}_m > \frac{T - \frac{1}{2} \beta_{m} T^{\frac{m}{M}}}{\frac{1}{2} \beta_{m} T^{\frac{m}{M}}}.
\end{align*}
Similar to the analysis in Case 2.1, we proceed with the following analysis.
In Case $m$.1, 
\begin{align*}
\frac{\widehat{\sigma}_m(0)}{\widehat{\sigma}_m(1) + \widehat{\sigma}_m(0)} T < \frac{1}{\frac{T - \frac{1}{2} \beta_m T^{\frac{m}{M}}}{\frac{1}{2} \beta_m T^{\frac{m}{M}}}+1} T = \frac{1}{2} \beta_m T^{\frac{m}{M}}.
\end{align*}
So Algorithm~\ref{alg:MStageANA} goes to Line~\ref{mrk:Case1} in the $m$-th stage experiment.
Then we have
\begin{align*}
(T(1), T(0)) = \bigg(T - \frac{1}{2} \beta_m T^{\frac{m}{M}}, \frac{1}{2} \beta_m T^{\frac{m}{M}}\bigg).
\end{align*}
We can express 
\begin{align*}
\frac{V(T(1), T(0) \vert \cE)}{V(T^*(1), T^*(0))} = & \ \frac{\sigma^2(1) + \sigma^2(0) + \frac{\frac{1}{2} \beta_m T^{\frac{m}{M}}}{T - \frac{1}{2} \beta_m T^{\frac{m}{M}}} \sigma^2(1) + \frac{T - \frac{1}{2} \beta_m T^{\frac{m}{M}}}{\frac{1}{2} \beta_m T^{\frac{m}{M}}} \sigma^2(0)}{(\sigma(1) + \sigma(0))^2}.
\end{align*}
Note that, 
\begin{multline}
\rho \cdot \sqrt{\frac{1-48^{\frac{1}{2}} C^2 \beta_{m-1}^{-\frac{1}{2}} T^{-\frac{m-1}{2M}} (\log{T})^{\frac{1}{2}}}{1+48^{\frac{1}{2}} C^2 \beta_{m-1}^{-\frac{1}{2}} T^{-\frac{m-1}{2M}} (\log{T})^{\frac{1}{2}}}} \leq \widehat{\rho}_{m-1} \leq \frac{T - \frac{1}{2} \beta_m T^{\frac{m}{M}}}{\frac{1}{2} \beta_m T^{\frac{m}{M}}} < \widehat{\rho}_m \\
\leq \rho \cdot \sqrt{\frac{1+48^{\frac{1}{2}} C^2 \beta_m^{-\frac{1}{2}} T^{-\frac{m}{2M}} (\log{T})^{\frac{1}{2}}}{1-48^{\frac{1}{2}} C^2 \beta_m^{-\frac{1}{2}} T^{-\frac{m}{2M}} (\log{T})^{\frac{1}{2}}}} < \rho \cdot \sqrt{\frac{1+48^{\frac{1}{2}} C^2 \beta_{m-1}^{-\frac{1}{2}} T^{-\frac{m-1}{2M}} (\log{T})^{\frac{1}{2}}}{1-48^{\frac{1}{2}} C^2 \beta_{m-1}^{-\frac{1}{2}} T^{-\frac{m-1}{2M}} (\log{T})^{\frac{1}{2}}}}, \label{eqn:TwoSidedBoundsm:NewHighProbBound}
\end{multline}
where the first and the fourth inequalities are due to \eqref{eqn:MStage:NewHighProbBound:ConfidenceBound1} and \eqref{eqn:MStage:NewHighProbBound:ConfidenceBound0};
the second and the third inequalities are due to the condition of Case $m$.1;
the last inequality is because $\beta_{m-1} T^{\frac{m-1}{M}} < \beta_m T^{\frac{m}{M}}$ so we have $48^{\frac{1}{2}} C^2 \beta_m^{-\frac{1}{2}} T^{-\frac{m}{2M}} (\log{T})^{\frac{1}{2}} < 48^{\frac{1}{2}} C^2 \beta_{m-1}^{-\frac{1}{2}} T^{-\frac{m-1}{2M}} (\log{T})^{\frac{1}{2}}$.

Then we have
\begin{align}
\frac{V(T(1), T(0) \vert \cE)}{V(T^*(1), T^*(0))} \leq & \ \frac{\sigma^2(1) + \sigma^2(0) + \sigma(1) \sigma(0) \left( \sqrt{\frac{1+48^{\frac{1}{2}} C^2 \beta_{m-1}^{-\frac{1}{2}} T^{-\frac{m-1}{2M}} (\log{T})^{\frac{1}{2}}}{1-48^{\frac{1}{2}} C^2 \beta_{m-1}^{-\frac{1}{2}} T^{-\frac{m-1}{2M}} (\log{T})^{\frac{1}{2}}}} + \sqrt{\frac{1-48^{\frac{1}{2}} C^2 \beta_{m-1}^{-\frac{1}{2}} T^{-\frac{m-1}{2M}} (\log{T})^{\frac{1}{2}}}{1+48^{\frac{1}{2}} C^2 \beta_{m-1}^{-\frac{1}{2}} T^{-\frac{m-1}{2M}} (\log{T})^{\frac{1}{2}}}} \right) }{(\sigma(1) + \sigma(0))^2} \nonumber \\
= & \ 1 + \frac{\sigma(1)\sigma(0)}{(\sigma(1) + \sigma(0))^2} \cdot \left(\frac{2}{\sqrt{1-48 C^4 \beta_{m-1}^{-1} T^{-\frac{m-1}{M}} \log{T}}} - 2\right) \nonumber \\
\leq & \ 1 + \frac{\sigma(1)\sigma(0)}{(\sigma(1) + \sigma(0))^2} \cdot \left(96 C^4 \beta_{m-1}^{-1} T^{-\frac{m-1}{M}} \log{T}\right), \label{eqn:BreakIntoTwoPartsm1:NewHighProbBound}
\end{align}
where the first inequality is due to Lemma~\ref{lem:h:rhohat}; the last inequality is due to Lemma~\ref{lem:AlgebraicTrick2:Refined}.

Note that, $\frac{\sigma(1) \sigma(0)}{(\sigma(1) + \sigma(0))^2} = \frac{\rho}{(\rho+1)^2}$ is a decreasing function when $\rho>1$.
Note also that,
\begin{align*}
\rho > \frac{T - \frac{1}{2} \beta_m T^{\frac{m}{M}}}{\frac{1}{2} \beta_m T^{\frac{m}{M}}} \cdot \sqrt{\frac{1-48^{\frac{1}{2}} C^2 \beta_m^{-\frac{1}{2}} T^{-\frac{m}{2M}} (\log{T})^{\frac{1}{2}}}{1+48^{\frac{1}{2}} C^2 \beta_m^{-\frac{1}{2}} T^{-\frac{m}{2M}} (\log{T})^{\frac{1}{2}}}} > \frac{1}{2} \frac{T - \frac{1}{2} \beta_m T^{\frac{m}{M}}}{\frac{1}{2} \beta_m T^{\frac{m}{M}}} > 1,
\end{align*}
where the first inequality is due to \eqref{eqn:TwoSidedBoundsm:NewHighProbBound}; the second inequality is due to Lemma~\ref{lem:AlgebraicTrick3:Refined}; the last inequality is due to Lemma~\ref{lem:AlgebraicTrick4:Refined}.

Then we have
\begin{align*}
\frac{\sigma(1) \sigma(0)}{(\sigma(1) + \sigma(0))^2} 
< \frac{\frac{1}{2} \frac{T - \frac{1}{2} \beta_m T^{\frac{m}{M}}}{\frac{1}{2} \beta_m T^{\frac{m}{M}}}}{\bigg( 1 + \frac{1}{2} \frac{T - \frac{1}{2} \beta_m T^{\frac{m}{M}}}{\frac{1}{2} \beta_m T^{\frac{m}{M}}} \bigg)^2} 
= \frac{\beta_m T^{\frac{m}{M}} (T - \frac{1}{2} \beta_m T^{\frac{m}{M}}) }{(T + \frac{1}{2} \beta_m T^{\frac{m}{M}})^2} 
\leq \frac{\beta_m T^{\frac{m}{M}}}{T}.
\end{align*}
Putting this into \eqref{eqn:BreakIntoTwoPartsm1:NewHighProbBound} we have that in Case $m$.1,
\begin{align*}
\frac{V(T(1), T(0) \vert \cE)}{V(T^*(1), T^*(0))} \leq 1 + \frac{96 C^4 \beta_m}{\beta_{m-1}} \cdot T^{-\frac{M-1}{M}} \log{T} = 1 + 96 \left(\frac{1000}{3}\right)^{-\frac{1}{M}} C^{\frac{4(M-1)}{M}} T^{-\frac{M-1}{M}} (\log{T})^{\frac{M-1}{M}}.
\end{align*}

\noindent\textbf{Case $\bm{m}$.2}:
In addition to the conditions in Case $m$ above, we also have
\begin{align*}
\frac{T - \frac{1}{2} \beta_{m+1} T^{\frac{m+1}{M}}}{\frac{1}{2} \beta_{m+1} T^{\frac{m+1}{M}}} < \widehat{\rho}_m \leq \frac{T - \frac{1}{2} \beta_{m} T^{\frac{m}{M}}}{\frac{1}{2} \beta_{m} T^{\frac{m}{M}}}.
\end{align*}
Similar to the analysis in Case 2.2, we proceed with the following analysis.
In Case $m$.2, 
\begin{align*}
\frac{1}{2} \beta_m T^{\frac{m}{M}} = \frac{1}{\frac{T - \frac{1}{2} \beta_m T^{\frac{m}{M}}}{\frac{1}{2} \beta_m T^{\frac{m}{M}}}+1} T \leq \frac{\widehat{\sigma}_m(0)}{\widehat{\sigma}_m(1) + \widehat{\sigma}_m(0)} T < \frac{1}{\frac{T - \frac{1}{2} \beta_{m+1} T^{\frac{m+1}{M}}}{\frac{1}{2} \beta_{m+1} T^{\frac{m+1}{M}}}+1} T < \frac{1}{2} \beta_{m+1} T^{\frac{m+1}{M}}.
\end{align*}
So Algorithm~\ref{alg:MStageANA} goes to Line~\ref{mrk:Case2} in the $m$-th stage experiment.
Then we have
\begin{align*}
(T(1), T(0)) = \bigg(\frac{\widehat{\sigma}_m(1)}{\widehat{\sigma}_m(1) + \widehat{\sigma}_m(0)} T, \frac{\widehat{\sigma}_m(0)}{\widehat{\sigma}_m(1) + \widehat{\sigma}_m(0)} T\bigg).
\end{align*}
We can then express
\begin{align*}
\frac{V(T(1), T(0) \vert \cE)}{V(T^*(1), T^*(0))} = & \ \frac{\sigma^2(1) + \sigma^2(0) + \frac{1}{\widehat{\rho}_m} \sigma^2(1) + \widehat{\rho}_m \sigma^2(0)}{(\sigma(1) + \sigma(0))^2}.
\end{align*}
Recall that, conditional on $\cE$, \eqref{eqn:MStage:NewHighProbBound:ConfidenceBound1} and \eqref{eqn:MStage:NewHighProbBound:ConfidenceBound0} lead to
\begin{align*}
\rho \cdot \sqrt{\frac{1-48^{\frac{1}{2}} C^2 \beta_m^{-\frac{1}{2}} T^{-\frac{m}{2M}} (\log{T})^{\frac{1}{2}}}{1+48^{\frac{1}{2}} C^2 \beta_m^{-\frac{1}{2}} T^{-\frac{m}{2M}} (\log{T})^{\frac{1}{2}}}} \leq \widehat{\rho}_m \leq \rho \sqrt{\frac{1+48^{\frac{1}{2}} C^2 \beta_m^{-\frac{1}{2}} T^{-\frac{m}{2M}} (\log{T})^{\frac{1}{2}}}{1-48^{\frac{1}{2}} C^2 \beta_m^{-\frac{1}{2}} T^{-\frac{m}{2M}} (\log{T})^{\frac{1}{2}}}}.
\end{align*}
So we have
\begin{align}
\frac{V(T(1), T(0) \vert \cE)}{V(T^*(1), T^*(0))} \leq & \ \frac{\sigma^2(1) + \sigma^2(0) + \sigma(1)\sigma(0) \bigg( \sqrt{\frac{1-48^{\frac{1}{2}} C^2 \beta_m^{-\frac{1}{2}} T^{-\frac{m}{2M}} (\log{T})^{\frac{1}{2}}}{1+48^{\frac{1}{2}} C^2 \beta_m^{-\frac{1}{2}} T^{-\frac{m}{2M}} (\log{T})^{\frac{1}{2}}}} + \sqrt{\frac{1+48^{\frac{1}{2}} C^2 \beta_m^{-\frac{1}{2}} T^{-\frac{m}{2M}} (\log{T})^{\frac{1}{2}}}{1-48^{\frac{1}{2}} C^2 \beta_m^{-\frac{1}{2}} T^{-\frac{m}{2M}} (\log{T})^{\frac{1}{2}}}} \bigg)}{(\sigma(1) + \sigma(0))^2} \nonumber \\
= & \ 1 + \frac{\sigma(1) \sigma(0)}{(\sigma(1) + \sigma(0))^2} \cdot \bigg( \frac{2}{\sqrt{1 - 48 C^4 \beta_m^{-1} T^{-\frac{m}{M}} \log{T}}} - 2 \bigg) \nonumber \\
\leq & \ 1 + \frac{\sigma(1) \sigma(0)}{(\sigma(1) + \sigma(0))^2} \cdot \left(96 C^4 \beta_m^{-1} T^{-\frac{m}{M}} \log{T}\right), \label{eqn:BreakIntoTwoPartsm2:NewHighProbBound}
\end{align}
where the first inequality is due to Lemma~\ref{lem:h:rhohat}; the last inequality is due to Lemma~\ref{lem:AlgebraicTrick2:Refined}.

Note that, $\frac{\sigma(1) \sigma(0)}{(\sigma(1) + \sigma(0))^2} = \frac{\rho}{(\rho+1)^2}$ is a decreasing function when $\rho>1$.
Note also that,
\begin{multline*}
\rho \geq \widehat{\rho}_m \cdot \sqrt{\frac{1-48^{\frac{1}{2}} C^2 \beta_m^{-\frac{1}{2}} T^{-\frac{m}{2M}} (\log{T})^{\frac{1}{2}}}{1+48^{\frac{1}{2}} C^2 \beta_m^{-\frac{1}{2}} T^{-\frac{m}{2M}} (\log{T})^{\frac{1}{2}}}} \\
> \frac{T - \frac{1}{2} \beta_{m+1} T^{\frac{m+1}{M}}}{\frac{1}{2} \beta_{m+1} T^{\frac{m+1}{M}}} \cdot \sqrt{\frac{1-48^{\frac{1}{2}} C^2 \beta_m^{-\frac{1}{2}} T^{-\frac{m}{2M}} (\log{T})^{\frac{1}{2}}}{1+48^{\frac{1}{2}} C^2 \beta_m^{-\frac{1}{2}} T^{-\frac{m}{2M}} (\log{T})^{\frac{1}{2}}}} > \frac{1}{2} \frac{T - \frac{1}{2} \beta_{m+1} T^{\frac{m+1}{M}}}{\frac{1}{2} \beta_{m+1} T^{\frac{m+1}{M}}} > 1,
\end{multline*}
where the first inequality is due to \eqref{eqn:MStage:NewHighProbBound:ConfidenceBound1} and \eqref{eqn:MStage:NewHighProbBound:ConfidenceBound0}; the second inequality is due to the condition of Case 2.2; the third inequality is due to Lemma~\ref{lem:AlgebraicTrick3:Refined}; the last inequality is due to Lemma~\ref{lem:AlgebraicTrick4:Refined}.
Then we have
\begin{align*}
\frac{\sigma(1) \sigma(0)}{(\sigma(1) + \sigma(0))^2}
< \frac{\frac{1}{2} \frac{T - \frac{1}{2} \beta_{m+1} T^{\frac{m+1}{M}}}{\frac{1}{2} \beta_{m+1} T^{\frac{m+1}{M}}}}{\bigg( 1 + \frac{1}{2} \frac{T - \frac{1}{2} \beta_{m+1} T^{\frac{m+1}{M}}}{\frac{1}{2} \beta_{m+1} T^{\frac{m+1}{M}}} \bigg)^2}
= \frac{\beta_{m+1} T^{\frac{m+1}{M}} (T - \frac{1}{2} \beta_{m+1} T^{\frac{m+1}{M}}) }{(T + \frac{1}{2} \beta_{m+1} T^{\frac{m+1}{M}})^2} 
\leq \frac{\beta_{m+1} T^{\frac{m+1}{M}}}{T}
\end{align*}
Putting this into \eqref{eqn:BreakIntoTwoPartsm2:NewHighProbBound} we have that in Case $m$.2,
\begin{align*}
\frac{V(T(1), T(0) \vert \cE)}{V(T^*(1), T^*(0))} \leq 1 + \frac{96 C^4 \beta_{m+1}}{\beta_m} \cdot T^{-\frac{M-1}{M}} \log{T} = 1 + 96 \left(\frac{1000}{3}\right)^{-\frac{1}{M}} C^{\frac{4(M-1)}{M}} T^{-\frac{M-1}{M}} (\log{T})^{\frac{M-1}{M}}.
\end{align*}

\noindent\underline{\textbf{Case ($\bm{M-1}$)}}:
\begin{align*}
\widehat{\rho}_l \leq \frac{T - \frac{1}{2} \beta_{l+1} T^{\frac{l+1}{M}}}{\frac{1}{2} \beta_{l+1} T^{\frac{l+1}{M}}}, \ \forall \ l \leq M-2.
\end{align*}
Due to the condition of Case ($M-1$), we immediately have
\begin{align*}
\frac{\widehat{\sigma}_{M-2}(0)}{\widehat{\sigma}_{M-2}(1) + \widehat{\sigma}_{M-2}(0)} T \geq \frac{1}{\frac{T - \frac{1}{2} \beta_{M-1} T^{\frac{M-1}{M}}}{\frac{1}{2} \beta_{M-1} T^{\frac{M-1}{M}}}+1} T = \frac{1}{2} \beta_{M-1} T^{\frac{M-1}{M}}.
\end{align*}
On the other hand, since
\begin{align*}
\widehat{\rho}_{M-2} \geq & \ \rho \sqrt{\frac{1-48^{\frac{1}{2}} C^2 \beta_{M-2}^{-\frac{1}{2}} T^{-\frac{M-2}{2M}} (\log{T})^{\frac{1}{2}}}{1+48^{\frac{1}{2}} C^2 \beta_{M-2}^{-\frac{1}{2}} T^{-\frac{M-2}{2M}} (\log{T})^{\frac{1}{2}}}} \\
\geq & \ \widehat{\rho}_1 \sqrt{\frac{1-48^{\frac{1}{2}} C^2 \beta_1^{-\frac{1}{2}} T^{-\frac{1}{2M}} (\log{T})^{\frac{1}{2}}}{1+48^{\frac{1}{2}} C^2 \beta_1^{-\frac{1}{2}} T^{-\frac{1}{2M}} (\log{T})^{\frac{1}{2}}}} \sqrt{\frac{1-48^{\frac{1}{2}} C^2 \beta_{M-2}^{-\frac{1}{2}} T^{-\frac{M-2}{2M}} (\log{T})^{\frac{1}{2}}}{1+48^{\frac{1}{2}} C^2 \beta_{M-2}^{-\frac{1}{2}} T^{-\frac{M-2}{2M}} (\log{T})^{\frac{1}{2}}}} \\
\geq & \sqrt{\frac{1-48^{\frac{1}{2}} C^2 \beta_1^{-\frac{1}{2}} T^{-\frac{1}{2M}} (\log{T})^{\frac{1}{2}}}{1+48^{\frac{1}{2}} C^2 \beta_1^{-\frac{1}{2}} T^{-\frac{1}{2M}} (\log{T})^{\frac{1}{2}}}} \sqrt{\frac{1-48^{\frac{1}{2}} C^2 \beta_{M-2}^{-\frac{1}{2}} T^{-\frac{M-2}{2M}} (\log{T})^{\frac{1}{2}}}{1+48^{\frac{1}{2}} C^2 \beta_{M-2}^{-\frac{1}{2}} T^{-\frac{M-2}{2M}} (\log{T})^{\frac{1}{2}}}} \\
> & \ \frac{1}{4} \\
\geq & \ \frac{\frac{1}{2}\beta_{M-1} T^{\frac{M-1}{M}}}{T - \frac{1}{2}\beta_{M-1} T^{\frac{M-1}{M}}},
\end{align*}
where the first and second inequalities are due to \eqref{eqn:MStage:NewHighProbBound:ConfidenceBound1} and \eqref{eqn:MStage:NewHighProbBound:ConfidenceBound0}; the third inequality is due to \eqref{eqn:WLOG:NewHighProbBound}; the fourth inequality is due to Lemma~\ref{lem:AlgebraicTrick3:Refined}; the last inequality is due to Lemma~\ref{lem:AlgebraicTrick4:Refined}.
Due to the above sequence of inequalities, we have $\frac{1}{\widehat{\rho}_{M-2}} \leq \frac{T - \frac{1}{2}\beta_{M-1} T^{\frac{M-1}{M}}}{\frac{1}{2}\beta_{M-1} T^{\frac{M-1}{M}}}$, which leads to
\begin{align*}
\frac{\widehat{\sigma}_{M-2}(1)}{\widehat{\sigma}_{M-2}(1) + \widehat{\sigma}_{M-2}(0)} T  \geq \frac{1}{1+\frac{T - \frac{1}{2} \beta_{M-1} T^{\frac{M-1}{M}}}{\frac{1}{2} \beta_{M-1} T^{\frac{M-1}{M}}}} T = \frac{1}{2} \beta_{M-1} T^{\frac{M-1}{M}}.
\end{align*}
So Algorithm~\ref{alg:MStageANA} goes to Line~\ref{mrk:Case3} in the $(M-2)$-th stage experiment. 
Then Algorithm~\ref{alg:MStageANA} goes to Line~\ref{mrk:LastStage} in the last stage.
We further distinguish two cases.

\noindent\textbf{Case ($\bm{M-1}$).1}:
In addition to the conditions in Case $(M-1)$ above, we also have
\begin{align*}
\widehat{\rho}_{M-1} > \frac{T - \frac{1}{2} \beta_{M-1} T^{\frac{M-1}{M}}}{\frac{1}{2} \beta_{M-1} T^{\frac{M-1}{M}}}.
\end{align*}
Similar to the analysis in Case $m$.1, we proceed with the following analysis.
In Case $(M-1)$.1, 
\begin{align*}
\frac{\widehat{\sigma}_{M-1}(0)}{\widehat{\sigma}_{M-1}(1) + \widehat{\sigma}_{M-1}(0)} T < \frac{1}{\frac{T - \frac{1}{2} \beta_{M-1} T^{\frac{M-1}{M}}}{\frac{1}{2} \beta_{M-1} T^{\frac{M-1}{M}}}+1} T = \frac{1}{2} \beta_{M-1} T^{\frac{M-1}{M}}.
\end{align*}
So Algorithm~\ref{alg:MStageANA} goes to Line~\ref{mrk:LastStage:Case1} in the $(M-1)$-th stage experiment, and we have
\begin{align*}
(T(1), T(0)) = \bigg(T - \frac{1}{2} \beta_{M-1} T^{\frac{M-1}{M}}, \frac{1}{2} \beta_{M-1} T^{\frac{M-1}{M}}\bigg).
\end{align*}
We can express 
\begin{align*}
\frac{V(T(1), T(0) \vert \cE)}{V(T^*(1), T^*(0))} = & \ \frac{\sigma^2(1) + \sigma^2(0) + \frac{\frac{1}{2} \beta_{M-1} T^{\frac{M-1}{M}}}{T - \frac{1}{2} \beta_{M-1} T^{\frac{M-1}{M}}} \sigma^2(1) + \frac{T - \frac{1}{2} \beta_{M-1} T^{\frac{M-1}{M}}}{\frac{1}{2} \beta_{M-1} T^{\frac{M-1}{M}}} \sigma^2(0)}{(\sigma(1) + \sigma(0))^2}.
\end{align*}
Note that, 
\begin{multline}
\rho \cdot \sqrt{\frac{1-48^{\frac{1}{2}} C^2 \beta_{M-2}^{-\frac{1}{2}} T^{-\frac{M-2}{2M}} (\log{T})^{\frac{1}{2}}}{1+48^{\frac{1}{2}} C^2 \beta_{M-2}^{-\frac{1}{2}} T^{-\frac{M-2}{2M}} (\log{T})^{\frac{1}{2}}}} \leq \widehat{\rho}_{M-2} \leq \frac{T - \frac{1}{2} \beta_{M-1} T^{\frac{M-1}{M}}}{\frac{1}{2} \beta_{M-1} T^{\frac{M-1}{M}}} < \widehat{\rho}_{M-1} \\
\leq \rho \cdot \sqrt{\frac{1+48^{\frac{1}{2}} C^2 \beta_{M-1}^{-\frac{1}{2}} T^{-\frac{M-1}{2M}} (\log{T})^{\frac{1}{2}}}{1-48^{\frac{1}{2}} C^2 \beta_{M-1}^{-\frac{1}{2}} T^{-\frac{M-1}{2M}} (\log{T})^{\frac{1}{2}}}} < \rho \cdot \sqrt{\frac{1+48^{\frac{1}{2}} C^2 \beta_{M-2}^{-\frac{1}{2}} T^{-\frac{M-2}{2M}} (\log{T})^{\frac{1}{2}}}{1-48^{\frac{1}{2}} C^2 \beta_{M-2}^{-\frac{1}{2}} T^{-\frac{M-2}{2M}} (\log{T})^{\frac{1}{2}}}}, \label{eqn:TwoSidedBoundsM-1:NewHighProbBound}
\end{multline}
where the first and the fourth inequalities are due to \eqref{eqn:MStage:NewHighProbBound:ConfidenceBound1} and \eqref{eqn:MStage:NewHighProbBound:ConfidenceBound0};
the second and the third inequalities are due to the conditions of Case $(M-1)$.1;
the last inequality is because $\beta_{M-2} T^{\frac{M-2}{M}} < \beta_{M-1} T^{\frac{M-1}{M}}$ so we have $48^{\frac{1}{2}} C^2 \beta_{M-1}^{-\frac{1}{2}} T^{-\frac{M-1}{2M}} (\log{T})^{\frac{1}{2}} < 48^{\frac{1}{2}} C^2 \beta_{M-2}^{-\frac{1}{2}} T^{-\frac{M-2}{2M}} (\log{T})^{\frac{1}{2}}$.

Then we have
\begin{align}
\frac{V(T(1), T(0) \vert \cE)}{V(T^*(1), T^*(0))} \leq & \ \frac{\sigma^2(1) + \sigma^2(0) + \sigma(1) \sigma(0) \left( \sqrt{\frac{1+48^{\frac{1}{2}} C^2 \beta_{M-2}^{-\frac{1}{2}} T^{-\frac{M-2}{2M}} (\log{T})^{\frac{1}{2}}}{1-48^{\frac{1}{2}} C^2 \beta_{M-2}^{-\frac{1}{2}} T^{-\frac{M-2}{2M}} (\log{T})^{\frac{1}{2}}}} + \sqrt{\frac{1-48^{\frac{1}{2}} C^2 \beta_{M-2}^{-\frac{1}{2}} T^{-\frac{M-2}{2M}} (\log{T})^{\frac{1}{2}}}{1+48^{\frac{1}{2}} C^2 \beta_{M-2}^{-\frac{1}{2}} T^{-\frac{M-2}{2M}} (\log{T})^{\frac{1}{2}}}} \right) }{(\sigma(1) + \sigma(0))^2} \nonumber \\
= & \ 1 + \frac{\sigma(1)\sigma(0)}{(\sigma(1) + \sigma(0))^2} \cdot \left(\frac{2}{\sqrt{1-48 C^4 \beta_{M-2}^{-1} T^{-\frac{M-2}{M}} \log{T}}} - 2\right) \nonumber \\
\leq & \ 1 + \frac{\sigma(1)\sigma(0)}{(\sigma(1) + \sigma(0))^2} \cdot \left(96 C^4 \beta_{M-2}^{-1} T^{-\frac{M-2}{M}} \log{T}\right), \label{eqn:BreakIntoTwoPartsM-11:NewHighProbBound}
\end{align}
where the first inequality is due to Lemma~\ref{lem:h:rhohat}; the last inequality is due to Lemma~\ref{lem:AlgebraicTrick2:Refined}.

Note that, $\frac{\sigma(1) \sigma(0)}{(\sigma(1) + \sigma(0))^2} = \frac{\rho}{(\rho+1)^2}$ is a decreasing function when $\rho>1$.
Note also that,
\begin{align*}
\rho > \frac{T - \frac{1}{2} \beta_{M-1} T^{\frac{M-1}{M}}}{\frac{1}{2} \beta_{M-1} T^{\frac{M-1}{M}}} \cdot \sqrt{\frac{1-48^{\frac{1}{2}} C^2 \beta_{M-1}^{-\frac{1}{2}} T^{-\frac{M-1}{2M}} (\log{T})^{\frac{1}{2}}}{1+48^{\frac{1}{2}} C^2 \beta_{M-1}^{-\frac{1}{2}} T^{-\frac{M-1}{2M}} (\log{T})^{\frac{1}{2}}}} > \frac{1}{2} \frac{T - \frac{1}{2} \beta_{M-1} T^{\frac{M-1}{M}}}{\frac{1}{2} \beta_{M-1} T^{\frac{M-1}{M}}} > 1,
\end{align*}
where the first inequality is due to \eqref{eqn:TwoSidedBoundsM-1:NewHighProbBound}; the second inequality is due to Lemma~\ref{lem:AlgebraicTrick3:Refined}; the last inequality is due to Lemma~\ref{lem:AlgebraicTrick4:Refined}.

Then we have
\begin{align*}
\frac{\sigma(1) \sigma(0)}{(\sigma(1) + \sigma(0))^2} 
< \frac{\frac{1}{2} \frac{T - \frac{1}{2} \beta_{M-1} T^{\frac{M-1}{M}}}{\frac{1}{2} \beta_{M-1} T^{\frac{M-1}{M}}}}{\bigg( 1 + \frac{1}{2} \frac{T - \frac{1}{2} \beta_{M-1} T^{\frac{M-1}{M}}}{\frac{1}{2} \beta_{M-1} T^{\frac{M-1}{M}}} \bigg)^2} 
= \frac{\beta_{M-1} T^{\frac{M-1}{M}} (T - \frac{1}{2} \beta_{M-1} T^{\frac{M-1}{M}}) }{(T + \frac{1}{2} \beta_{M-1} T^{\frac{M-1}{M}})^2} 
\leq \frac{\beta_{M-1} T^{\frac{M-1}{M}}}{T}.
\end{align*}
Putting this into \eqref{eqn:BreakIntoTwoPartsM-11:NewHighProbBound} we have that in Case $(M-1)$.1,
\begin{align*}
\frac{V(T(1), T(0) \vert \cE)}{V(T^*(1), T^*(0))} \leq 1 + \frac{96 C^4 \beta_{M-1}}{\beta_{M-2}} \cdot T^{-\frac{M-1}{M}} \log{T} = 1 + 96 \left(\frac{1000}{3}\right)^{-\frac{1}{M}} C^{\frac{4(M-1)}{M}} T^{-\frac{M-1}{M}} (\log{T})^{\frac{M-1}{M}}.
\end{align*}

\noindent\textbf{Case ($\bm{M-1}$).2}:
In addition to the conditions in Case $(M-1)$ above, we also have
\begin{align*}
\widehat{\rho}_{M-1} \leq \frac{T - \frac{1}{2} \beta_{M-1} T^{\frac{M-1}{M}}}{\frac{1}{2} \beta_{M-1} T^{\frac{M-1}{M}}}.
\end{align*}
Due to the condition of Case ($M-1$).2, we immediately have
\begin{align*}
\frac{\widehat{\sigma}_{M-1}(0)}{\widehat{\sigma}_{M-1}(1) + \widehat{\sigma}_{M-1}(0)} T \geq \frac{1}{\frac{T - \frac{1}{2} \beta_{M-1} T^{\frac{M-1}{M}}}{\frac{1}{2} \beta_{M-1} T^{\frac{M-1}{M}}}+1} T = \frac{1}{2} \beta_{M-1} T^{\frac{M-1}{M}}.
\end{align*}
On the other hand, since
\begin{align*}
\widehat{\rho}_{M-1} \geq & \ \rho \sqrt{\frac{1-48^{\frac{1}{2}} C^2 \beta_{M-1}^{-\frac{1}{2}} T^{-\frac{M-1}{2M}} (\log{T})^{\frac{1}{2}}}{1+48^{\frac{1}{2}} C^2 \beta_{M-1}^{-\frac{1}{2}} T^{-\frac{M-1}{2M}} (\log{T})^{\frac{1}{2}}}} \\
\geq & \ \widehat{\rho}_1 \sqrt{\frac{1-48^{\frac{1}{2}} C^2 \beta_1^{-\frac{1}{2}} T^{-\frac{1}{2M}} (\log{T})^{\frac{1}{2}}}{1+48^{\frac{1}{2}} C^2 \beta_1^{-\frac{1}{2}} T^{-\frac{1}{2M}} (\log{T})^{\frac{1}{2}}}} \sqrt{\frac{1-48^{\frac{1}{2}} C^2 \beta_{M-1}^{-\frac{1}{2}} T^{-\frac{M-1}{2M}} (\log{T})^{\frac{1}{2}}}{1+48^{\frac{1}{2}} C^2 \beta_{M-1}^{-\frac{1}{2}} T^{-\frac{M-1}{2M}} (\log{T})^{\frac{1}{2}}}} \\
\geq & \ \sqrt{\frac{1-48^{\frac{1}{2}} C^2 \beta_1^{-\frac{1}{2}} T^{-\frac{1}{2M}} (\log{T})^{\frac{1}{2}}}{1+48^{\frac{1}{2}} C^2 \beta_1^{-\frac{1}{2}} T^{-\frac{1}{2M}} (\log{T})^{\frac{1}{2}}}} \sqrt{\frac{1-48^{\frac{1}{2}} C^2 \beta_{M-1}^{-\frac{1}{2}} T^{-\frac{M-1}{2M}} (\log{T})^{\frac{1}{2}}}{1+48^{\frac{1}{2}} C^2 \beta_{M-1}^{-\frac{1}{2}} T^{-\frac{M-1}{2M}} (\log{T})^{\frac{1}{2}}}} \\
> & \ \frac{1}{4} \\
\geq & \ \frac{\frac{1}{2}\beta_{M-1} T^{\frac{M-1}{M}}}{T - \frac{1}{2}\beta_{M-1} T^{\frac{M-1}{M}}},
\end{align*}
where the first and second inequalities are due to \eqref{eqn:MStage:NewHighProbBound:ConfidenceBound1} and \eqref{eqn:MStage:NewHighProbBound:ConfidenceBound0}; the third inequality is due to \eqref{eqn:WLOG:NewHighProbBound}; the fourth inequality is due to Lemma~\ref{lem:AlgebraicTrick3:Refined}; the last inequality is due to Lemma~\ref{lem:AlgebraicTrick4:Refined}.
Due to the above sequence of inequalities, we have $\frac{1}{\widehat{\rho}_{M-1}} \leq \frac{T - \frac{1}{2}\beta_{M-1} T^{\frac{M-1}{M}}}{\frac{1}{2}\beta_{M-1} T^{\frac{M-1}{M}}}$, which leads to
\begin{align*}
\frac{\widehat{\sigma}_{M-1}(1)}{\widehat{\sigma}_{M-1}(1) + \widehat{\sigma}_{M-1}(0)} T  \geq \frac{1}{1+\frac{T - \frac{1}{2} \beta_{M-1} T^{\frac{M-1}{M}}}{\frac{1}{2} \beta_{M-1} T^{\frac{M-1}{M}}}} T = \frac{1}{2} \beta_{M-1} T^{\frac{M-1}{M}}.
\end{align*}
So Algorithm~\ref{alg:MStageANA} goes to Line~\ref{mrk:LastStage:Case2} in the $(M-1)$-th stage experiment. 
Then we have
\begin{align*}
(T(1), T(0)) = \bigg(\frac{\widehat{\sigma}_{M-1}(1)}{\widehat{\sigma}_{M-1}(1) + \widehat{\sigma}_{M-1}(0)} T, \frac{\widehat{\sigma}_{M-1}(0)}{\widehat{\sigma}_{M-1}(1) + \widehat{\sigma}_{M-1}(0)} T\bigg).
\end{align*}
We can then express
\begin{align*}
\frac{V(T(1), T(0) \vert \cE)}{V(T^*(1), T^*(0))} = & \ \frac{\sigma^2(1) + \sigma^2(0) + \frac{1}{\widehat{\rho}_{M-1}} \sigma^2(1) + \widehat{\rho}_{M-1} \sigma^2(0)}{(\sigma(1) + \sigma(0))^2}.
\end{align*}
Recall that, conditional on $\cE$, \eqref{eqn:MStage:NewHighProbBound:ConfidenceBound1} and \eqref{eqn:MStage:NewHighProbBound:ConfidenceBound0} lead to
\begin{align*}
\rho \cdot \sqrt{\frac{1-48^{\frac{1}{2}} C^2 \beta_{M-1}^{-\frac{1}{2}} T^{-\frac{M-1}{2M}} (\log{T})^{\frac{1}{2}}}{1+48^{\frac{1}{2}} C^2 \beta_{M-1}^{-\frac{1}{2}} T^{-\frac{M-1}{2M}} (\log{T})^{\frac{1}{2}}}} \leq \widehat{\rho}_{M-1} \leq \rho \sqrt{\frac{1+48^{\frac{1}{2}} C^2 \beta_{M-1}^{-\frac{1}{2}} T^{-\frac{M-1}{2M}} (\log{T})^{\frac{1}{2}}}{1-48^{\frac{1}{2}} C^2 \beta_{M-1}^{-\frac{1}{2}} T^{-\frac{M-1}{2M}} (\log{T})^{\frac{1}{2}}}}.
\end{align*}
So we have
\begin{align*}
\frac{V(T(1), T(0) \vert \cE)}{V(T^*(1), T^*(0))} \leq & \ \frac{\sigma^2(1) + \sigma^2(0) + \sigma(1)\sigma(0) \bigg( \sqrt{\frac{1-48^{\frac{1}{2}} C^2 \beta_{M-1}^{-\frac{1}{2}} T^{-\frac{M-1}{2M}} (\log{T})^{\frac{1}{2}}}{1+48^{\frac{1}{2}} C^2 \beta_{M-1}^{-\frac{1}{2}} T^{-\frac{M-1}{2M}} (\log{T})^{\frac{1}{2}}}} + \sqrt{\frac{1+48^{\frac{1}{2}} C^2 \beta_{M-1}^{-\frac{1}{2}} T^{-\frac{M-1}{2M}} (\log{T})^{\frac{1}{2}}}{1-48^{\frac{1}{2}} C^2 \beta_{M-1}^{-\frac{1}{2}} T^{-\frac{M-1}{2M}} (\log{T})^{\frac{1}{2}}}} \bigg)}{(\sigma(1) + \sigma(0))^2} \\
= & \ 1 + \frac{\sigma(1) \sigma(0)}{(\sigma(1) + \sigma(0))^2} \cdot \bigg( \frac{2}{\sqrt{1 - 48 C^4 \beta_{M-1}^{-1} T^{-\frac{M-1}{M}} \log{T}}} - 2 \bigg) \\
\leq & \ 1 + \frac{\sigma(1) \sigma(0)}{(\sigma(1) + \sigma(0))^2} \cdot \left(96 C^4 \beta_{M-1}^{-1} T^{-\frac{M-1}{M}} \log{T}\right) \\
\leq & \ 1 + 24 C^4 \beta_{M-1}^{-1} T^{-\frac{M-1}{M}} \log{T},
\end{align*}
where the first inequality is due to Lemma~\ref{lem:h:rhohat}; the second inequality is due to Lemma~\ref{lem:AlgebraicTrick2:Refined}; the last inequality is because $\frac{\sigma(1) \sigma(0)}{(\sigma(1) + \sigma(0))^2} \leq \frac{1}{4}$.

Finally, using the definition of $\beta_{M-1} = \frac{400}{3} C^4 \log{T} \cdot (\frac{1000}{3} C^4 \log{T})^{-\frac{M-1}{M}}$,
\begin{multline*}
24 C^4 \beta_{M-1}^{-1} T^{-\frac{M-1}{M}} \log{T} = \frac{9}{50} \cdot \left(\frac{T}{\frac{1000}{3}C^4\log{T}}\right)^{-\frac{M-1}{M}} \\
= \frac{9}{50} \cdot \frac{1000}{3} \cdot \left(\frac{1000}{3}\right)^{-\frac{1}{M}} C^{\frac{4(M-1)}{M}} T^{-\frac{M-1}{M}} (\log{T})^{\frac{M-1}{M}} \\
< 96 \left(\frac{1000}{3}\right)^{-\frac{1}{M}} C^{\frac{4(M-1)}{M}} T^{-\frac{M-1}{M}} (\log{T})^{\frac{M-1}{M}},
\end{multline*}
where the last inequality is because $\frac{9}{50} \cdot \frac{1000}{3} = 60 < 96$.
So we have
\begin{align*}
\frac{V(T(1), T(0) \vert \cE)}{V(T^*(1), T^*(0))} \leq 1 + 96 \cdot \left(\frac{1000}{3}\right)^{-\frac{1}{M}} C^{\frac{4(M-1)}{M}} T^{-\frac{M-1}{M}} (\log{T})^{\frac{M-1}{M}}.
\end{align*}

To conclude, in all cases, we have shown that
\begin{align*}
\frac{V(T(1), T(0) \vert \cE)}{V(T^*(1), T^*(0))} \leq 1 + 96 \cdot \left(\frac{1000}{3}\right)^{-\frac{1}{M}} C^{\frac{4(M-1)}{M}} T^{-\frac{M-1}{M}} (\log{T})^{\frac{M-1}{M}}.
\end{align*}
\hfill \halmos
\endproof

\subsubsection{Completing the proof of Corollary~\ref{coro:MStageANA}.}
\proof{Proof of Corollary~\ref{coro:MStageANA}.}
We first show Algorithm~\ref{alg:MStageANA} is feasible under the parameters as defined in Corollary~\ref{coro:MStageANA}.
To start, it is easy to see $1 < \beta_1 T^{\frac{1}{M}}$.
Then for any $m \leq M-2$, 
\begin{align*}
\beta_m T^{\frac{m}{M}} = 
\frac{400}{3} C^4 \log{T} \cdot \left(\frac{T}{\frac{1000}{3} C^4 \log{T}}\right)^{\frac{m}{M}}
< \frac{400}{3} C^4 \log{T} \cdot \left(\frac{T}{\frac{1000}{3} C^4 \log{T}}\right)^{\frac{m+1}{M}}
= \beta_{m+1} T^{\frac{m+1}{M}},
\end{align*}
where the inequality is due to Lemma~\ref{lem:AlgebraicTrick:Basic}.
Finally, 
\begin{align*}
\beta_{M-1} T^{\frac{M-1}{M}} 
= \frac{400}{3} C^4 \log{T} \cdot \left(\frac{T}{\frac{1000}{3} C^4 \log{T}}\right)^{\frac{M-1}{M}}
\leq \frac{400}{3} C^4 \log{T} \cdot \left(\frac{T}{\frac{1000}{3} C^4 \log{T}}\right)
= \frac{2}{5} T
\leq T.
\end{align*}
Combining all above we know Algorithm~\ref{alg:MStageANA} is feasible, i.e., $1 < \beta_1 T^{\frac{1}{M}} < ... < \beta_{M-1} T^{\frac{M-1}{M}} < T$.

Next, due to Lemma~\ref{lem:NewHighProbBound}, conditional on $\cE$ that happens with probability at least $1 - \frac{4}{T^2}$, 
\begin{align}
\frac{V(T(1), T(0)\vert \cE)}{V(T^*(1), T^*(0))} \leq 1 + 96 \cdot \left(\frac{1000}{3}\right)^{-\frac{1}{M}} C^{\frac{4(M-1)}{M}} T^{-\frac{M-1}{M}} (\log{T})^{\frac{M-1}{M}}. \label{eqn:coro:MStageANA:1}
\end{align}

On the other hand, on the low probability event $\overline{\cE}$ that happens with probability at most $\frac{4}{T^2}$,
\begin{align}
\frac{V(T(1), T(0) \vert \overline{\cE})}{V(T^*(1), T^*(0))} = & \ \frac{T}{T - \frac{1}{2} \beta_1 T^{\frac{1}{M}}} \cdot \frac{\sigma^2(1)}{(\sigma(1) + \sigma(0))^2} + \frac{T}{\frac{1}{2} \beta_1 T^{\frac{1}{M}}} \cdot \frac{\sigma^2(0)}{(\sigma(1) + \sigma(0))^2} \nonumber \\
\leq & \ \max\left\{\frac{T}{T - \frac{1}{2} \beta_1 T^{\frac{1}{M}}}, \frac{T}{\frac{1}{2} \beta_1 T^{\frac{1}{M}}}\right\} \nonumber \\
= & \ 2 \beta_1^{-1} T^{1-\frac{1}{M}}, \label{eqn:coro:MStageANA:2}
\end{align}
where the inequality is due to Lemma~\ref{lem:g:rho}.

So overall we have
\begin{align*}
& \ \sup_{\cF \in \sP^{[C]}} \ \frac{\bE[V(T(1), T(0))]}{V(T^*(1), T^*(0))} \\
\leq & \ \left(1 - \frac{4}{T^2}\right) \left( 1 + 96 \cdot \left(\frac{1000}{3}\right)^{-\frac{1}{M}} C^{\frac{4(M-1)}{M}} T^{-\frac{M-1}{M}} (\log{T})^{\frac{M-1}{M}} \right) + \frac{4}{T^2} \cdot 2 \beta_1^{-1} T^{1-\frac{1}{M}} \\
\leq & 1 + 96 \cdot \left(\frac{1000}{3}\right)^{-\frac{1}{M}} C^{\frac{4(M-1)}{M}} T^{-\frac{M-1}{M}} (\log{T})^{\frac{M-1}{M}} + \frac{3}{50} C^{-4} (\log{T})^{-1} \left(\frac{T}{\frac{1000}{3}C^4\log{T}}\right)^{-\frac{1}{M}} T^{-1} \\
\leq & \ 1 + \left( 96 + \frac{3}{50} \frac{1}{(C^4 \log{T})^2} \right) \left(\frac{1000}{3}\right)^{-\frac{1}{M}} C^{\frac{4(M-1)}{M}} T^{-\frac{M-1}{M}} (\log{T})^{\frac{M-1}{M}} \\
< & \ 1 + 97 \left(\frac{1000}{3}\right)^{-\frac{1}{M}} C^{\frac{4(M-1)}{M}} T^{-\frac{M-1}{M}} (\log{T})^{\frac{M-1}{M}},
\end{align*}
where the first inequality is using the total law of probability, and upper bounding the two parts using \eqref{eqn:coro:MStageANA:1} and \eqref{eqn:coro:MStageANA:2};
the second probability is upper bounding $1 - \frac{4}{T^2}$ by 1;
the third inequality is because $\left(\frac{T}{\frac{1000}{3}C^4\log{T}}\right)^{-\frac{1}{M}} \leq \left(\frac{T}{\frac{1000}{3}C^4\log{T}}\right)^{\frac{1}{M}}$;
the last inequality is because $C^4 \log{T} \geq 1$.
\hfill \halmos
\endproof

\subsection{Proof of Corollary~\ref{coro:2Stage:ArmPulls}}
\label{sec:proof:coro:2Stage:ArmPulls}

\proof{Proof of Corollary~\ref{coro:2Stage:ArmPulls}.}
Our analysis of the two-stage adaptive Neyman allocation (Algorithm~\ref{alg:2StageANA}) will be based on the following two events.
\begin{align*}
\cE_1(1) = & \ \bigg\{ \left| \widehat{\sigma}^2_1(1) - \sigma^2(1) \right| < 2^{\frac{1}{2}} T^{-\frac{1}{4} + \frac{\eps}{2}} \sigma^2(1)\bigg\}, \\
\cE_1(0) = & \ \bigg\{ \left| \widehat{\sigma}^2_1(0) - \sigma^2(0) \right| < 2^{\frac{1}{2}} T^{-\frac{1}{4} + \frac{\eps}{2}} \sigma^2(0)\bigg\}.
\end{align*}
Denote $\cE = \cE_1(1) \cap \cE_1(0)$. Then $\Pr(\cE) = \Pr(\cE_1(1) \cap \cE_1(0)) \geq 1 - \Pr(\overline{\cE}_1(1)) - \Pr(\overline{\cE}_1(0))$.
We further have
\begin{align*}
\Pr(\cE) = & \ 1 - \Pr\left( \vert \widehat{\sigma}^2_1(1) - \sigma^2(1) \vert \geq 2^{\frac{1}{2}} T^{-\frac{1}{4} + \frac{\eps}{2}} \sigma^2(1) \right) - \Pr\left( \vert \widehat{\sigma}^2_1(0) - \sigma^2(0) \vert \geq 2^{\frac{1}{2}} T^{-\frac{1}{4} + \frac{\eps}{2}} \sigma^2(0) \right) \\
\geq & \ 1 - \frac{\kappa(1) \sigma^4(1)}{2 T^{-\frac{1}{2} + \eps} \sigma^4(1) T_1(1)} - \frac{\kappa(0) \sigma^4(0)}{2 T^{-\frac{1}{2} + \eps} \sigma^4(0) T_1(0)} \\
= & \ 1 - \frac{\kappa(1) + \kappa(0)}{T^{\eps}},
\end{align*}
where the inequality is due to Lemma~\ref{lem:LightTail}.

Conditional on the event $\cE$, we have
\begin{align*}
\sigma^2(1) \left( 1 - 2^{\frac{1}{2}} T^{-\frac{1}{4} + \frac{\eps}{2}} \right) \ \leq \ \widehat{\sigma}^2_1(1) \ \leq \ \sigma^2(1) \left( 1 + 2^{\frac{1}{2}} T^{-\frac{1}{4} + \frac{\eps}{2}} \right), \\
\sigma^2(0) \left( 1 - 2^{\frac{1}{2}} T^{-\frac{1}{4} + \frac{\eps}{2}} \right) \ \leq \ \widehat{\sigma}^2_1(0) \ \leq \ \sigma^2(0) \left( 1 + 2^{\frac{1}{2}} T^{-\frac{1}{4} + \frac{\eps}{2}} \right). 
\end{align*}
Conditional on event $\cE$, and given that $\sigma(1), \sigma(0) > 0$, we have $\widehat{\sigma}^2_1(1), \widehat{\sigma}^2_1(0) > 0$.
Denote $\rho = \frac{\sigma(1)}{\sigma(0)}$ and $\widehat{\rho} = \frac{\widehat{\sigma}_1(1)}{\widehat{\sigma}_1(0)}$.
Without loss of generality, assume $\widehat{\rho}_1 \geq 1$.
We distinguish two cases.

\noindent \textbf{Case 1}:
\begin{align*}
\widehat{\rho}_1 > \frac{T - \frac{1}{2} T^{\frac{1}{2}}}{\frac{1}{2} T^{\frac{1}{2}}}.
\end{align*}
In this case, 
\begin{align*}
(T(1), T(0)) = \bigg(T - \frac{1}{2} T^{\frac{1}{2}}, \frac{1}{2} T^{\frac{1}{2}}\bigg).
\end{align*}
We further distinguish two cases.

First, 
\begin{align*}
\rho < \frac{T - \frac{1}{2} T^{\frac{1}{2}}}{\frac{1}{2} T^{\frac{1}{2}}}.
\end{align*}
In this case, $\frac{\rho}{\rho+1} < \frac{T - \frac{1}{2} T^{\frac{1}{2}}}{T} = \frac{T(1)}{T}$. 
So $\vert \frac{T(1)}{T} - \frac{\rho}{\rho+1} \vert = \frac{T(1)}{T} - \frac{\rho}{\rho+1}$.
Conditional on event $\cE$, we have
\begin{align*}
\frac{T - \frac{1}{2} T^{\frac{1}{2}}}{\frac{1}{2} T^{\frac{1}{2}}} < \widehat{\rho}_1 \leq \rho \sqrt{ \frac{1+2^{\frac{1}{2}} T^{-\frac{1}{4} + \frac{\eps}{2}}}{1-2^{\frac{1}{2}} T^{-\frac{1}{4} + \frac{\eps}{2}}} }
\end{align*}
So we have
\begin{multline*}
\frac{T(1)}{T} - \frac{\rho}{\rho+1} = \frac{T - \frac{1}{2} T^{\frac{1}{2}}}{T} - \frac{\frac{T - \frac{1}{2} T^{\frac{1}{2}}}{\frac{1}{2} T^{\frac{1}{2}}} \sqrt{ \frac{1-2^{\frac{1}{2}} T^{-\frac{1}{4} + \frac{\eps}{2}}}{1+2^{\frac{1}{2}} T^{-\frac{1}{4} + \frac{\eps}{2}}} }}{1 + \frac{T - \frac{1}{2} T^{\frac{1}{2}}}{\frac{1}{2} T^{\frac{1}{2}}} \sqrt{ \frac{1-2^{\frac{1}{2}} T^{-\frac{1}{4} + \frac{\eps}{2}}}{1+2^{\frac{1}{2}} T^{-\frac{1}{4} + \frac{\eps}{2}}} }} \\
\leq \frac{T - \frac{1}{2} T^{\frac{1}{2}}}{T} \Big( 1 - \sqrt{ \frac{1-2^{\frac{1}{2}} T^{-\frac{1}{4} + \frac{\eps}{2}}}{1+2^{\frac{1}{2}} T^{-\frac{1}{4} + \frac{\eps}{2}}} } \Big) \leq 1 - \sqrt{ \frac{1-2^{\frac{1}{2}} T^{-\frac{1}{4} + \frac{\eps}{2}}}{1+2^{\frac{1}{2}} T^{-\frac{1}{4} + \frac{\eps}{2}}} } \leq 2^{\frac{1}{2}} T^{-\frac{1}{4} + \frac{\eps}{2}},
\end{multline*}
where the first inequality is because $\sqrt{ \frac{1-\delta}{1+\delta} } \leq 1$;
the second inequality is because $\sqrt{ \frac{1-\delta}{1+\delta} } \geq 1 - \delta$.

Second, 
\begin{align*}
\rho \geq \frac{T - \frac{1}{2} T^{\frac{1}{2}}}{\frac{1}{2} T^{\frac{1}{2}}}.
\end{align*}
In this case, $\frac{\rho}{\rho+1} \geq \frac{T - \frac{1}{2} T^{\frac{1}{2}}}{T} = \frac{T(1)}{T}$. 
So $\vert \frac{T(1)}{T} - \frac{\rho}{\rho+1} \vert = \frac{\rho}{\rho+1} - \frac{T(1)}{T}$.
So we have
\begin{align*}
\frac{\rho}{\rho+1} - \frac{T(1)}{T} \leq 1 - \frac{T - \frac{1}{2} T^{\frac{1}{2}}}{T} = \frac{1}{2} T^{-\frac{1}{2}}.
\end{align*}

\noindent \textbf{Case 2}:
\begin{align*}
1 \leq \widehat{\rho}_1 \leq \frac{T - \frac{1}{2} T^{\frac{1}{2}}}{\frac{1}{2} T^{\frac{1}{2}}}.
\end{align*}
In this case, 
\begin{align*}
(T(1), T(0)) = \bigg( \frac{\widehat{\sigma}_1(1)}{\widehat{\sigma}_1(1)+\widehat{\sigma}_1(0)}T, \frac{\widehat{\sigma}_1(0)}{\widehat{\sigma}_1(1)+\widehat{\sigma}_1(0)}T \bigg).
\end{align*}

Conditional on event $\cE$, we have
\begin{align*}
\rho \sqrt{ \frac{1-2^{\frac{1}{2}} T^{-\frac{1}{4} + \frac{\eps}{2}}}{1+2^{\frac{1}{2}} T^{-\frac{1}{4} + \frac{\eps}{2}}} } \leq \widehat{\rho}_1 \leq \rho \sqrt{ \frac{1+2^{\frac{1}{2}} T^{-\frac{1}{4} + \frac{\eps}{2}}}{1-2^{\frac{1}{2}} T^{-\frac{1}{4} + \frac{\eps}{2}}} }.
\end{align*}
So we have
\begin{multline*}
\Big\vert \frac{T(1)}{T} - \frac{\rho}{\rho+1} \Big\vert = \Big\vert \frac{\widehat{\rho}_1}{\widehat{\rho}_1+1} - \frac{\rho}{\rho+1} \Big\vert = \frac{\vert \widehat{\rho}_1 - \rho \vert}{(\widehat{\rho}_1+1)(\rho+1)} \leq \frac{\rho}{\rho+1} \Big\vert \frac{\widehat{\rho}_1}{\rho} - 1 \Big\vert \leq \Big\vert \frac{\widehat{\rho}_1}{\rho} - 1 \Big\vert \\
\leq \max\Big\{ \sqrt{ \frac{1+2^{\frac{1}{2}} T^{-\frac{1}{4} + \frac{\eps}{2}}}{1-2^{\frac{1}{2}} T^{-\frac{1}{4} + \frac{\eps}{2}}} } - 1, 1 - \sqrt{ \frac{1-2^{\frac{1}{2}} T^{-\frac{1}{4} + \frac{\eps}{2}}}{1+2^{\frac{1}{2}} T^{-\frac{1}{4} + \frac{\eps}{2}}} } \Big\} \leq 2^{\frac{1}{2}} T^{-\frac{1}{4} + \frac{\eps}{2}},
\end{multline*}
where the last inequality is because $\sqrt{\frac{1+\delta}{1-\delta}} \leq \frac{1}{1-\delta}$ so $\sqrt{\frac{1+\delta}{1-\delta}}-1 \leq \frac{\delta}{1-\delta} \leq \delta$, and because $\sqrt{\frac{1-\delta}{1+\delta}} \geq 1-\delta$.

To sum up, consolidating all the above cases, we always have
\begin{align*}
\Big\vert\frac{T(1)}{T} - \frac{\rho}{\rho+1}\Big\vert \leq 2^{\frac{1}{2}} T^{-\frac{1}{4} + \frac{\eps}{2}},
\end{align*}
which is on the order of $O\Big(T^{-\frac{1}{4} + \frac{\eps}{2}}\Big)$.
\hfill \halmos
\endproof

\subsection{Proof of Corollary~\ref{coro:MStage:ArmPulls}}
\label{sec:proof:coro:MStage:ArmPulls}

\proof{Proof of Corollary~\ref{coro:MStage:ArmPulls}.}

We proceed with the similar clean event analysis as in Theorem~\ref{thm:MStageANA}.
Suppose there are two length-$T$ arrays for the treated and the control, respectively, with each value being an independent and identically distributed copy of the representative random variables $Y(1)$ and $Y(0)$, respectively.
When Algorithm~\ref{alg:MStageANA} suggests to conduct an $m$-th stage experiment parameterized by $(T_m(1), T_m(0))$, the observations from the $m$-th stage experiment are generated by reading the next $T_m(1)$ values from the treated array, and the next $T_m(0)$ values from the control array.

Even though Algorithm~\ref{alg:MStageANA} adaptively determines the number of treated and control units, it is always the first few values of of the two arrays that are read.
For any $m \leq M-1$, let $\widehat{\psi}^2_m(1)$ and $\widehat{\psi}^2_{m}(0)$ be the sample variance estimators obtained from reading the first $\frac{\beta_m}{2}T^{\frac{m}{M}}$ values in the treated array and control array, respectively.
Depending on the execution of Algorithm~\ref{alg:MStageANA}, only a few of the sample variance estimators $\widehat{\sigma}^2_m(1)$ or $\widehat{\sigma}^2_m(0)$ are calculated. 
When one sample variance estimator $\widehat{\sigma}^2_m(1)$ or $\widehat{\sigma}^2_m(0)$ is calculated following Algorithm~\ref{alg:MStageANA}, it is equivalent to reading the corresponding $\widehat{\psi}^2_m(1)$ or $\widehat{\psi}^2_m(0)$ from the array.

Define the following events.
For any $m \leq M-1$, define
\begin{align*}
\cE_m(1) = & \ \bigg\{ \left| \widehat{\psi}^2_m(1) - \sigma^2(1) \right| < 2^{\frac{1}{2}} \beta_m^{-\frac{1}{2}} T^{-\frac{m}{2M} + \frac{\eps}{2}} \sigma^2(1)\bigg\}, \\
\cE_m(0) = & \ \bigg\{ \left| \widehat{\psi}^2_m(0) - \sigma^2(0) \right| < 2^{\frac{1}{2}} \beta_m^{-\frac{1}{2}} T^{-\frac{m}{2M} + \frac{\eps}{2}} \sigma^2(0)\bigg\}.
\end{align*}
Denote the intersect of all above events as $\cE$, i.e., 
\begin{align*}
\cE = \bigcap_{m=1}^{M-1} \left(\cE_m(1) \cap \cE_m(0)\right).
\end{align*}
Then due to union bound, 
\begin{align*}
\Pr(\cE) \geq 1 - \sum_{m=1}^{M-1} \Pr(\overline{\cE}_m(1)) - \sum_{m=1}^{M-1} \Pr(\overline{\cE}_m(0)).
\end{align*}
We further have
\begin{align*}
& \Pr(\cE) \\
= & \ 1 - \sum_{m=1}^{M-1} \Pr\left( \vert \widehat{\psi}^2_m(1) - \sigma^2(1) \vert \geq 2^{\frac{1}{2}} \beta_m^{-\frac{1}{2}} T^{-\frac{m}{2M} + \frac{\eps}{2}} \sigma^2(1) \right) - \sum_{m=1}^{M-1} \Pr\left( \vert \widehat{\psi}^2_m(0) - \sigma^2(0) \vert \geq 2^{\frac{1}{2}} \beta_m^{-\frac{1}{2}} T^{-\frac{m}{2M} + \frac{\eps}{2}} \sigma^2(0) \right) \\
\geq & \ 1 - \sum_{m=1}^{M-1} \frac{\kappa(1) \sigma^4(1)}{2 \beta_m^{-1} T^{-\frac{m}{M} + \eps} \sigma^4(1) \frac{1}{2} \beta_m T^{\frac{m}{M}}} - \sum_{m=1}^{M-1} \frac{\kappa(0) \sigma^4(0)}{2 \beta_m^{-1} T^{-\frac{m}{M} + \eps} \sigma^4(0) \frac{1}{2} \beta_m T^{\frac{m}{M}}} \\
= & \ 1 - \sum_{m=1}^{M-1} \frac{\kappa(1) + \kappa(0)}{T^{\eps}} \\
= & \ 1 - (M-1) \frac{\kappa(1) + \kappa(0)}{T^{\eps}},
\end{align*}
where the inequality is due to Lemma~\ref{lem:LightTail}.

Conditional on the event $\cE$, we have, for any $m \leq M-1$,
\begin{align*}
\sigma^2(1) \left( 1 - 2^{\frac{1}{2}} \beta_m^{-\frac{1}{2}} T^{-\frac{m}{2M} + \frac{\eps}{2}} \right) \ \leq \ \widehat{\psi}^2_m(1) \ \leq \ \sigma^2(1) \left( 1 + 2^{\frac{1}{2}} \beta_m^{-\frac{1}{2}} T^{-\frac{m}{2M} + \frac{\eps}{2}} \right), \\
\sigma^2(0) \left( 1 - 2^{\frac{1}{2}} \beta_m^{-\frac{1}{2}} T^{-\frac{m}{2M} + \frac{\eps}{2}} \right) \ \leq \ \widehat{\psi}^2_m(0) \ \leq \ \sigma^2(0) \left( 1 + 2^{\frac{1}{2}} \beta_m^{-\frac{1}{2}} T^{-\frac{m}{2M} + \frac{\eps}{2}} \right). 
\end{align*}

Since $\sigma(1), \sigma(0) > 0$, we can denote $\rho = \frac{\sigma(1)}{\sigma(0)}$.
For any $m \leq M-1$, when $\widehat{\sigma}^2_m(1)$ and $\widehat{\sigma}^2_m(0)$ are calculated during Algorithm~\ref{alg:MStageANA}, $\widehat{\sigma}^2_m(1) = \widehat{\psi}^2_m(1)$ and $\widehat{\sigma}^2_m(0) = \widehat{\psi}^2_m(0)$.
Conditional on the event $\cE$, and given that $\sigma(1), \sigma(0) > 0$, we have $\widehat{\sigma}^2_m(1), \widehat{\sigma}^2_m(0) > 0$.
Then we can denote $\widehat{\rho}_m = \frac{\widehat{\sigma}_m(1)}{\widehat{\sigma}_m(0)}$.

Suppose Algorithm~\ref{alg:MStageANA} terminates at an iteration indicated by $m$.
Without loss of generality, assume $\widehat{\rho}_m \geq 1$.
We distinguish two cases.

\noindent \textbf{Case 1}:
\begin{align*}
\widehat{\rho}_m > \frac{T - \frac{1}{2} \beta_m T^{\frac{m}{M}}}{\frac{1}{2} \beta_m T^{\frac{m}{M}}}.
\end{align*}
In this case, 
\begin{align*}
(T(1), T(0)) = \bigg(T - \frac{1}{2} \beta_m T^{\frac{m}{M}}, \frac{1}{2} \beta_m T^{\frac{m}{M}}\bigg).
\end{align*}
We further distinguish two cases.

First, 
\begin{align*}
\rho < \frac{T - \frac{1}{2} \beta_m T^{\frac{m}{M}}}{\frac{1}{2} \beta_m T^{\frac{m}{M}}}.
\end{align*}
In this case, $\frac{\rho}{\rho+1} < \frac{T - \frac{1}{2} \beta_m T^{\frac{m}{M}}}{T} = \frac{T(1)}{T}$. 
So $\vert \frac{T(1)}{T} - \frac{\rho}{\rho+1} \vert = \frac{T(1)}{T} - \frac{\rho}{\rho+1}$.
Conditional on event $\cE$, we have
\begin{align*}
\frac{T - \frac{1}{2} \beta_m T^{\frac{m}{M}}}{\frac{1}{2} \beta_m T^{\frac{m}{M}}} < \widehat{\rho}_m \leq \rho \sqrt{ \frac{1+2^{\frac{1}{2}} \beta_m^{-\frac{1}{2}} T^{-\frac{m}{2M} + \frac{\eps}{2}}}{1-2^{\frac{1}{2}} \beta_m^{-\frac{1}{2}} T^{-\frac{m}{2M} + \frac{\eps}{2}}} }
\end{align*}
So we have
\begin{multline*}
\frac{T(1)}{T} - \frac{\rho}{\rho+1} = \frac{T - \frac{1}{2} \beta_m T^{\frac{m}{M}}}{T} - \frac{\frac{T - \frac{1}{2} \beta_m T^{\frac{m}{M}}}{\frac{1}{2} \beta_m T^{\frac{m}{M}}} \sqrt{ \frac{1-2^{\frac{1}{2}} \beta_m^{-\frac{1}{2}} T^{-\frac{m}{2M} + \frac{\eps}{2}}}{1+2^{\frac{1}{2}} \beta_m^{-\frac{1}{2}} T^{-\frac{m}{2M} + \frac{\eps}{2}}} }}{1 + \frac{T - \frac{1}{2} \beta_m T^{\frac{m}{M}}}{\frac{1}{2} \beta_m T^{\frac{m}{M}}} \sqrt{ \frac{1-2^{\frac{1}{2}} \beta_m^{-\frac{1}{2}} T^{-\frac{m}{2M} + \frac{\eps}{2}}}{1+2^{\frac{1}{2}} \beta_m^{-\frac{1}{2}} T^{-\frac{m}{2M} + \frac{\eps}{2}}} }} \\
\leq \frac{T - \frac{1}{2} \beta_m T^{\frac{m}{M}}}{T} \Big( 1 - \sqrt{ \frac{1-2^{\frac{1}{2}} \beta_m^{-\frac{1}{2}} T^{-\frac{m}{2M} + \frac{\eps}{2}}}{1+2^{\frac{1}{2}} \beta_m^{-\frac{1}{2}} T^{-\frac{m}{2M} + \frac{\eps}{2}}} } \Big) \leq 1 - \sqrt{ \frac{1-2^{\frac{1}{2}} \beta_m^{-\frac{1}{2}} T^{-\frac{m}{2M} + \frac{\eps}{2}}}{1+2^{\frac{1}{2}} \beta_m^{-\frac{1}{2}} T^{-\frac{m}{2M} + \frac{\eps}{2}}} } \leq 2^{\frac{1}{2}} \beta_m^{-\frac{1}{2}} T^{-\frac{m}{2M} + \frac{\eps}{2}},
\end{multline*}
where the first inequality is because $\sqrt{ \frac{1-\delta}{1+\delta} } \leq 1$;
the second inequality is because $\sqrt{ \frac{1-\delta}{1+\delta} } \geq 1 - \delta$.
Because $\beta_{m} T^{\frac{m}{M}} < \beta_{m+1} T^{\frac{m+1}{M}}$, we always have
\begin{align*}
\Big\vert\frac{T(1)}{T} - \frac{\rho}{\rho+1}\Big\vert \leq 2^{\frac{1}{2}} \beta_1^{-\frac{1}{2}} T^{-\frac{1}{2M} + \frac{\eps}{2}}.
\end{align*}

Second, 
\begin{align*}
\rho \geq \frac{T - \frac{1}{2} \beta_m T^{\frac{m}{M}}}{\frac{1}{2} \beta_m T^{\frac{m}{M}}}.
\end{align*}
In this case, $\frac{\rho}{\rho+1} \geq \frac{T - \frac{1}{2} \beta_m T^{\frac{m}{M}}}{T} = \frac{T(1)}{T}$. 
So $\vert \frac{T(1)}{T} - \frac{\rho}{\rho+1} \vert = \frac{\rho}{\rho+1} - \frac{T(1)}{T}$.
So we have
\begin{align*}
\frac{\rho}{\rho+1} - \frac{T(1)}{T} \leq 1 - \frac{T - \frac{1}{2} \beta_m T^{\frac{m}{M}}}{T} = \frac{1}{2} \beta_m T^{-\frac{M-m}{M}}.
\end{align*}
Because $\beta_{m} T^{\frac{m}{M}} < \beta_{m+1} T^{\frac{m+1}{M}}$, we always have
\begin{align*}
\Big\vert\frac{T(1)}{T} - \frac{\rho}{\rho+1}\Big\vert \leq \frac{1}{2} \beta_{M-1} T^{-\frac{1}{M}}.
\end{align*}

\noindent \textbf{Case 2}:
\begin{align*}
\frac{T - \frac{1}{2} \beta_{m+1} T^{\frac{m+1}{M}}}{\frac{1}{2} \beta_{m+1} T^{\frac{m+1}{M}}} < \widehat{\rho}_m \leq \frac{T - \frac{1}{2} \beta_m T^{\frac{m}{M}}}{\frac{1}{2} \beta_m T^{\frac{m}{M}}}.
\end{align*}
In this case, 
\begin{align*}
(T(1), T(0)) = \bigg( \frac{\widehat{\sigma}_m(1)}{\widehat{\sigma}_m(1)+\widehat{\sigma}_m(0)}T, \frac{\widehat{\sigma}_m(0)}{\widehat{\sigma}_m(1)+\widehat{\sigma}_m(0)}T \bigg).
\end{align*}

Conditional on event $\cE$, we have
\begin{align*}
\rho \sqrt{ \frac{1-2^{\frac{1}{2}} \beta_m^{-\frac{1}{2}} T^{-\frac{m}{2M} + \frac{\eps}{2}}}{1+2^{\frac{1}{2}} \beta_m^{-\frac{1}{2}} T^{-\frac{m}{2M} + \frac{\eps}{2}}} } \leq \widehat{\rho}_m \leq \rho \sqrt{ \frac{1+2^{\frac{1}{2}} \beta_m^{-\frac{1}{2}} T^{-\frac{m}{2M} + \frac{\eps}{2}}}{1-2^{\frac{1}{2}} \beta_m^{-\frac{1}{2}} T^{-\frac{m}{2M} + \frac{\eps}{2}}} }.
\end{align*}
So we have
\begin{multline*}
\Big\vert \frac{T(1)}{T} - \frac{\rho}{\rho+1} \Big\vert = \Big\vert \frac{\widehat{\rho}_m}{\widehat{\rho}_m+1} - \frac{\rho}{\rho+1} \Big\vert = \frac{\vert \widehat{\rho}_m - \rho \vert}{(\widehat{\rho}_m+1)(\rho+1)} \leq \frac{\rho}{\rho+1} \Big\vert \frac{\widehat{\rho}_m}{\rho} - 1 \Big\vert \leq \Big\vert \frac{\widehat{\rho}_m}{\rho} - 1 \Big\vert \\
\leq \max\Big\{ \sqrt{ \frac{1+2^{\frac{1}{2}} \beta_m^{-\frac{1}{2}} T^{-\frac{m}{2M} + \frac{\eps}{2}}}{1-2^{\frac{1}{2}} \beta_m^{-\frac{1}{2}} T^{-\frac{m}{2M} + \frac{\eps}{2}}} } - 1, 1 - \sqrt{ \frac{1-2^{\frac{1}{2}} \beta_m^{-\frac{1}{2}} T^{-\frac{m}{2M} + \frac{\eps}{2}}}{1+2^{\frac{1}{2}} \beta_m^{-\frac{1}{2}} T^{-\frac{m}{2M} + \frac{\eps}{2}}} } \Big\} \leq 2^{\frac{1}{2}} \beta_m^{-\frac{1}{2}} T^{-\frac{m}{2M} + \frac{\eps}{2}},
\end{multline*}
where the last inequality is because $\sqrt{\frac{1+\delta}{1-\delta}} \leq \frac{1}{1-\delta}$ so $\sqrt{\frac{1+\delta}{1-\delta}}-1 \leq \frac{\delta}{1-\delta} \leq \delta$, and because $\sqrt{\frac{1-\delta}{1+\delta}} \geq 1-\delta$.
Because $\beta_{m} T^{\frac{m}{M}} < \beta_{m+1} T^{\frac{m+1}{M}}$, we always have
\begin{align*}
\Big\vert\frac{T(1)}{T} - \frac{\rho}{\rho+1}\Big\vert \leq 2^{\frac{1}{2}} \beta_1^{-\frac{1}{2}} T^{-\frac{1}{2M} + \frac{\eps}{2}}.
\end{align*}

To sum up, consolidating all the above cases, we always have
\begin{align*}
\Big\vert\frac{T(1)}{T} - \frac{\rho}{\rho+1}\Big\vert \leq 3^{-\frac{1}{2}} 15^{\frac{1}{2M}} T^{-\frac{1}{2M} + \frac{\eps}{2}},
\end{align*}
which is on the order of $O\Big(T^{-\frac{1}{2M} + \frac{\eps}{2}}\Big)$.
\hfill \halmos
\endproof

\section{Additional Simulations Using Synthetic Data}
\label{sec:append:AdditionalSynthetic}

\begin{figure}[!tb]
\centering
\caption{Normalized mean squared error with respect to sample size when $\sigma(1) / \sigma(0)=1$}
\label{fig:MSE:sigma=1}
\includegraphics[width=0.75\textwidth]{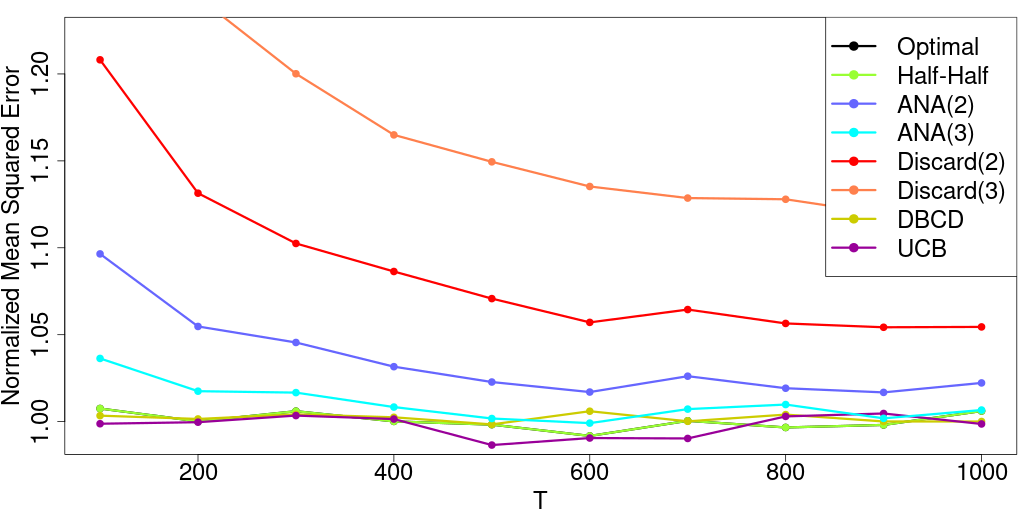}
\end{figure}

\begin{figure}[!tb]
\centering
\caption{Normalized proxy mean squared error with respect to sample size when $\sigma(1) / \sigma(0)=1$}
\label{fig:proxyMSE:sigma=1}
\includegraphics[width=0.75\textwidth]{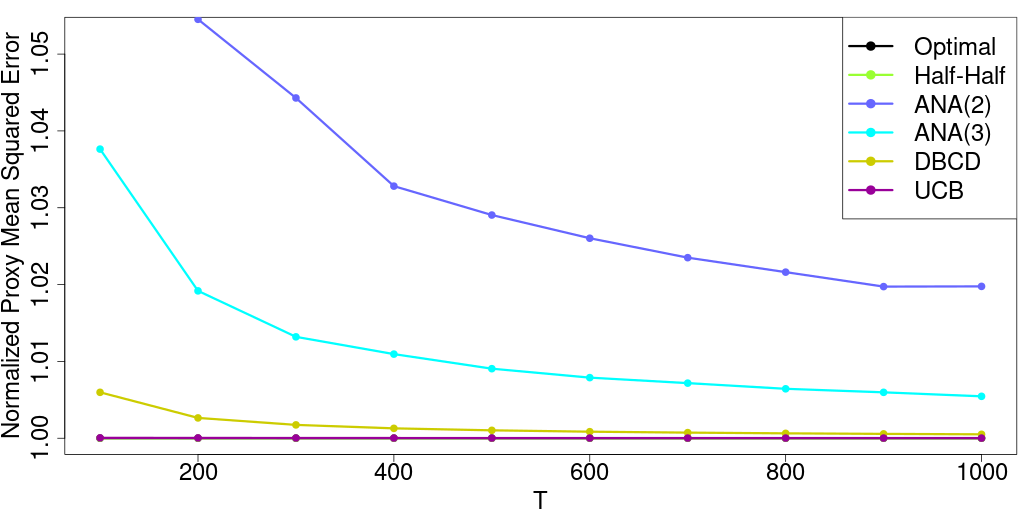}
\end{figure}

\begin{figure}[!tb]
\centering
\caption{Gap between $\bE[V(T(1),T(0))]$, $\Var(\widehat{\tau})$, and $\bE[(\widehat{\tau} - \tau)^2]$ when $\sigma(1) / \sigma(0)=1$}
\label{fig:Gap:sigma=1}
\includegraphics[width=0.75\textwidth]{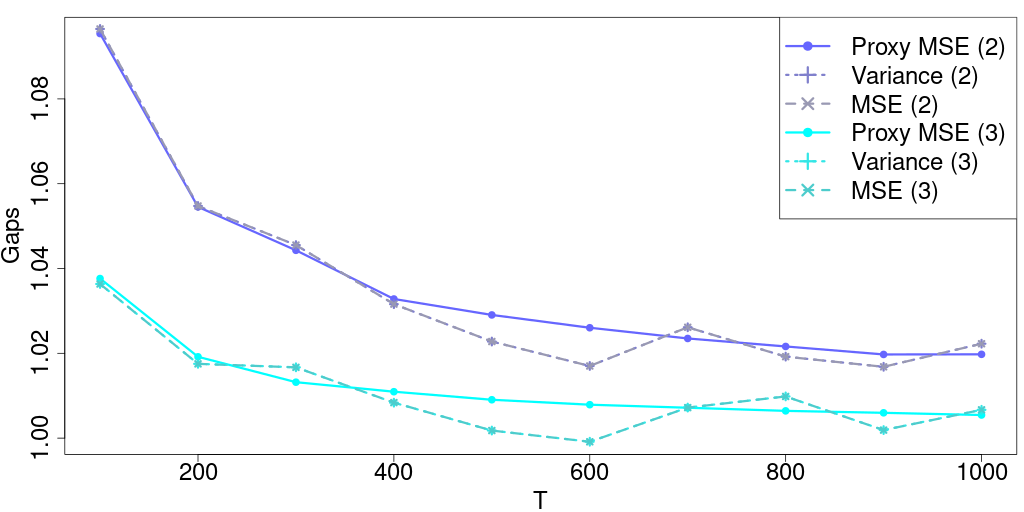}
\end{figure}

\begin{figure}[!tb]
\centering
\caption{Normalized mean squared error with respect to sample size when $\sigma(1) / \sigma(0)=10$}
\label{fig:MSE:sigma=10}
\includegraphics[width=0.75\textwidth]{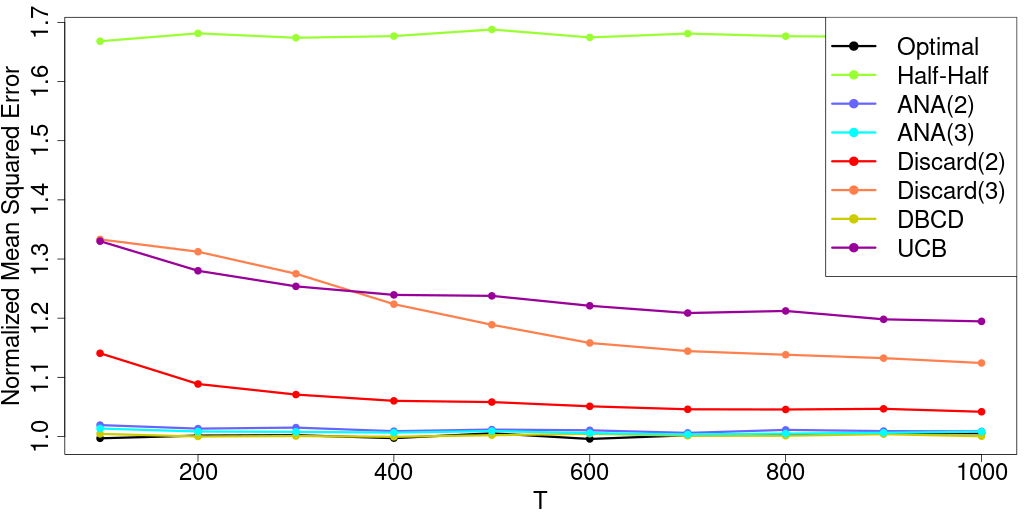}
\end{figure}

\begin{figure}[!tb]
\centering
\caption{Normalized proxy mean squared error with respect to sample size when $\sigma(1) / \sigma(0)=10$}
\label{fig:proxyMSE:sigma=10}
\includegraphics[width=0.75\textwidth]{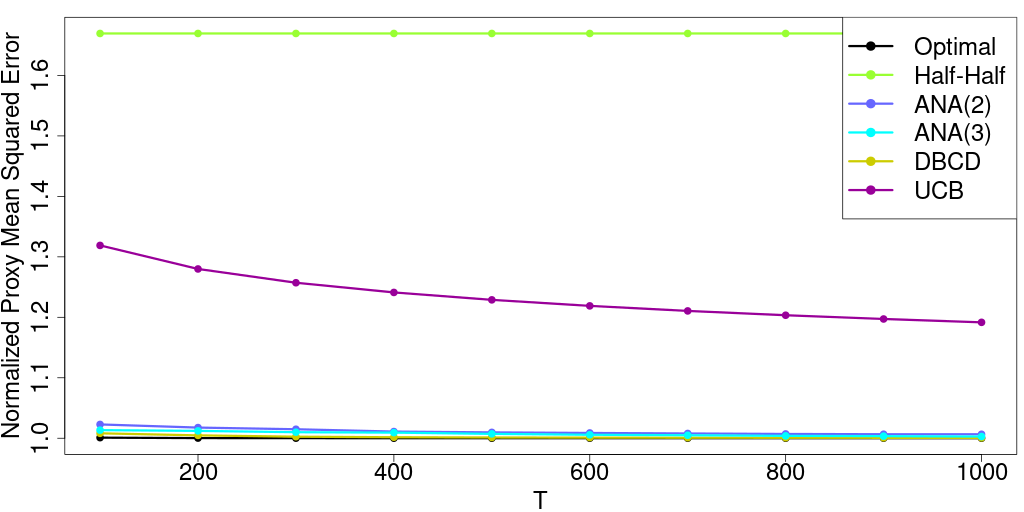}
\end{figure}

\begin{figure}[!tb]
\centering
\caption{Gap between $\bE[V(T(1),T(0))]$, $\Var(\widehat{\tau})$, and $\bE[(\widehat{\tau} - \tau)^2]$ when $\sigma(1) / \sigma(0)=10$}
\label{fig:Gap:sigma=10}
\includegraphics[width=0.75\textwidth]{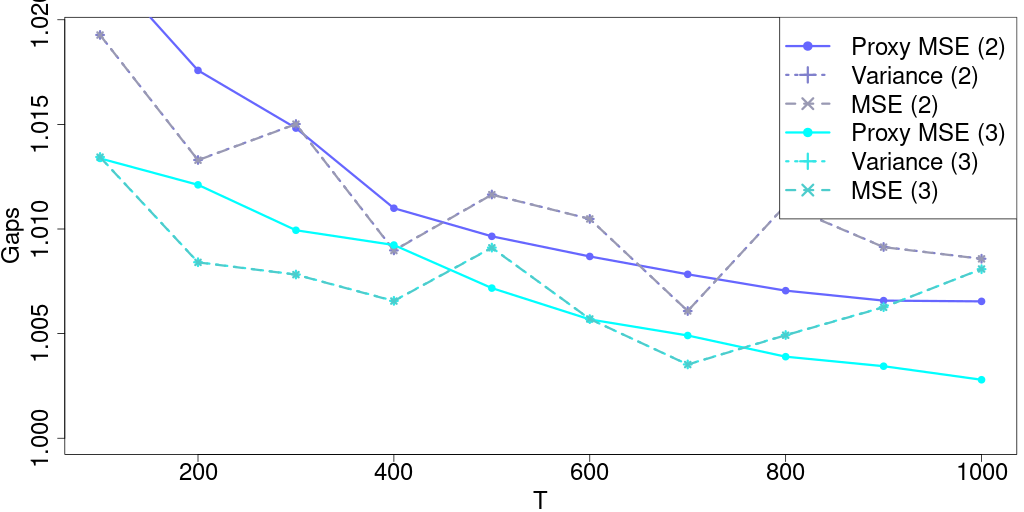}
\end{figure}

In this section, we present additional simulations using the same data generating process as we have studied in Section~\ref{sec:Synthetic}.
The only difference is that we set $\sigma(1) = 1$ or $10$ in this section.
See Figures~\ref{fig:MSE:sigma=1} --~\ref{fig:Gap:sigma=10}.
The results in this section essentially follow the same pattern as we have seen in Section~\ref{sec:Synthetic}.
One major difference comes from Figures~\ref{fig:MSE:sigma=1} and~\ref{fig:proxyMSE:sigma=1} when $\sigma(1) = 1$, in which case the two treatments are equally optimal.
In this case, the half-half allocation is also the optimal allocation, and the the upper confidence bound algorithm has better performance.

\clearpage

\end{document}